\documentclass[a4paper,fleqn]{cas-dc}
\renewcommand{\printorcid}{} 

\usepackage[numbers]{natbib}
\usepackage{subfig}
\usepackage{algorithm,algpseudocode}
\usepackage{tabularx}
\usepackage{stfloats}
\usepackage{esvect}

\def\tsc#1{\csdef{#1}{\textsc{\lowercase{#1}}\xspace}}
\tsc{WGM}
\tsc{QE}
\tsc{EP}
\tsc{PMS}
\tsc{BEC}
\tsc{DE}

\begin{document}
\let\WriteBookmarks\relax
\def\floatpagepagefraction{1}
\def\textpagefraction{.001}
\shorttitle{Prediction of the motion of chest internal points using an RNN trained with RTRL for latency compensation in lung radiotherapy}
\shortauthors{Pohl, Uesaka, Demachi and Chhatkuli}

\title [mode = title]{Prediction of the motion of chest internal points using a recurrent neural network trained with real-time recurrent learning for latency compensation in lung cancer radiotherapy}                

\author[1]{Michel Pohl}
\cormark[1]
\ead[url]{michel.pohl@centrale-marseille.fr}

\address[1]{The University of Tokyo, Graduate School of Engineering, Department of Bioengineering, Tokyo, Japan}

\author[1,2]{Mitsuru Uesaka}

\address[2]{The University of Tokyo, Graduate School of Engineering, Department of Nuclear Engineering and Management, Tokyo, Japan}

\author[2]{Kazuyuki Demachi}

\author[3]{Ritu {Bhusal Chhatkuli}}

\address[3]{National Institute for Quantum and Radiological Science and Technology, Chiba, Japan}

\cortext[cor1]{Corresponding author}

\begin{abstract}
During the radiotherapy treatment of patients with lung cancer, the radiation delivered to healthy tissue around the tumor needs to be minimized, which is difficult because of respiratory motion and the latency of linear accelerator (LINAC) systems. In the proposed study, we first use the Lucas-Kanade pyramidal optical flow algorithm to perform deformable image registration (DIR) of chest computed tomography (CT) scan images of four patients with lung cancer. We then track three internal points close to the lung tumor based on the previously computed deformation field and predict their position with a recurrent neural network (RNN) trained using real-time recurrent learning (RTRL) and gradient clipping. The breathing data is quite regular, sampled at approximately 2.5Hz, and includes artificially added drift in the spine direction. The amplitude of the motion of the tracked points ranged from 12.0mm to 22.7mm. Finally, we propose a simple method for recovering and predicting three-dimensional (3D) tumor images from the tracked points and the initial tumor image, based on a linear correspondence model and the Nadaraya-Watson non-linear regression. The root-mean-square (RMS) error, maximum error and jitter corresponding to the RNN prediction on the test set were smaller than the same performance measures obtained with linear prediction and least mean squares (LMS). In particular, the maximum prediction error associated with the RNN, equal to 1.51mm, is respectively 16.1\% and 5.0\% lower than the error given by a linear predictor and LMS. The average prediction time per time step with RTRL is equal to 119ms, which is less than the 400ms marker position sampling time.  The tumor position in the predicted images appears visually correct, which is confirmed by the high mean cross-correlation between the original and predicted images, equal to 0.955. The standard deviation of the Gaussian kernel and the number of layers in the optical flow algorithm were the parameters having the most significant impact on registration performance. Their optimization led respectively to a 31.3\% and 36.2\% decrease in the registration error. Using only a single layer proved to be detrimental to the registration quality because tissue motion in the lower part of the lung has a high amplitude relative to the resolution of the CT scan images. The random initialization of the hidden units and the number of these hidden units were found to be the most important factors affecting the performance of the RNN. Increasing the number of hidden units from 15 to 250 led to a 56.3\% decrease in the prediction error on the cross-validation data. Similarly, optimizing the standard deviation of the initial Gaussian distribution of the synaptic weights $\sigma_{init}^{RNN}$ led to a 28.4\% decrease in the prediction error on the cross-validation data, with the error minimized for $\sigma_{init}^{RNN} = 0.02$ with the four patients.
\end{abstract}




\maketitle

\section{Introduction}

\subsection{\normalsize Lung cancer and respiratory motion}

Lung and bronchus cancer is estimated to represent 12.7\% of all new cancer cases with 229,000 expected new cases in 2020 in the United States, according to the National Cancer Institute. Comparatively, in the same year, 136,000 deaths are expected to occur, making up 22.4\% of all cancer deaths \cite{NIC2020LungBronchusCancer}. 

Nearly half of patients suffering from cancer benefit from radiation therapy during the course of their treatment. Usually, a certain amount of normal tissue surrounding the tumor receives irradiation as well, due to normal movements of the organs causing tumor displacements, such as breathing in the case of lung cancer. Lung tumors therefore exhibit a rather cyclic motion, with some changes in frequency and amplitude over time. It has previously been reported that such motion can be up to 5cm \cite{chen2001fluoroscopic}. Phase shift as well as intrafractional baseline shift and drift can be observed. The term "shift" refers to sudden changes in the mean tumor position whereas the term "drift" refers to continuous changes, during a single treatment. Baseline drifts of $1.65 \pm 5.95$ mm (mean position $\pm$ standard deviation), $1.50 \pm 2.54$ mm, and $0.45 \pm 2.23$ mm have respectively been reported concerning the spine axis, dorsoventral axis, and left-right direction in \cite{takao2016intrafractional}. Noise is naturally present and can be partly caused by cardiac or gastrointestinal movements. The tumor shape is not rigid and deforms over time to a certain extent. Moreover, the motion of lung tumors can vary across patients and fractions \cite{verma2010survey, ehrhardt20134d}. 

\subsection{\normalsize Systems for lung tumor tracking}

In image-guided radiotherapy (IGRT), several methods have been designed to track the three-dimensional (3D) position of the tumor in real-time as accurately as possible. Indeed, concerning the respiratory movement, it has been mentioned that "a systematic tracking error of 2mm can be significant" \cite{murphy2004tracking} in terms of dose delivery and safety. One of these modalities is called beam gating and consists of turning on and off a static beam according to the recorded tumor position. Conversely, in beam tracking, the radiation beam follows the tumor and conforms to its position as it moves.

Visualizing clearly a lung tumor in 3D during radiotherapy treatment is difficult. Therefore, one often records some surrogate signals and uses a correspondence model to infer the tumor location from them. Such signals may be the position of internal or external markers. Internal fiducial markers are small metallic objects implanted in the lung, close to the tumor, prior to the radiotherapy treatment, and whose position can be measured by fluoroscopic imaging systems, such as CyberKnife's orthogonal X-ray imaging sources and flat-panel detectors \cite{khankan2017demystifying}. In contrast, external fiducial markers are objects attached to the patient's chest whose position can be recorded using an infrared tracking system.

\subsection{\normalsize Prediction methods for latency compensation}
\label{subsection:intro_prediction_methods}

\begin{figure}
	\centering
		\includegraphics[width=.55\columnwidth]{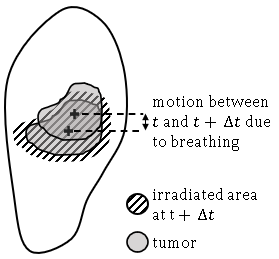}
	\caption{Excessive irradiation of healthy lung tissue due to an overall system delay $ \Delta t$ not compensated. The area irradiated, represented here using diagonal stripes, is larger than the tumor size, to take into consideration effects such as variation of the tumor shape during the treatment. }
	\label{fig:lung_latency}
\end{figure}

Current treatment systems generally suffer from an inherent time latency due to image acquisition and processing, communication delays, and preparation of the radiation delivery system. A latency of around 300ms has been reported for a robotic arm mounted linear accelerator (LINAC) in \cite{schweikard2000robotic}. Concerning a gantry mounted multileaf collimator (MLC) based LINAC, Shirato et al. reported a latency of 90ms \cite{shirato2000physical}, whereas Poulsen et al. reported latencies from 350ms to 1400ms for sampling intervals between 150ms and 1000ms \cite{poulsen2010detailed}. Verma et al. summarize the situation as follows: "For most radiation treatments, the latency will be more than 100ms, and can be up to two seconds" \cite{verma2010survey}. If this latency is not compensated, it may lead to errors in the estimation of the tumor position, and thus to serious damage to healthy tissue and ineffective irradiation of the tumor (Fig. \ref{fig:lung_latency}).

Various methods have been proposed to predict respiratory motion. When using online methods, as opposed to offline methods, the prediction coefficients are updated with each new training example, which enables continuous adaptation to natural changes in the breathing characteristics. Adaptive linear filters such as least mean squares (LMS) have been applied to prediction in radiotherapy as early as 2004 \cite{vedam2004predicting}. However, Murphy remarked that the performance of adaptive filters deteriorates significantly when the response time, that is to say, the time interval in advance for which the prediction is made, also called the look-ahead time, exceeds 200ms \cite{murphy2004tracking}. Artificial neural networks (ANNs) require more computational power but proved to have a better performance than linear adaptive filters, as the latency time becomes higher and the breathing signals non-stationary and complex \cite{verma2010survey}. Sharp et al. confirmed this result when predicting the position of an implanted marker sampled at imaging rates from 1Hz to 30Hz with a latency varying from 33ms to 1s, using a feedforward ANN with one hidden layer, trained with the conjugate gradient method \cite{sharp2004prediction}. Also, Goodband et al. compared different online training methods for multilayer perceptron ANNs in radiation therapy \cite{goodband2008comparison}. They predicted the motion of a marker block resting on the chest with a sampling frequency of 30Hz. For a latency time of 400ms, the lowest root-mean-square error (RMSE) was achieved using a feedforward ANN with one hidden layer, trained with a variation of the conjugate gradient method.

Recurrent neural networks (RNNs) are a specific type of ANN suited for temporal series processing, featuring a feedback loop enabling storage of information over time. RNNs have been applied in many areas such as meteorology to predict wind speed \cite{balluff2020meteorological} and air quality \cite{athira2018deepairnet}, and finance to predict stock prices \cite{selvin2017stock} and currency exchange rates \cite{hazazi13extended}. Concerning lung radiotherapy, Kai et al. used an RNN with a single hidden layer, trained with back-propagation through time (BPTT), for the prediction of the position of an implanted marker \cite{kai2018prediction}. Also, an online training approach based on extended Kalman filtering (EKF) has been applied to an RNN with a single hidden layer for the prediction of breathing data from the Cyberknife system \cite{lee2011respiratory}. In this work, we propose and investigate a standard online training algorithm for RNNs called real-time recurrent learning (RTRL) and apply it to the prediction of lung tumor position.

\subsection{\normalsize Chest image registration}

In the proposed study, we artificially track arbitrary internal points near the tumor by calculating the deformation or displacement vector field (DVF) in the whole chest in computed tomography (CT) scan images. This internal correspondence calculation process, known as deformable image registration (DIR), has been extensively studied for various applications in radiotherapy, such as tumor tracking, correction of the irradiation plan relative to the patient position on the couch, and ventilation imaging for lung function estimation. The different  DIR algorithms can be classified into two categories. The first category is referred to as feature-based registration \cite{ehrhardt20134d}. In feature-based registration, highly structured image regions such as vertebrae, ribs, the lung surface, the bronchial and vascular tree are first matched by algorithms such as the iterative closest point (ICP) \cite{besl1992method}, and a dense deformation field is subsequently calculated using interpolation methods such as B-splines \cite{mcclelland2006continuous}. In contrast, intensity-based deformable registration methods consist in calculating directly the entire global deformation using only image intensity information without performing segmentation or feature extraction beforehand. The Lucas-Kanade optical flow \cite{lucas1981iterative}, the Horn-Schunck optical flow \cite{horn1981determining}, and the different variants of the "Demons algorithm" \cite{thirion1995fast, thirion1998image} are examples in that category. Computing large displacements with these methods can be difficult. To cope with this problem, an approach referred to interchangeably as "coarse to fine strategy", "pyramidal implementation", or "multi-resolution scheme" can be used. It consists of iteratively calculating and refining the DVF of gradually more detailed versions of the images to be matched.

\subsection{\normalsize Contributions of the proposed study}

\begin{figure}
	\centering
		\includegraphics[width=.85\columnwidth]{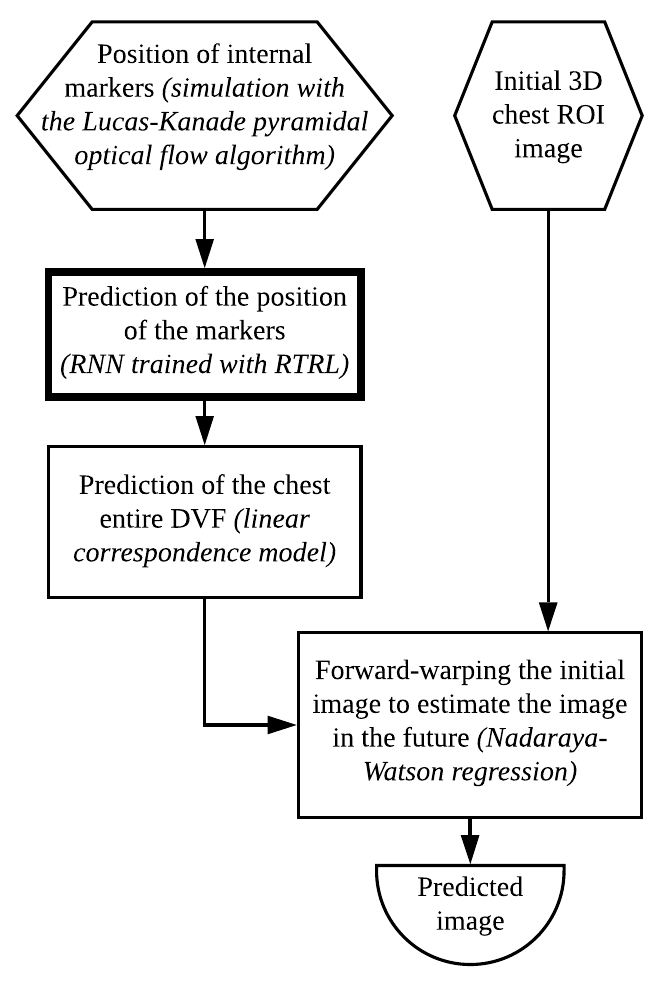}
	\caption{Overview of the proposed prediction algorithm}
	\label{fig:overview_algo}
\end{figure}

The main contributions of this study are the following. First, we discuss in detail parameter optimization of the iterative and pyramidal version of the Lucas-Kanade optical flow algorithm in the context of DIR of chest CT scan images. That algorithm has often been used in the context of chest imaging \cite{xu2008lung, akino2014evaluation, dhont2019multi}, but there are no studies about proper selection of the parameters for accurate registration of chest CT scan images, to the extent of our knowledge. Secondly, this is the first application of RNNs trained with the RTRL algorithm to predict breathing signals and compensate for the inherent latency of treatment systems in radiotherapy. The optimal choice of the RNN parameters is discussed thoroughly. In contrast to the related studies about marker position prediction with ANNs mentioned in Section \ref{subsection:intro_prediction_methods}, our study describes the simultaneous prediction of the position of three markers \cite{sharp2004prediction, goodband2008comparison, kai2018prediction}, rather than the position of one marker only. Finally, we propose a simple method to reconstruct and predict 3D lung tumor images given only the trajectory of internal markers and an initial 3D image of that tumor (Fig. \ref{fig:overview_algo}).

\section{Materials and methods}

\subsection{\normalsize Chest image data} \label{subsection:chest image data}

\begin{figure}
    \captionsetup[subfigure]{labelformat=empty}
    \centering
    \subfloat{{\includegraphics[width=.18\columnwidth]{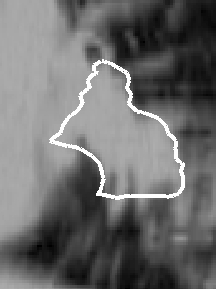} }}%
    \subfloat{{\includegraphics[width=.18\columnwidth]{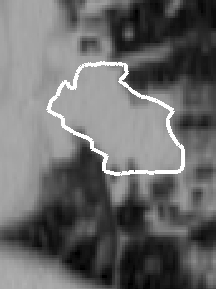} }}%
    \subfloat{{\includegraphics[width=.18\columnwidth]{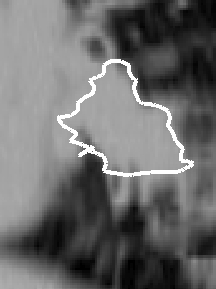} }}%
    \subfloat{{\includegraphics[width=.18\columnwidth]{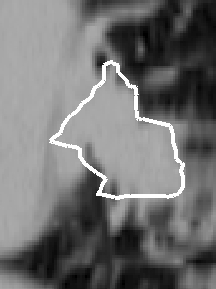} }}%
    \subfloat{{\includegraphics[width=.18\columnwidth]{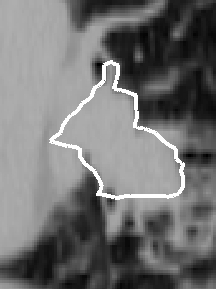} }}%
  
    \subfloat[\text{\normalsize $t=t_1$}]{{\includegraphics[width=.18\columnwidth]{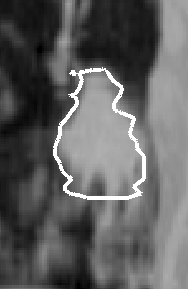} }}%
    \subfloat[\text{\normalsize $t=t_3$}]{{\includegraphics[width=.18\columnwidth]{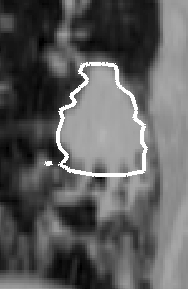} }}%
    \subfloat[\text{\normalsize $t=t_5$}]{{\includegraphics[width=.18\columnwidth]{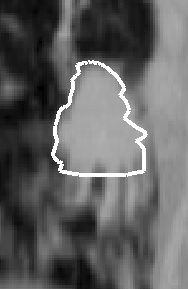} }}%
    \subfloat[\text{\normalsize $t=t_7$}]{{\includegraphics[width=.18\columnwidth]{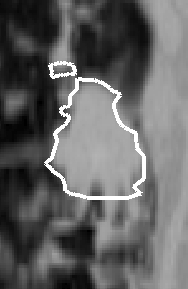} }}%
    \subfloat[\text{\normalsize $t=t_9$}]{{\includegraphics[width=.18\columnwidth]{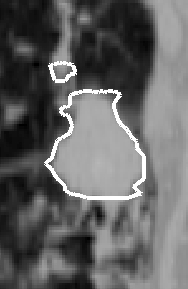} }}%
    \caption{Sagittal (top line) and coronal (bottom line) cross-sections of the 3D ROI of patient 2 at different phases of the breathing cycle. The coordinates of the cross-sections are the same as in Fig. \ref{fig:org_im}. The tumor was delineated by a physician in each image.}
    \label{fig:org_im_patient2}
\end{figure}

\begin{figure*}
    \captionsetup[subfigure]{labelformat=empty}
    \centering
    \subfloat[Patient 1 - sagittal]{{\includegraphics[height=2.7cm]{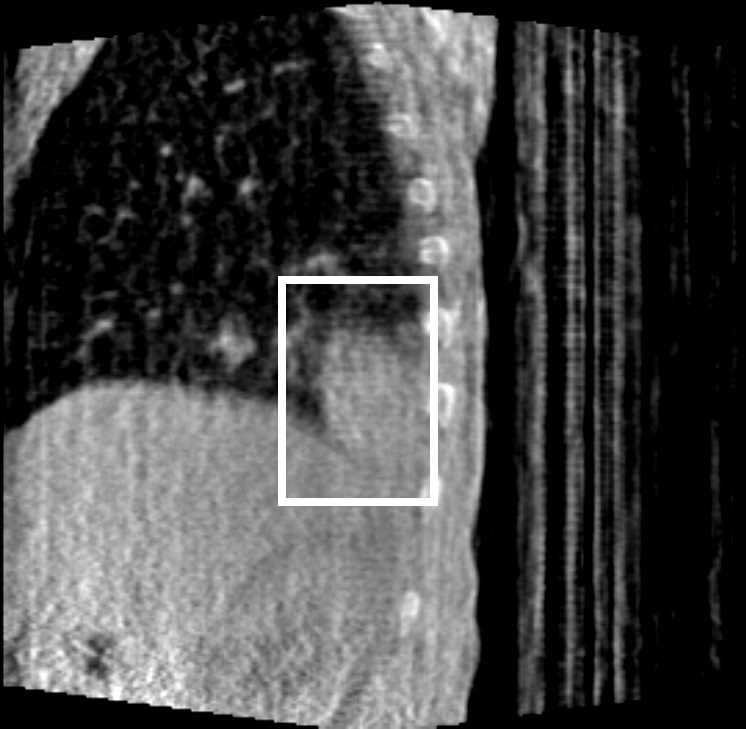} }}%
    \subfloat[Patient 2 - sagittal]{{\includegraphics[height=2.7cm]{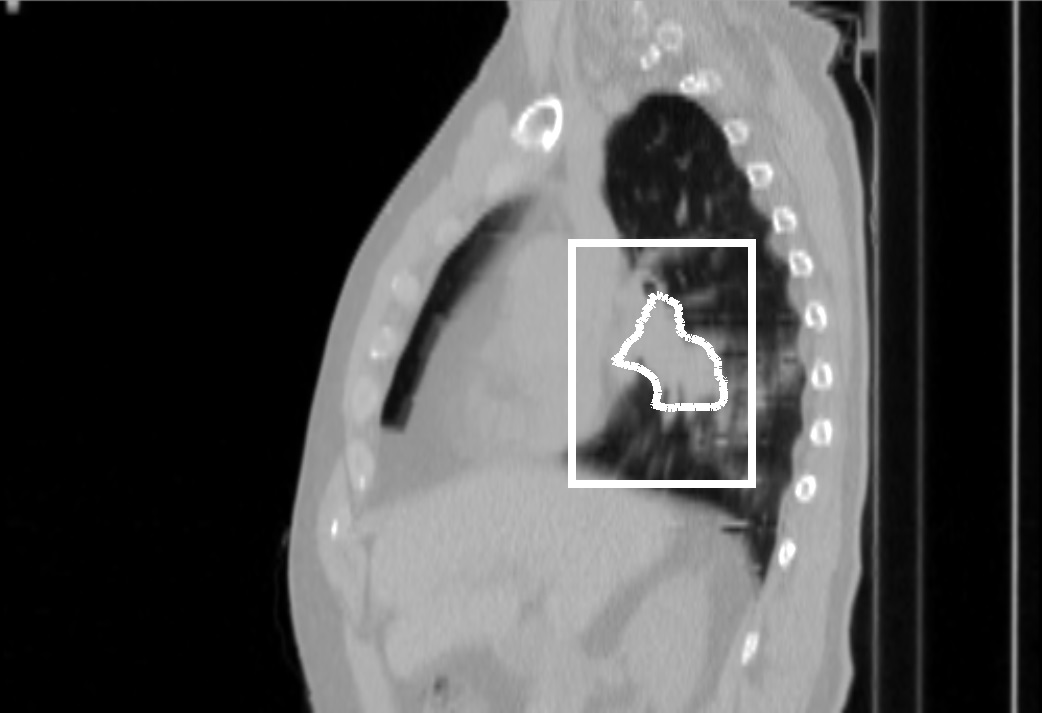} }}%
    \subfloat[Patient 3 - sagittal]{{\includegraphics[height=2.7cm]{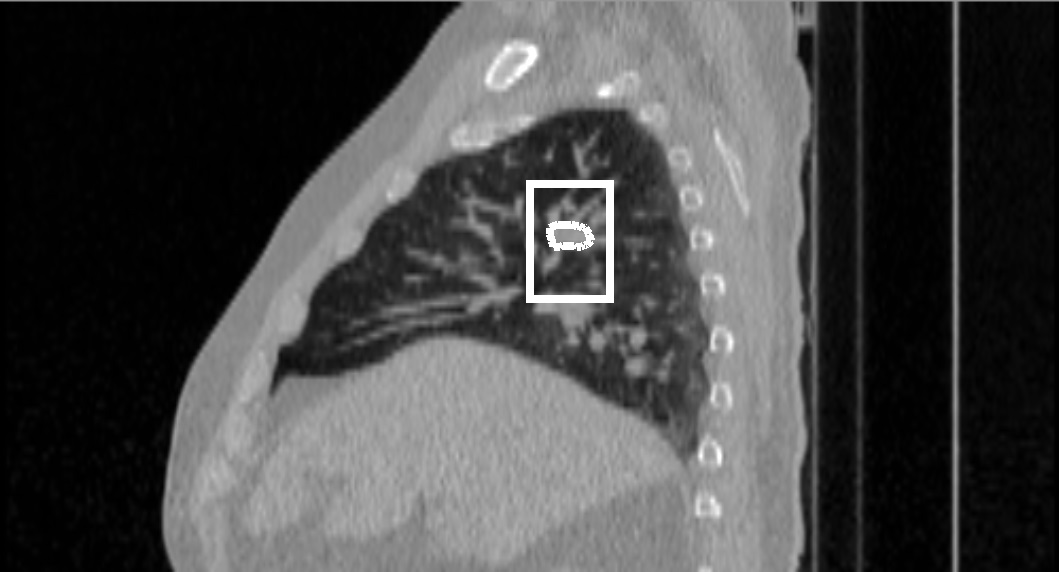} }}%
    \subfloat[Patient 4 - sagittal]{{\includegraphics[height=2.7cm]{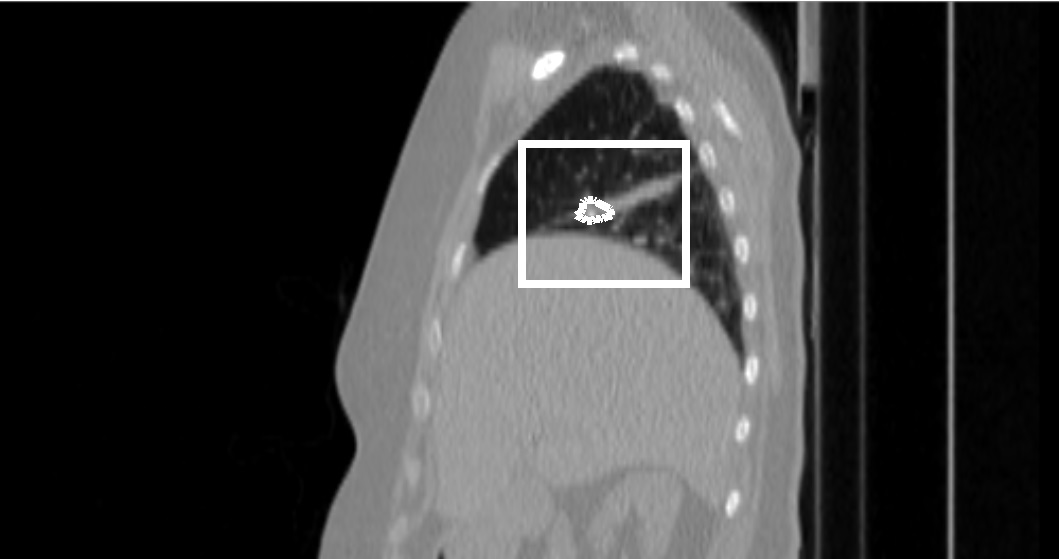} }}%
  
    \subfloat[Patient 1 - coronal]{{\includegraphics[height=2.7cm]{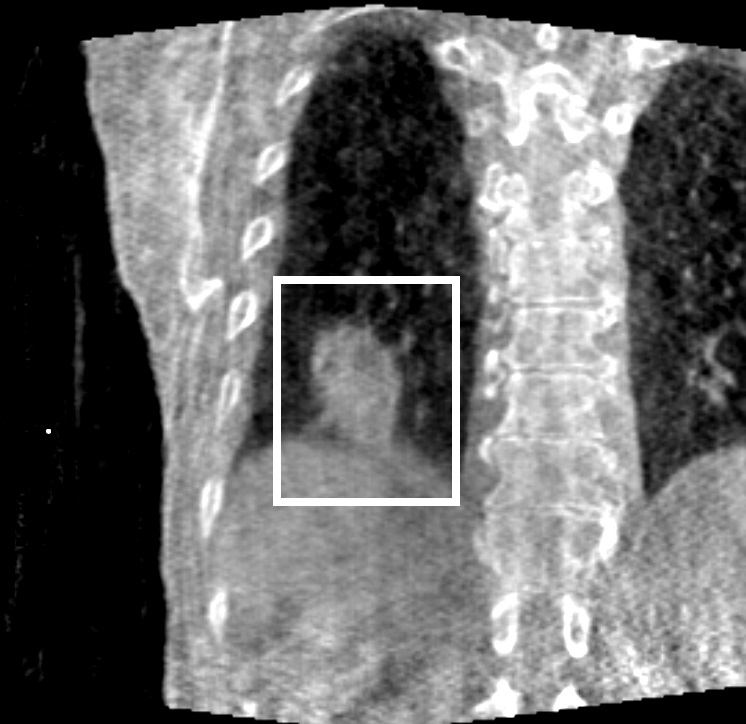} }}%
    \subfloat[Patient 2 - coronal]{{\includegraphics[height=2.7cm]{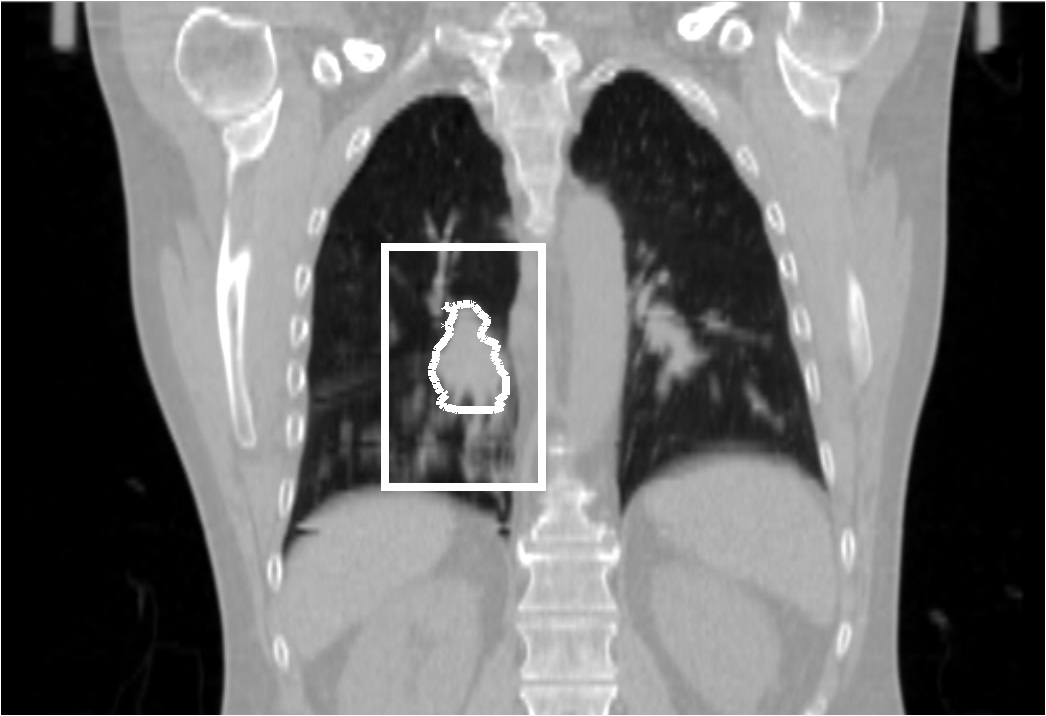} }}%
    \subfloat[Patient 3 - coronal]{{\includegraphics[height=2.7cm]{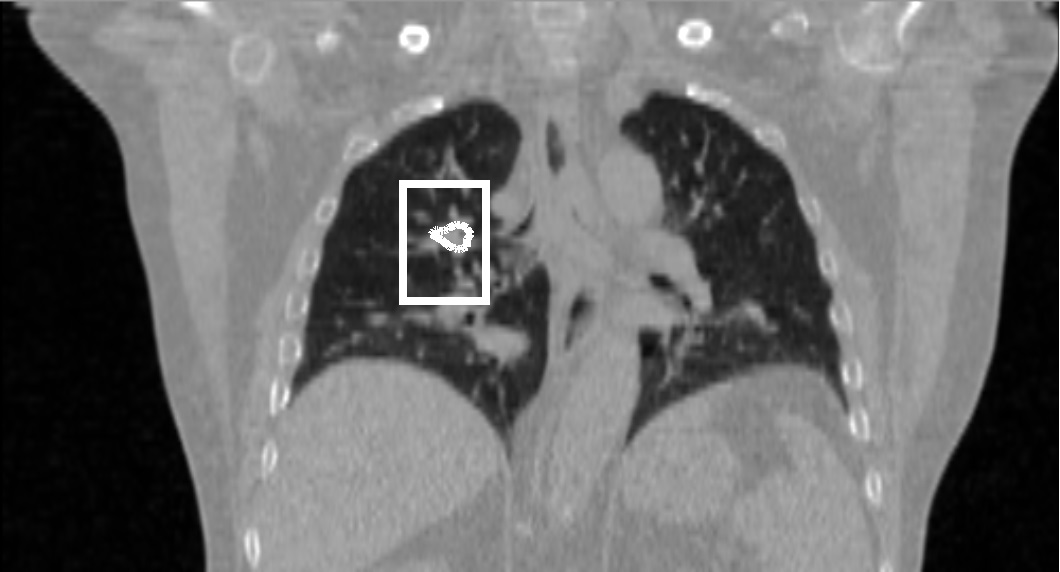} }}%
    \subfloat[Patient 4 - coronal]{{\includegraphics[height=2.7cm]{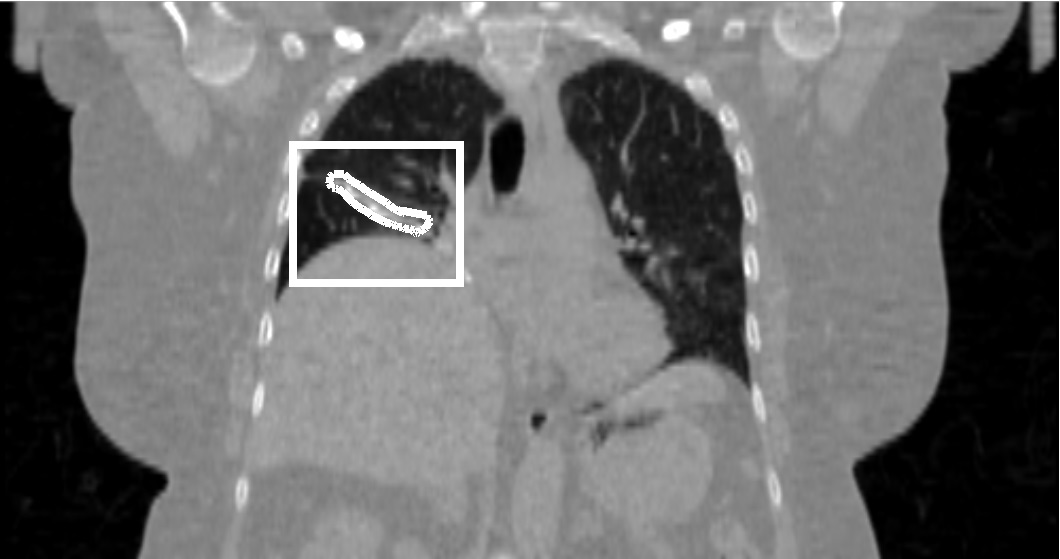} }}%
    \caption{Sagittal (top line) and coronal (bottom line) cross-sections of the 3D ROI of each patient at \text{\normalsize $t=t_1$}. The tumor of each of patients 2, 3, and 4 was delineated by a physician.}
    \label{fig:org_im}
\end{figure*}

The data used in this study consists of chest 3D 16-bit image sequences of 4 patients with lung cancer. Each of the 4 sequences consists of ten 3D images of the chest at different phases of the breathing process. The first sequence is a 4D-CBCT (four-dimensional cone-beam computed tomography) sequence acquired by the Elekta Synergy XVI system in the University of Tokyo Hospital and the three remaining sequences are 4DCT (four-dimensional computed tomography) sequences acquired by a 16-slice helical CT simulator (Brilliance Big Bore, Philips Medical System) in Virginia Commonwealth University Massey Cancer Center. 

Each sequence was resampled using trilinear interpolation such that 1 voxel corresponds to $1mm^3$. For each sequence, a 3D region of interest (ROI) encompassing the tumor was selected (Figs. \ref{fig:org_im_patient2}, \ref{fig:org_im}) and the size of each of them is recorded in Table \ref{table:ROI_size_param}. Then, each sequence was extended to $ N = 2400$ images\footnotemark by introducing a breathing drift in the z-direction (the spine axis). Indeed, it has been reported that the axis along which the respiratory drift is the greatest is the craniocaudal axis \cite{takao2016intrafractional}. More precisely, $I(\cdot, t_k)$, the image at time $t_k$, where $k \in \{1, ..., 2400\}$, results from the translation along the z-axis defined in Eq. \ref{eq:init_im_translation}.%
\begin{equation} \label{eq:init_im_translation}
 I(\vec{x}, t_k) = I \left(\vec{x} + A sin\left(\frac{2 \pi t_k}{T}\right) \vec{e_z}, t_{k \; mod \; 10} \right)
\end{equation} 

\footnotetext{Prior to the extension of the sequences, the 10 original images were permuted for each patient so that each series begins at a phase where the tumor is approximately located at its center position, with regards to the overall cyclic breathing motion. This is performed in order to increase the accuracy of the optical flow registration that follows.
}

\begin{table*}
\begin{tabular*}{\tblwidth}{@{} LLLLL@{} }
\toprule
Patient & 1 & 2 & 3 & 4 \\
\midrule
ROI size (in $mm^3$) & $65 \times 56 \times 82 $ & 
			           $76 \times 87 \times 116 $ & 
			   		   $41 \times 39 \times 56 $ &
			           $80 \times 79 \times 67 $ \\
$T$ (in s)         & 400 & 320 & 800 & 480 \\
$A$ (in mm)        & 2.0 & 1.5 & 4.0 & 2.5 \\
\bottomrule
\end{tabular*}
\caption{Description of the ROI size and motion parameters, defined in Eq. \ref{eq:init_im_translation}, for each patient.}
\label{table:ROI_size_param}
\end{table*}

In this equation, $\vec{x}$ refers to a selected voxel in the image $I(\cdot, t_k)$, $\vec{e_z}$ is a unit vector in the z-direction, and $A$ and $T$ are respectively the amplitude and the period of the added sinusoidal drift (see Table \ref{table:ROI_size_param}). The voxel intensity values on the right side of Eq. \ref{eq:init_im_translation} are computed using trilinear interpolation. Finally, Poisson noise with parameter $\lambda = 1000$ is added to the extended sequences, given that this type of noise is prevalent in CT scan imaging \cite{boas2012ct, diwakar2018review}. Because the average breathing cycle of an adult lasts 4s \cite{barrett2019ganong}, we can assume that the interval of time between each image is equal to 400ms, or in other words, that the sampling rate is equal to 2.5Hz.

\subsection{\normalsize Chest image registration} \label{subsection:chest image registration}

First, the pyramidal and iterative Lucas-Kanade optical flow algorithm (Algorithm \ref{alg:PILK-OF}) is used to calculate $\vec{u}(\cdot, t)$, the DVF between the first image (at time $t_1$) and the image at time $t$, which approximately satisfies Eq. \ref{eq:DVF_def}. 
\begin{equation} \label{eq:DVF_def}
 I(\vec{x}, t_1) = I(\vec{x} + \vec{u}(\vec{x}, t), t)
\end{equation} 

In the pyramidal and iterative Lucas-Kanade optical flow algorithm, a multiresolution representation of the two images to be registered, $I(\cdot, t_1)$ and $I(\cdot, t)$, is first computed. For this purpose, an initial low-pass Gaussian filter of standard deviation $\sigma_{init}$ is first applied to both of them. Given the representations of $I(\cdot, t_1)$ and $I(\cdot, t)$ at the layer $l$, denoted by $I_l(\cdot, t_1)$ and $I_{l}(\cdot, t)$, these representations have another low-pass Gaussian filter of standard deviation $\sigma_{sub}$ applied to them. They are then subsampled by a factor 2 to create their representations at the layer $l+1$, $I_{l+1}(\cdot, t_1)$ and $I_{l+1}(\cdot, t)$. Indeed, prior Gaussian filtering has been shown to increase the accuracy of the resulting computed optical flow in general \cite{sharmin2012optimal}.

The displacement vector at a given voxel $\vec{x_0}$ and layer $l$ between $t_1$ and $t$ is the argument $\vec{v_0}$ that minimizes the energy $E(\vec{v})$ in Eq. \ref{eq:DVF_energy_min}.
\begin{equation} \label{eq:DVF_energy_min}
 E(\vec{v}) = \sum_{\vec{x}} K_{\sigma_{LK}}(\| \vec{x} - \vec{x_0} \|_2) \left[ 
 \vec{\nabla}{I_l} (\vec{x}, t_1) \cdot \vec{v} + \frac{\partial I_l}{\partial t} (\vec{x}, t_1)
 \right] ^2
\end{equation} 

In that equation, $\vec{\nabla}$ refers to the spatial gradient operator, calculated here by applying the Scharr filter \cite{wikiSobelOperator, levkine2012prewitt}. Furthermore, $K_{\sigma_{LK}}$ refers to the probability density function of a centered normal distribution of standard deviation $\sigma_{LK}$ (Eq. \ref{eq:normal_gaussian}). 
\begin{equation} \label{eq:normal_gaussian}
	K_{\sigma}(x) = \frac{1}{\sqrt{2 \pi \sigma^2}} exp \left( - \frac{x^2}{2 \sigma^2}\right)	
\end{equation} 

The minimization of $E(\vec{v})$ is iterated to decrease the residual error, and the displacement field calculated at the layer $l$ is propagated at the layer $l-1$ to give a first approximation of the displacement field at the layer $l-1$. The algorithm is detailed in \cite{bouguet2001pyramidal, fleet2006optical}.

\begin{algorithm}
\caption{Pyramidal Iterative Lucas-Kanade Optical Flow}
\label{alg:PILK-OF}
\begin{algorithmic}
\State \textbf{Input} :
\State $I$ : initial image at time $t_1$
\State $J$ : image at an arbitrary time $t$
\State
\State \textbf{Parameters} :
\State $\sigma_{init}$, $\sigma_{sub}$, $\sigma_{LK}$ : standard deviation of various Gaussian filters
\State $n_{layers}$ : number of layers
\State $n_{iter}$ : number of iterations
\State
\State \textbf{Pyramidal representation of $I$ and $J$}
\State In what follows $\mathcal{G}(\cdot, \sigma)$ designates the isotropic Gaussian filter operator with standard deviation $\sigma$, and $\mathcal{S}_2(\cdot)$ the subsampling operator by a factor 2, defined by $S_2(I)(\vec{x}) = I(2\vec{x})$
\State $I_1 := \mathcal{G}(I, \sigma_{init})$ (initial filtering)
\State $J_1 := \mathcal{G}(J, \sigma_{init})$
\For {$l = 1, ..., n_{layers}-1$}
\State $I_{l+1} := \mathcal{S}_2(\mathcal{G}(I_l, \sigma_{sub}))$
\State $J_{l+1} := \mathcal{S}_2(\mathcal{G}(J_l, \sigma_{sub}))$
\EndFor
\State $g_{n_{layers}} := 0$ (DVF guess initialization)
\State
\State \textbf{Computation of the DVF}
\For {$l = n_{layers}, ..., 1$}
\For {$x \in I_l$}
\State $G(x) {\footnotesize:=} \sum_{v} K_{\sigma_{LK}}(\|x - v\|_2)
\begin{bmatrix}
I_x^2  & I_xI_y & I_xI_z\\
I_xI_y & I_y^2  & I_yI_z \\
I_xI_z & I_yI_z & I_z^2
\end{bmatrix}$
\State where $I_x$ (resp. $I_y$, $I_z$) is the partial derivative of $I_l$ in the $x$-direction (resp. $y$ and $z$ directions) at voxel $v$ and $K_{\sigma_{LK}}$ is defined in Eq. \ref{eq:normal_gaussian}
\EndFor
\State $r_l^0 := 0$ (DVF refinement initialization)
\For {$i = 1, ..., n_{iter}$}
\For {$x \in I_l$}
\State $\delta I^i(x) := I_l(x) - J_l (x + g_l(x) + r_l^{i-1}(x))$
\EndFor
\For {$x \in I_l$}
\State $ b(x) := \sum_{v} K_{\sigma_{LK}}(\|x - v\|_2)
\begin{bmatrix}
\delta I^i(v) I_x(v) \\
\delta I^i(v) I_y(v) \\
\delta I^i(v) I_z(v) 
\end{bmatrix}$
\State $ r_l^i(x) := r_l^{i-1}(x) + G(x)^{-1}b(x)$
\EndFor
\EndFor
\If {l > 1}
\For {$x \in I_{l-1}$}
\State $g_{l-1}(x) := 2 (g_l(x/2) + r_l^{n_{iter}}(x/2)) $
\EndFor
\EndIf
\EndFor
\State 
\State \textbf{Output :} 3D displacement field $u(x)$
\State $u(x) := g_1(x) + r_1^{n_{iter}}(x) $ 

\end{algorithmic}
\end{algorithm}

\subsection{\normalsize Prediction of the position of internal points}
\label{subsection:method_RNN_prediction_pts}

After the computation of the optical flow, $r=3$ internal points $\vec{x_1}, ..., \vec{x_r}$ are selected close to the tumor in the initial image at $t=t_1$. They are considered to be points of known position during the treatment. It is reported in \cite{harley2010fiducial} that internal markers are usually implanted near or inside the tumor and that their number is generally 3 or 4. 

\begin{figure}
	\centering
		\includegraphics[width=\columnwidth]{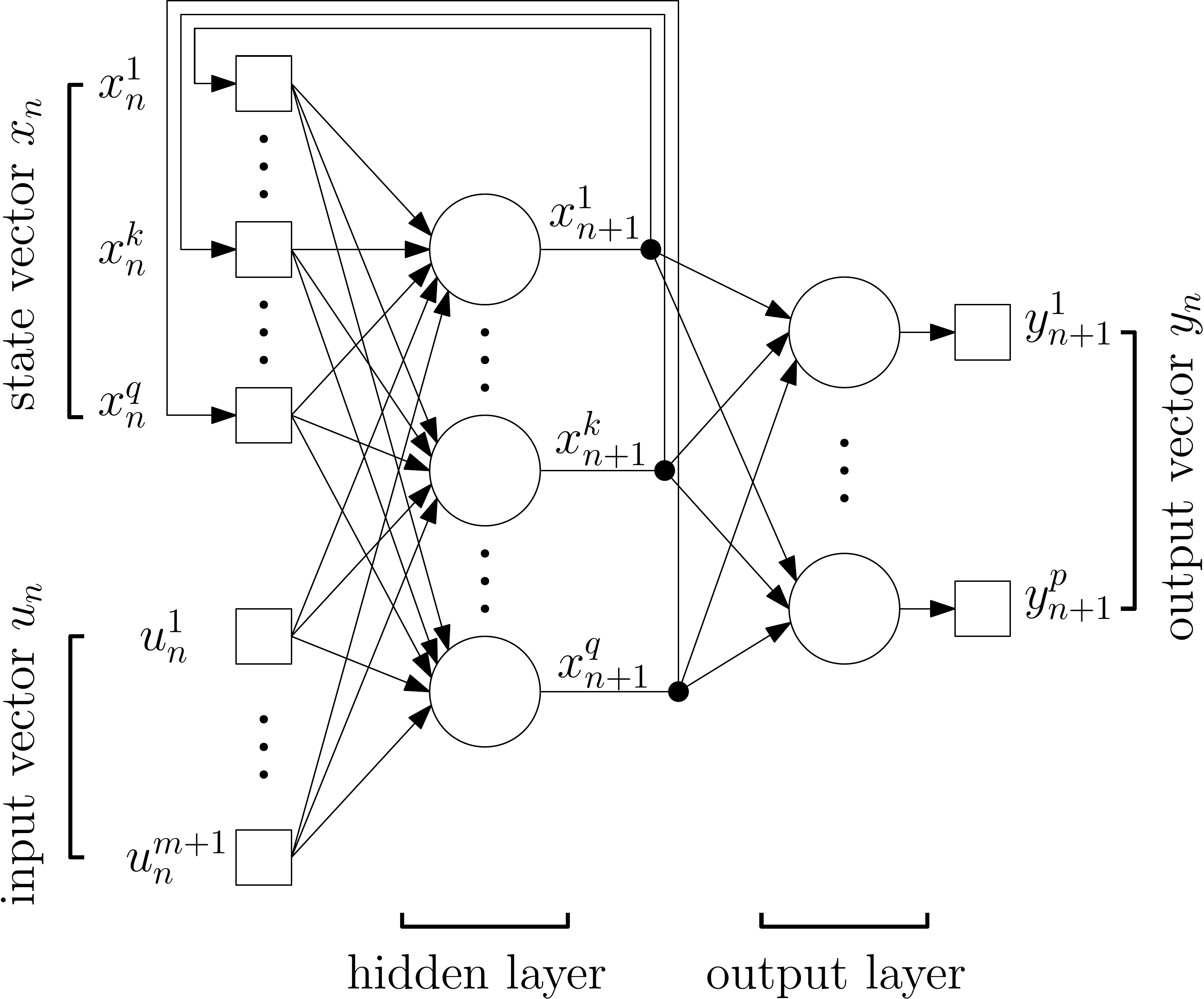}
	\caption{Structure of the RNN predicting the markers' position. The input vector \text{\normalsize $u_n$}, corresponding to the positions in the past, and the output vector \text{\normalsize $y_{n+1}$}, corresponding to the predicted positions, are defined in Eq. \ref{eq:RNN_in_out_def}.}.
	\label{fig:rnn_structure}
\end{figure}

We predict the motion of these $r$ points using an RNN. The input $u_n$ of the RNN is a vector of size $3rL+1$, where $L$ represents the signal history length (SHL): the time interval in the past, the information of which is used for making one prediction. $u_n$ consists of the concatenation of the displacement vectors $\vec{u}(\vec{x_p}, t_n)$, ..., $\vec{u}(\vec{x_p}, t_{n+L-1})$ for each point $p \in [1,...,r]$ (Eq. \ref{eq:RNN_in_out_def}). An additional $1$ was added to account for a bias unit. Each time-series $\big(u_d(\vec{x_p},t_n)$ $\big)_{n=1,...,N}$, for $d = x,y$, and $z$, and $p \in [1, ..., r]$, is normalized prior to being used as an input (Eq. \ref{eq:u_n_normalization}), in order to facilitate the learning process\footnotemark. The output $y_{n+1}$ of the RNN is a vector of size $3r$ consisting of the position of these $r$ points at the time $t_{n+L}$ (Eq. \ref{eq:RNN_in_out_def}). In particular, this means that the positions of all the markers are predicted simultaneously. Specifically, not only information concerning marker 1 but also the positions of markers 2,..., $r=3$ are used to predict the position of that first marker, which may help in mitigating the influence of noise.

\footnotetext{The relationships $\mathbb{E}\big((u_d(\vec{x_p},t_n))_{n=1,...,N}\big)=0$ and $\mathrm{Var}\big((u_d(\vec{x_p},t_n))_{n=1,...,N}\big)=1$ only hold true on the training set, that is to say for $N = N_{train} = 2000$. Indeed, in a practical case scenario, we cannot compute the mean and variance using future data. On the cross-validation set and the training set, we respectively subtract and divide each series by their mean $\mu_{d,p}$ and standard deviation $\sigma_{d,p}$ computed on the training set, before processing by the RNN.}

\begin{equation} \label{eq:RNN_in_out_def}
u_n
=
\begin{pmatrix}
1 \\
u_x(\vec{x_1}, t_n)\\
u_y(\vec{x_1}, t_n)\\
u_z(\vec{x_1}, t_n)\\
...\\
u_z(\vec{x_r}, t_n)\\
u_x(\vec{x_1}, t_{n+1})\\
...\\
u_z(\vec{x_r}, t_{n+L-1})
\end{pmatrix}
\qquad
y_{n+1}
=
\begin{pmatrix}
u_x(\vec{x_1}, t_{n+L})\\
u_y(\vec{x_1}, t_{n+L})\\
u_z(\vec{x_1}, t_{n+L})\\
...\\
u_z(\vec{x_r}, t_{n+L})
\end{pmatrix}
\end{equation}

\begin{equation} \label{eq:u_n_normalization}
\forall (d,p) \in \{x,y,z\} \times \{1,...,r\}, 
\left\{
\begin{matrix}
\mathbb{E}\big((u_d(\vec{x_p},t_n))_{n=1,...,N}\big)=0 \\
\mathrm{Var}\big((u_d(\vec{x_p},t_n))_{n=1,...,N}\big)=1
\end{matrix}
\right.
\end{equation}

The RNN architecture can be visualized in Fig. \ref{fig:rnn_structure}. It has one hidden layer which computes $q$ internal states $x_{n+1}^1$, ..., $x_{n+1}^q$ (scalar values) from the input $u_n$ and the internal states $x_n^1$, ..., $x_n^q$. The RNN output layer computes the output vector $y_{n+1}$ from the internal states $x_{n+1}^1$, ..., $x_{n+1}^q$. 

The system state vector $x_{n+1} = [x_{n+1}^1$, ..., $x_{n+1}^q]^T$ is calculated according to the measurement equation (left part of Eq. \ref{eq:RNN_system_measurement}) using the synaptic weight matrices $W_{a,n}$ and $W_{b,n}$, and a non-linear activation function $\Phi : \mathbf{R}^q \rightarrow \mathbf{R}^q$. The output vector $y_{n}$ is calculated by multiplying the synaptic weight matrix $W_{c,n}$ by the system states $x_n$, as described in the linear measurement equation (right part of Eq. \ref{eq:RNN_system_measurement}). In this research, we chose the hyperbolic tangent function as the activation function $\Phi$ (Eq. \ref{eq:non_linearity}).

\begin{equation} \label{eq:RNN_system_measurement}
 x_{n+1} = \Phi(W_{a,n} x_n + W_{b,n} u_n)
\qquad
 y_{n} = W_{c,n} x_n
\end{equation}

\begin{equation}\label{eq:non_linearity}
\Phi \left(
\begin{matrix}
a_1\\
...\\
a_q
\end{matrix}
\right) 
=
\begin{pmatrix}
\phi (a_1)\\
...\\
\phi (a_q)
\end{pmatrix}
\quad
\text{where}
\quad
\phi(a) = tanh(a)
\end{equation}

The RNN is trained using the RTRL algorithm (Algorithm \ref{alg:RNN-RTRL}). Prior to the learning process, each synaptic weight is initialized according to a normal distribution of standard deviation $\sigma_{init}^{RNN}$. RTRL is an online learning method, and so the weight matrices $W_{a,n}$, $W_{b,n}$, and $W_{c,n}$ are updated at every time step to take into account the recent changes in the breathing pattern of the patient. Given the predicted positions of the markers $y_n$ and the real position of the markers $d_n$, we can compute the instantaneous error vector $e_n$ and instantaneous error function $E_n$ as in Eq. \ref{eq:error_vector_and_function}.  

\begin{equation} \label{eq:error_vector_and_function}
 e_{n} = d_n - y_n
\qquad
 E_{n} = \frac{1}{2} \|e_n\|_2^2
\end{equation}

The weight matrix $W_{k,n+1}$ at time $n+1$, where $k = a,b$ or $c$, is computed from the corresponding weight matrix $W_{k,n}$ at time $n$ by performing a single gradient descent update. However, RNNs updated by the gradient rule may be unstable, and as proposed in \cite{pascanu2013difficulty}, we prevent large weight updates by clipping the gradient norm to address instability. Specifically, given the learning rate $\eta$ and a threshold $\theta$, we update each weight matrix $W_{k,n}$, where $k = a,b$ or $c$, according to Eq. \ref{eq:RNN_weights_update_rule}. Details concerning the calculation of the terms ${\partial E_n}/{\partial W_{k,n}}$ can be found in \cite{haykin2009neural}, whose description was extended in this work to encompass RNNs with a multidimensional output vector. The RNN main characteristics are summarized in Table \ref{table:RNN_configuration}. The RTRL computation complexity is $\mathcal{O}(q^2 (q+m) (q+p))$. 

\begin{table}
\begin{tabular*}{\tblwidth}{@{} LL@{} }
\toprule
RNN characteristic &  \\
\midrule
Output layer size       & $p = 3 r $ \\
Input layer size        & $m = 3 r L$\\
Number of hidden layers & 1 \\
Size of the hidden layer & $q$ \\
Activation function $\phi$ &  Hyperbolic tangent \\
Training algorithm &  RTRL (online learning)\\
Optimization method &  Stochastic gradient descent \\ 
                    & with gradient clipping \\
                    & (learn. rate $\eta$ and clip. threshold $\theta$) \\
Weights initialization & Gaussian with std. dev. $\sigma_{init}^{RNN}$ \\
Input data normalization & Yes (online)\\
Cross-validation metric & MAE (Eq. \ref{eq:MAE_def}) \\
Nb. of runs for evaluation & 10 \\
\bottomrule
\end{tabular*}
\caption{Configuration of the RNN for predicting the motion of the internal points, as described in Section \ref{subsection:method_RNN_prediction_pts}. $r$ refers to the number of internal points selected and $L$ to the SHL.}
\label{table:RNN_configuration}
\end{table} 

\begin{figure*}[b]
\begin{equation} \label{eq:RNN_weights_update_rule}
\resizebox{0.96\hsize}{!}{
\mbox{\normalsize\(
W_{k,n+1} = W_{k,n} - \rho \dfrac{\partial E_n}{\partial W_{k,n}}
\text{ where }
\rho = 
\left\{
\begin{matrix}
\eta & \text{if }  
\sqrt{\left\|  \dfrac{\partial E_n}{\partial W_{a,n}}\right\|_2^2
+\left\|\dfrac{\partial E_n}{\partial W_{b,n}}\right\|_2^2
+\left\|\dfrac{\partial E_n}{\partial W_{c,n}}\right\|_2^2} \leq \theta\\
\frac{\displaystyle{\eta \theta}}{
\sqrt{\left\|  \dfrac{\partial E_n}{\partial W_{a,n}}\right\|_2^2
+\left\|\dfrac{\partial E_n}{\partial W_{b,n}}\right\|_2^2
+\left\|\dfrac{\partial E_n}{\partial W_{c,n}}\right\|_2^2}
} & \text{if } 
\sqrt{\left\|  \dfrac{\partial E_n}{\partial W_{a,n}}\right\|_2^2
+\left\|\dfrac{\partial E_n}{\partial W_{b,n}}\right\|_2^2
+\left\|\dfrac{\partial E_n}{\partial W_{c,n}}\right\|_2^2} > \theta\\
\end{matrix}
\right.
\)}
}
\end{equation}
\end{figure*}

\begin{algorithm}
\caption{RTRL with gradient clipping}
\label{alg:RNN-RTRL}
\begin{algorithmic}
\State \textbf{Parameters} :
\State $L$ : signal history length
\State $r$ : number of internal points considered
\State $m = 3rL$ dimension of the input space
\State $q$ : dimension of the state space
\State $p = 3r$ dimension of the output space
\State $\eta$ : learning rate
\State $\theta$ : gradient threshold
\State $\sigma_{init}^{RNN}$ : standard deviation of the initial weights
\State
\State \textbf{Initialization}
\State $W_{a,n=1}$ : $q \times q$ matrix initialized randomly according to a Gaussian distribution with standard deviation $\sigma_{init}^{RNN}$
\State $W_{b,n=1}$ : $q \times (m+1)$ matrix initialized randomly according to a Gaussian distribution with standard deviation $\sigma_{init}^{RNN}$
\State $W_{c,n=1}$ : $p \times q$ matrix initialized randomly according to a Gaussian distribution with standard deviation $\sigma_{init}^{RNN}$
\State State vector $x_{n=1} := 0_{q \times 1}$ 
\For{$j = 1,..., q$}
\State $\Lambda_{j,n=1} := 0_{q \times (q+m+1)}$ 
\EndFor
\State
\State \textbf{Learning and prediction}
\For{$n = 1,2,...$}
\State $y_n := W_{c,n}x_n$ (prediction)
\State $e_n := d_n - y_n$ (error vector update)
\For{$j = 1,..., q$} (gradient calculation)
\State $w_{j,n} := \begin{bmatrix} 
w_{a,j,n} \\
w_{b,j,n}
\end{bmatrix}$ 
{\footnotesize where}
$\begin{matrix}
W_{a,n} = [w_{a,1,n}, ..., w_{a,q,n}]^T\\
W_{b,n} = [w_{b,1,n}, ..., w_{b,q,n}]^T
\end{matrix}$
\State $\Delta w_{j,n} := \Lambda_{j,n}^T W_{c,n}^T e_n$ 
\EndFor
\State $\Delta W_{c,n} := e_n \otimes x_n$ 
\State $\kappa := \sqrt{\|\Delta w_{1,n}\|_2^2 + ... + \|\Delta w_{q,n}\|_2^2 + \|\Delta W_{c,n}\|_2^2}$
\If{$ \kappa > \theta $} (gradient clipping)
\For{$j = 1,..., q$}
\State $\Delta w_{j,n} := \frac{\theta}{\kappa} \Delta w_{j,n}$ 
\EndFor
\State $\Delta W_{c,n} := \frac{\theta}{\kappa} \Delta W_{c,n}$ 
\EndIf
\State $W_{c,n+1} := W_{c,n} + \eta \Delta W_{c,n}$ (gradient update)
\State $\xi_n := \begin{bmatrix} 
x_n \\
u_n
\end{bmatrix}$ 
\State $\Phi_n := diag(\phi'(w_{1,n}^T\xi_n), ..., \phi'(w_{q,n}^T\xi_n))$
\For{$j = 1,..., q$}
\State $w_{j,n+1} := w_{j,n} + \eta \Delta w_{j,n}$ (gradient update)
\State $ U_{j,n} := \begin{bmatrix} 
0 \\
\xi_n^T \\
0
\end{bmatrix}
\leftarrow j^{th} \mbox{ row} $
\State $ \Lambda_{j,n+1} := \Phi_n [ W_{a,n} \Lambda_{j,n} + U_{j,n}]$ 
\EndFor  
\State $ W_{a,n+1} := [w_{a,1,n+1}, ..., w_{a,q,n+1}]^T$ 
\State $ W_{b,n+1} := [w_{b,1,n+1}, ..., w_{b,q,n+1}]^T$ 
\State $x_{n+1} := \Phi(W_{a,n}x_n + W_{b,n}u_n)$ (hidden states update) 
\EndFor

\end{algorithmic}
\end{algorithm}

\subsection{\normalsize Application to chest image prediction}

In what follows, we propose a simple method to predict future 3D images of the ROI based on marker position prediction as described in Section \ref{subsection:method_RNN_prediction_pts}. First, we assume that the motion of each voxel is linked to the motion of the markers via a linear relationship, which indirectly models the connectivity between the tissues (Eq. \ref{eq:spatial_lin_reg}). The coefficients $\gamma_p(\vec{x})$ are calculated using linear regression.

\begin{equation}\label{eq:spatial_lin_reg}
\vec{u}(\vec{x}, t) = \sum_{p = 1}^r \gamma_p(\vec{x}) \vec{u}(\vec{x_p}, t)
\end{equation}

\begin{figure}
	\centering
		\includegraphics[width=.7\columnwidth]{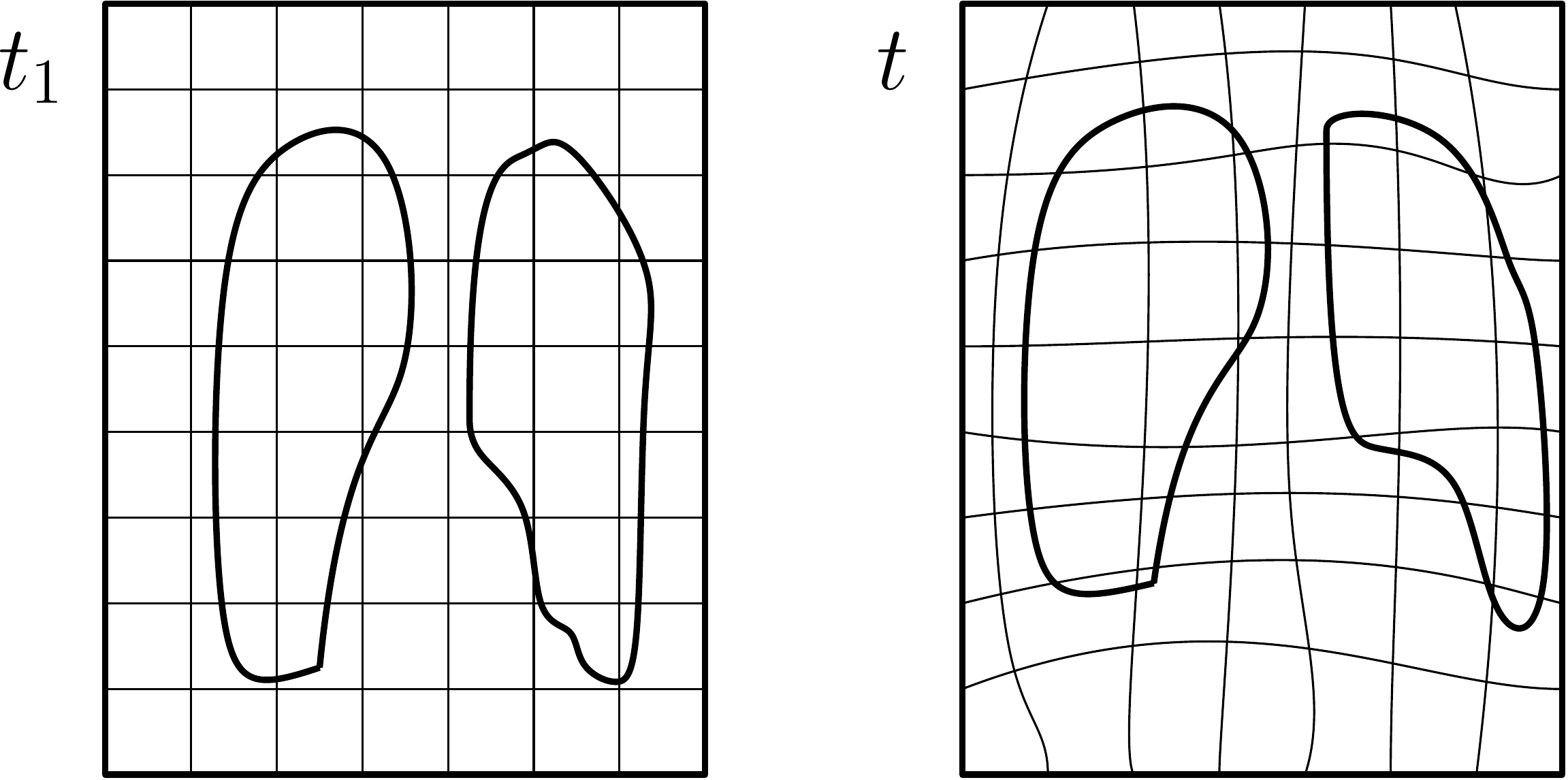}
	\caption{Warping the initial lung image at \text{\normalsize $t=t_1$} to estimate the lung image at \text{\normalsize $t$}}
	\label{fig:lung_warping}
\end{figure}

Given the position of the markers at time $t_{1}$, ..., $t_{n}$, their position at time $t_{n+1}$ can be predicted using the RNN, and the whole DVF at $t_{n+1}$, $\vec{u}(\cdot, t_{n+1})$, can then be recovered using Eq. \ref{eq:spatial_lin_reg}. In order to estimate the image at time $t_{n+1}$, we can warp the initial image $I(\cdot, t_1)$ by the field $\vec{u}(\cdot, t_{n+1})$ (Fig. \ref{fig:lung_warping}). This relies on the assumption that the image at $t_{n+1}$ can be approximately reconstructed via warping the image at $t_1$. 

\begin{figure}
	\centering
		\includegraphics[width=0.8\columnwidth]{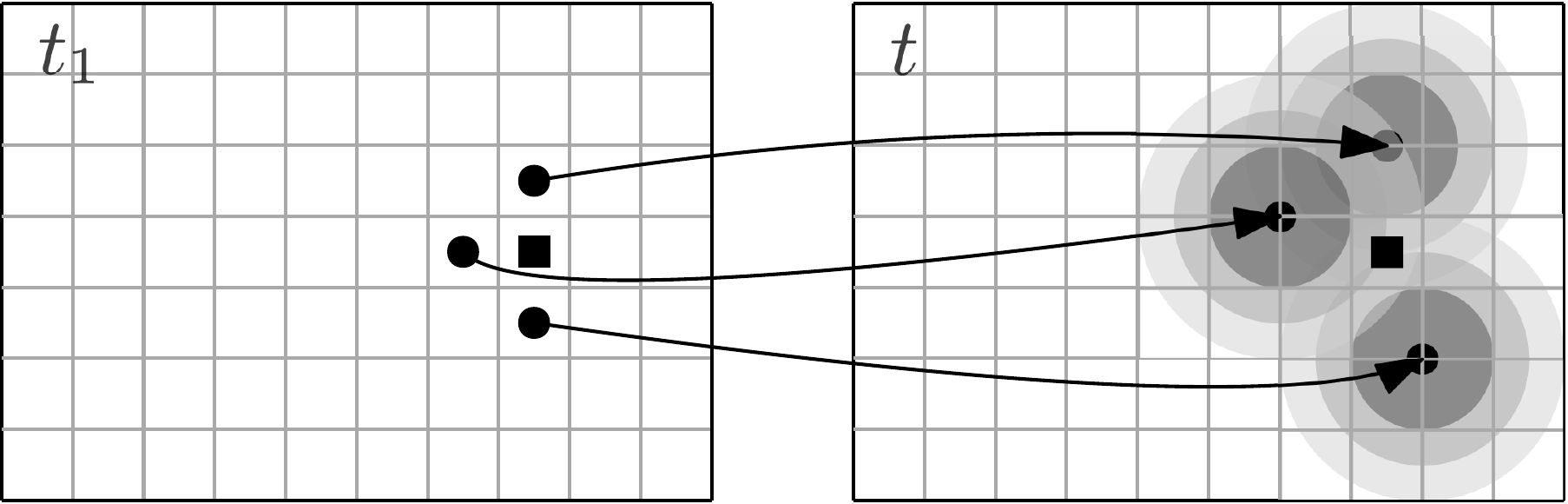}
	\caption{Warping the initial image at \text{\normalsize $t=t_1$} using Nadaraya-Watson regression with a Gaussian kernel. 
	The closer a point at \text{\normalsize $t=t_1$} arrives next to the square point at \text{\normalsize $t$}, the more it contributes to the intensity of that square point at \text{\normalsize $t$}.}
	\label{fig:warp_vectors_principle}
\end{figure}

In order to estimate the image at time $t$ from the DVF at time $t_1$, we use the Nadaraya-Watson non-parametric regression method, described in Fig. \ref{fig:warp_vectors_principle} and Eq. \ref{eq:Nadaraya_Watson}. The modified kernel $\widetilde{K}$ used in that equation is a variant of the Gaussian kernel $K$ defined in Eq. \ref{eq:normal_gaussian}. $\sigma_w$ represents the standard deviation of the new kernel $\widetilde{K}$ and $h$ represents the window size of the kernel calculation. Imposing an arbitrary window size $h$ is necessary because the calculations would be slow otherwise\footnotemark. However, this may lead to some voxels in the destination image $I(\cdot, t)$ not having any corresponding voxel in the source image $I(\cdot, t_1)$. Therefore, $h$ needs to be chosen appropriately large. Furthermore, $\sigma_w$ needs to be selected such that the images do not appear either too blurry or with too many artifacts, such as inappropriate impainting due to voxels in the destination image having only one antecedent voxel. Theoretical details about the Nadaraya-Watson statistical estimator can be found in \cite{tsybakov2008introduction}. The computational complexity of image warping is $\mathcal{O}(V h^3)$ where $V$ is the volume (in voxels) of the image considered.

\footnotetext{When calculating the optical flow, $\widetilde{K}$ was also used instead of $K$ to process the data reasonably fast, but we did not introduce this notation for two reasons. First, it is generally assumed that there is a window when using a Gaussian kernel so that was implicit. Secondly, adjusting the size of the window is particularly important when reconstructing images, because of the problem of voxels without antecedent. }

\begin{equation} \label{eq:Nadaraya_Watson}
I_{NW}(\vec{x},t) = \frac{
\sum_{\vec{p}} I(\vec{p}, t_1)
\widetilde{K}_{\sigma_w, h}\big( 
\| \vec{x} - (\vec{p} + \vec{u}(\vec{p}, t))  \|_2
\big)
}
{
\sum_{\vec{p}}
\widetilde{K}_{\sigma_w, h}\big( 
\| \vec{x} - (\vec{p} + \vec{u}(\vec{p}, t))  \|_2
\big)
}
\end{equation}

\begin{equation} \label{eq:modified_kernel}
\widetilde{K}_{\sigma_w, h} (x) = 
\left\{
\begin{matrix}
K_{\sigma_w}(x) & \text{if } |x| < h & \text{(cf Eq. \ref{eq:normal_gaussian})} \\
0             & \text{otherwise}   &
\end{matrix}
\right.
\end{equation}

\section{Results and discussion}

\subsection{\normalsize Chest image registration}

In order to determine the parameters giving the most accurate DVF for each image sequence, we calculated the registration error defined in Eq. \ref{eq:OF_error_def}, for the following set of parameters, on the initial ROI sequences of $n=10$ images :
\begin{enumerate}[\textbullet]
	\item $\sigma_{init} \in \{0.2, 0.5, 1.0, 2.0\}$
	\item $\sigma_{sub} \in \{0.2, 0.5, 1.0, 2.0\}$
	\item $\sigma_{LK} \in \{1.0, 2.0, 3.0, 4.0\}$
	\item number of layers $n_{layers} \in \{1, 2, 3, 4\}$
	\item number of iterations $n_{iter} \in \{1, 2, 3\}$
\end{enumerate}

\begin{equation} \label{eq:OF_error_def}
 e_{DVF} = \sqrt{\frac{1}{(n-1)|I|}\sum_{k=2}^{n} \sum_{\vec{x}} \big[ I(\vec{x}, t_1) - I(\vec{x}+\vec{u}(\vec{x}, t_k), t_k) \big] ^2} 
\end{equation} 

\begin{figure*}
    \centering
    \subfloat[Registration error as a function of \text{\normalsize $\sigma_{init}$}]{{\includegraphics[width=5cm]{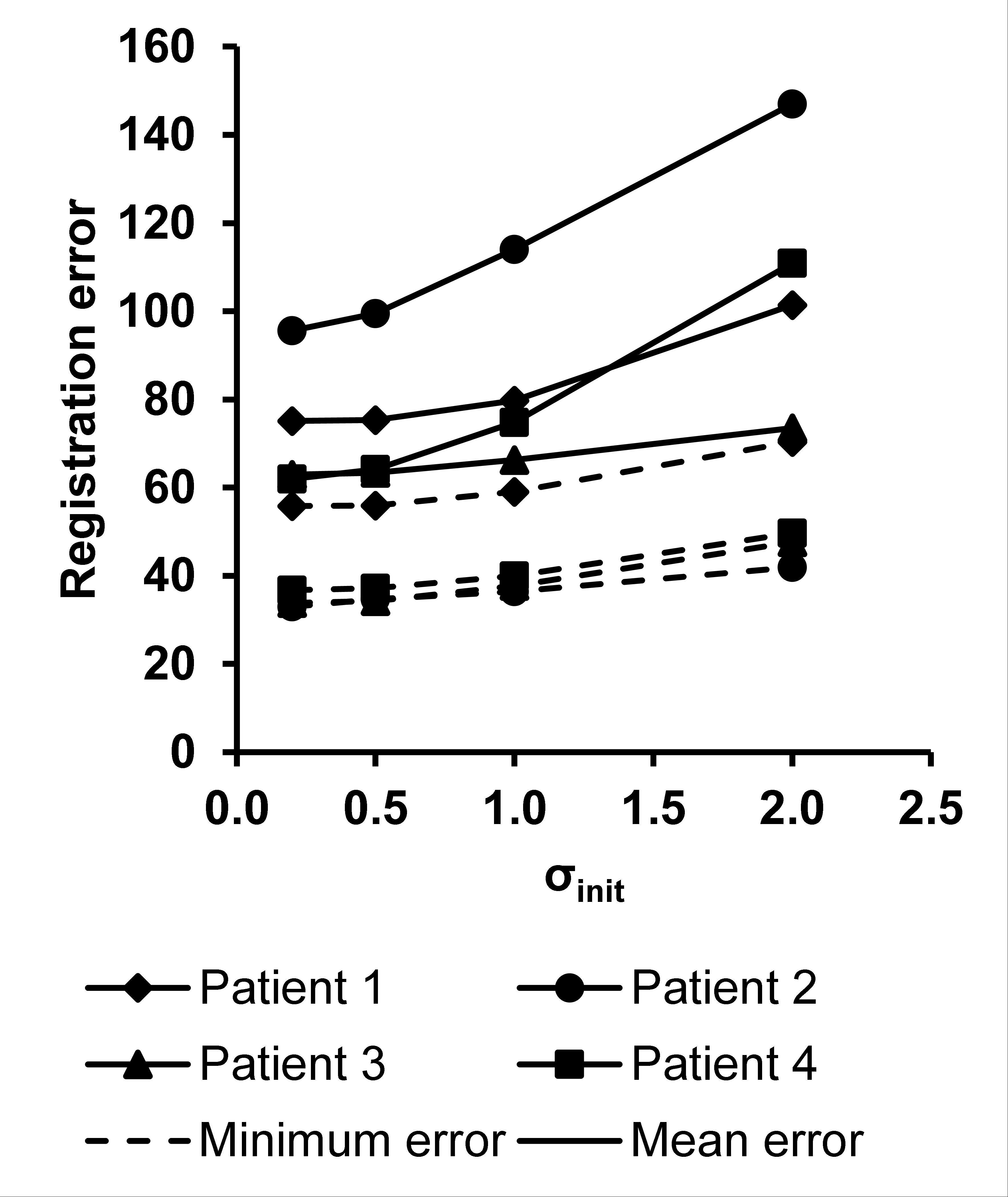} \label{subfig:DVF_err_sg_init}}}%
    \qquad
    \subfloat[Registration error as a function of \text{\normalsize $\sigma_{sub}$}]{{\includegraphics[width=5cm]{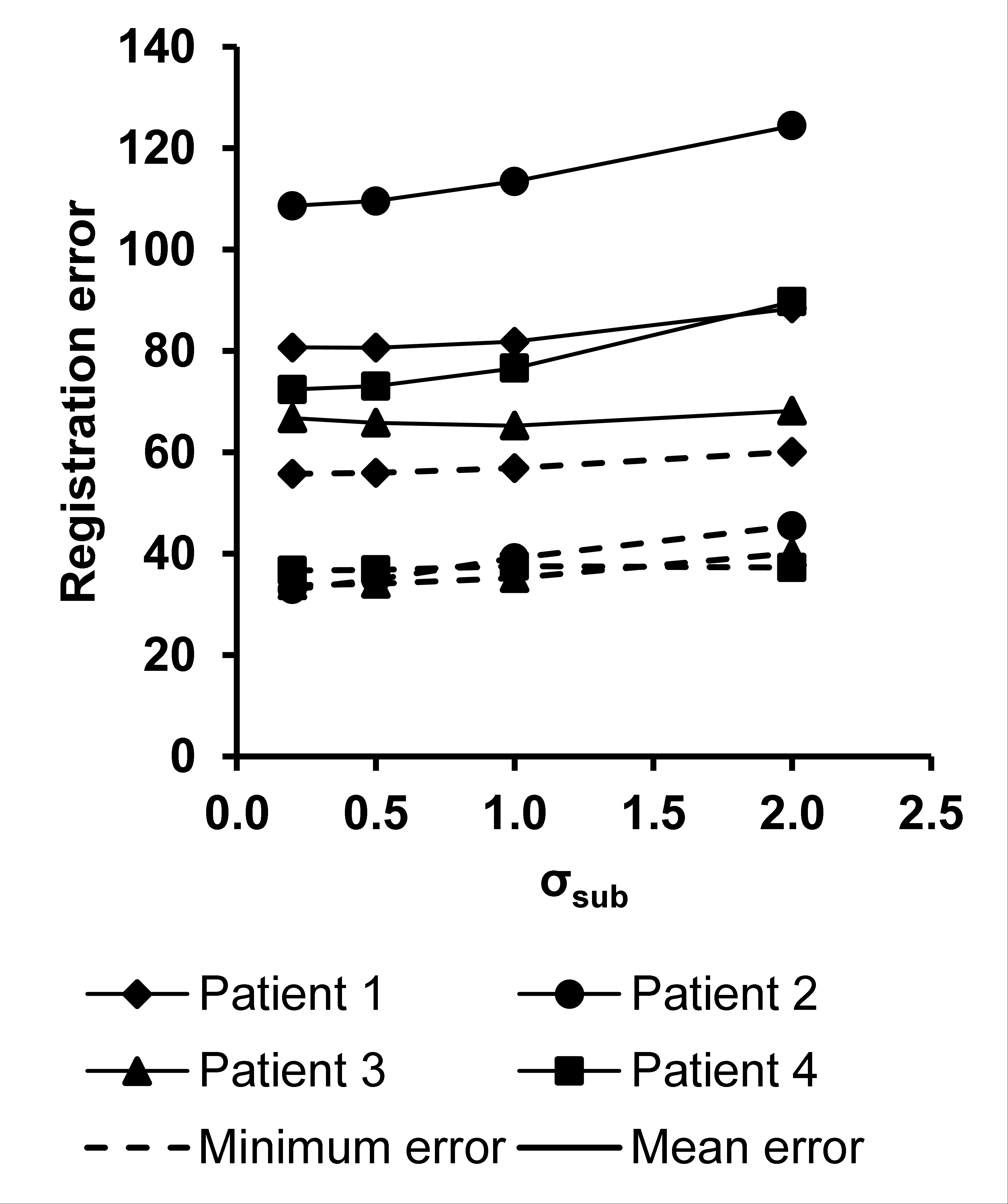} \label{subfig:DVF_err_sg_sub}}}%
    \qquad
    \subfloat[Registration error as a function of \text{\normalsize $\sigma_{LK}$}]{{\includegraphics[width=5cm]{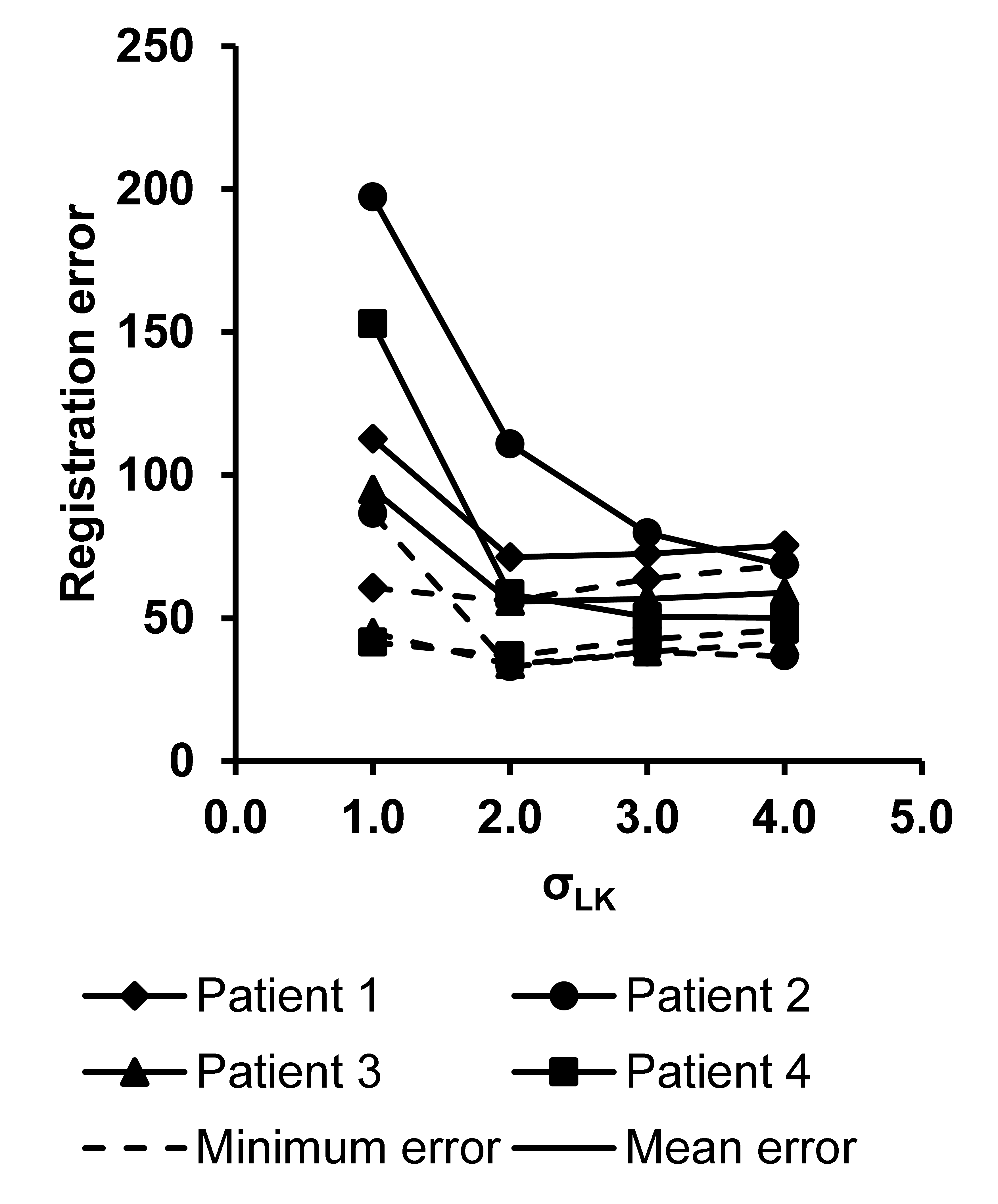} \label{subfig:DVF_err_sg_LK}}}%
    \qquad
    \subfloat[Registration error as a function of the number of layers]{{\includegraphics[width=5cm]{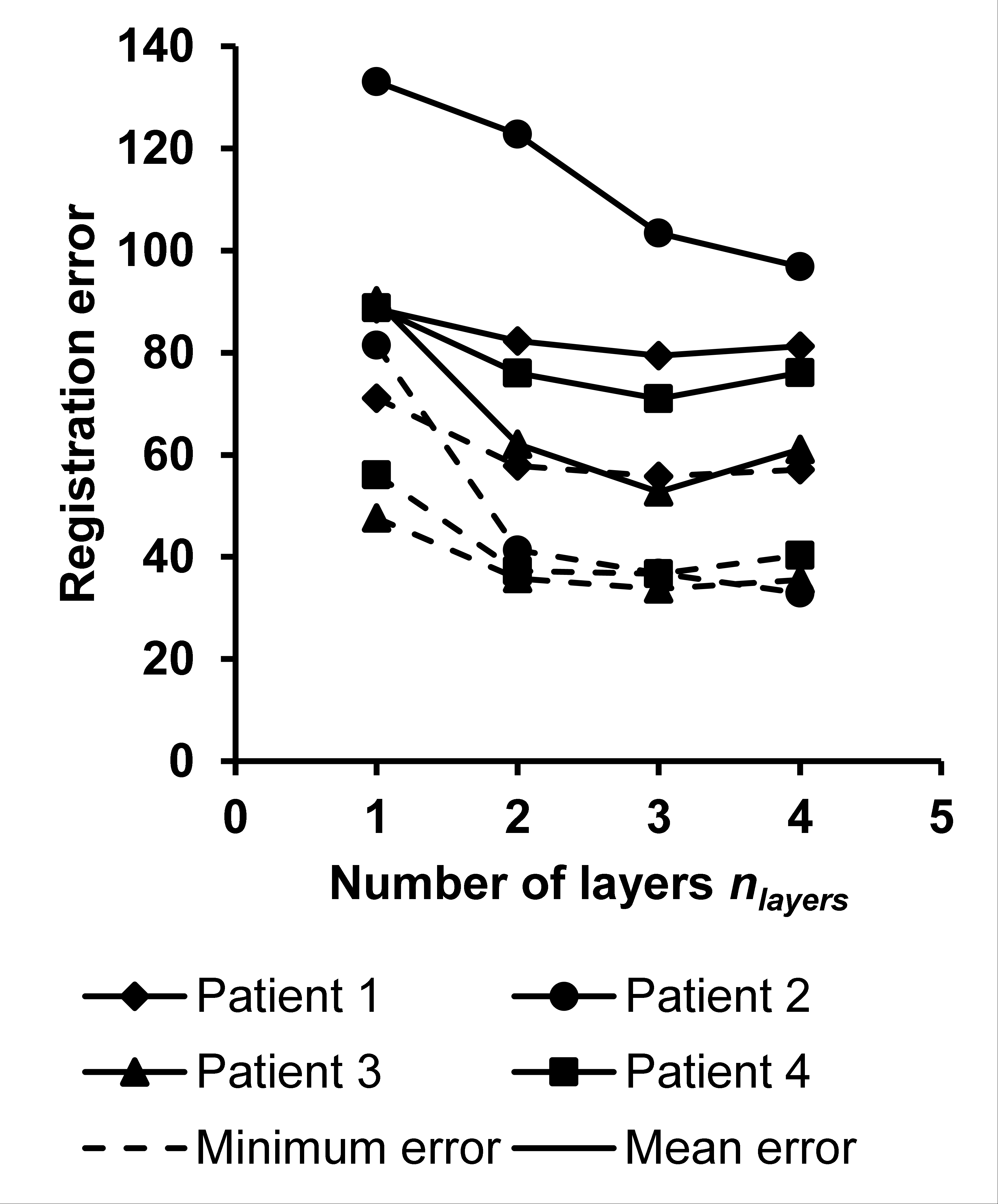} \label{subfig:DVF_err_nlayers}}}%
    \qquad
    \subfloat[Registration error as a function of the number of iterations]{{\includegraphics[width=5cm]{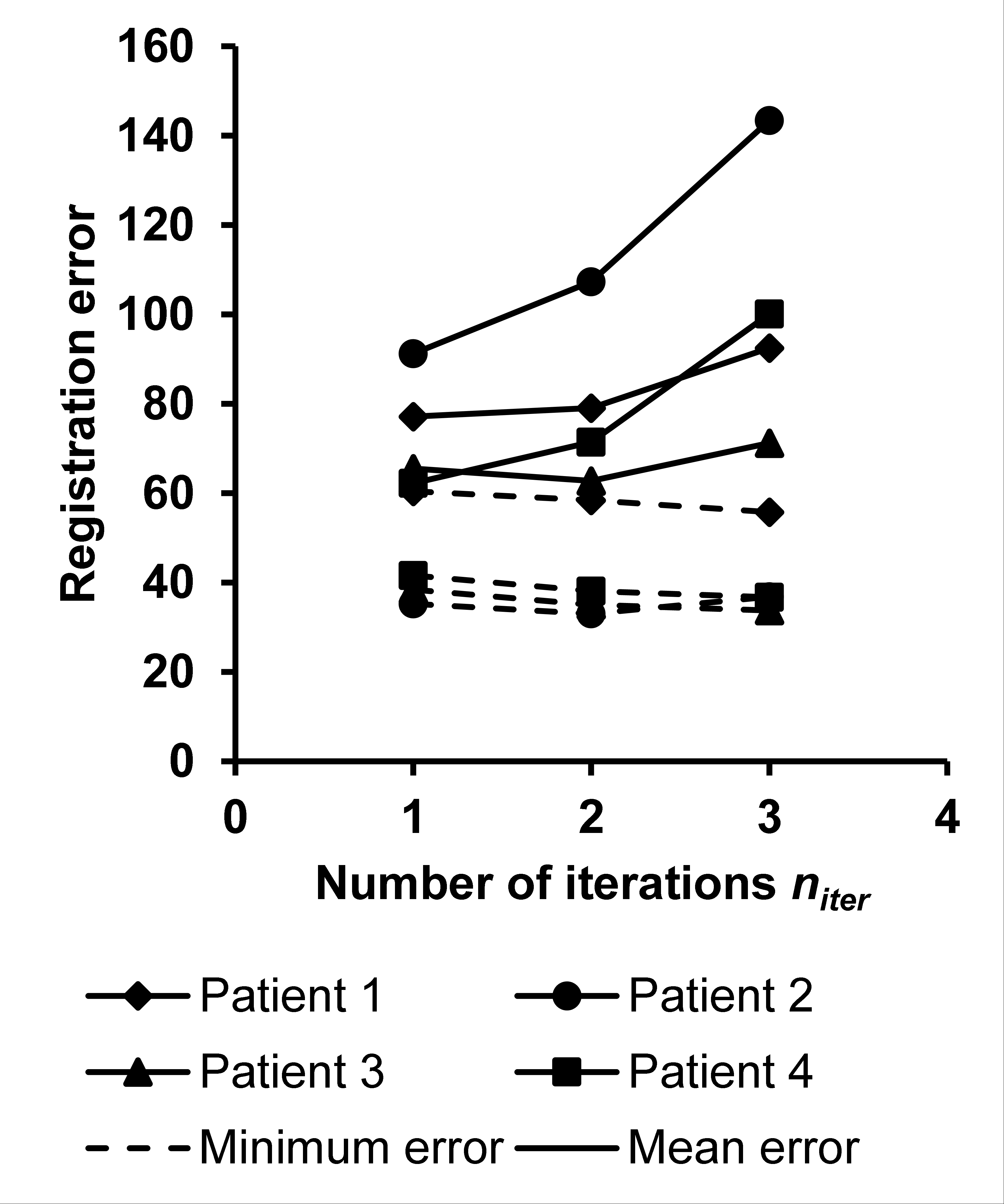} \label{subfig:DVF_err_niter}}}%
    \caption{Registration error $e_{DVF}$ as a function of the parameters of the Lucas-Kanade iterative and pyramidal optical flow algorithm (cf Eq. \ref{eq:OF_error_def}). The minimum error refers to the minimum of the registration error across every parameter, and the mean error refers to the registration error averaged over the four parameters not studied in each graph.}%
    \label{fig:OF_grid_search_eval}%
\end{figure*}


\begin{figure}[ht!]
    \centering
    \includegraphics[width=.9\columnwidth]{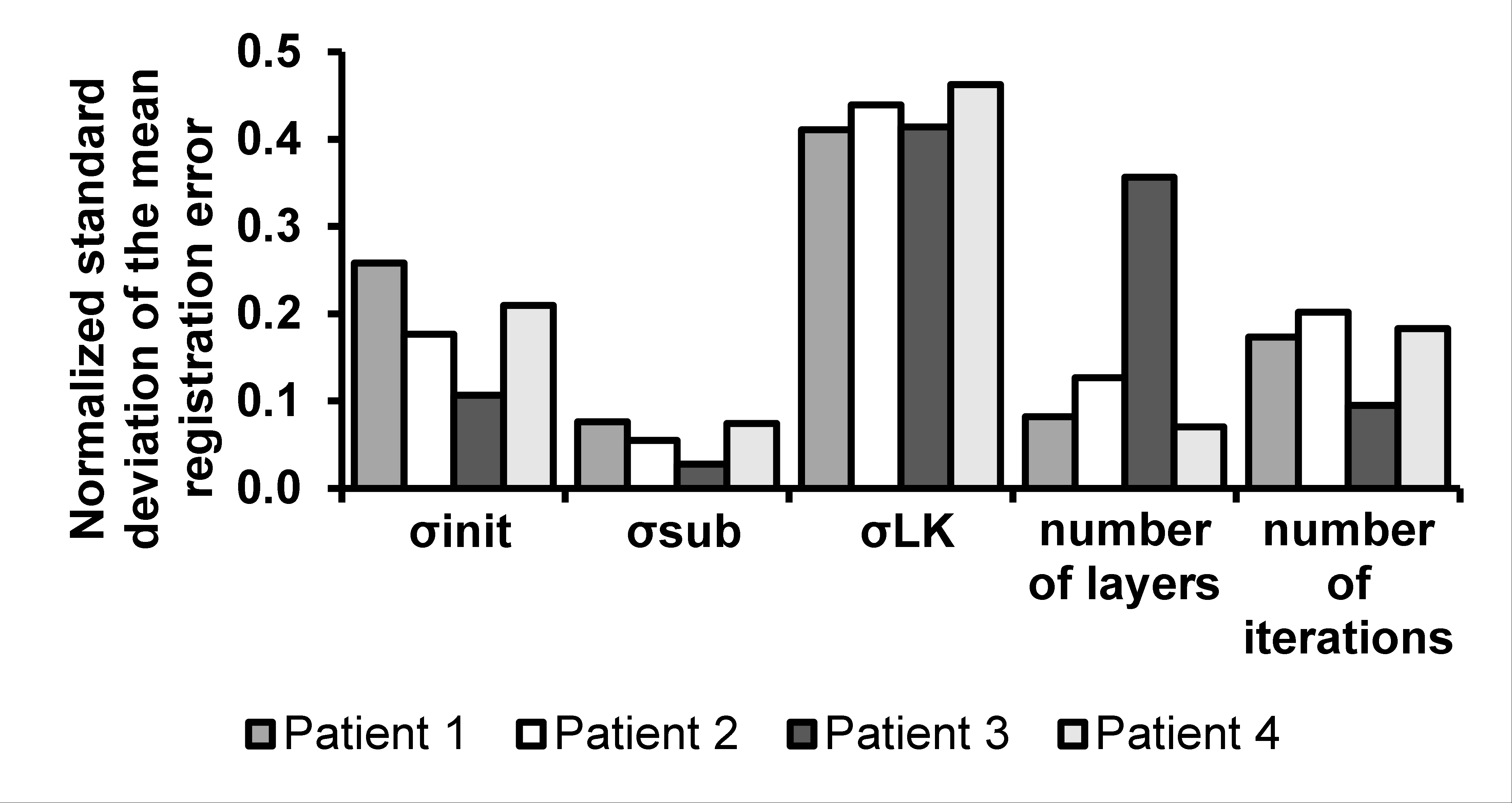}
    \includegraphics[width=.9\columnwidth]{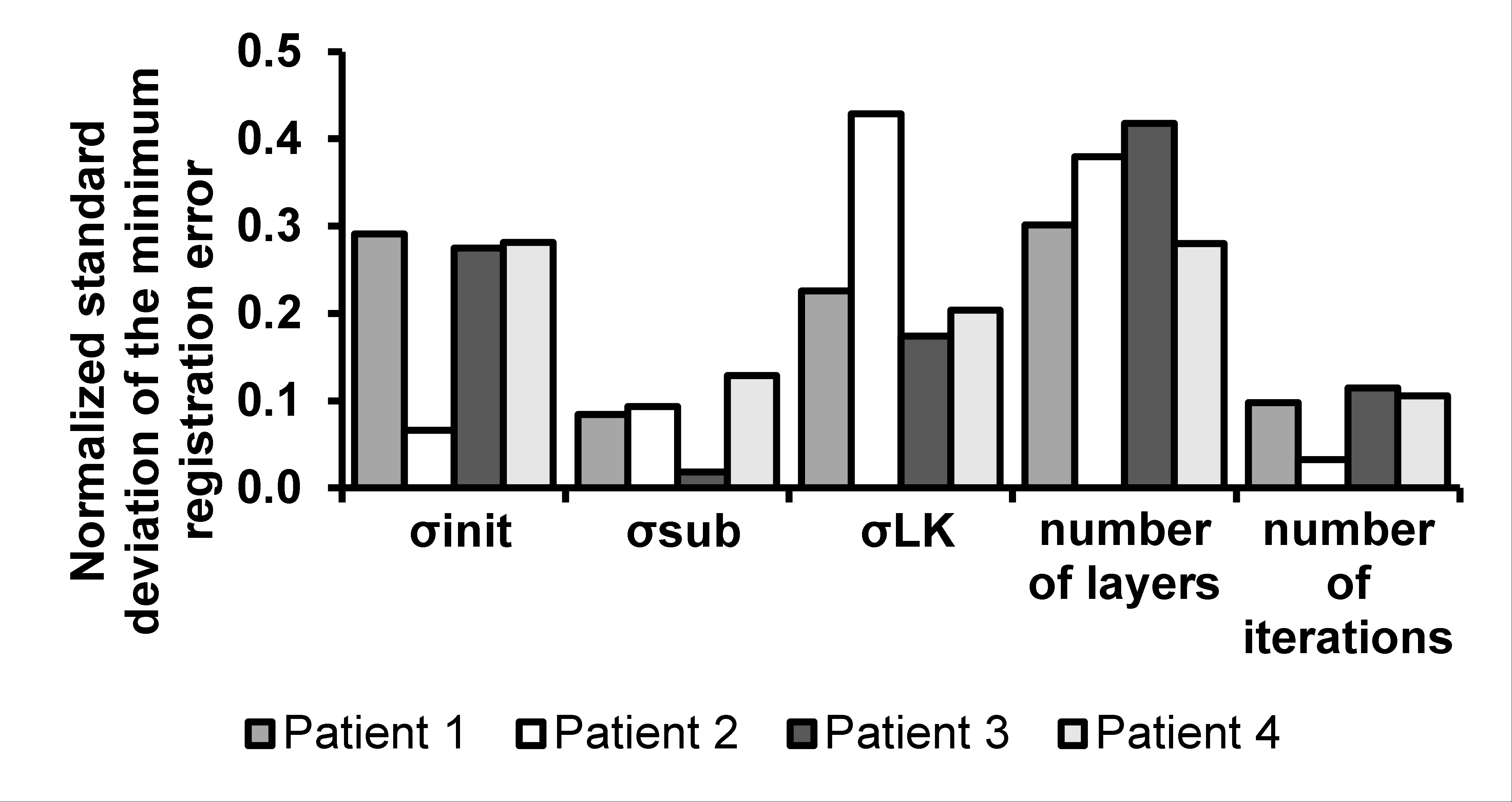}
    \caption{Relative influence of the parameters of the Lucas-Kanade optical flow algorithm on the registration error.}
    \label{fig:OF_grid_search_eval_std_dev}%
\end{figure}

The results of this grid search optimization are displayed in Fig. \ref{fig:OF_grid_search_eval} and Fig. \ref{fig:OF_grid_search_eval_std_dev}. Fig. \ref{fig:OF_grid_search_eval} shows for each parameter two different types of errors. The first one is the mean registration error: the registration error averaged over every other parameter. The second type is the minimum registration error, which represents the minimum error over the entire set of parameters. Both the minimum error and mean error increase for every patient when $\sigma_{init}$ increases (Fig. \ref{subfig:DVF_err_sg_init}), which means that initial filtering had a detrimental effect on the accuracy of the registration, because the initial images were not very noisy. Similarly, the registration minimum error increases with $\sigma_{sub}$, except for patient 3 (Fig. \ref{subfig:DVF_err_sg_sub}). Both errors as a function of $\sigma_{LK}$ are either decreasing or strictly convex, except for the minimum error of patient 2 (Fig. \ref{subfig:DVF_err_sg_LK}). Setting $\sigma_{LK} = 1.0$ (lowest value tested) leads to large mean registration errors. Likewise, the errors associated with $n_{layers}$ are either decreasing or strictly convex (Fig. \ref{subfig:DVF_err_nlayers}). Using only one layer (simple Lucas-Kanade algorithm) entails large errors, because the motion of the chest has a high amplitude relative to the imaging resolution. This supports the previous claims in the literature that a multiresolution scheme is generally needed for accurate registration of chest CT scan images \cite{xu2008lung, zhang2008use}. Increasing $n_{iter}$ results in a decrease in the minimum error, except for patient 2, and an increase in the mean error, except for patient 3 (Fig. \ref{subfig:DVF_err_niter}). For all the patients, $\sigma_{init} = 0.2$, $\sigma_{sub} = 0.2$, and $\sigma_{LK} = 2.0$ led to the highest displacement field accuracy. The registration was the most accurate using $n_{layers} = 3$ and $n_{iter} = 3$ for patients 1, 3, and 4, and using $n_{layers} = 4$ and $n_{iter} = 2$ for patient 2. 

\begin{figure}
    \captionsetup[subfigure]{labelformat=empty}
    \centering
    \subfloat[\text{\normalsize $t=t_{2209}$}]{{\includegraphics[width=.4\columnwidth]{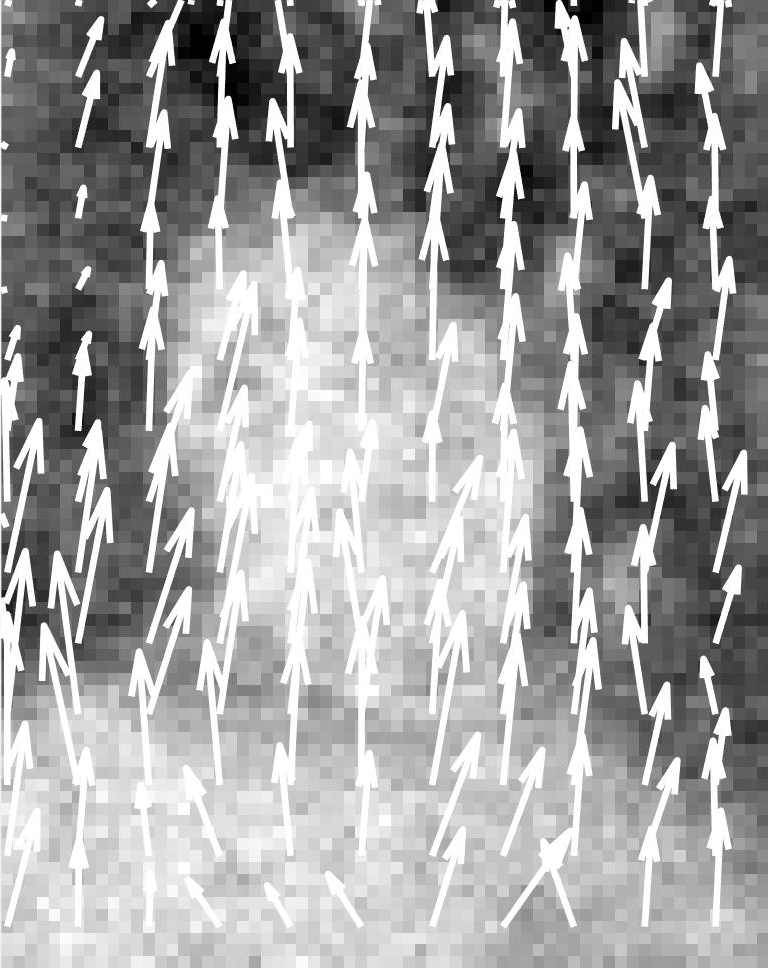} }}%
    \quad
    \subfloat[\text{\normalsize $t=t_{2374}$}]{{\includegraphics[width=.4\columnwidth]{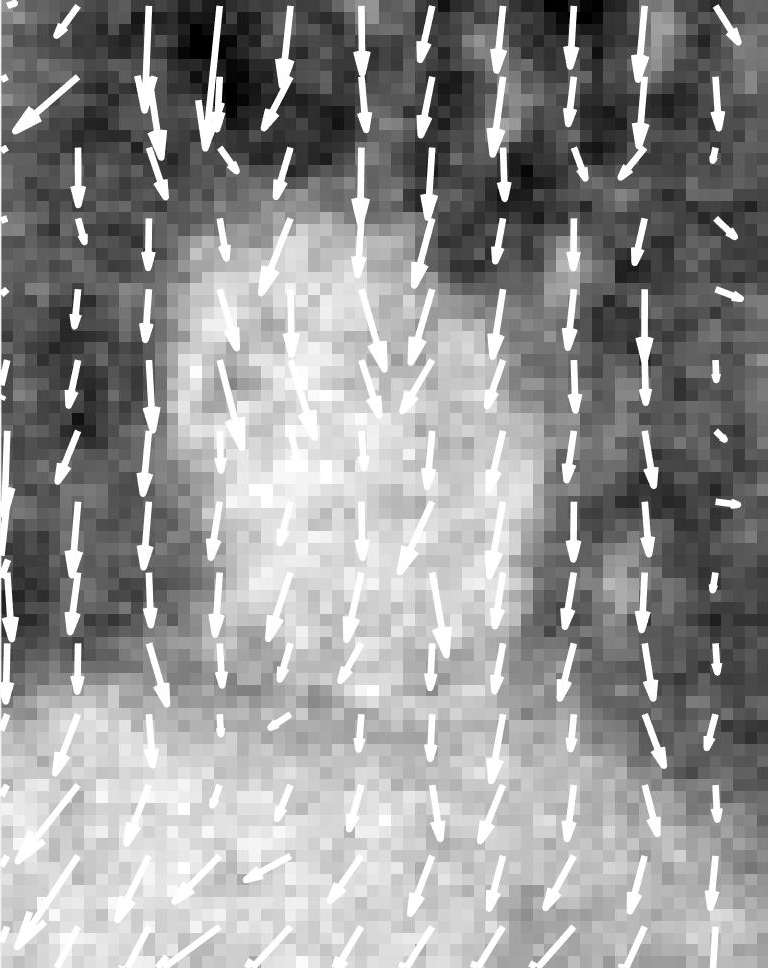} }}%
    \caption{Displacement vector field in the ROI for patient 1 at \text{\normalsize $t=t_{2209}$} (end of expiration) and \text{\normalsize $t=t_{2374}$} (end of inspiration) projected in a coronal cross-section displayed in the background at \text{\normalsize $t=t_1$} (same coordinates as in Fig. \ref{fig:org_im}). The origins of each of the displayed two-dimensional (2D) displacement vectors are separated from each other by 6 voxels.}
    \label{fig:OFvectors}
\end{figure}

The normalized standard deviation of the mean error and minimum error relative to each parameter is reported in Fig. \ref{fig:OF_grid_search_eval_std_dev}. "Normalization" means that for each patient, the sum of all the contributions was set to be equal to 1 by multiplying them by a proportionality coefficient. $\sigma_{LK}$ is the parameter that contributes the most to the variation in the mean error. $\sigma_{LK}$ and $n_{layers}$ are the two parameters that have the highest influence on the minimum registration error, and this emphasizes the importance of using more than one layer when performing lung image registration. The minimum registration error varied with $\sigma_{LK}$ from 68.5 to 55.8 for patient 1, from 86.4 to 33.0 for patient 2, from 45.9 to 36.7 for patient 3, and from 44.7 to 33.8 for patient 4. In other words, optimizing $\sigma_{LK}$ led to a 31.3\% average decrease in the minimum registration error. Similarly, carefully selecting $n_{layers}$ led to a 36.2\% average decrease in the minimum registration error. 

\begin{figure}
    \captionsetup[subfigure]{labelformat=empty}
    \centering  
    \subfloat[Coronal view]{{\includegraphics[height=5cm]{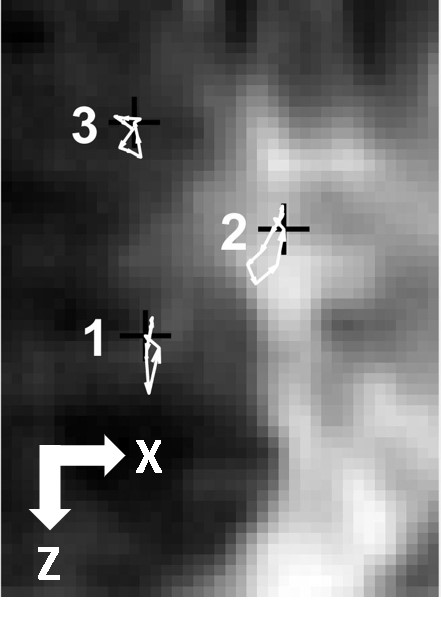} }}%
    \quad
    \subfloat[Sagittal view]{{\includegraphics[height=5cm]{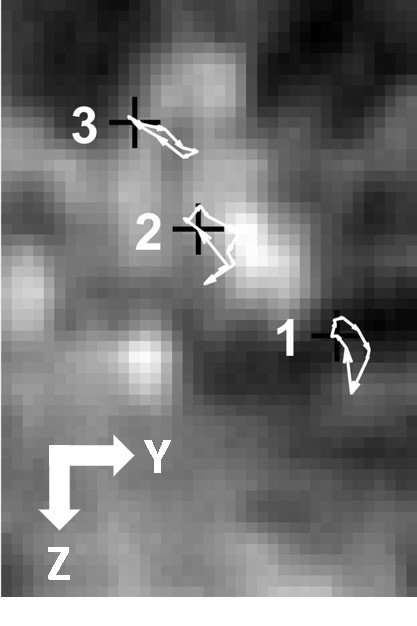} }}%
    \caption{Trajectories of the internal points, between \text{\normalsize $t=t_1$} and \text{\normalsize $t=t_{10}$}, for patient 3, calculated using the optical flow algorithm and displayed on top of the average intensity projection (AIP) of the ROI at \text{\normalsize $t=t_1$}. The position of these internal points at time \text{\normalsize $t=t_1$} is denoted by a black cross marker.}
    \label{fig:traj}
\end{figure}

\begin{figure}%
    \centering
    \includegraphics[width=0.8\columnwidth]{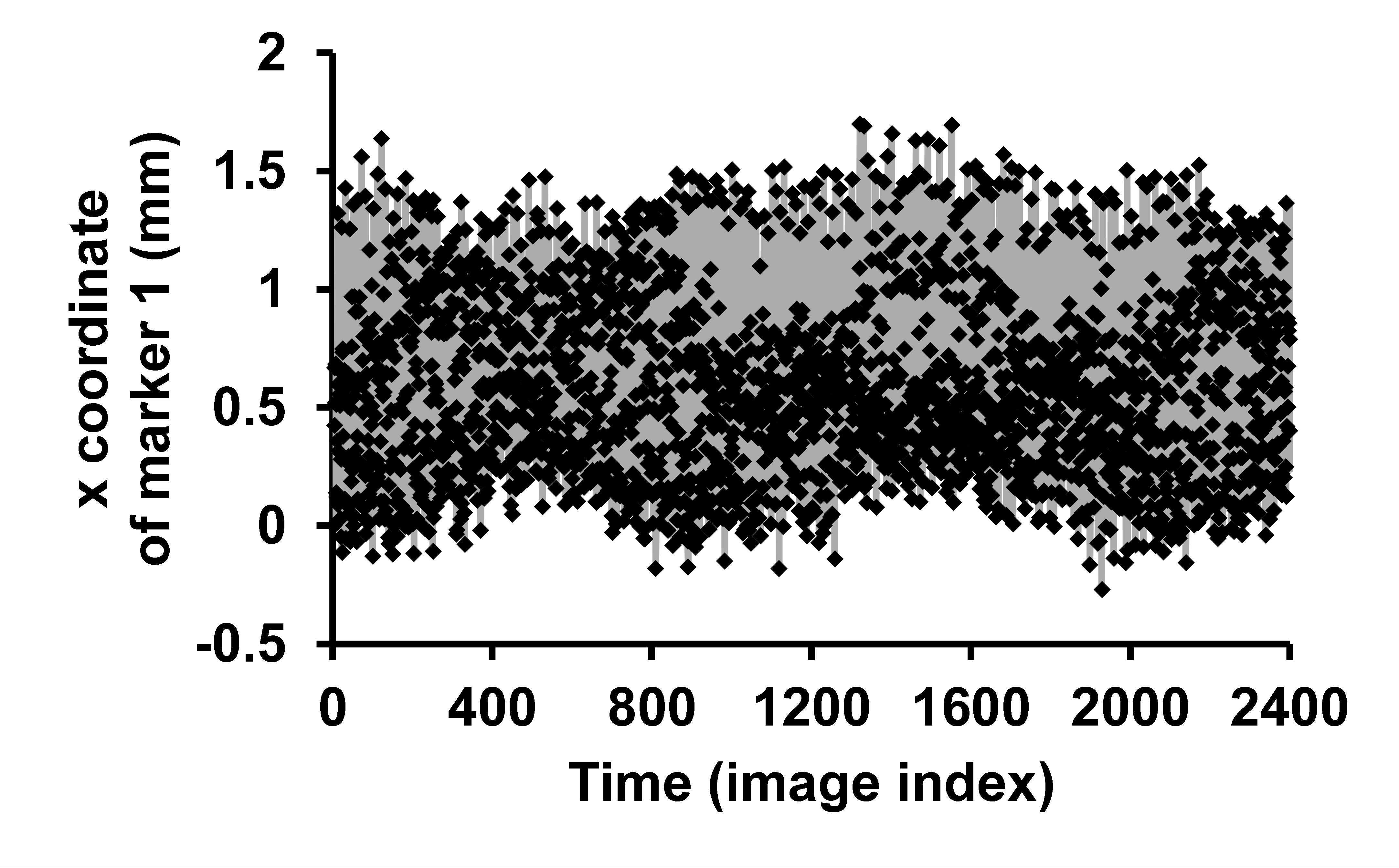}
    \includegraphics[width=0.8\columnwidth]{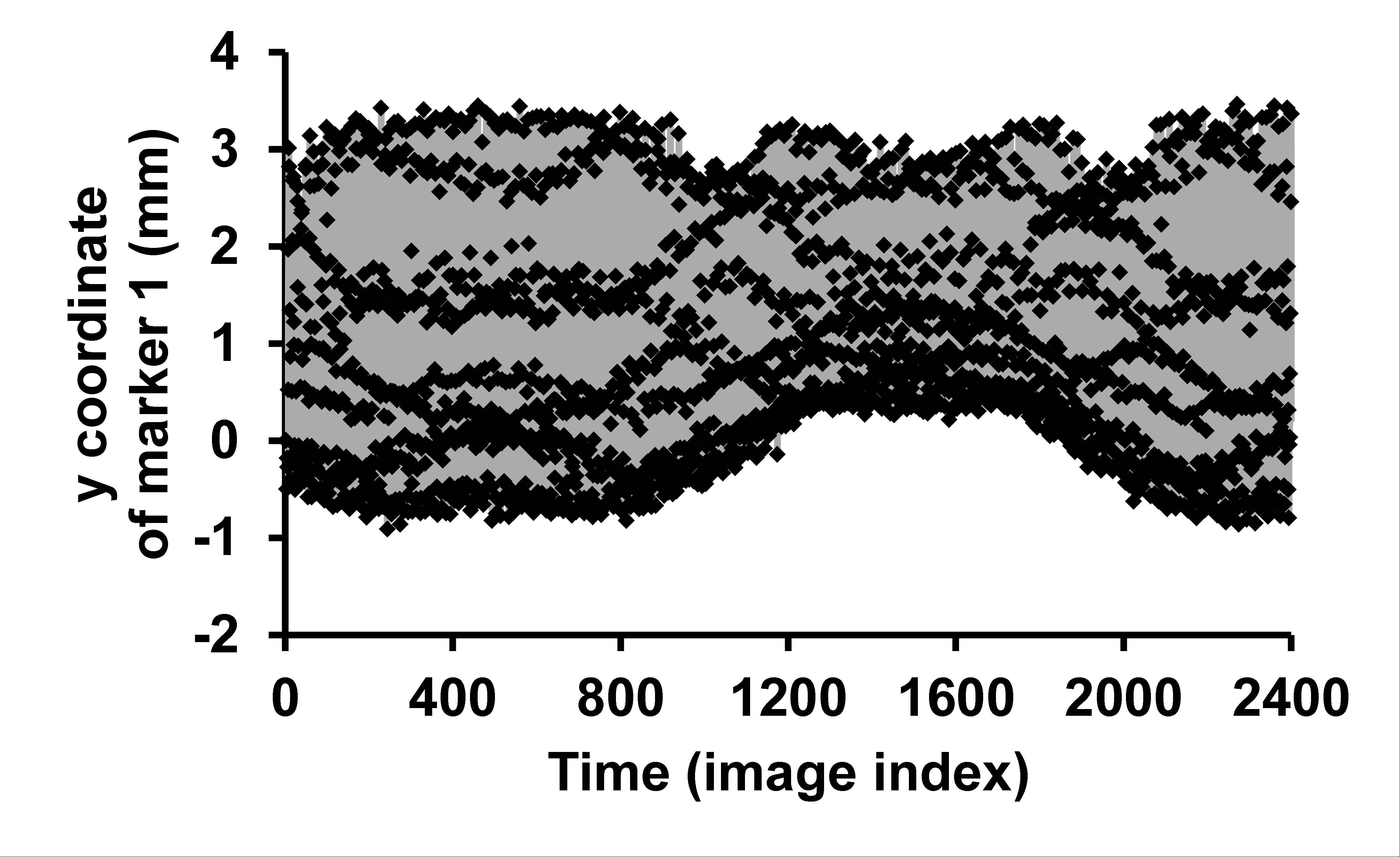}
    \includegraphics[width=0.8\columnwidth]{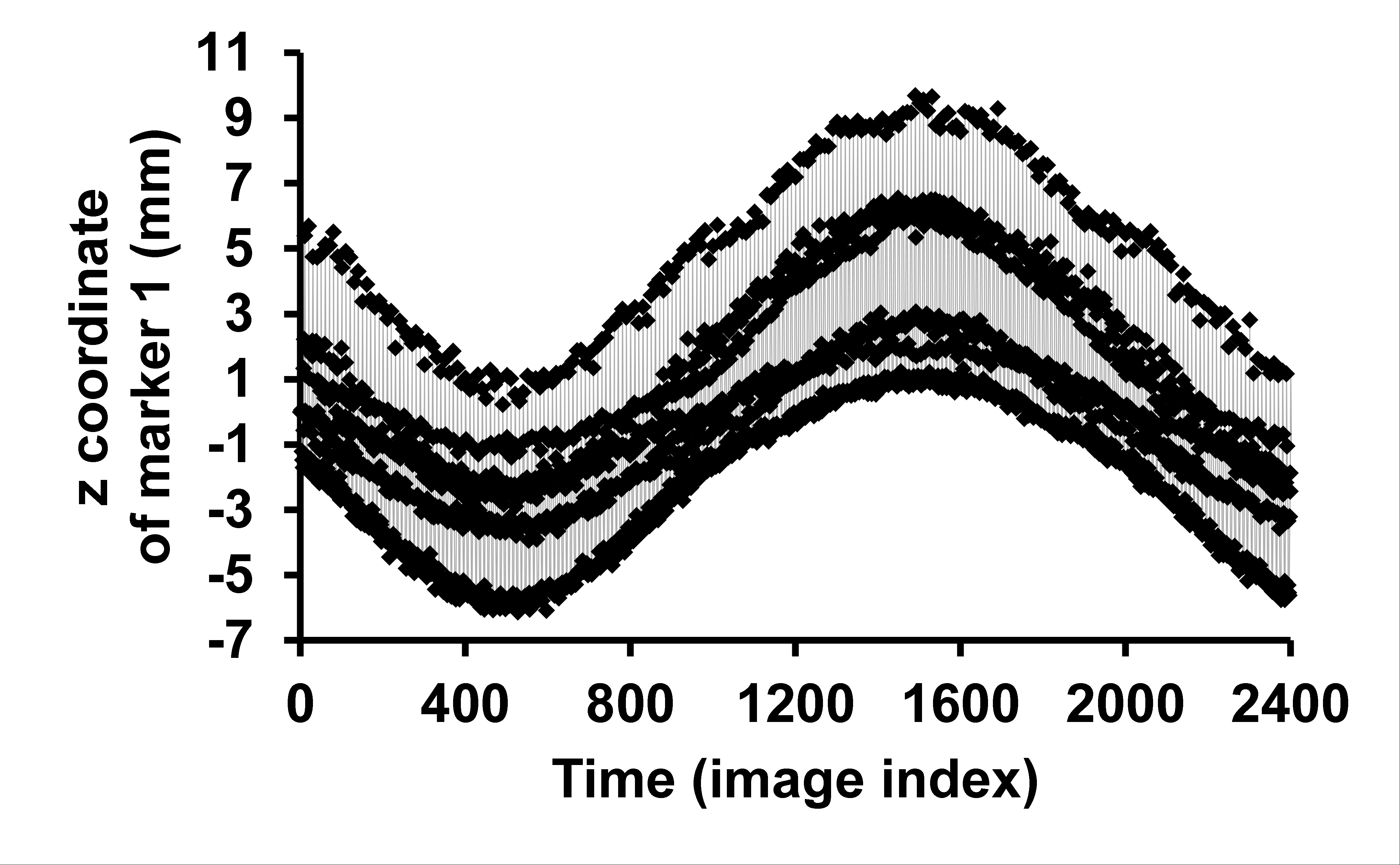}      
    \caption{Motion of marker 1 of patient 3. The dot at time t corresponds to the signal sampled at time t. The axes are the same as in Fig. \ref{fig:traj}. The data is divided into 3 sets, namely the training set, between \text{\normalsize $t=t_{1}$} and \text{\normalsize $t=t_{2000}$}, the cross-validation set, between \text{\normalsize $t=t_{2001}$} and \text{\normalsize $t=t_{2200}$}, and the test set, between \text{\normalsize $t=t_{2201}$} and \text{\normalsize $t=t_{2400}$}.}%
    \label{fig:patient3_extended_time_series}%
\end{figure}

\begin{figure*}
    \centering
    \includegraphics[width=5.4cm, height=3.7cm]{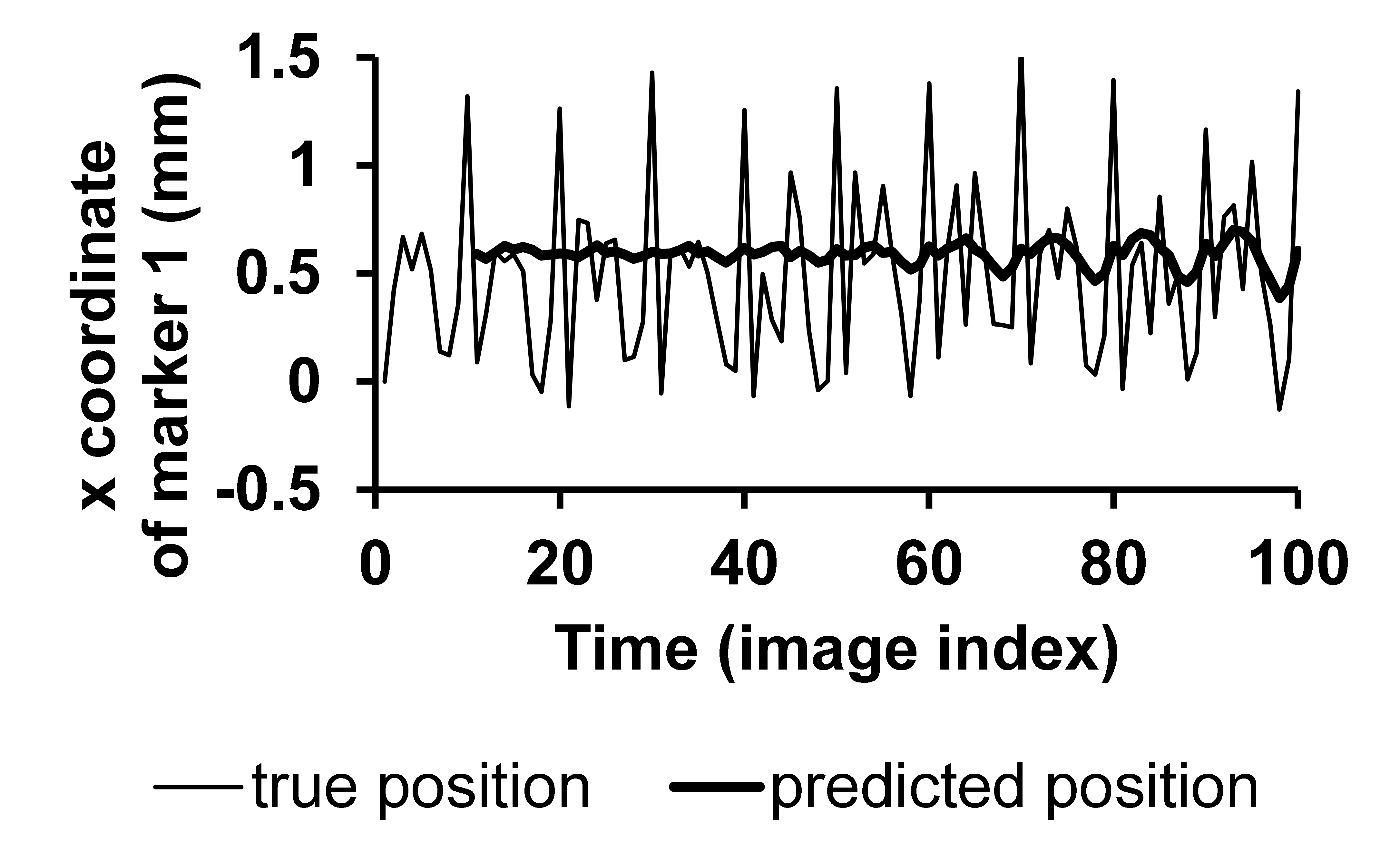}
    \quad
    \includegraphics[width=5.4cm, height=3.7cm]{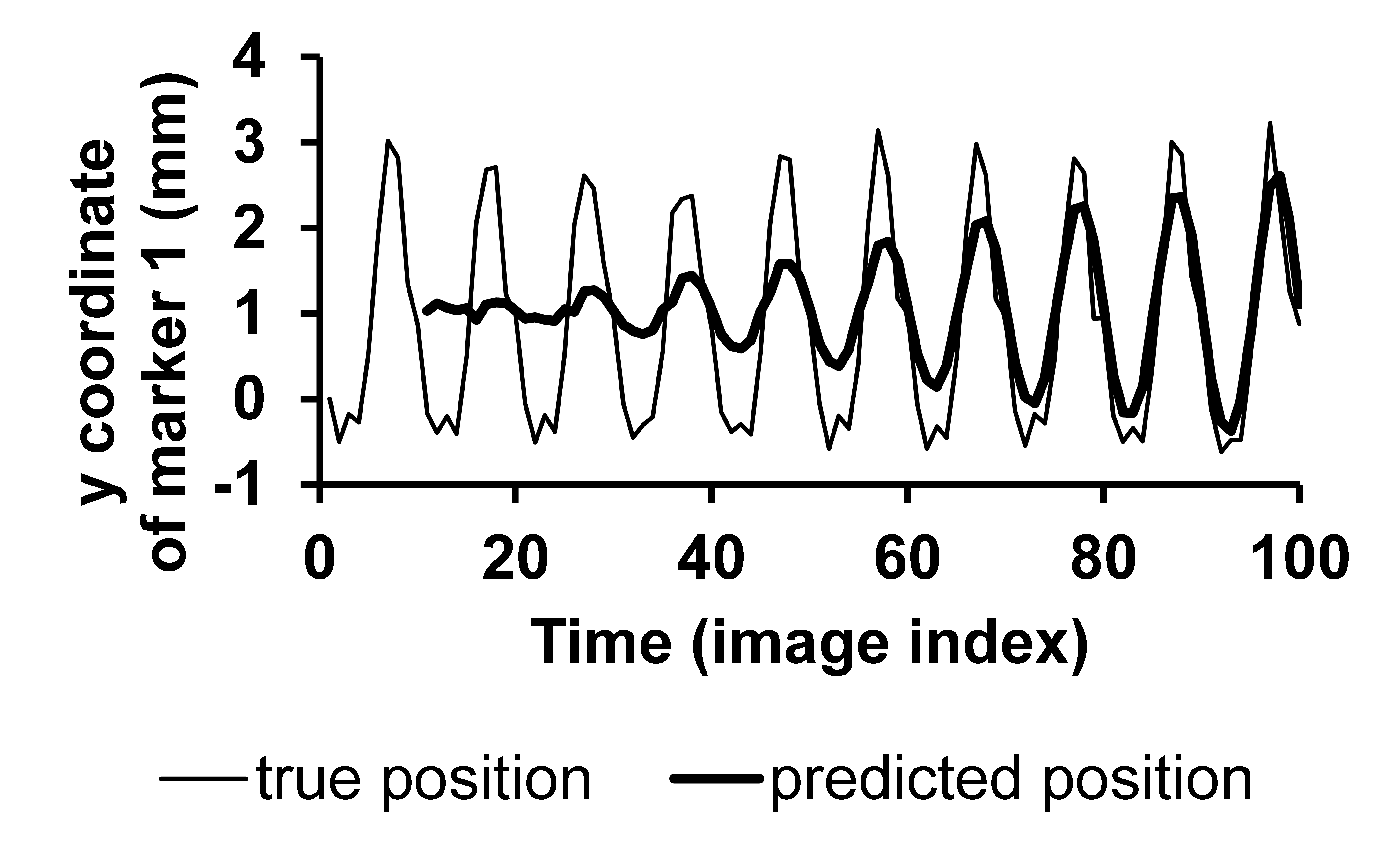}
	\quad    
    \includegraphics[width=5.4cm, height=3.7cm]{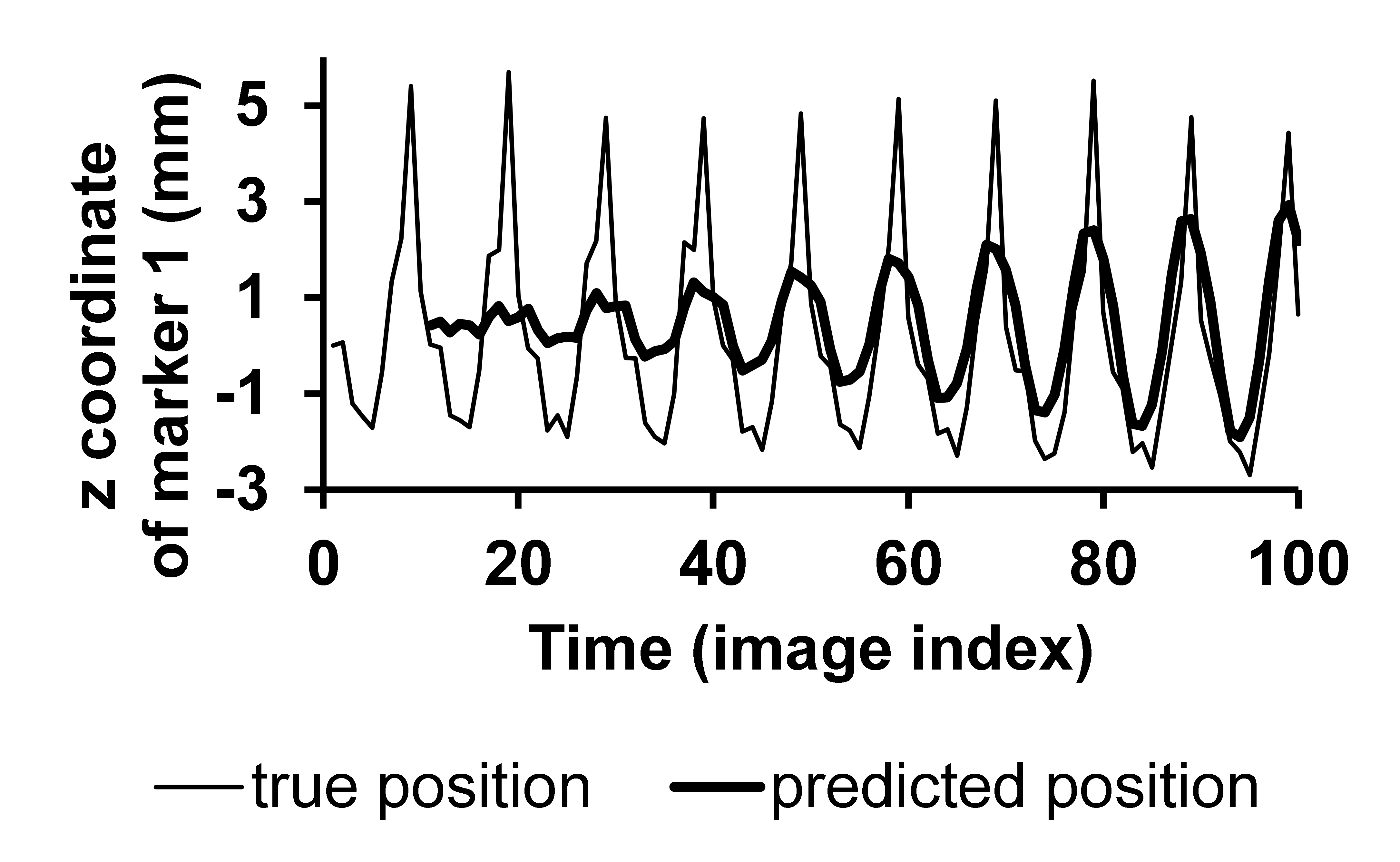}
    \quad
    \includegraphics[width=5.4cm, height=3.7cm]{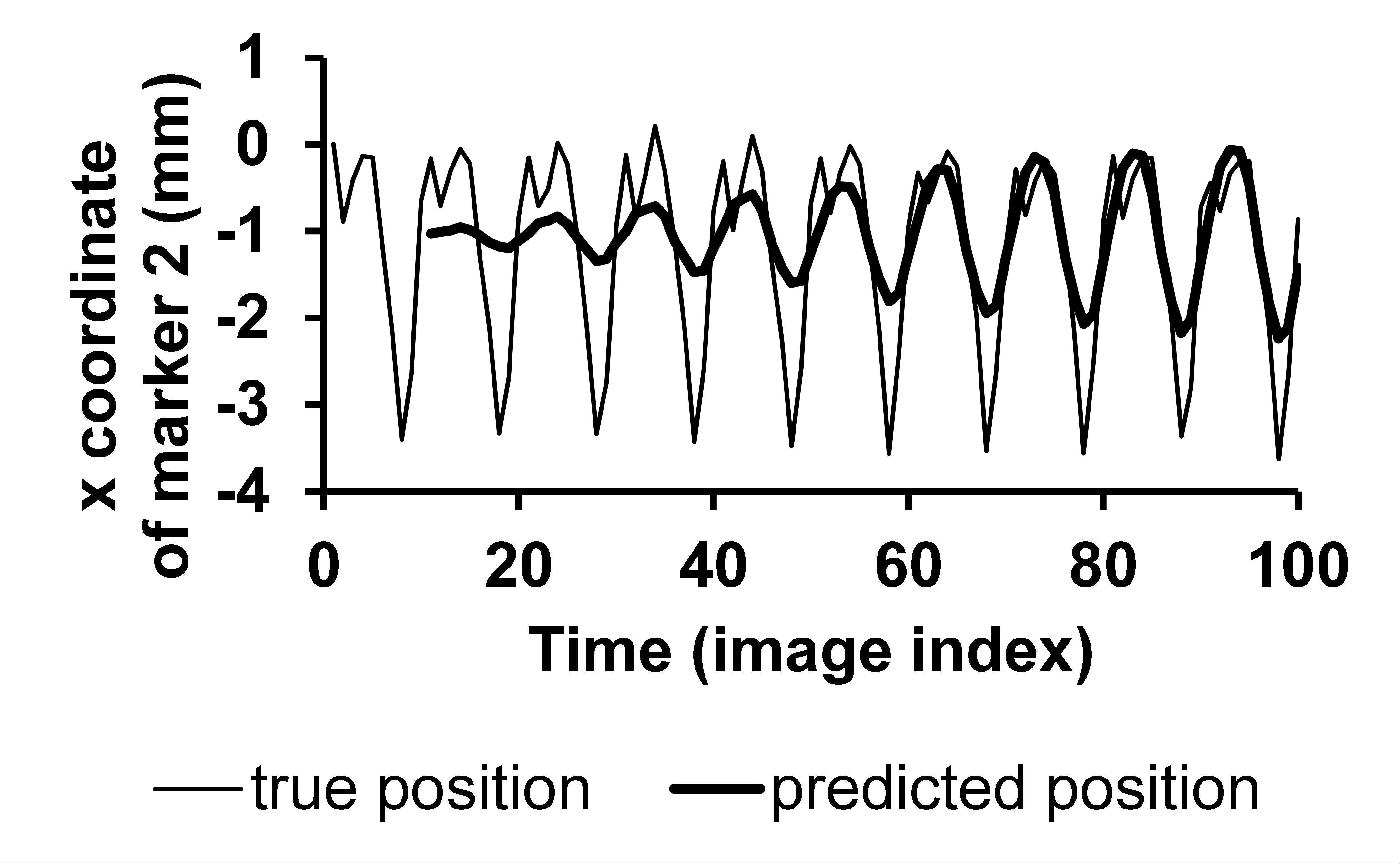}
    \quad
    \includegraphics[width=5.4cm, height=3.7cm]{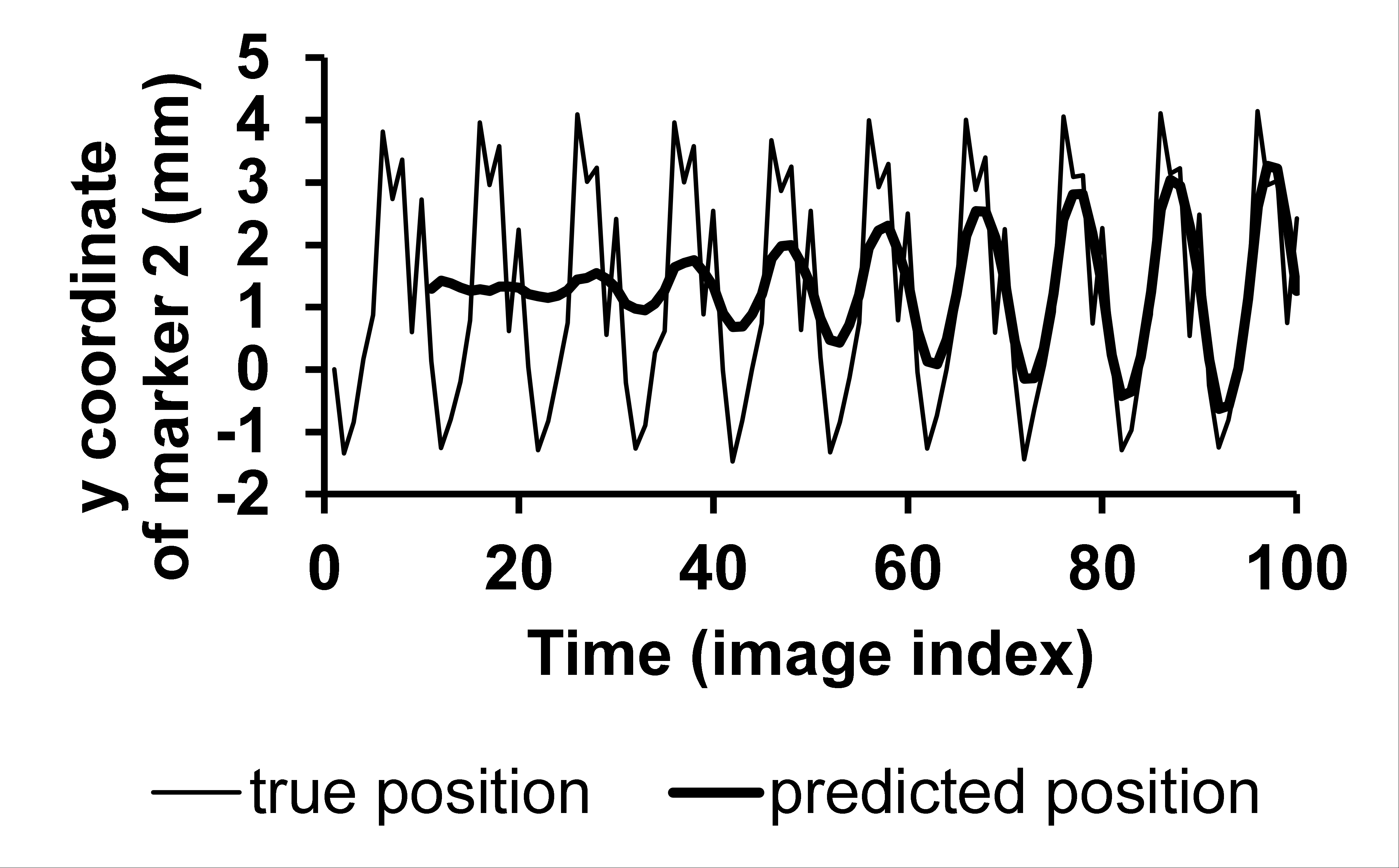}
	\quad   
    \includegraphics[width=5.4cm, height=3.7cm]{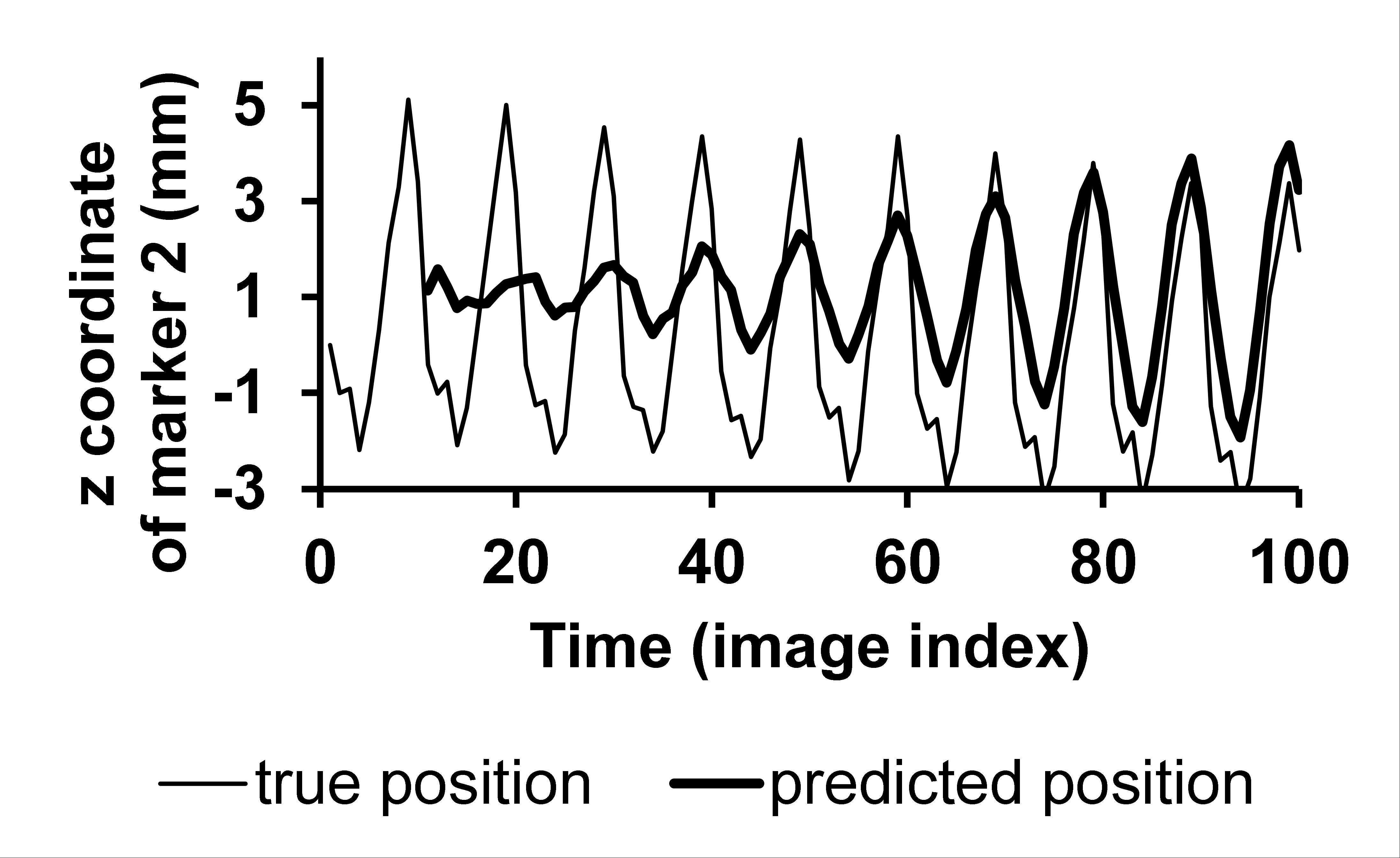}
    \quad
    \includegraphics[width=5.4cm, height=3.7cm]{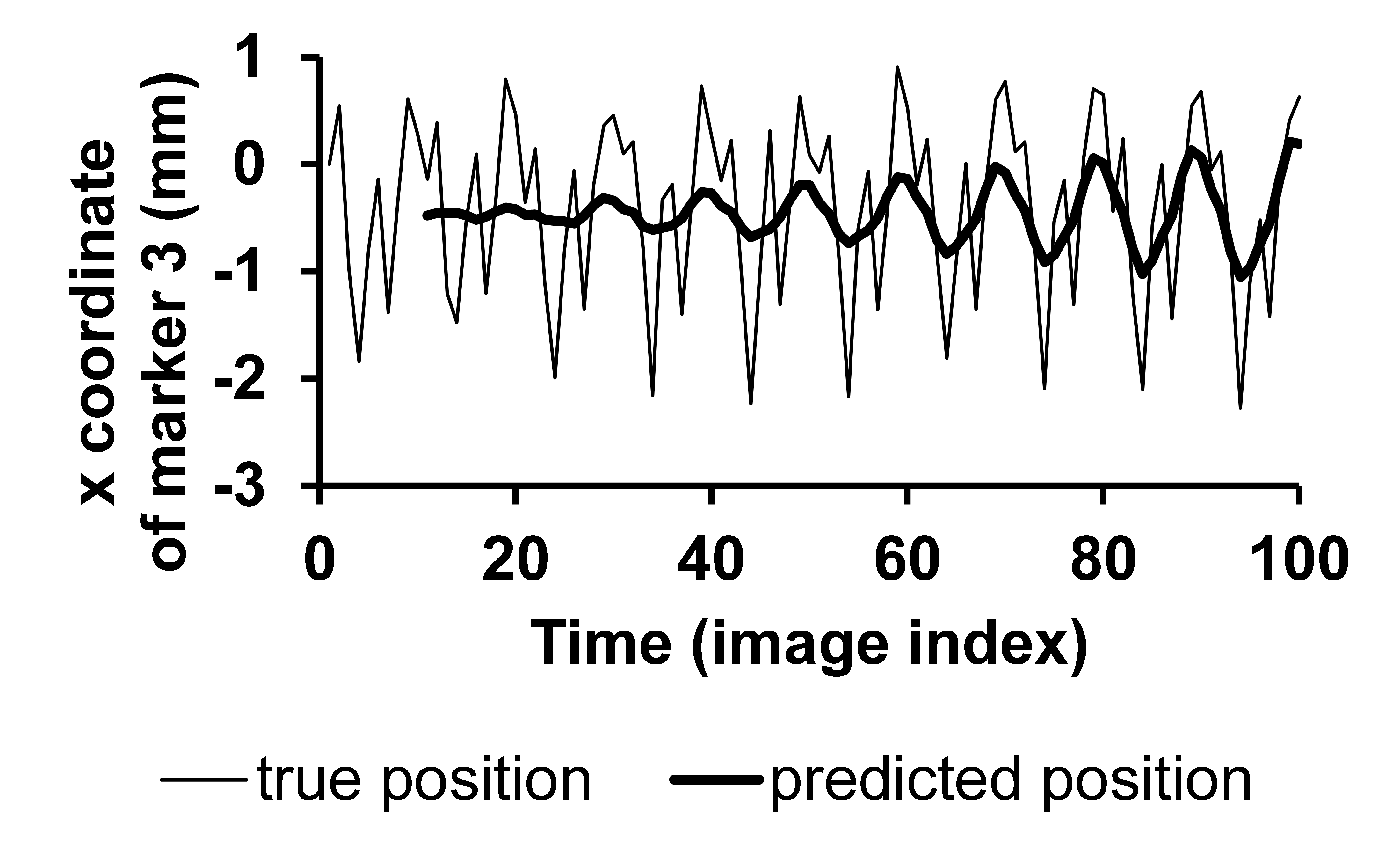}
    \quad
    \includegraphics[width=5.4cm, height=3.7cm]{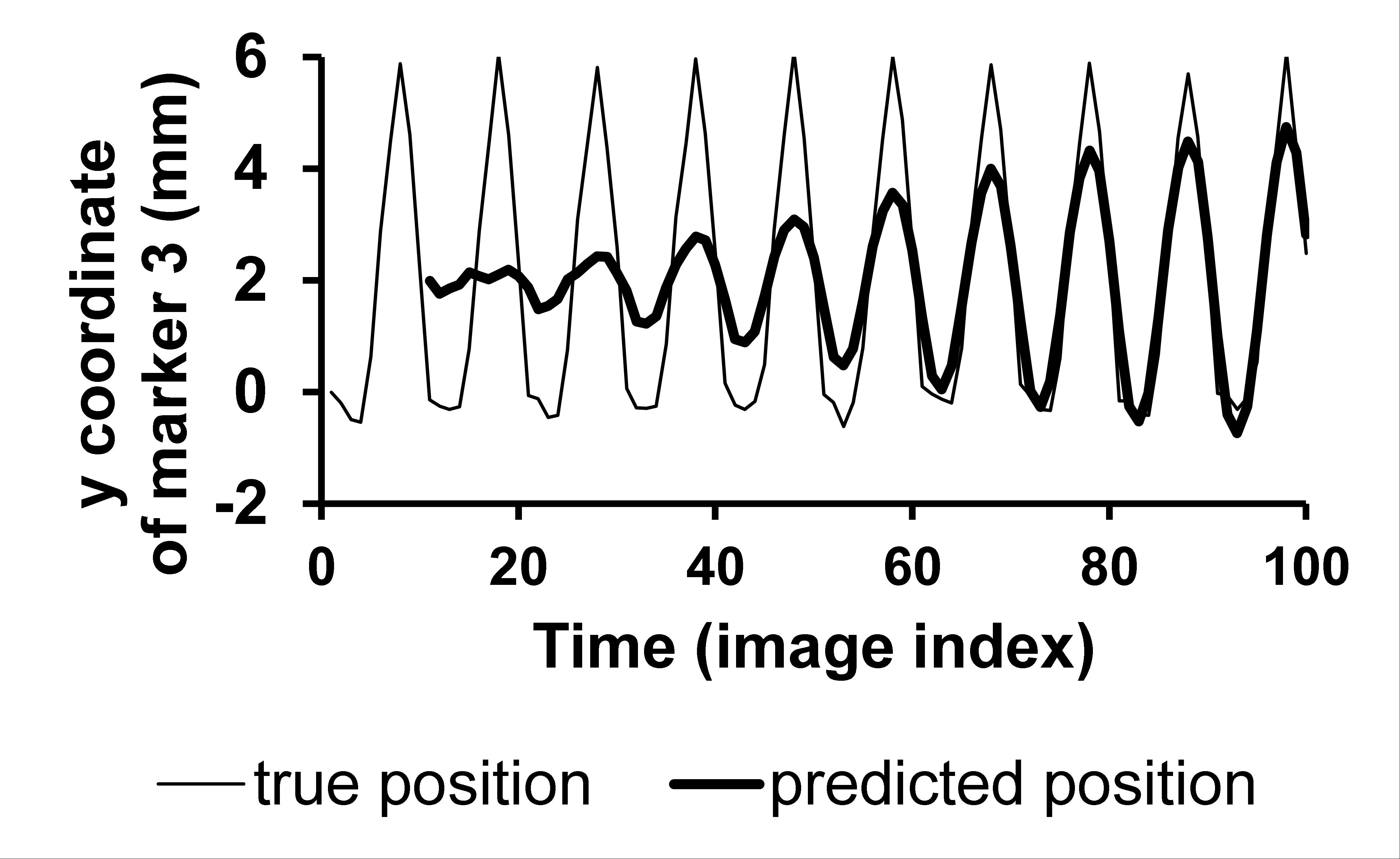}
	\quad
    \includegraphics[width=5.4cm, height=3.7cm]{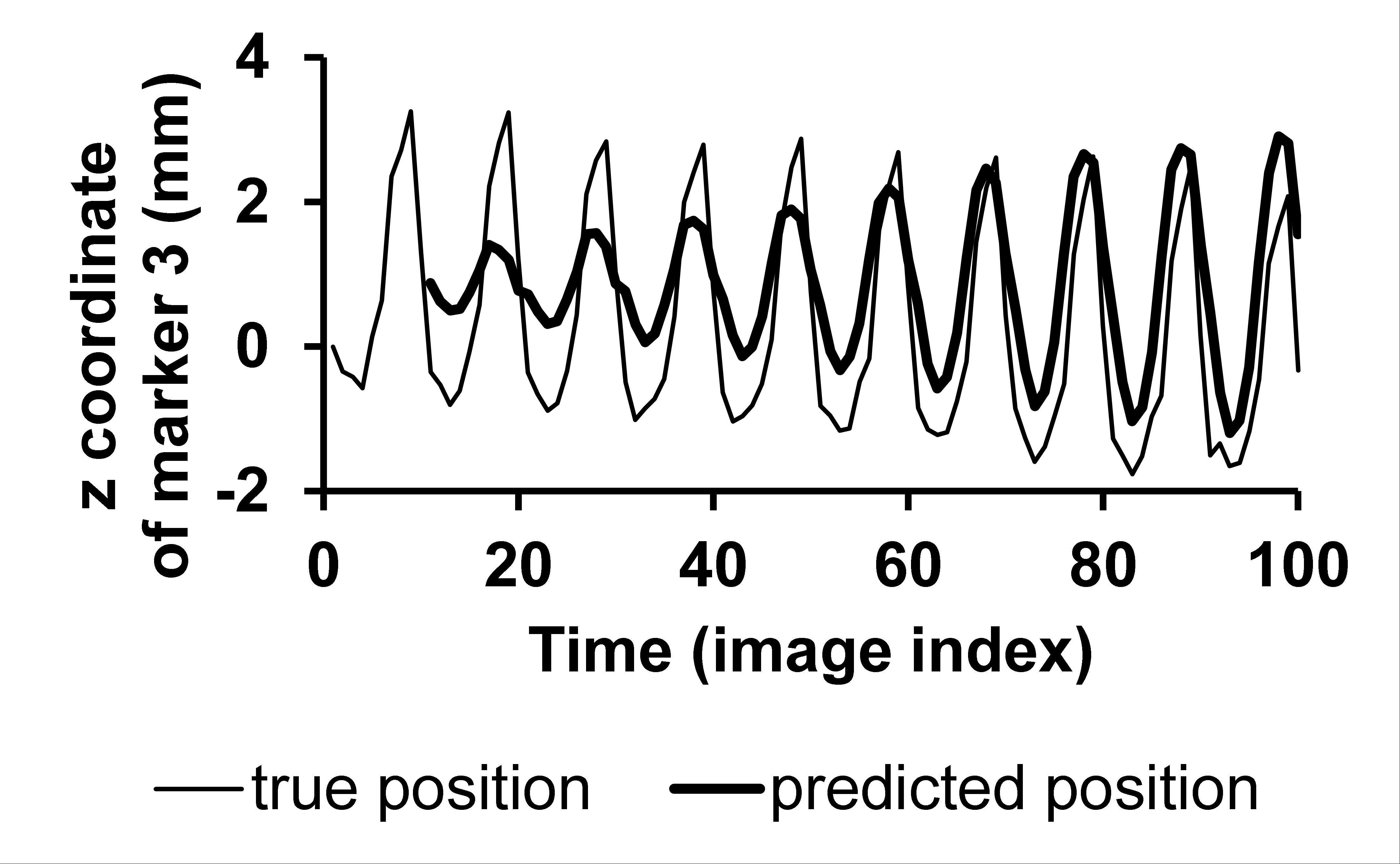}       
    \caption{RNN training for predicting the position of the markers of patient 3, displayed between \text{\normalsize $t=t_1$} and \text{\normalsize $t=t_{100}$}.  The axes are the same as in Fig. \ref{fig:traj}.}%
	\label{fig:pred_t1_t100}%
\end{figure*}

The deformation vectors in the lungs mainly point downwards during inspiration and upwards during expiration (Fig. \ref{fig:OFvectors} and Appendix \ref{appendix:DVF_display}). The trajectories of the selected points of each patient also reflect the up and down motion of the lung structures (Fig. \ref{fig:traj} and Appendix \ref{appendix:traj_display}). These points move predominantly along the z-direction (spine axis) but other directions can be non-negligible. Marker 3 of patient 3 is the only marker for which the motion in the z-direction is not the most significant. Indeed, its motion amplitude during one breathing cycle in the y-direction (dorsoventral direction) is approximately equal to 6.5mm, whereas it is approximately equal to 3.5mm along the z-direction (Fig. \ref{fig:pred_t1_t100}). The amplitude of the motion of each marker between $t_1$ and $t_{2400}$ is reported in Table \ref{table:motion amplitude}.

The optical flow algorithm optimized on the first breathing cycle (10 images) captured relatively well the z component of the motion, including the artificial drift, on the entire sequence of 2,400 images, despite the added noise (Fig. \ref{fig:patient3_extended_time_series} and Appendix \ref{appendix:entire_motion_dataset_patient3}).  

\begin{table}
\begin{tabular*}{\tblwidth}{@{} LLLLL@{} }
\toprule
Patient number & 1 & 2 & 3 & 4 \\
\midrule
Marker 1 & 19.4 & 19.7 & 15.8 & 13.2 \\
Marker 2 &  22.7 & 17.5 & 17.7 & 13.7 \\
Marker 3 &  21.9 & 16.7 & 12.0 & 13.6 \\
\bottomrule
\end{tabular*}
\caption{Amplitude of the motion of the selected internal points, in mm, between \text{\normalsize $t=t_1$} and \text{\normalsize $t=t_{2400}$}}
\label{table:motion amplitude}
\end{table}

\begin{figure*}
    \centering
    \subfloat[Prediction error as a function of the gradient threshold]{\includegraphics[width=5cm]{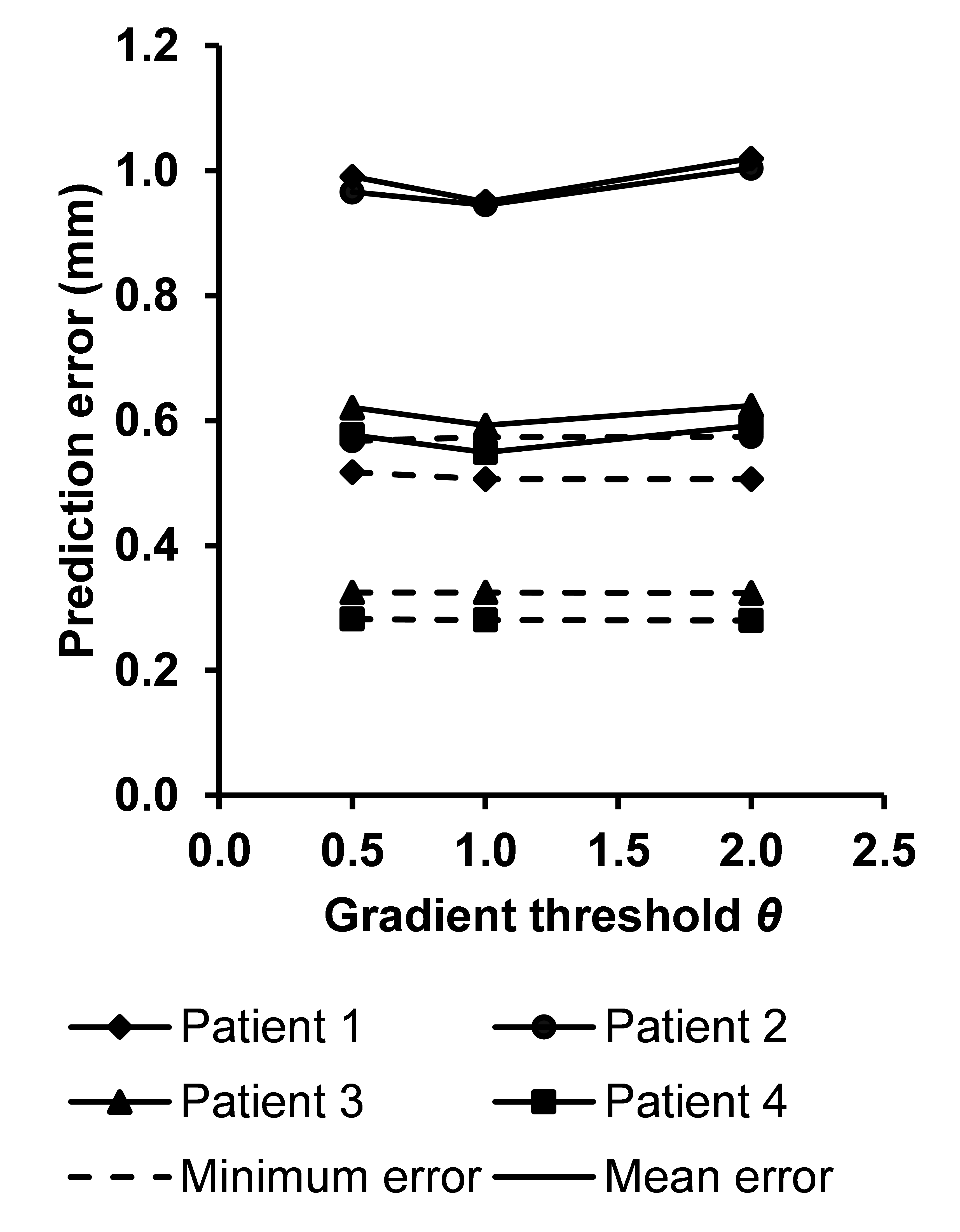} \label{RNN_pred_error_grad_thresh}}
    \quad
    \subfloat[Prediction error as a function of the learning rate]{\includegraphics[width=5cm]{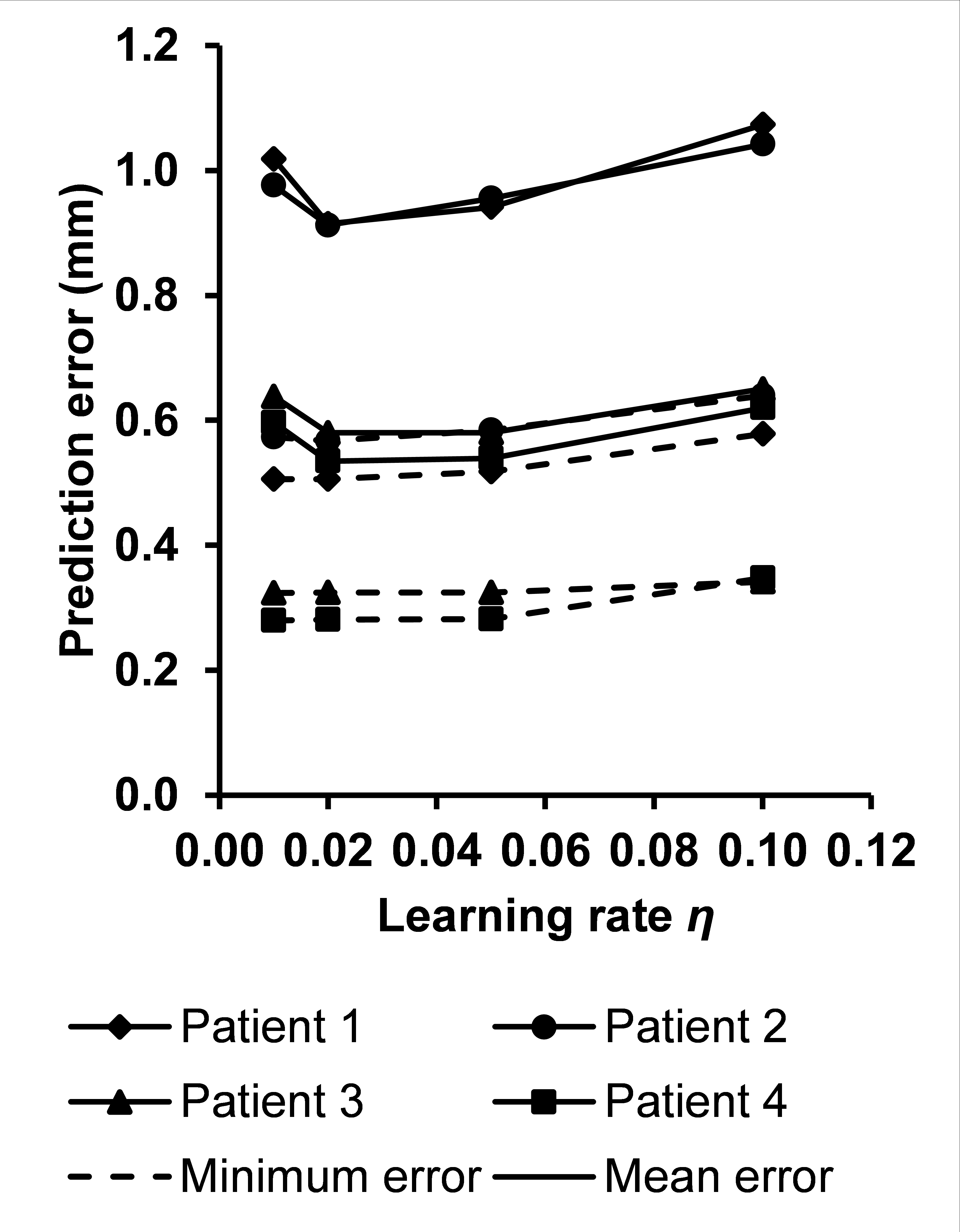} \label{RNN_pred_error_learn_rate}}
    \quad
    \subfloat[Prediction error as a function of the standard deviation of the initial weights distribution]{\includegraphics[width=5cm]{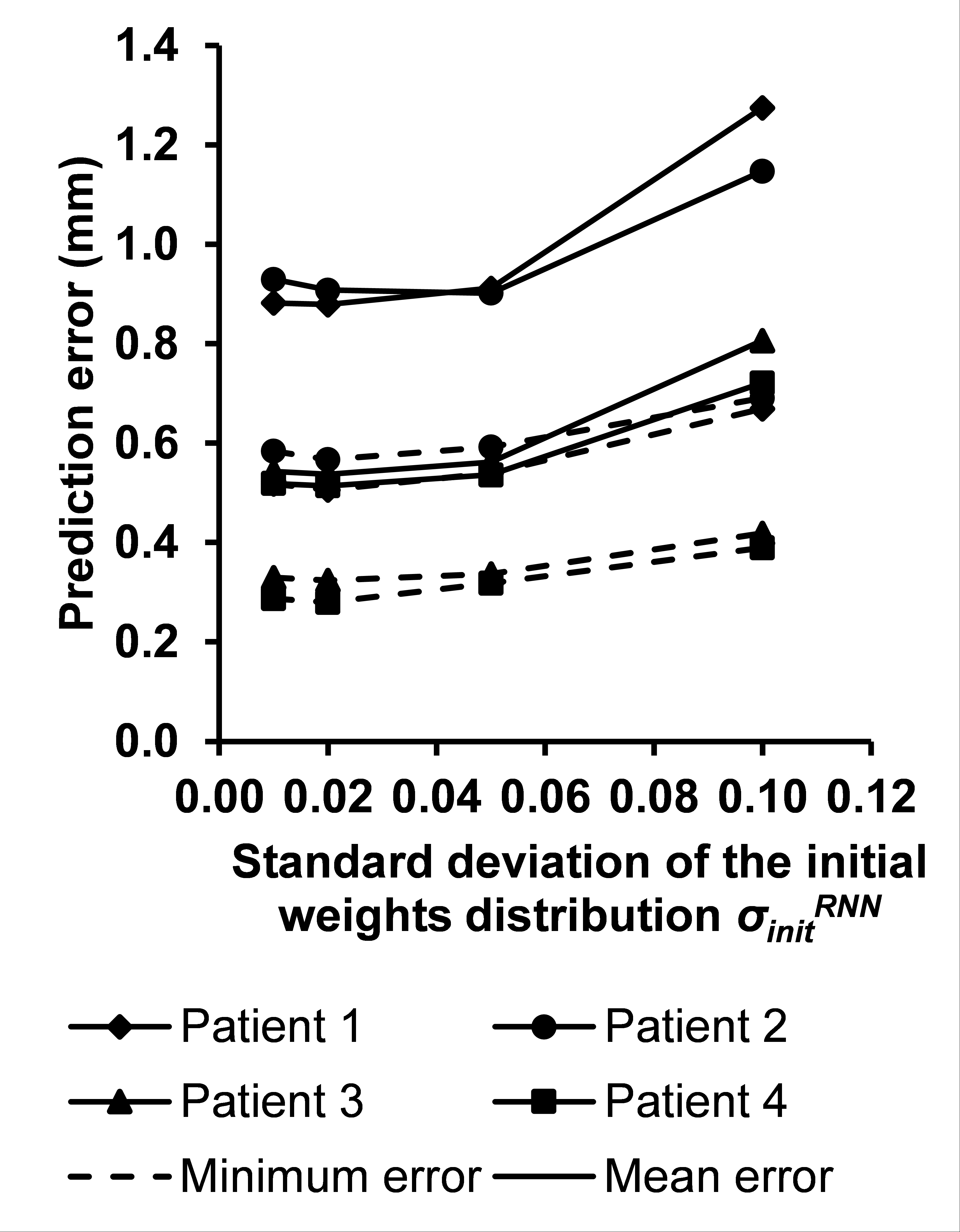} \label{RNN_pred_error_init_weights}}
    \quad
    \subfloat[Prediction error as a function of the signal history length]{\includegraphics[width=5cm]{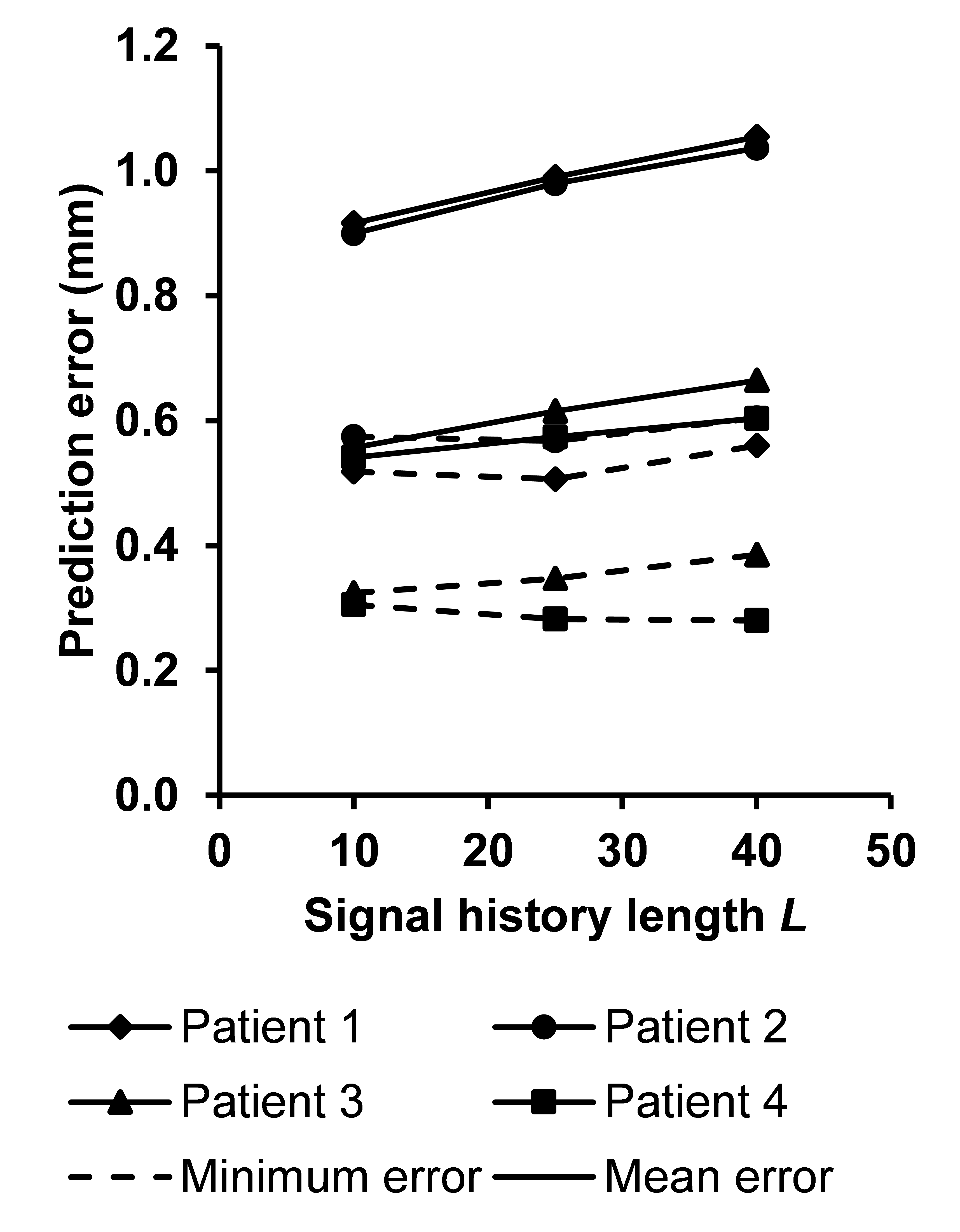} \label{RNN_pred_error_SHL}}
    \quad
    \subfloat[Prediction error as a function of the number of hidden units]{\includegraphics[width=5cm]{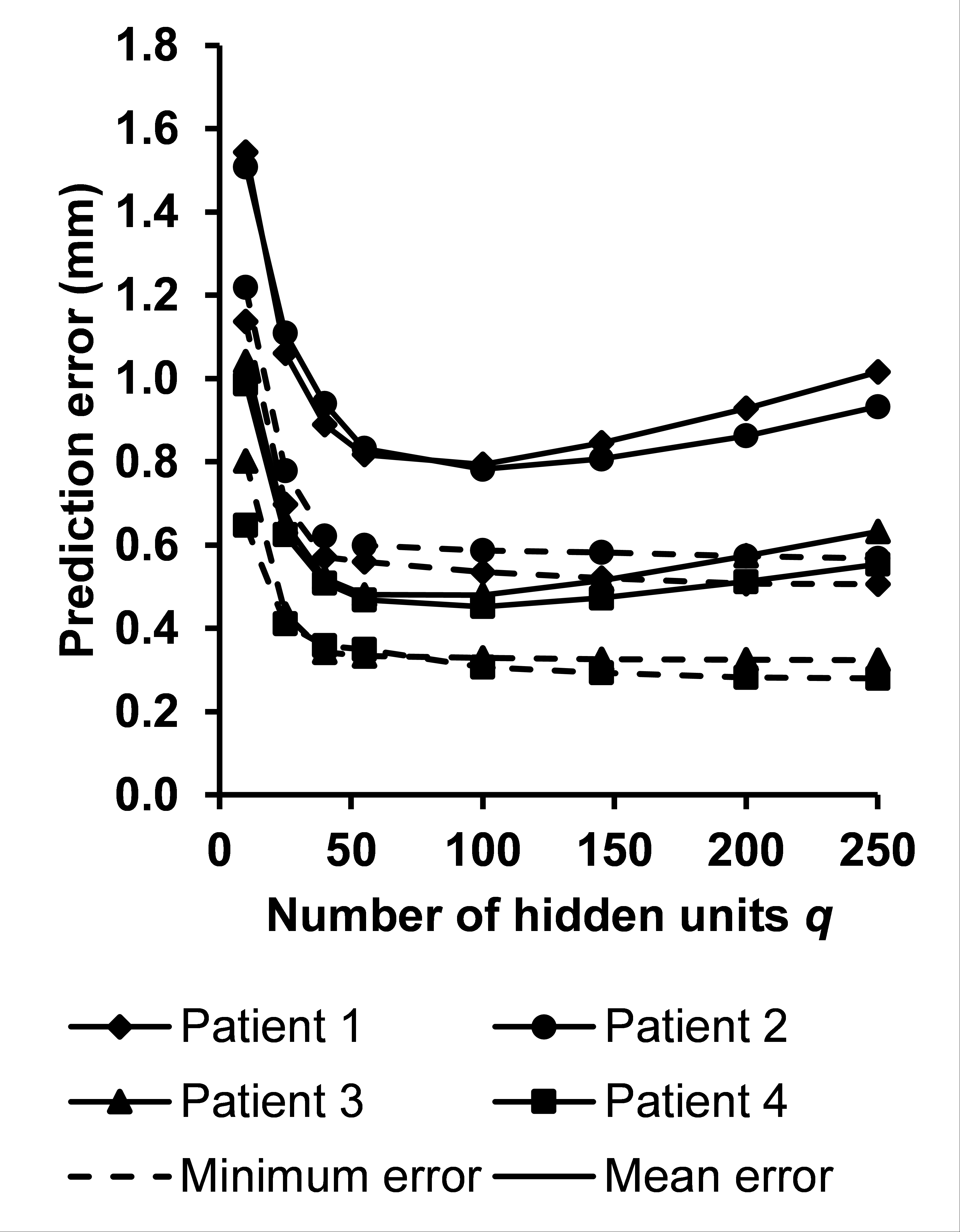} \label{RNN_pred_error_nb_hidden_units}}
    \caption{Prediction error $e_{MAE}$ calculated on the cross-validation set between $t=t_{2001}$ and $t=t_{2200}$, as a function of the RNN parameters (Eq. \ref{eq:MAE_def}). The minimum error corresponds to the minimum of $e_{MAE}$ across all parameters, and for a given graph, the mean error corresponds to the average of $e_{MAE}$ over the four parameters not studied in the graph. Each type of error is averaged over 10 runs to take into account the random initialization of the initial weights.}
    \label{fig:RNN_grid_search_pred_error}
\end{figure*}

\subsection{\normalsize Prediction of the position of internal points}

The parameters intervening in the RTRL learning algorithm have also been optimized by performing a grid search, with the following range of parameters :
\begin{enumerate}[\textbullet]
	\item gradient threshold $\theta \in \{0.5, 1.0, 2.0\}$
	\item learning rate $\eta \in \{0.01, 0.02, 0.05, 0.10\}$
	\item weights std. deviation $\sigma_{init}^{RNN} \in \{0.01, 0.02, 0.05, 0.10\}$
	\item signal history length $L \in \{10, 25, 40\}$
	\item \begin{flushleft}	nb. of hidden units $q \in \{10, 25, 40, 55, 100, 145, 200, 250\}$ \end{flushleft}
\end{enumerate}

Fig. \ref{fig:RNN_grid_search_pred_error} details how the prediction mean average error (MAE) on the cross-validation set between $t_{2001}$ and $t_{2200}$, defined in Eq. \ref{eq:MAE_def}, is affected by the choice of these parameters. 
\begin{equation} \label{eq:MAE_def}
 e_{MAE} = \frac{1}{200 r} \sum_{k = 2001}^{2200} \sum_{p =1}^r
 	\Big\| \vv{M_{true}^p(t_k)M_{pred}^p(t_k)} \Big\|_2
\end{equation} 

In this equation, $M_{true}^p(t_k)$ is the 3D position of the $p^{th}$ marker at the instant $t_k$, calculated by the optical flow registration algorithm, $M_{pred}^p(t_k)$ is the predicted position of that marker at the same instant, and $\| \cdot \|_2$ refers to the euclidean norm. In order to take into consideration the random initialization of the initial synaptic weights, the MAE was averaged over 10 runs\footnotemark. Each graph in Fig. \ref{fig:RNN_grid_search_pred_error} describes the influence of one parameter and for each graph, two types of errors are displayed. The first one is the mean error: the MAE averaged over all the other parameters not studied in the graph. The second one is the minimum error: the minimum of the MAE across all the parameters. The mean prediction error as a function of $\eta$ presents a bell shape (Fig. \ref{RNN_pred_error_learn_rate}). Both errors are maximum for $\eta = 0.10$ and we found the lowest minimum errors for $\eta = 0.01$ or $\eta = 0.02$, depending on the patient index. The mean error varies with $\sigma_{init}^{RNN}$ from 1.27mm to 0.88mm for patient 1, from 1.14mm to 0.90mm for patient 2, from 0.81mm to 0.54mm for patient 3, and from 0.72mm to 0.51mm for patient 4 (Fig. \ref{RNN_pred_error_init_weights}). In other words, optimizing $\sigma_{init}^{RNN}$ led to a 28.4\% average decrease in the mean error. Both error curves are strictly convex because when the initial weights are too low, many time steps are required to grow them using the gradient descent updating rule, and when they are too high, they are difficult to control. Both errors were maximum for $\sigma_{init}^{RNN} = 0.1$ and attained their minimum for $\sigma_{init}^{RNN} = 0.02$, except the mean error of patient 2 which was minimized for $\sigma_{init}^{RNN} = 0.05$. The mean prediction error increases with the SHL, but the variation of the minimum error with the SHL was dependent on the patient index (Fig. \ref{RNN_pred_error_SHL}). Finally, the prediction error strongly decreases when $q$ increases (Fig. \ref{RNN_pred_error_nb_hidden_units}). The minimum error for $q=10$, equal to 1.14mm, 1.22mm, 0.80mm, and 0.65mm respectively for patients 1,2,3 and 4, dropped down to 0.51mm, 0.57mm, 0.32mm, and 0.28mm for $q=250$, which corresponds to a 56.3\% error decrease on average. It is thus recommended to set a high value of $q$ while keeping in mind that this may also result in a relatively high computing time. The mean error as a function of $q$ is strictly convex and increases from $q=100$ to $q=250$.

\footnotetext{\label{RNN_instability_footnote} The RNN is updated in real-time and is thus prone to numerical errors when updating the synaptic weights. The errors are actually averaged over the runs among the 10 runs for which no numerical error occurred. We performed in total 46,080 prediction runs over the four patients and all the parameters. 46 prediction runs resulted in numerical errors, which corresponds to a 0.0998\% occurrence rate of numerical errors.}

\begin{figure}%
    \centering
    \subfloat[Influence of each RNN parameter on the prediction mean error]{{\includegraphics[width=.95\columnwidth]{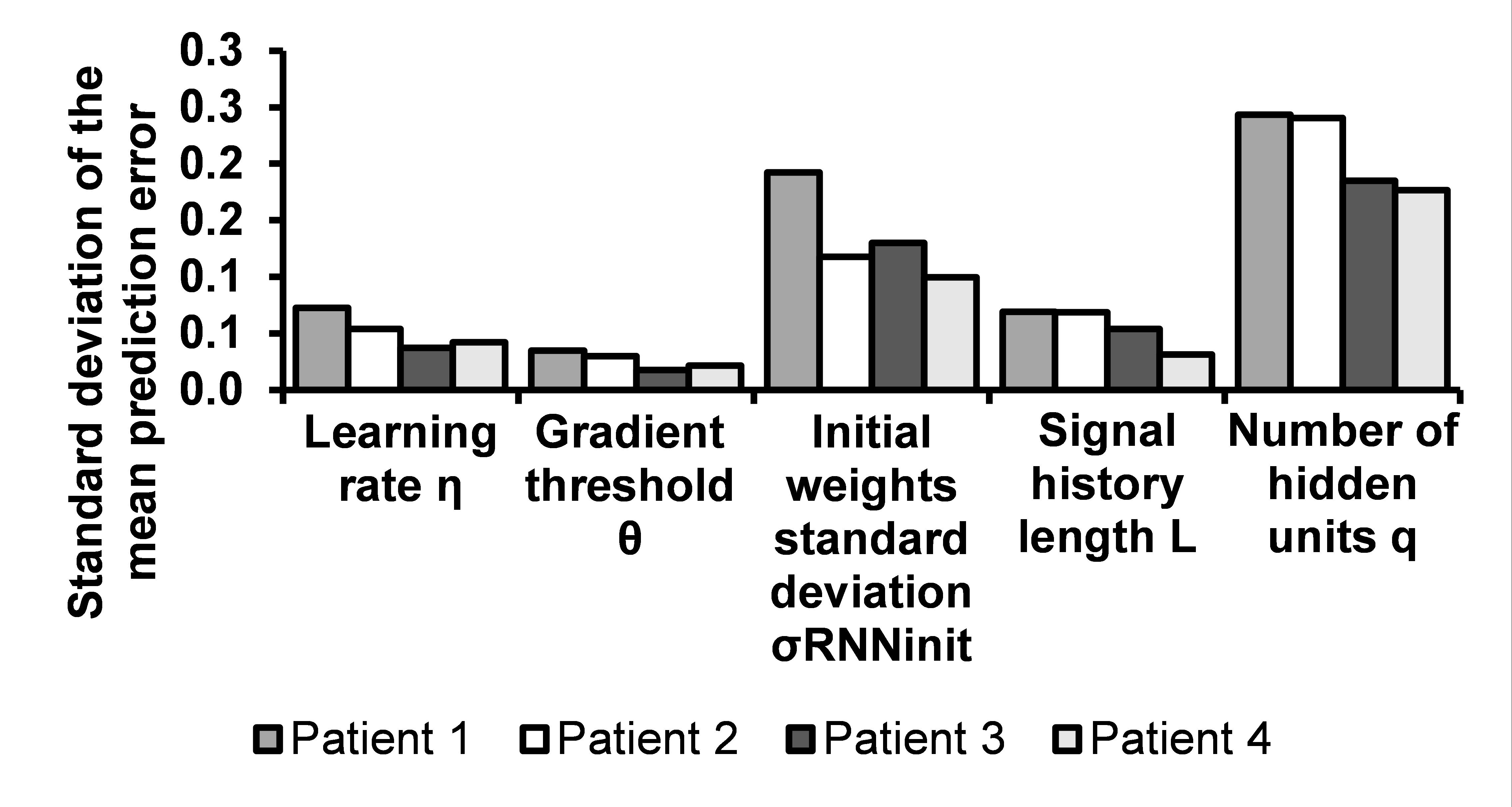} }}%
    \quad
    \subfloat[Influence of each RNN parameter on the prediction minimum error]{{\includegraphics[width=.95\columnwidth]{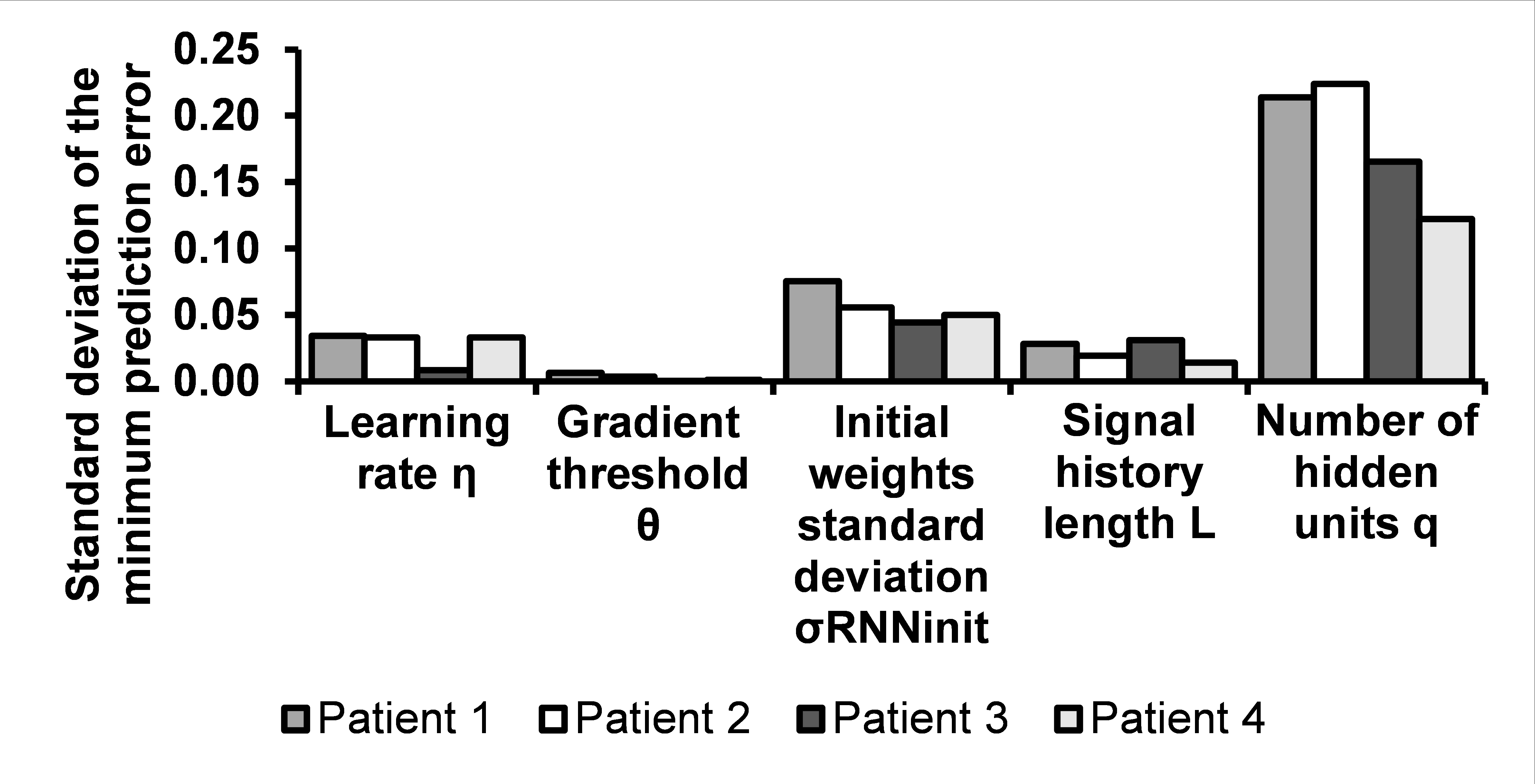} }}%
    \caption{Relative influence of each of the RNN parameters on the prediction performance on the cross-validation set}%
    \label{fig:RNN_param_influence}%
\end{figure}

\begin{figure}
	\centering
		\includegraphics[width=.8\columnwidth]{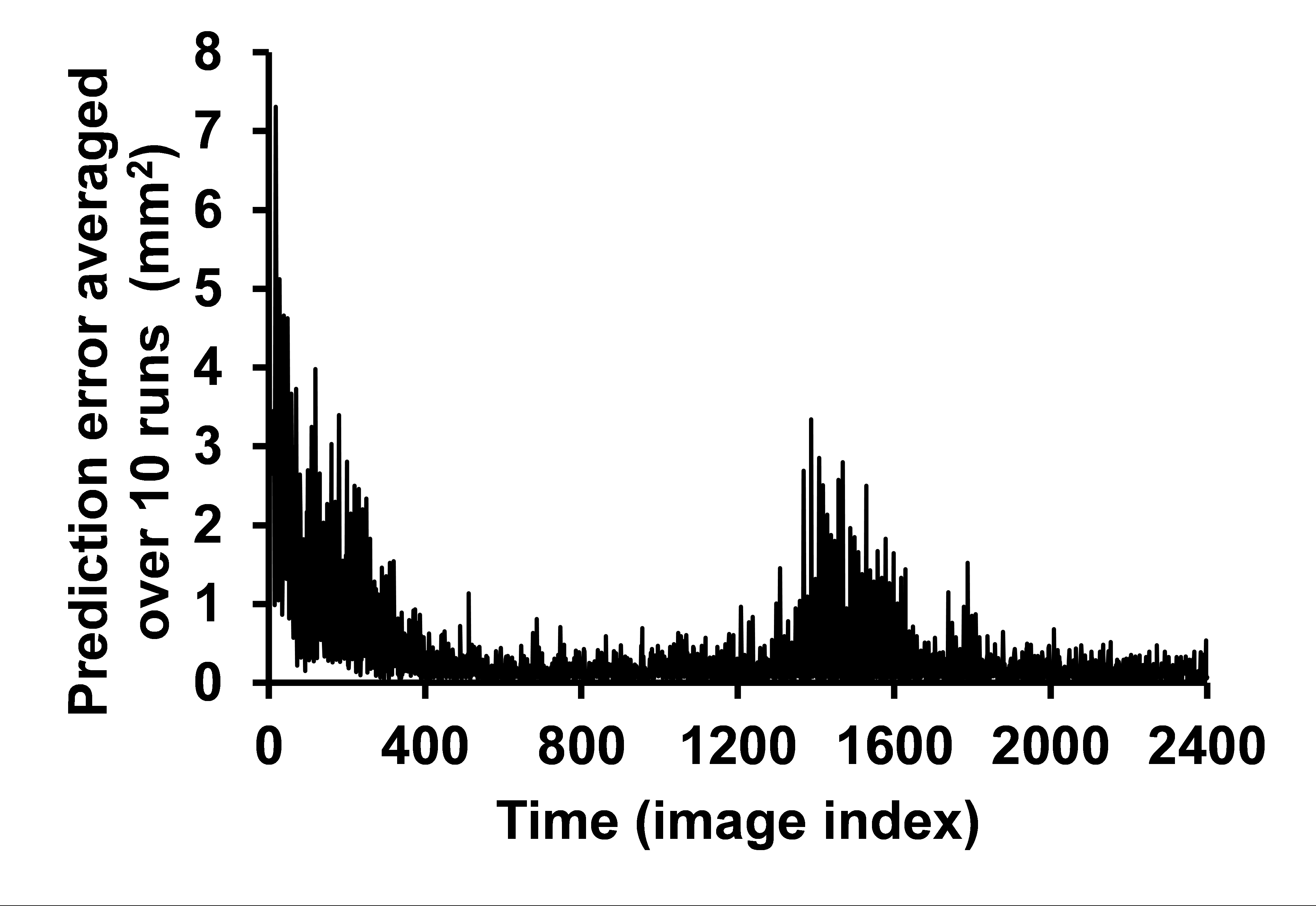}
	\caption{RNN loss function $E_n$ on the normalized data for patient 3 (cf Eq. \ref{eq:error_vector_and_function})}
	\label{fig:loss_error}
\end{figure}

The standard deviation of the mean prediction error and the minimum prediction error, relative to each parameter, is reported in Fig. \ref{fig:RNN_param_influence}. We observe that both $\sigma_{init}^{RNN}$ and $q$ are the parameters having the strongest impact on prediction accuracy. It would be interesting to evaluate the RNN trained with RTRL using less repetitive temporal data and reevaluate the importance of the SHL in that case.

\begin{table*}[width=.9\textwidth,cols=7,pos=h]
\begin{tabular*}{\tblwidth}{@{} LLLLLLL@{} }
\toprule
Error type & Prediction & Patient 1 & Patient 2 & Patient 3 & Patient 4 & Error averaged \\
 & method & & & & & over the 4 patients\\
\midrule
\midrule
Max error  & RNN with RTRL     & 1.82 $\pm$ 0.06     & 1.65 $\pm$ 0.04   & 1.16 $\pm$ 0.03   & 1.42 $\pm$ 0.06   & 1.51  \\
(in mm)    & Linear prediction & 1.96                & 2.30              & 1.65              & 1.30              & 1.80  \\
  	       & LMS               & 2.07                & 1.69              & 1.40              & 1.21              & 1.59  \\
           & No prediction     & 9.11                & 5.98              & 4.66              & 4.60              & 6.09  \\
\midrule
RMSE       & RNN with RTRL     & 0.529 $\pm$ 0.005   & 0.585 $\pm$ 0.003 & 0.338 $\pm$ 0.002 & 0.324 $\pm$ 0.002 & 0.444 \\
(in mm)    & Linear prediction & 0.512               & 0.610             & 0.333             & 0.341             & 0.449 \\
           & LMS               & 0.595               & 0.661             & 0.360             & 0.344             & 0.490 \\
           & No prediction     & 4.29                & 3.23              & 2.25              & 2.08              & 2.96  \\
\midrule
nRMSE      & RNN with RTRL     & 0.0829 $\pm$ 0.0007 & 0.118 $\pm$ 0.001 & 0.121 $\pm$ 0.001 & 0.109 $\pm$ 0.001 & 0.108 \\
(no unit)  & Linear regression & 0.080               & 0.124             & 0.121             & 0.115             & 0.110 \\
           & LMS               & 0.0932              & 0.133             & 0.129             & 0.116             & 0.118 \\
           & No prediction     & 0.671               & 0.651             & 0.807             & 0.701             & 0.708 \\
\midrule
Jitter     & RNN with RTRL     & 3.72              & 2.83              & 1.96              & 1.86              & 2.59  \\
(in mm)    & Linear regression & 3.69              & 2.86              & 1.98              & 1.82              & 2.59  \\
           & LMS               & 3.74              & 2.91              & 2.01              & 1.88              & 2.63  \\
           & No prediction     & 3.76              & 2.96              & 2.03              & 1.87              & 2.66  \\
\bottomrule
\end{tabular*}
\caption{RNN prediction performance computed on the test data, between \text{\normalsize $t=t_{2201}$} and \text{\normalsize $t=t_{2400}$}, in comparison with other methods. Each cell indicates the maximum error, RMSE, nRMSE or jitter associated with the prediction of the position of the markers (Eq. \ref{eq:max_pred_error_def}, Eq. \ref{eq:pred_RMSE_def}, Eq. \ref{eq:pred_nRMSE_def} and Eq. \ref{eq:jitter_def}). The error and 95\% mean confidence interval mentioned for the RNN are calculated using 10 random initializations and assuming that the error distribution is Gaussian (Eq. \ref{eq:max_err_conf_interval} and Eq. \ref{eq:RMS_err_conf_interval}). The confidence half range associated with the jitter measure has not been provided as the former is low compared with the latter (order of magnitude $10^{-3}$ mm).}
\label{table:RNN_lin_reg_comparison}
\end{table*}

\begin{table}
\begin{tabular*}{\tblwidth}{@{} LL@{} }
\toprule
Prediction algorithm & Calculation time per time step (in ms)\\
\midrule
RNN with RTRL     & 119.1 \\
Linear regression & 0.0052\\
LMS               & 0.318 \\                   
\bottomrule
\end{tabular*}
\caption{Time performance of the RNN in comparison with other prediction methods (Dell Intel Core i9-9900K 3.60GHz CPU NVidia GeForce RTX 2080 SUPER GPU 32Gb RAM with Matlab).}
\label{table:RNNtime_perf}
\end{table}

\begin{figure*}%
    \centering
    \includegraphics[width=5.4cm, height=3.7cm]{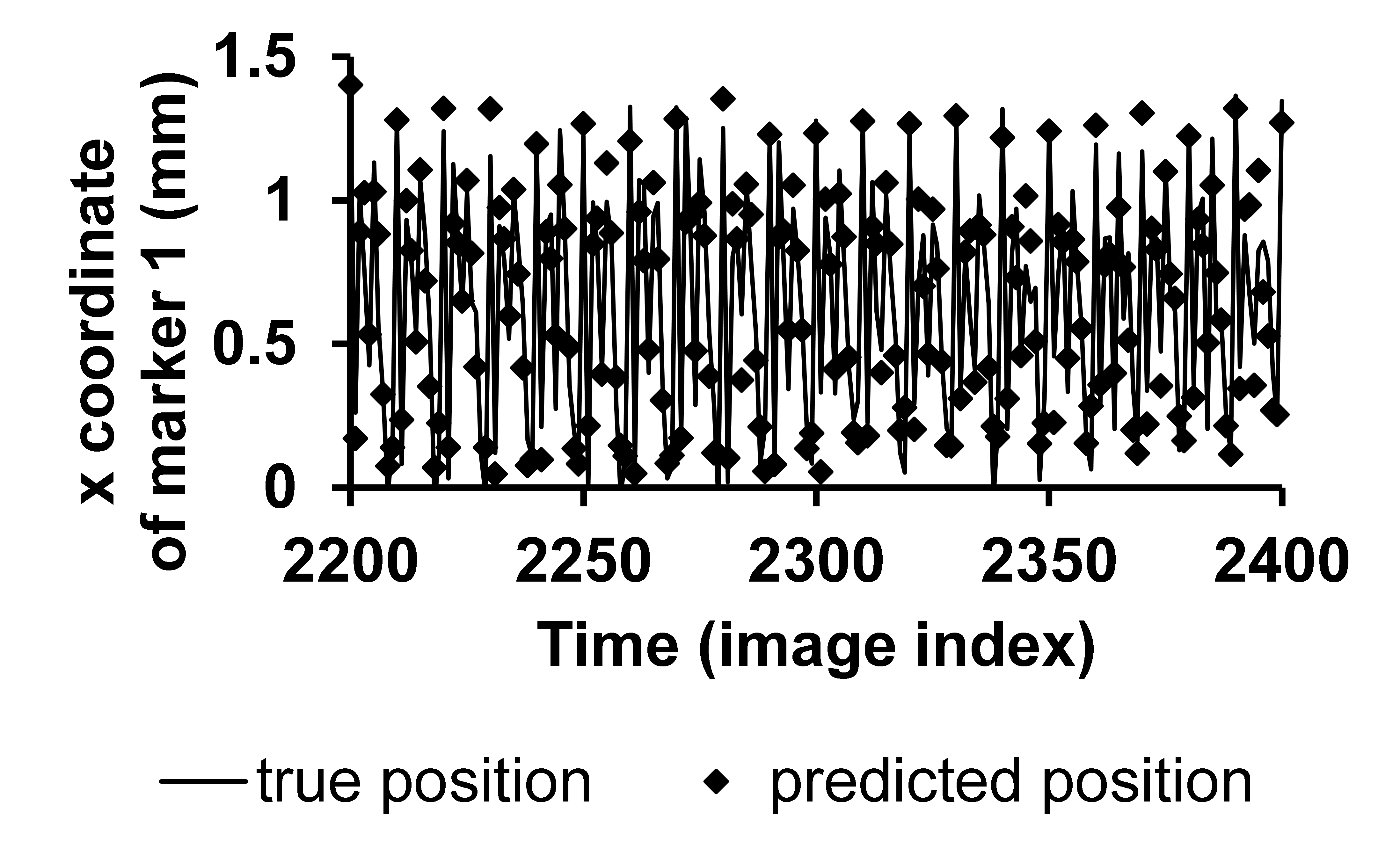}
    \includegraphics[width=5.4cm, height=3.7cm]{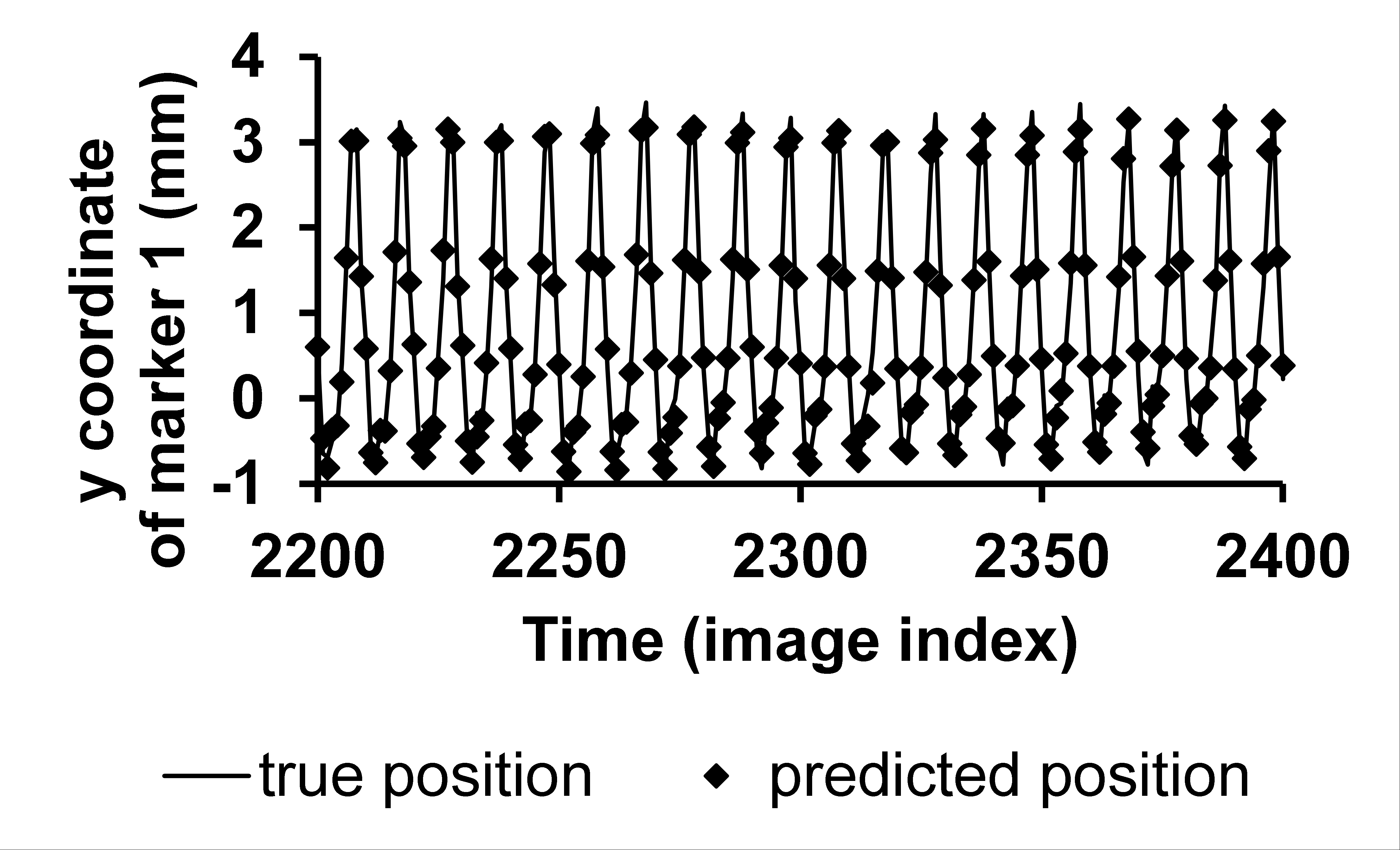}
    \includegraphics[width=5.4cm, height=3.7cm]{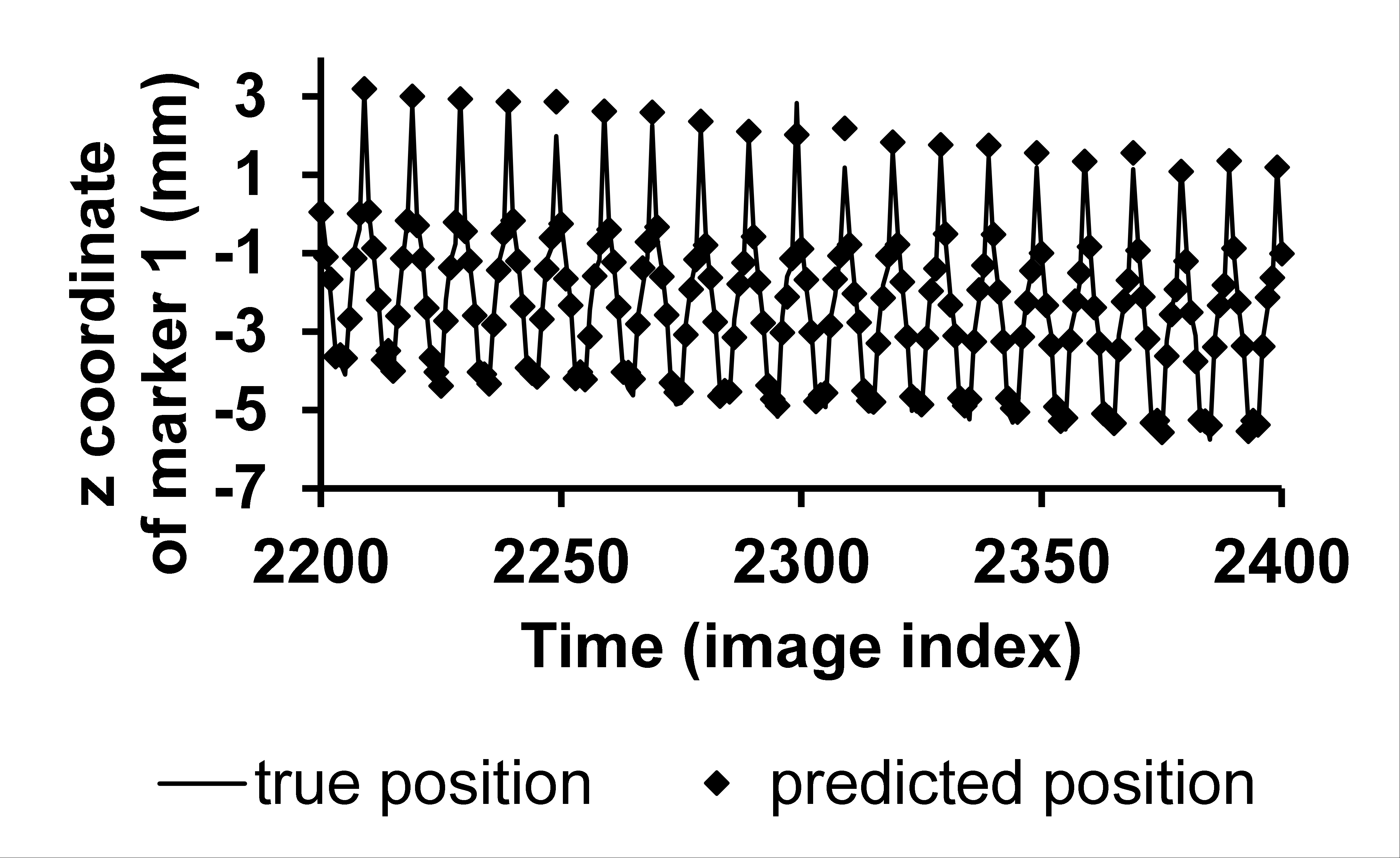}
    \quad
    \includegraphics[width=5.4cm, height=3.7cm]{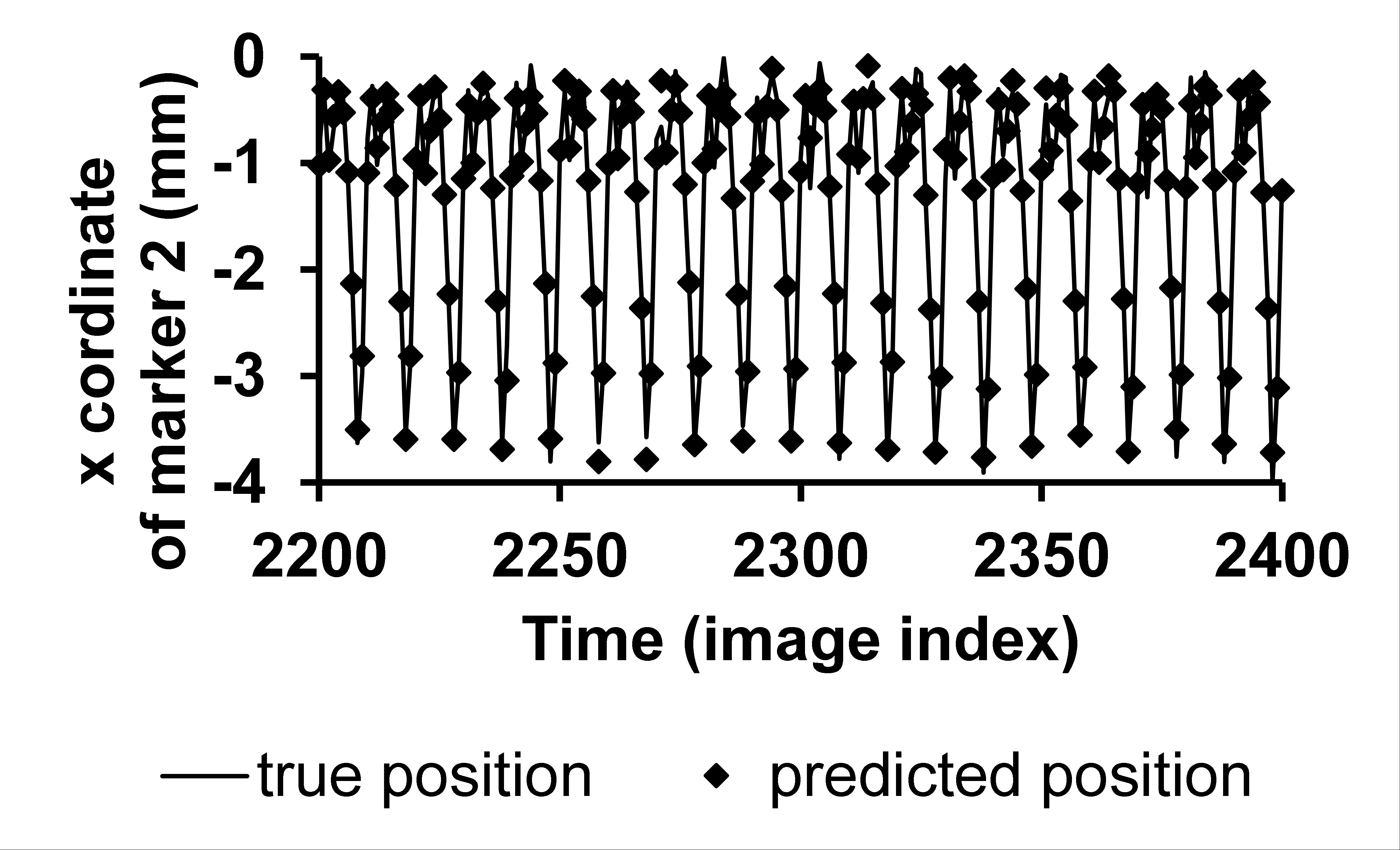}
    \includegraphics[width=5.4cm, height=3.7cm]{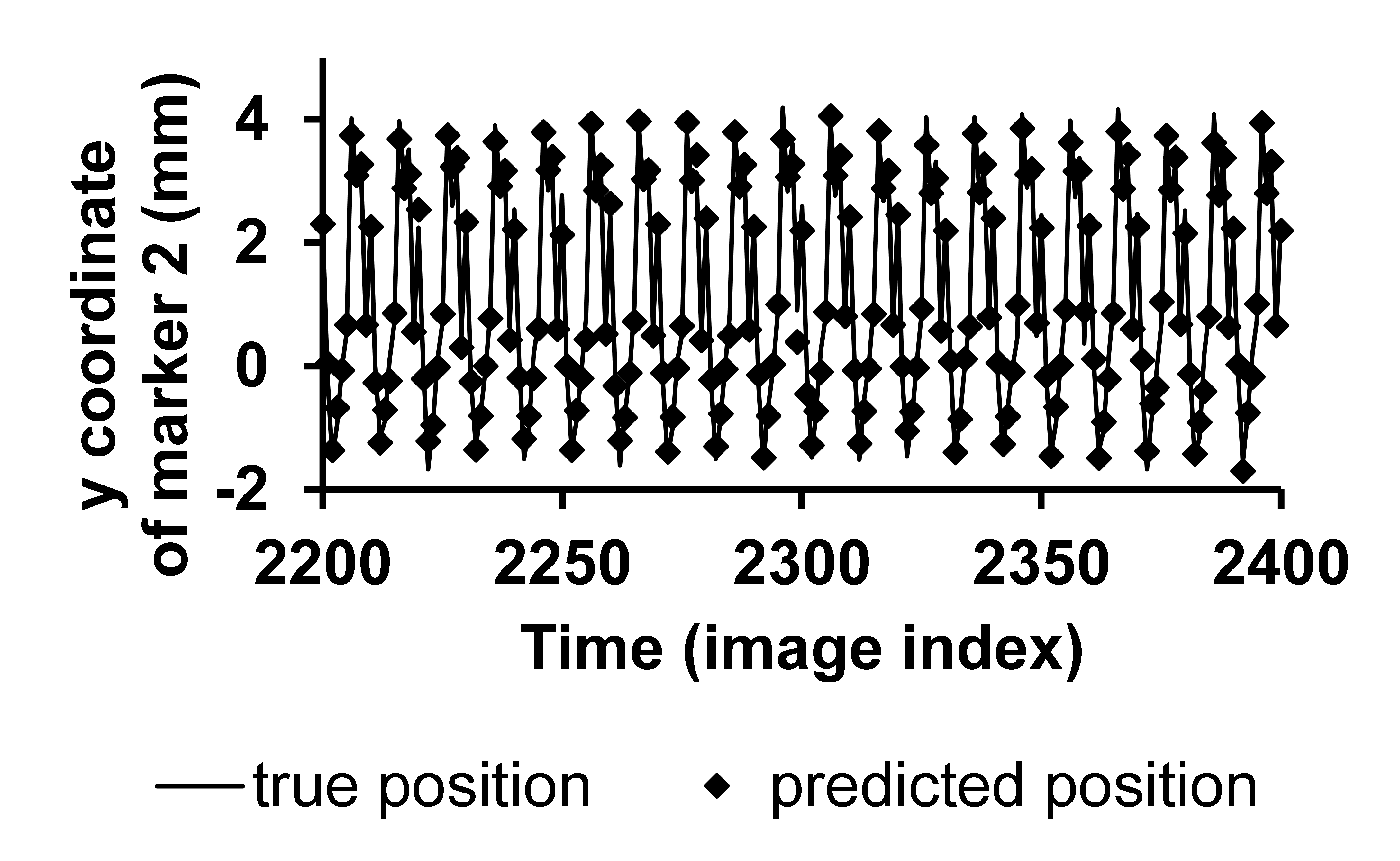}
    \includegraphics[width=5.4cm, height=3.7cm]{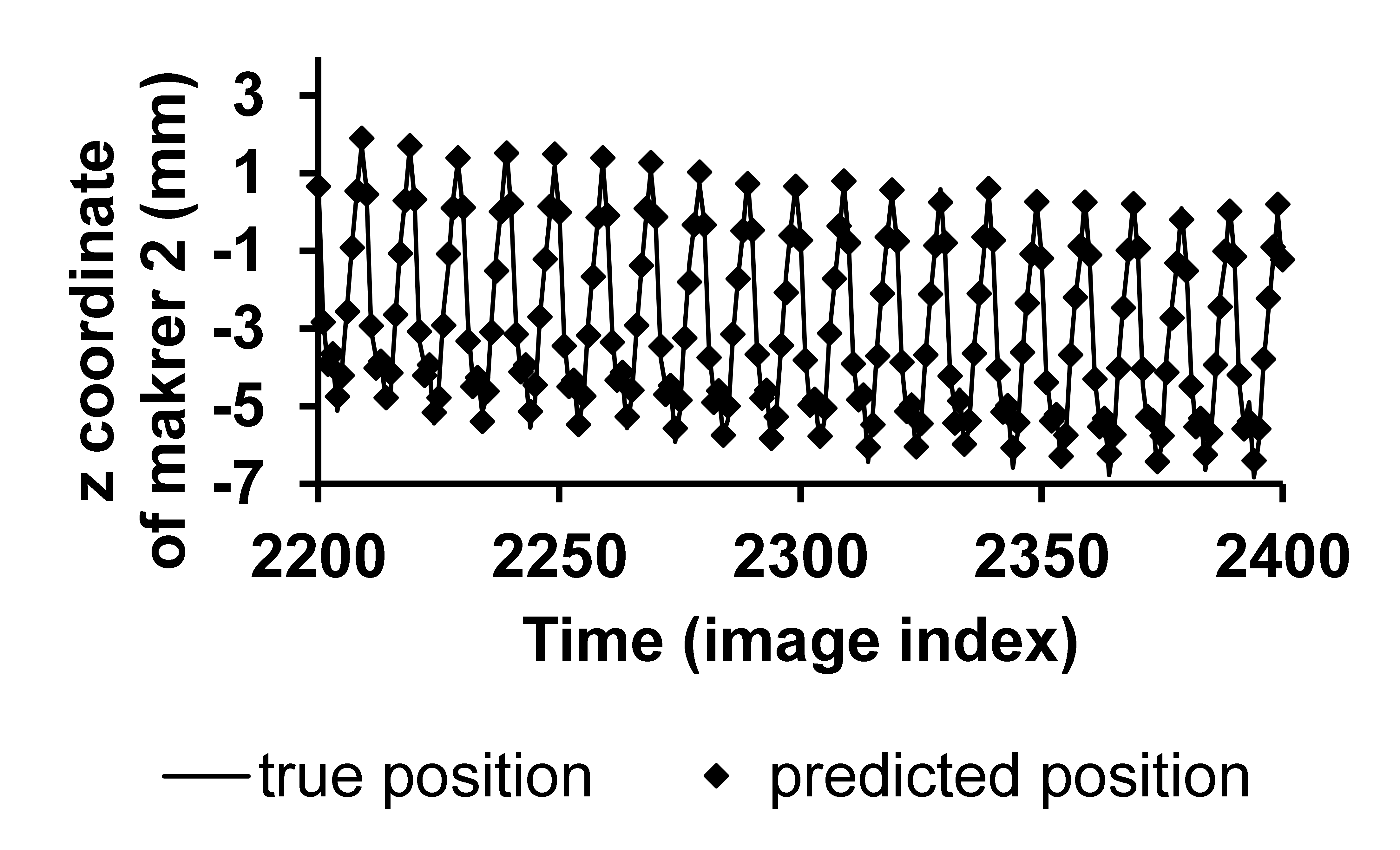}
    \quad
    \includegraphics[width=5.4cm, height=3.7cm]{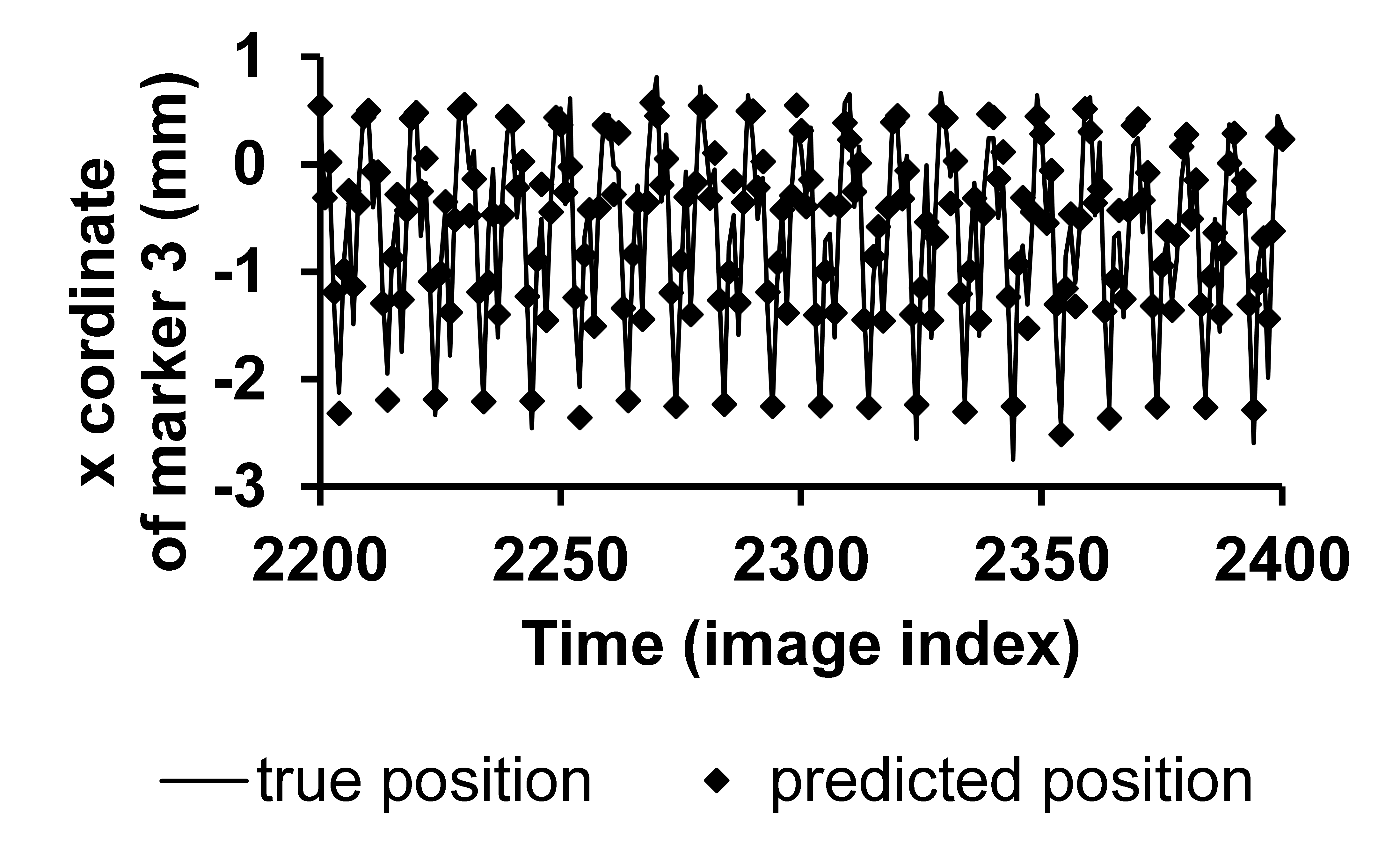}
    \includegraphics[width=5.4cm, height=3.7cm]{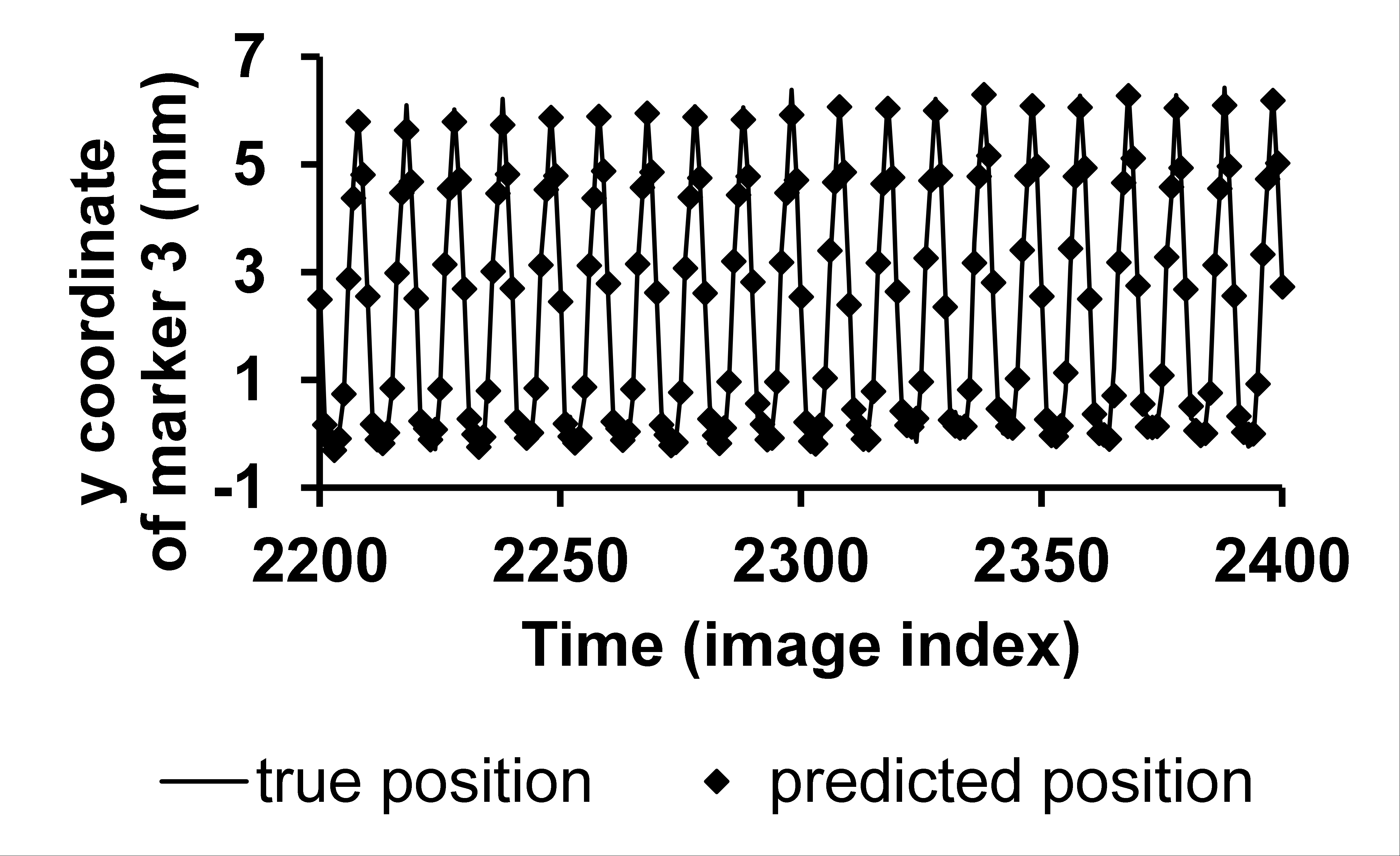}
    \includegraphics[width=5.4cm, height=3.7cm]{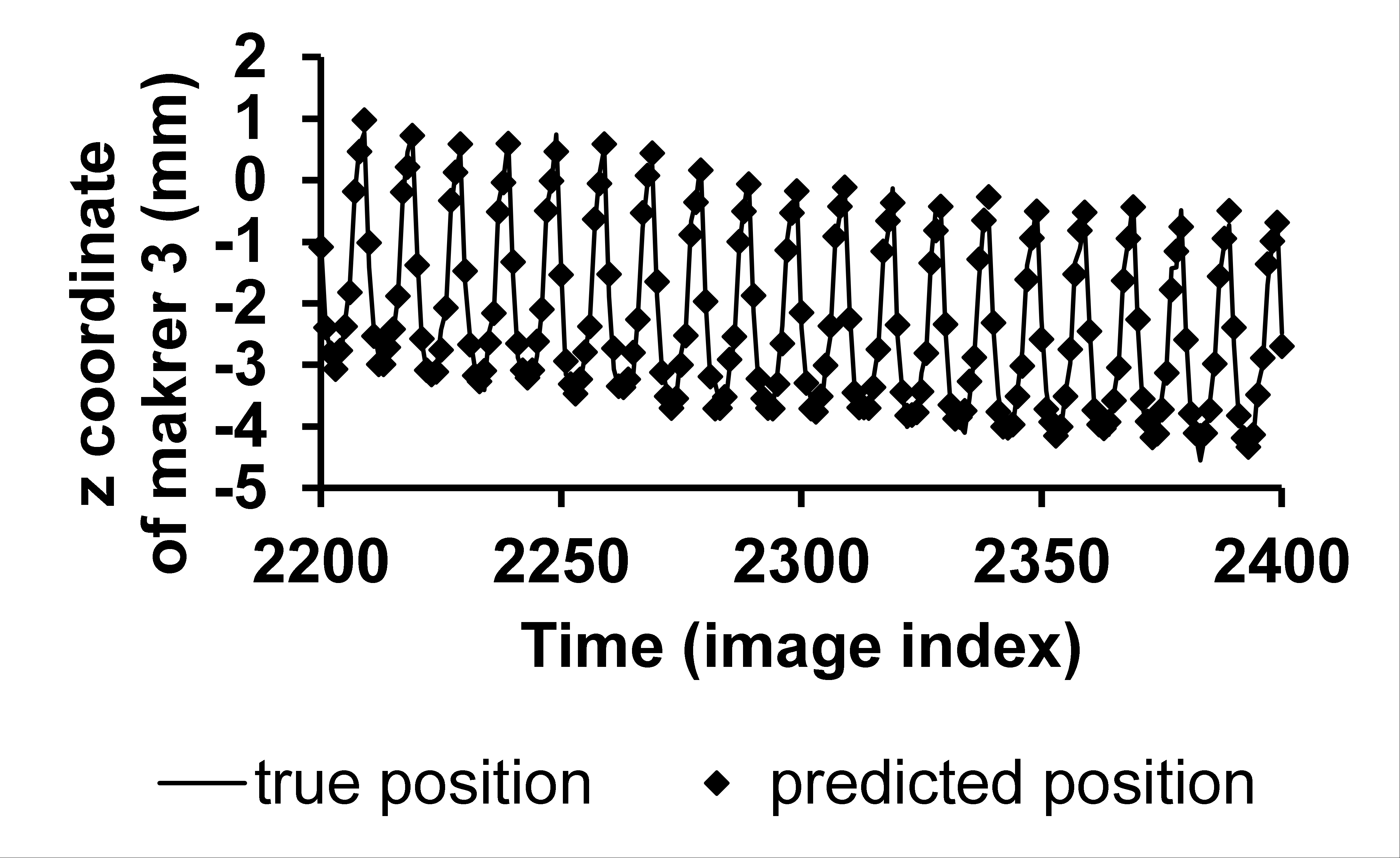}       
    \caption{Prediction of the position of the markers of patient 3 on the test data. The axes are the same as in Fig. \ref{fig:traj}.}%
    \label{fig:patient3_pred_t2201_t2400}%
\end{figure*}

The parameters that achieved the lowest (minimum) MAE error on the cross-validation set without leading to any numerical error have been used for evaluation on the test data between $t_{2201}$ and $t_{2400}$. For every patient, we set $q = 250$ and $\sigma_{init}^{RNN} = 0.02$. The value of $\eta$ was set to 0.01 for patients 3 and 4, and 0.02 for patients 1 and 2. Table \ref{table:RNN_lin_reg_comparison} shows the performance of the RNN on that test data, using the parameters selected as mentioned beforehand, in terms of the maximum prediction error, RMSE, and normalized RMSE, defined respectively in Eq. \ref{eq:max_pred_error_def}, Eq. \ref{eq:pred_RMSE_def} and Eq. \ref{eq:pred_nRMSE_def}. In Eq. \ref{eq:pred_nRMSE_def}, $\mu_{true}^p$ designates the mean position of all observations of point $p$ on the test set. 

\begin{equation}\label{eq:max_pred_error_def}
 e_{max} = \underset{k = 2201, ..., 2400}{max} \underset{p =1, ..., r}{max}
 	\Big\| \vv{M_{true}^p(t_k)M_{pred}^p(t_k)} \Big\|_2
\end{equation}
\begin{equation} \label{eq:pred_RMSE_def}
 e_{RMS} = \sqrt{\frac{1}{200 r} \sum_{k = 2201}^{2400} \sum_{p =1}^r
 	\Big\| \vv{M_{true}^p(t_k)M_{pred}^p(t_k)} \Big\|_2^2}
\end{equation} 

\begin{equation} \label{eq:pred_nRMSE_def}
 e_{nRMS} = \frac{\sqrt{\sum_{k = 2201}^{2400} \sum_{p =1}^r
 						\Big\| \vv{M_{true}^p(t_k)M_{pred}^p(t_k)} \Big\|_2^2}}
 				 {\sqrt{\sum_{k = 2201}^{2400} \sum_{p =1}^r
 						\Big\| \vv{M_{true}^p(t_k)\mu_{true}^p} \Big\|_2^2}}
\end{equation}

Furthermore, we evaluated the jitter of each prediction method on the test data. Jitter measures how oscillatory the predicted signal is (Eq. \ref{eq:jitter_def}). Prediction with low jitter is desirable since it makes control of the treatment robot easier. The jitter measure $J$ is minimized when the prediction is constant, thus there is a trade-off between accuracy and jitter.

\begin{equation} \label{eq:jitter_def}
 J = \frac{1}{199 r} \sum_{k = 2201}^{2399} \sum_{p =1}^r
	\Big\| \vv{M_{pred}^p(t_{k+1})M_{pred}^p(t_k)} \Big\|_2
\end{equation} 

Because the RNN is evaluated using 10 runs with random weight initialization, not only the errors $e_{max}$ and $e_{RMS}$ are calculated, but also the corresponding 95\% mean confidence intervals $I_{max}$  and $I_{RMS}$ (assuming that both $ e_{max} $ and $e_{RMS}$ follow a Gaussian distribution) defined in Eq. \ref{eq:max_err_conf_interval} and Eq. \ref{eq:RMS_err_conf_interval}, where $ \sigma_{max} $ and $\sigma_{RMS}$ are the corresponding standard deviations of $ e_{max} $ and $e_{RMS}$ over the 10 runs\footnotemark. The performance of the RNN was compared with the point-wise and coordinate-wise linear predictor defined in Eq. \ref{eq:linear_pred_def}. In that equation, $p$ is the tracked point index, $d$ represents the x,y, or z component of the 3D displacement $\vec{u}(\vec{x_p}, t)$, $(a_k^{d,p})$ are regression constants, and $L_{lin}$ is the SHL, arbitrarily set to $L_{lin} = 10$. We also compared the RNN with the LMS filter (Algorithm \ref{alg:LMS}) \cite{haykin2014adaptive}, for which we selected a SHL of $L_{LMS} = 10$ and a learning rate $\eta_{LMS} = 0.01$. The time series input data for the LMS algorithm was also normalized as described in Section \ref{subsection:method_RNN_prediction_pts}.

\footnotetext{Because numerical errors may happen (cf footnote \ref{RNN_instability_footnote}), $\sigma_{max}$ and $\sigma_{RMS}$ are actually the standard deviations over the runs among the 10 runs for which no numerical error occurred. Also, in Eq. \ref{eq:max_err_conf_interval} and Eq. \ref{eq:RMS_err_conf_interval}, $\sqrt{10}$ should also be replaced by $\sqrt{n_0}$, where $n_0$ is the number of runs among the 10 runs for which no numerical error occurred, to be precise.}

\begin{equation} \label{eq:max_err_conf_interval}
I_{max} = \left[ 
e_{max} - \frac{1.96 \sigma_{max}}{\sqrt{10}},
e_{max} + \frac{1.96 \sigma_{max}}{\sqrt{10}}
\right]
\end{equation}

\begin{equation} \label{eq:RMS_err_conf_interval}
I_{RMS} = \left[ 
e_{RMS} - \frac{1.96 \sigma_{RMS}}{\sqrt{10}},
e_{RMS} + \frac{1.96 \sigma_{RMS}}{\sqrt{10}}
\right]
\end{equation}

\begin{equation} \label{eq:linear_pred_def}
	\begin{aligned}
	u_d^{pred}(\vec{x_p}, t_{n+L_{lin}}) = a_0^{d,p} + \sum_{k = 1}^{L_{lin}} a_k^{d,p}u_d(\vec{x_p}, t_{n+L_{lin}-k}) \\ d=x,y,z \quad p=1,2,3
	\end{aligned}
\end{equation}

\begin{algorithm}
\caption{Least mean squares}
\label{alg:LMS}
\begin{algorithmic}
\State \textbf{Parameters} :
\State $L$ : signal history length
\State $r$ : number of internal points considered
\State $m = 3rL$ dimension of the input space
\State $p = 3r$ dimension of the output space
\State $\eta$ : learning rate
\State
\State \textbf{Initialization}
\State $W_{n=1} = 0_{p \times (m+1)}$
\State
\State \textbf{Learning and prediction}
\For{$n = 1,2,...$}
\State $y_{n} := W_n u_n$ (prediction)
\State $W_{n+1} := W_n + \eta (d_n - y_n) u_n^T$ (weights update)
\EndFor
\end{algorithmic}
\end{algorithm}

The RNN achieves a lower maximum and RMS prediction error as well as a lower jitter (averaged over the 4 patients) than linear prediction and LMS (Table \ref{table:RNN_lin_reg_comparison}). In particular, the maximum prediction error corresponding to the RNN, averaged over the 4 patients, equal to 1.51mm, is respectively 16.1\% and 5.0\% lower than the maximum error corresponding to linear prediction and LMS, equal to 1.80mm and 1.59mm. Furthermore, the averaged maximum prediction error and the averaged RMSE given by the RNN are respectively approximately 4 times and 7 times lower than the corresponding errors given by a system without prediction, defined by $\overrightarrow{u_{pred}}( \cdot , t_{n+1}) = \vec{u}( \cdot , t_n)$. The prediction errors are higher for patients 1 and 2, which correlates with the higher motion amplitude of these patients' markers (Table \ref{table:motion amplitude}). Concerning the prediction with the RNN, the maximum tracking error for each patient is below the 2mm threshold recommended by Murphy \cite{murphy2004tracking}. By contrast, the maximum error with linear prediction corresponding to patient 2 and maximum error with LMS corresponding to patient 1 exceeded that threshold.

The average calculation time per time step (time for performing one prediction) of the RNN obtained with GPU programming was equal to 119.1ms  (Dell Intel Core i9-9900K 3.60GHz CPU NVidia GeForce RTX 2080 SUPER GPU 32Gb RAM with Matlab), which is lower than the marker position sampling time, approximately equal to 400ms (Table  \ref{table:RNNtime_perf}).  

Both the maximum error and the RMS error achieved with the RNN in our study are lower than the corresponding errors reported in related studies about prediction in radiotherapy \cite{sharp2004prediction, goodband2008comparison, kai2018prediction} mentioned in Section \ref{subsection:intro_prediction_methods} (cf Table \ref{table:comparison_with litterature}). However, comparison with the prediction methods in the literature is difficult because the datasets, sampling rates, and look-ahead time vary between the studies. The regularity of the breathing motion as well as the low motion amplitudes in our dataset are factors that may have contributed to lower prediction errors in our research. 

\begin{table*}[width=.95\textwidth,cols=10,pos=h]
\begin{tabular*}{\tblwidth}{@{} LLLLLLLLLL@{} }
\toprule
Author   & Network & Training           & Signal       & Sampling & Nb. of   & Signal     & Response & Signal  & Prediction       \\
         &         & method             & predicted    & rate     & patients & amplitude  & time     & history & error            \\
\midrule
\midrule
Sharp    & MLP     & Conjugate          & 1 implanted  & 3 Hz     & 14       & 9.1mm      & 1) 400ms & -       & 1) RMSE 3.5mm    \\
         &         & gradient (offline) & marker       &          &          & to 31.6mm  & 2) 1.0s  &         & 2) RMSE 5.5mm    \\
Goodband & MLP     & Conjugate          & 1 external   & 30 Hz    & 24       & 8mm        & 400ms    & 133ms   & Max error        \\ 
         &         & gradient (online)  & marker       &          &          & to 60mm    &          &         & 5.027mm          \\
Lee      & RNN     & HEKF (online)      & Cyberknife   & 26 Hz    & -        & Normalized & 500ms    & -       & nRMSE from       \\  
         &         &                    & data         &          &          & to max 1   &          &         & 0.040 to 0.193   \\
Kai      & RNN     & BPTT (offline)     & 1 implanted  & 30 Hz    & 7        & -          & 1.0s     & 4.0s    & RMSE from        \\
         &         &                    & marker       &          &          &            &          &         & 0.48mm to 1.37mm \\
\midrule
Proposed & RNN     & RTRL (online)      & 3 implanted  & 2.5 Hz   & 4        & 12.0mm     & 400ms    & up to   & Max error 1.51mm \\
work     &         &                    & markers      &          &          & to 22.7mm  &          & 16.0s   & RMSE 0.444mm     \\
         &         &                    & (simulation) &          &          &            &          &         & nRMSE 0.108      \\
\bottomrule
\end{tabular*}
\caption{Comparison of the prediction performance of RNNs trained with RTRL with previous ANNs models proposed for prediction in radiotherapy (studies \cite{sharp2004prediction}, \cite{goodband2008comparison}, \cite{hazazi13extended} and \cite{kai2018prediction}, introduced in Section \ref{subsection:intro_prediction_methods}). MLP stands for "multilayer perceptron".}
\label{table:comparison_with litterature}
\end{table*}

During the beginning of the learning process, the predicted values oscillate around the mean position signal and adjust progressively to reach the actual signal (Fig. \ref{fig:pred_t1_t100}). This is illustrated by the loss function decreasing for small values of the time index (Fig. \ref{fig:loss_error}). The error loss function of patient 3 rises again between $t_{1200}$ and $t_{1800}$ when variations in the marker motion pattern appear (cf Fig. \ref{fig:patient3_extended_time_series} and Appendix \ref{appendix:entire_motion_dataset_patient3}). The predicted values on the test data follow closely the original motion signal. The breathing drift, corresponding to a decreasing trend in the z position of the markers on the test data for patient 3, is also well captured by the RNN (Fig. \ref{fig:patient3_pred_t2201_t2400}).

\subsection{\normalsize Chest image prediction}

\begin{figure*}%
    \centering
    \captionsetup[subfigure]{justification=centering}   
    \captionsetup[subfigure]{labelformat=empty} 
    \subfloat[Patient 1 \\* predicted \\* \text{\normalsize $t=t_{2209}$}]{{\includegraphics[height=2.8cm]{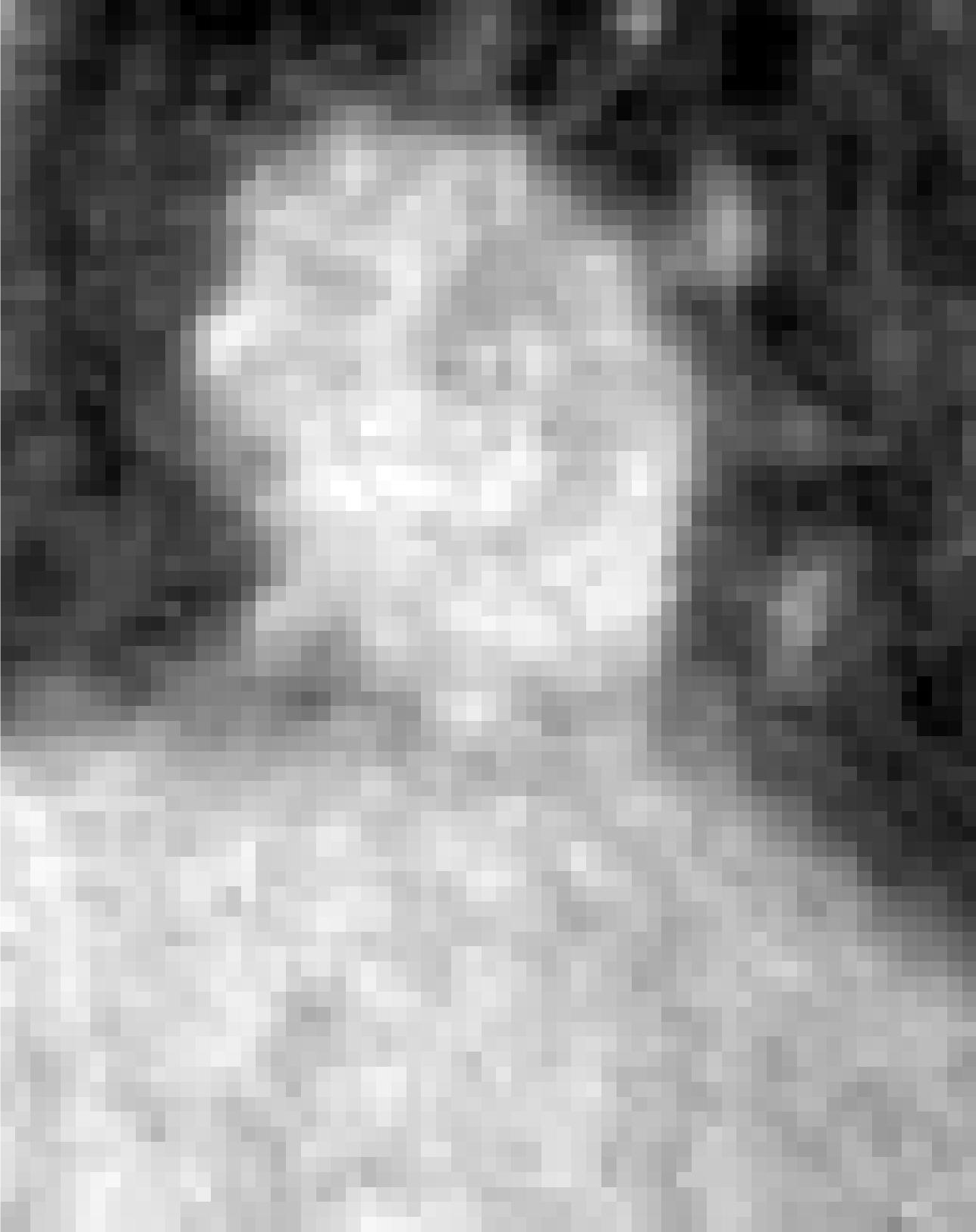} }}%
    \subfloat[Patient 1 \\* original \\* \text{\normalsize $t=t_{2209}$}]{{\includegraphics[height=2.8cm]{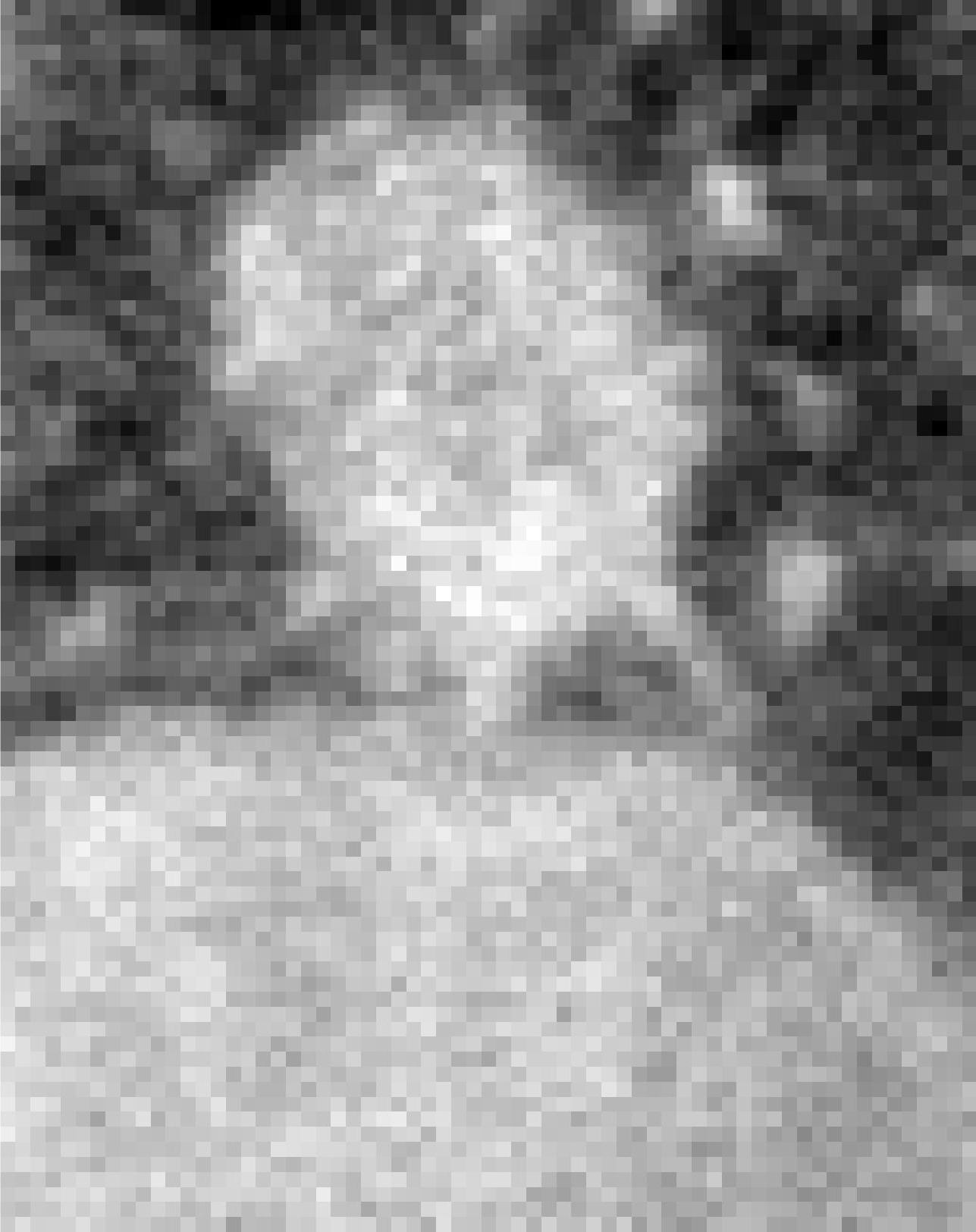} }}%
    \subfloat[Patient 1 \\* predicted \\* \text{\normalsize $t=t_{2374}$}]{{\includegraphics[height=2.8cm]{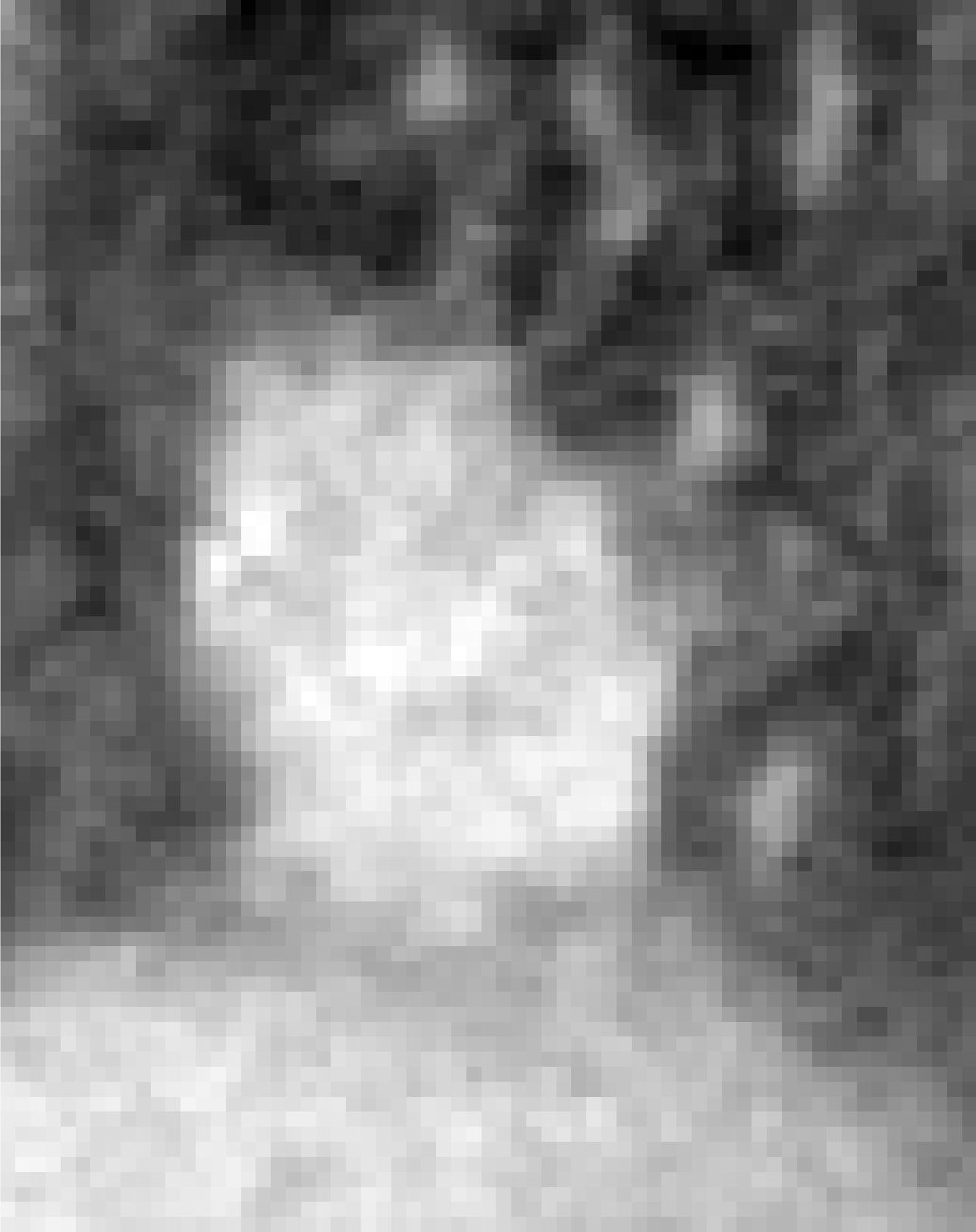} }}%
    \subfloat[Patient 1 \\* original \\* \text{\normalsize $t=t_{2374}$}]{{\includegraphics[height=2.8cm]{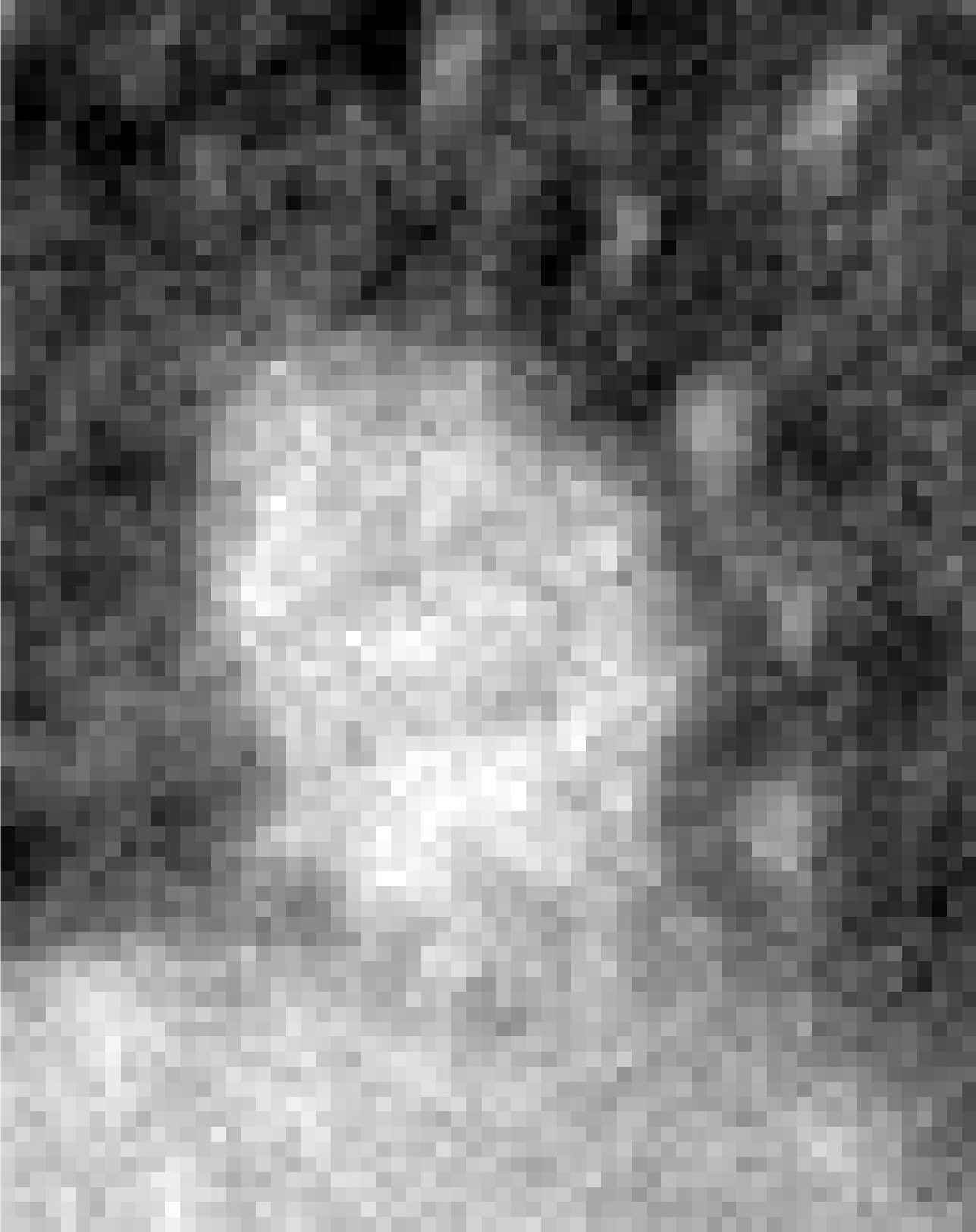} }}%
    \subfloat[Patient 2 \\* predicted \\* \text{\normalsize $t=t_{2209}$}]{{\includegraphics[height=2.8cm]{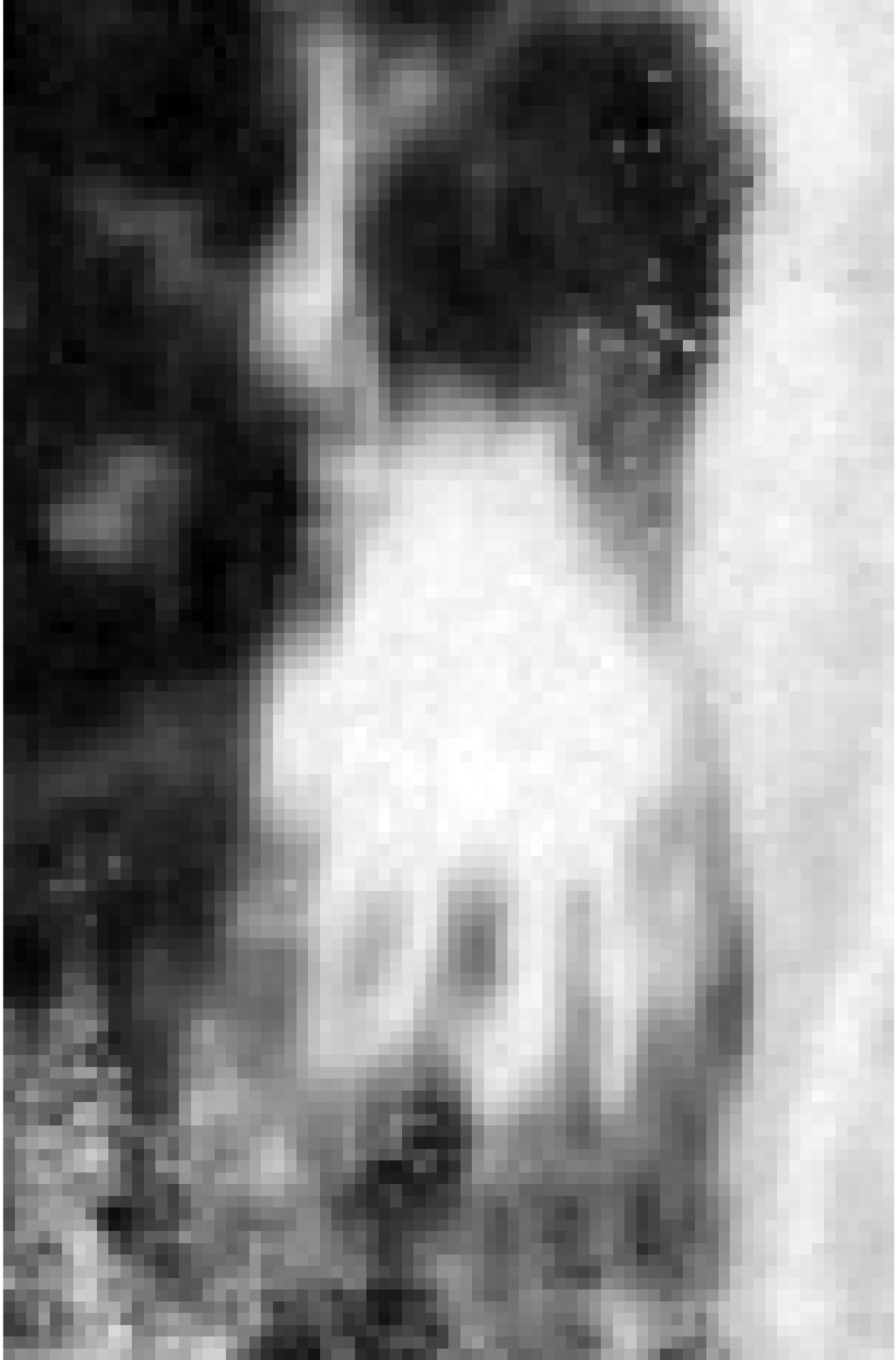} }}%
    \subfloat[Patient 2 \\* original \\* \text{\normalsize $t=t_{2209}$}]{{\includegraphics[height=2.8cm]{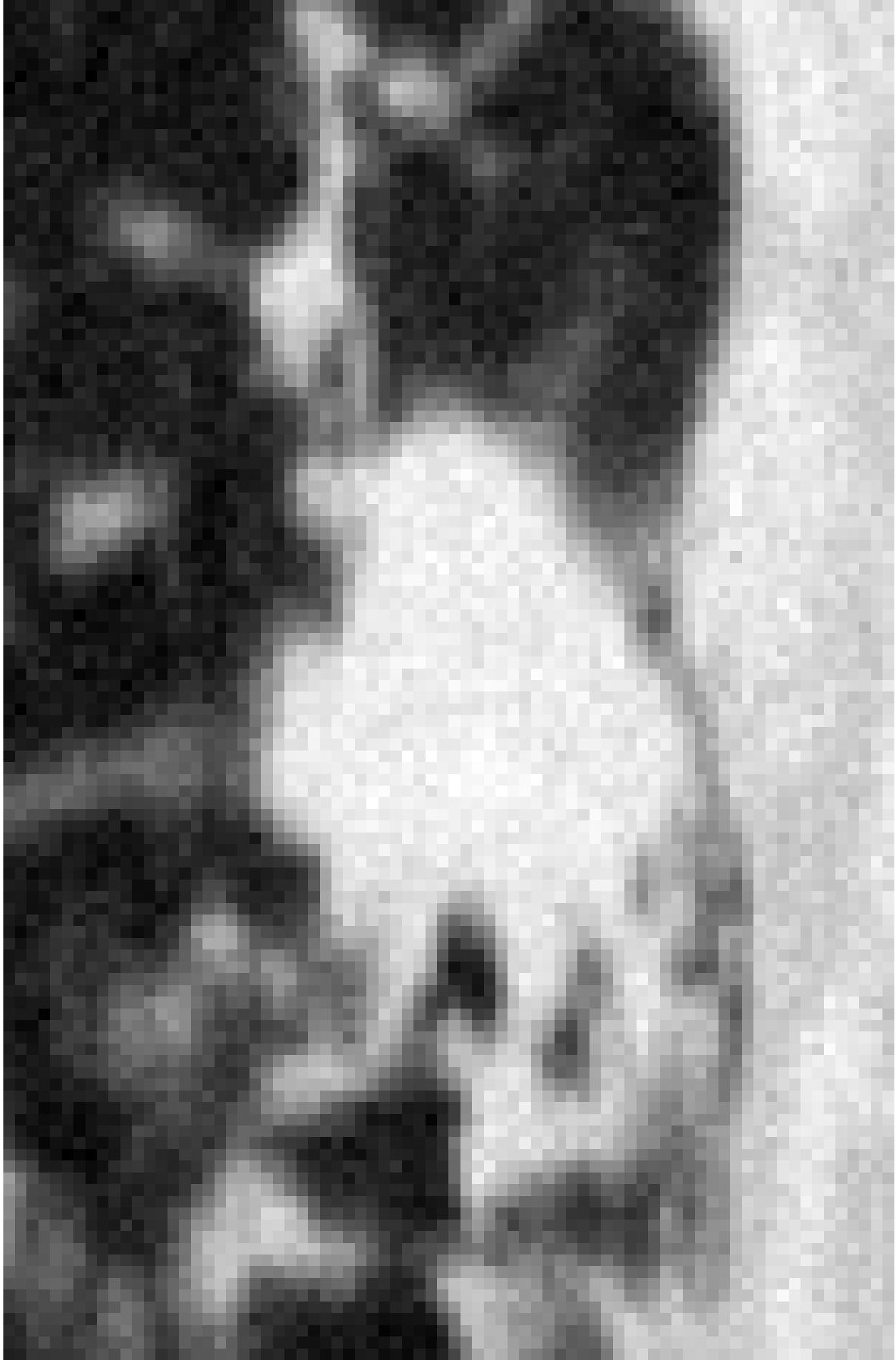} }}%
    \subfloat[Patient 2 \\* predicted \\* \text{\normalsize $t=t_{2374}$}]{{\includegraphics[height=2.8cm]{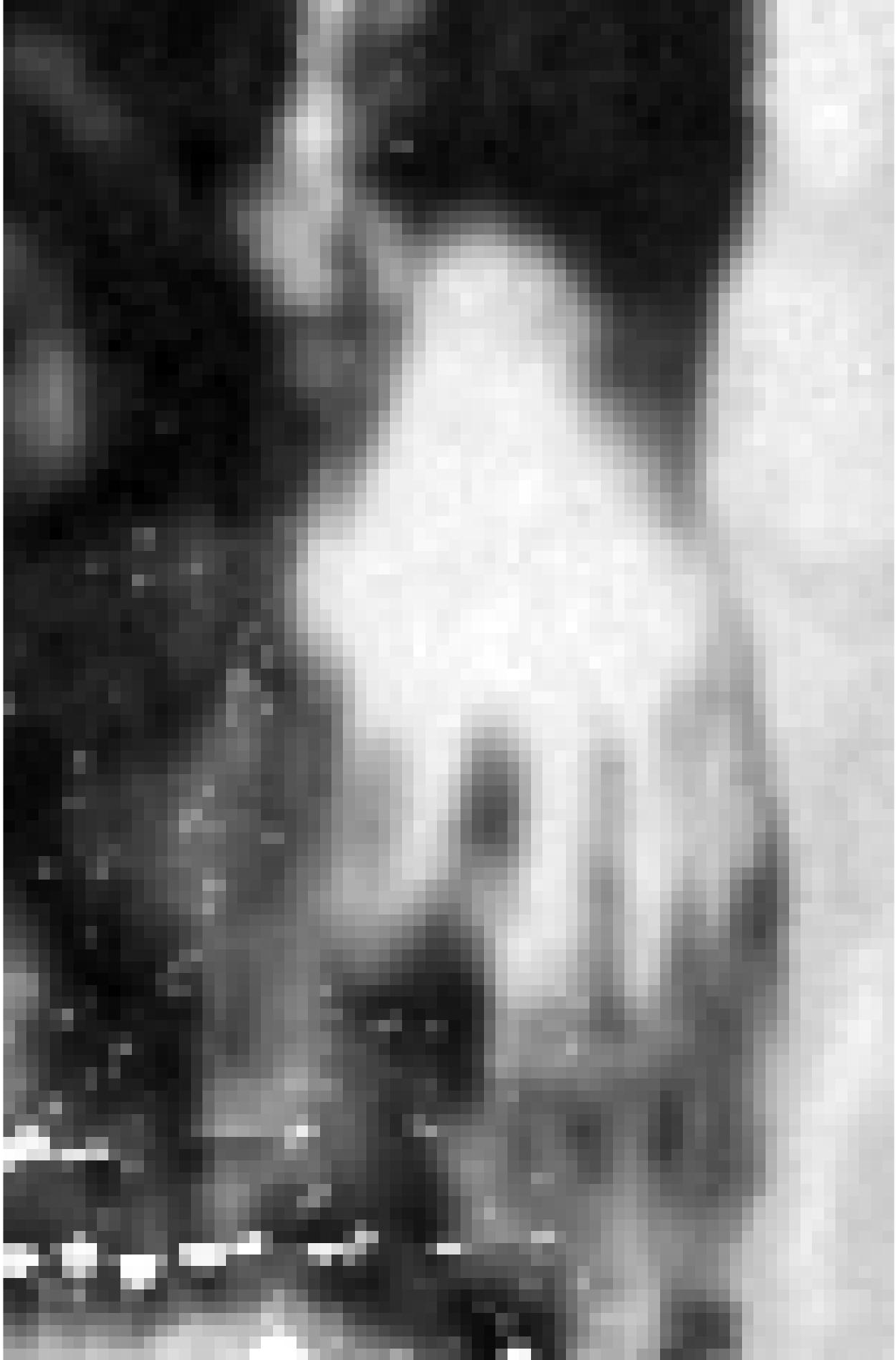} }}%
    \subfloat[Patient 2 \\* original \\* \text{\normalsize $t=t_{2374}$}]{{\includegraphics[height=2.8cm]{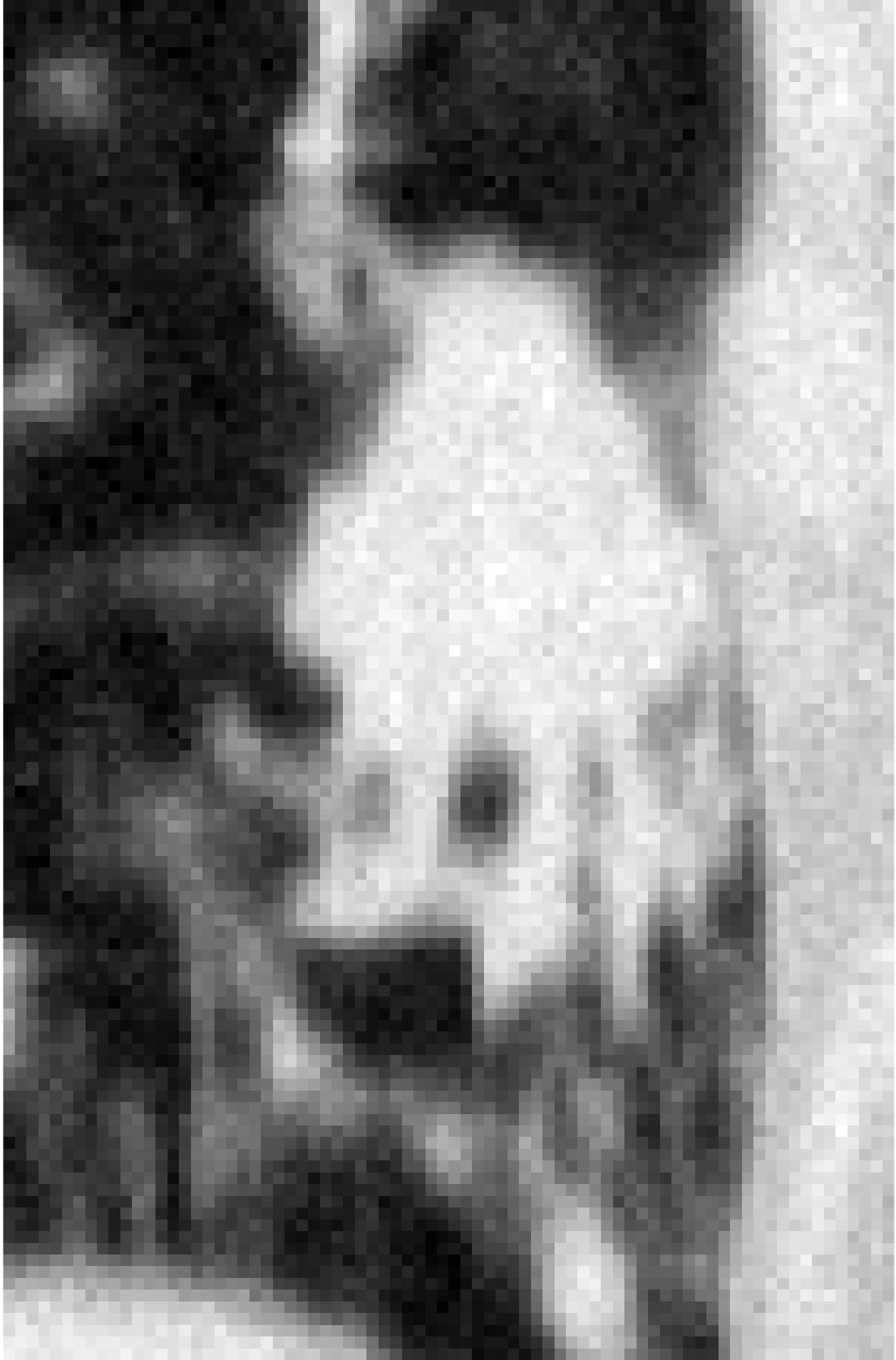} }}%
    \quad
    \subfloat[Patient 3 \\* predicted \\* \text{\normalsize $t=t_{2209}$}]{{\includegraphics[height=2.1cm]{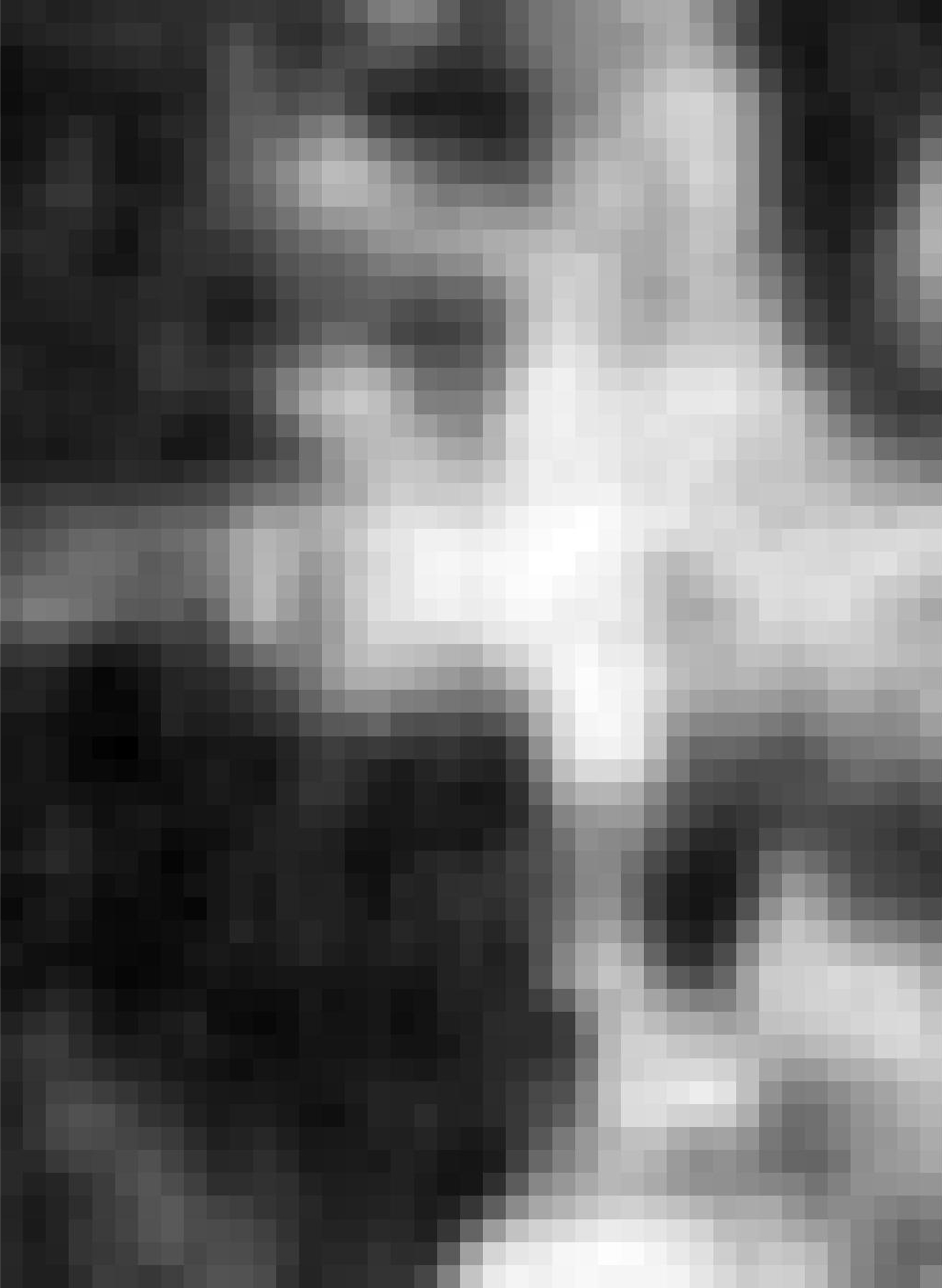} }}%
    \subfloat[Patient 3 \\* original \\* \text{\normalsize $t=t_{2209}$}]{{\includegraphics[height=2.1cm]{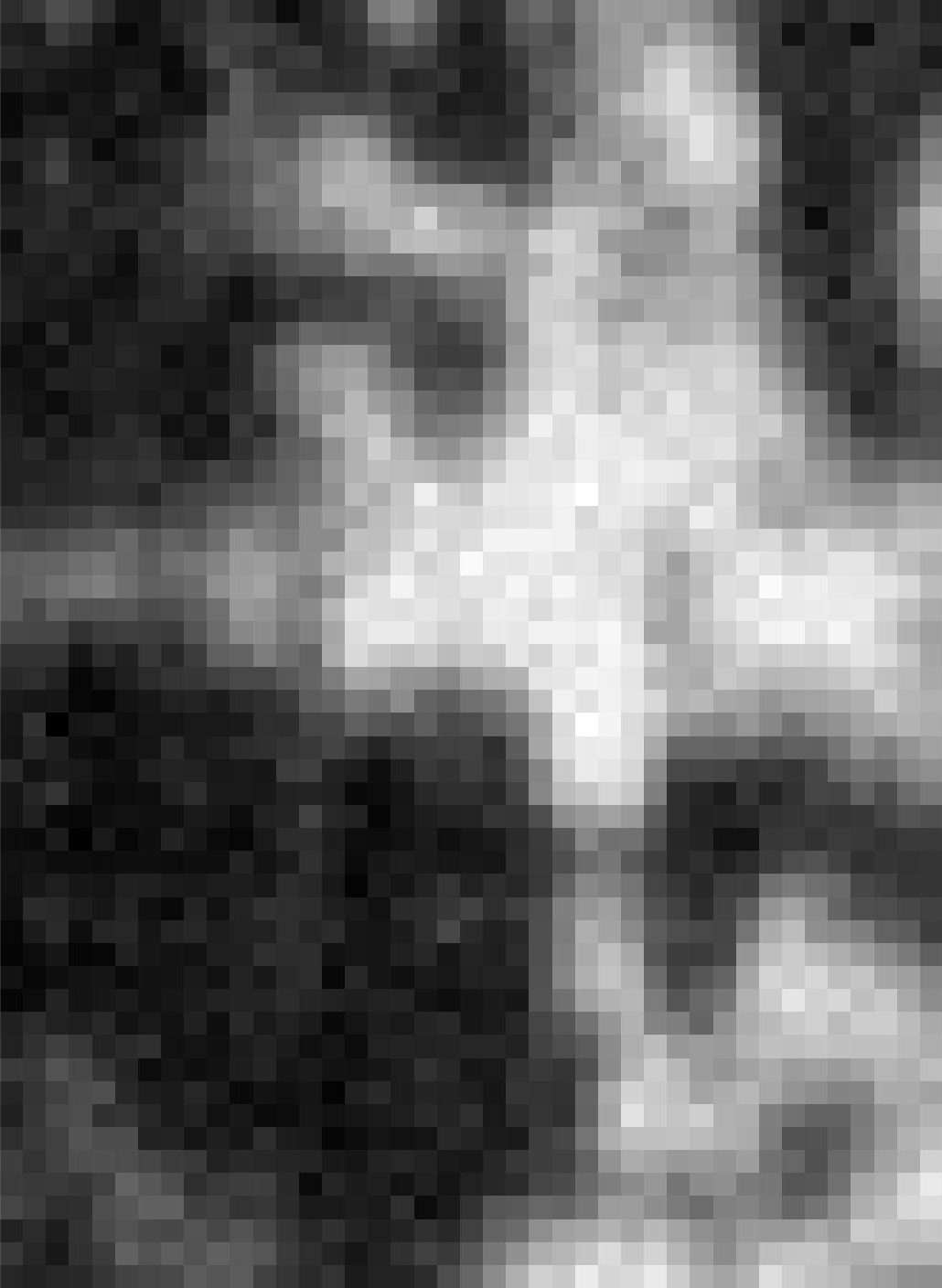} }}%
    \subfloat[Patient 3 \\* predicted \\* \text{\normalsize $t=t_{2374}$}]{{\includegraphics[height=2.1cm]{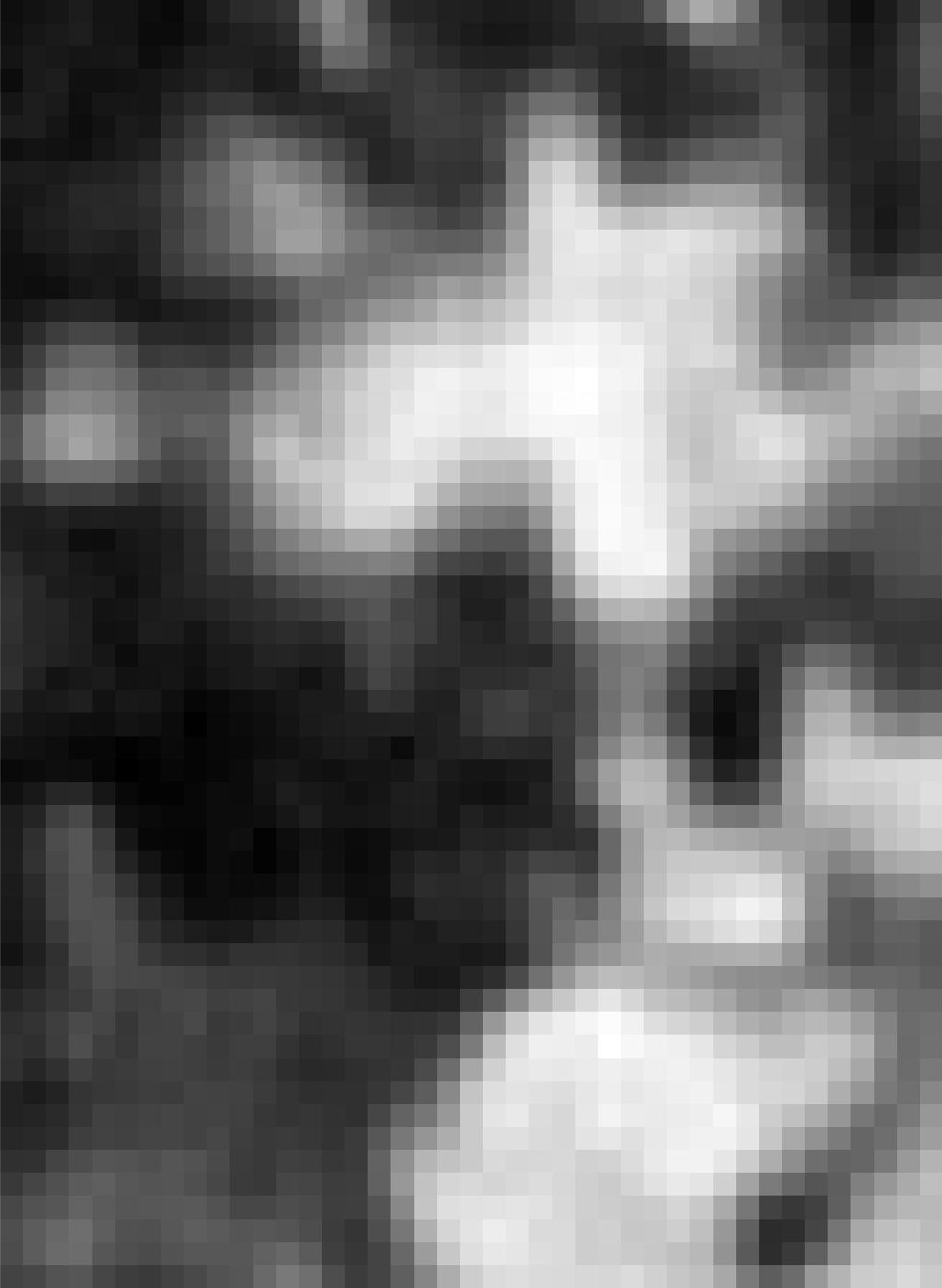} }}%
    \subfloat[Patient 3 \\* original \\* \text{\normalsize $t=t_{2374}$}]{{\includegraphics[height=2.1cm]{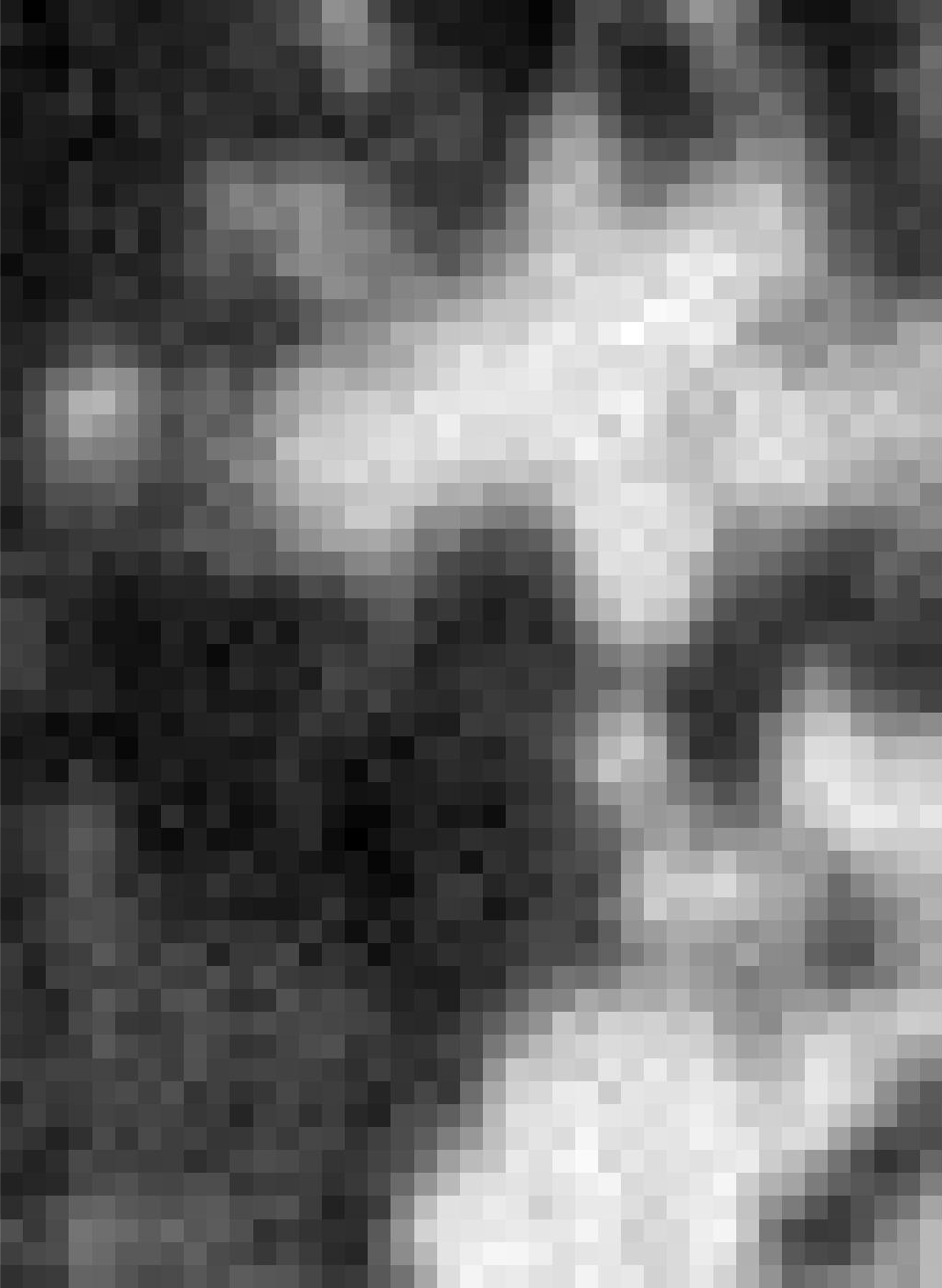} }}%
    \subfloat[Patient 4 \\* predicted \\* \text{\normalsize $t=t_{2209}$}]{{\includegraphics[height=2.1cm]{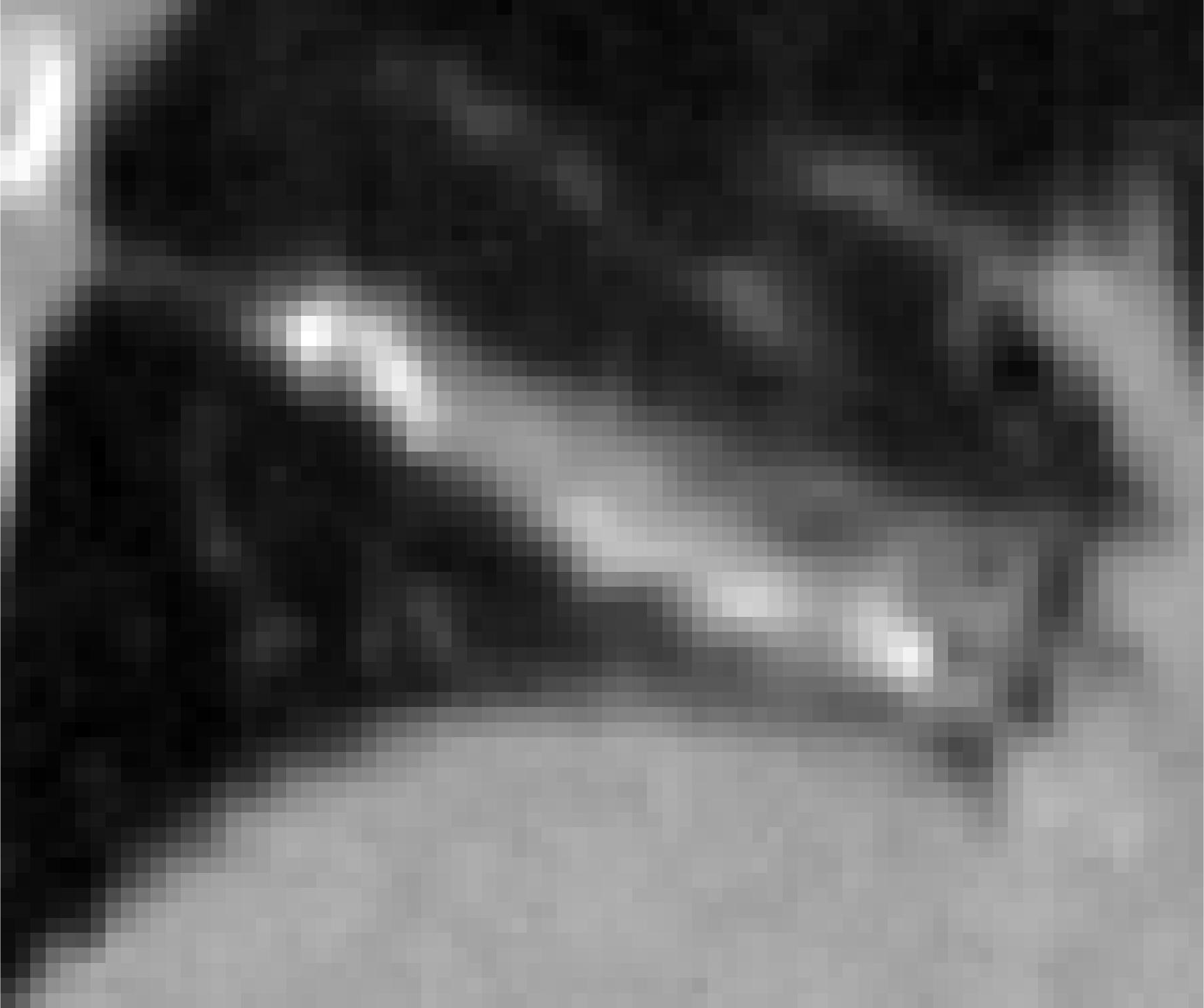} }}%
    \subfloat[Patient 4 \\* original \\* \text{\normalsize $t=t_{2209}$}]{{\includegraphics[height=2.1cm]{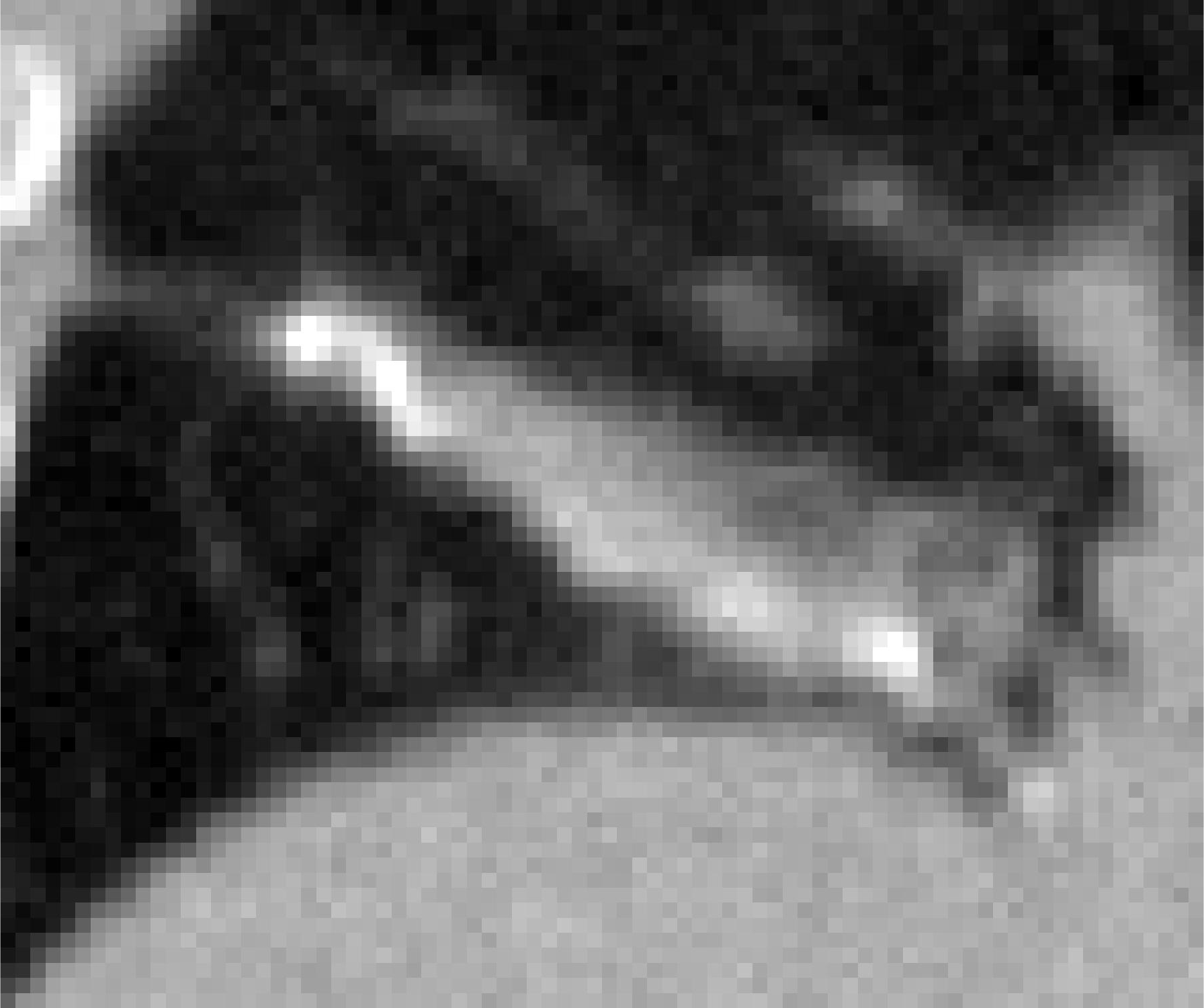} }}%
    \subfloat[Patient 4 \\* predicted \\* \text{\normalsize $t=t_{2374}$}]{{\includegraphics[height=2.1cm]{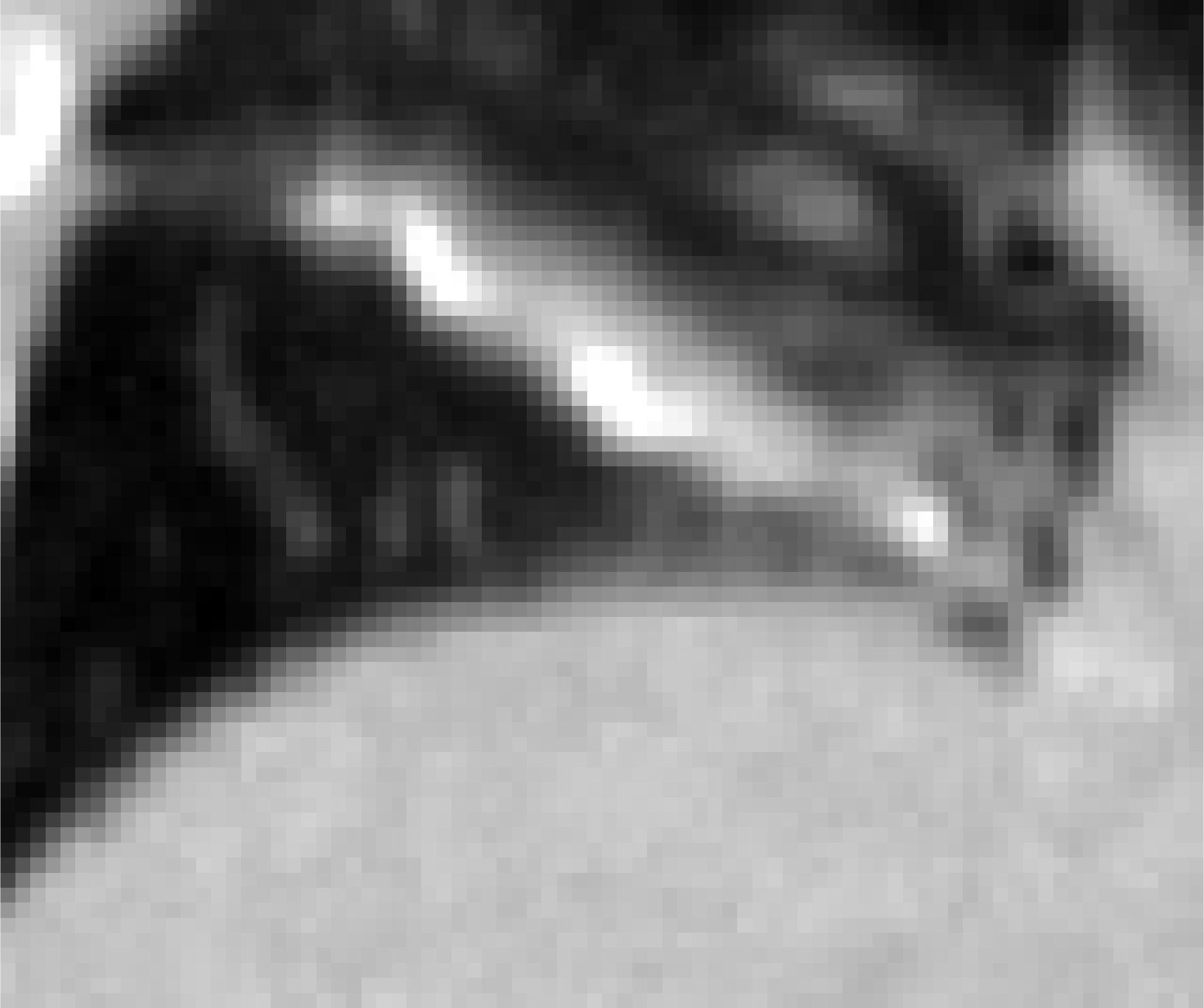} }}%
    \subfloat[Patient 4 \\* original \\* \text{\normalsize $t=t_{2374}$}]{{\includegraphics[height=2.1cm]{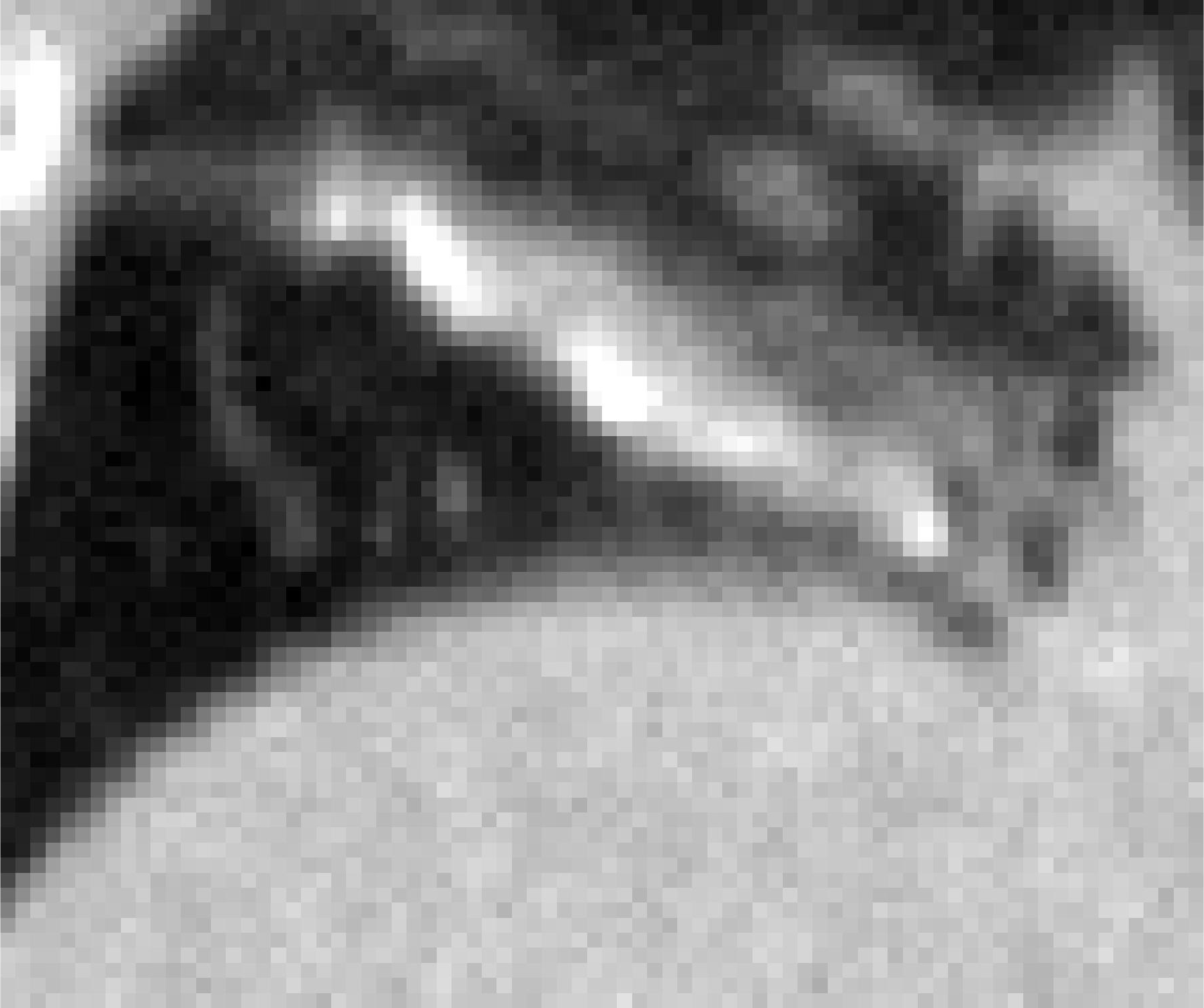} }}%
    \caption{Original and predicted ROI coronal cross-sections (same coordinates as in Fig. \ref{fig:org_im}), at \text{\normalsize $t=t_{2209}$} (end of expiration for patient 1 and end of inspiration for the other patients) and \text{\normalsize $t=t_{2374}$} (opposite case).}%
    \label{fig:pred_cor_cc}%
\end{figure*}

\begin{table*}[width=.9\textwidth,cols=6,pos=h]
\begin{tabular*}{\tblwidth}{@{} LLLLLL@{} }
\toprule
DVF used for warping & Patient 1 & Patient 2 & Patient 3 & Patient 4 & Average over all patients\\
\midrule
Initial DVF &  0.960 &  0.982 &  0.976 & 0.992 & 0.978\\
DVF from markers &  0.923 &  0.957 &  0.959 & 0.979 & 0.954\\
Predicted DVF from markers &  0.923 & 0.958 & 0.959 & 0.978 & 0.955\\
\bottomrule
\end{tabular*}
\caption{Precision of the displacement vector field (DVF) calculated at each step of the image prediction process. Each cell in the table corresponds to the cross-correlation between the initial ROI images, that is, the images from the sequence constructed in Section \ref{subsection:chest image data}, and the warped initial image at \text{\normalsize $t=t_1$}, averaged over the test data. The first line corresponds to the average for $k \in \{2201, ..., 2400\}$ of the cross-correlation between the initial image at time \text{\normalsize $t_k$} and the initial image at time \text{\normalsize $t=t_1$} warped with the DVF directly calculated using the optical flow algorithm (cf Section \ref{subsection:chest image registration}). The second line corresponds to the average for $k \in \{2201, ..., 2400\}$ of the mean cross-correlation between the initial image at time \text{\normalsize $t_k$}, and the initial image at time \text{\normalsize $t_1$} warped with the DVF calculated from the markers' position using the linear correspondence model (Eq. \ref{eq:spatial_lin_reg}), without prediction. Most importantly, the last line corresponds to the mean cross-correlation between the predicted and the initial images.}
\label{tb:cc_results}
\end{table*}

We chose the window size $h = 3$ and the standard deviation $\sigma_w = 0.5$ based on the visual quality of the resulting images, to warp $I(\cdot, t_1)$ using the Nadaraya-Watson estimator (cf Eq. \ref{eq:Nadaraya_Watson}). The position of the tumor on the predicted images is almost the same as on the initial images (Fig. \ref{fig:pred_cor_cc} and Appendix \ref{appendix:predicted_images}). The predicted images are less noisy due to the Gaussian filtering inherent to the warping process. However, some structures like blood vessels may have an unclear or imprecise position, or even be absent in the predicted images, such as the vessel on the bottom left of the tumor in the predicted coronal cross-section of patient 1 at $t=t_{2209}$ (Fig. \ref{fig:pred_cor_cc}). Artifacts consisting of trails of white dots and blurring appeared below the tumor of patient 2. These white trails may have appeared due to an inexact DVF and target voxels with only one antecedent voxel in the initial image at $t_1$. Moreover, points without antecedent voxels appeared for patient 4 (lower right corner of the sagittal cross-section and AIP at the end-of-exhale point), which resulted in voxels impainted in black by default (Figs. \ref{fig:pred_sag_cc}, \ref{fig:pred_sag_AIP}).

The efficiency of the proposed image prediction algorithm is confirmed by the high cross-correlation between the predicted and original images averaged over the test data and the four patients, equal to 0.955 (Table \ref{tb:cc_results}). The cross-correlation $\rho(I,J)$ between two images or vectors I and J is defined by Eq. \ref{eq:cc_def}, where $cov(I,J)$ is the covariance between I and J, and $\sigma(I)$ and $\sigma(J)$ designate respectively the standard deviation of $I$ and $J$. %

\begin{equation}
\rho(I,J) = \frac{cov(I,J)}{\sigma(I) \sigma(J)}
\label{eq:cc_def}
\end{equation} %

We also observe from Table \ref{tb:cc_results} that the step most hampering the image prediction process is not the prediction of the markers' location, but the reconstruction of the entire DVF from the linear correspondence model (cf Eq. \ref{eq:spatial_lin_reg}). We chose a simple correspondence model because it is not the main focus of the study. However, this model can be improved to take into account effects such as hysteresis and phase offset \cite{ehrhardt20134d}. 

\section{Conclusion}

This is the first study of RNNs trained with RTRL for latency compensation in lung cancer radiotherapy, to the extent of our knowledge. RNNs are ANNs that are well suited for time-series prediction and the RTRL online learning method enables the predictor to continuously adapt to changes in the patient breathing patterns. Gradient clipping was performed to minimize the likelihood of a numerical error while continually updating the synaptic weights. The image data used in this study consisted of four patients' temporal series of 10 3D chest CT scan images. Each of them was artificially extended into a series of 2,400 images by simulating the natural drift process while breathing. The sampling time is equal to approximately 400ms. Comparatively, it has been reported that the time delay of radiotherapy treatment systems ranges from 100ms up to 2s. The positions of internal points near the tumor of lung cancer patients, derived from the Lucas-Kanade pyramidal and iterative optical flow algorithm, were predicted with a 400ms response time. The amplitude of the motion of these points varied from 12.0mm to 22.7mm. The RMS error, maximum error, and jitter on the test set were all smaller than the corresponding performance measures given by linear prediction and LMS. In particular, the maximum prediction error given by the RNN trained with RTRL was equal to 1.51mm, which is respectively 16.1\% and 5.0\% lower than the maximum prediction error given by linear prediction and LMS (table \ref{table:RNN_lin_reg_comparison}). In comparison, the maximum error and RMS error resulting from the prediction with the RNN were respectively 4 times and 7 times lower than the same errors resulting from a system without prediction. Furthermore, when performing prediction with the RNN, the maximum tracking error for each patient was below the 2mm threshold suggested by Murphy \cite{murphy2004tracking}. The average calculation time per time step of the RNN was equal to 119.1ms (Dell Intel Core i9-9900K 3.60GHz CPU NVidia GeForce RTX 2080 SUPER GPU 32Gb RAM with Matlab), which is lower than the marker position sampling time, equal to 400ms. Finally, we combined prediction of the position of internal points using the RNN with a linear correspondence model and forward-warping using Nadaraya-Watson non-linear regression to perform 3D chest image prediction. The mean cross-correlation between the initial and predicted images is equal to 0.955 (table \ref{tb:cc_results}), and the overall tumor position in the predicted images appears to be visually correct. 

This research gives valuable insight concerning proper parameter adjustment for maximizing prediction performance with RNNs trained with RTRL in the context of radiotherapy. We performed grid search and found that $q = 250$ hidden units and an initial standard deviation of the synaptic weights equal to $\sigma_{init}^{RNN} = 0.02$ were optimal on the cross-validation set for all patients. These two parameters were the parameters having the largest impact on the prediction error on the cross-validation set. Optimizing $q$ and $\sigma_{init}^{RNN}$ respectively led to a decrease of 56.3\% and 28.4\% in the MAE. The minimum prediction error is a convex function of $\sigma_{init}^{RNN}$ and decreases when $q$ increases. However, the general variation of that prediction error as a function of the SHL was different from patient to patient, hence the optimal value of the SHL also varied among the patients. 

This is also the first detailed study of the pyramidal iterative Lucas-Kanade optical flow algorithm applied to lung CT scan images providing details about the precise influence of each parameter on the registration error. The pyramidal iterative Lucas-Kanade optical flow is a classical DIR algorithm, but proper parameter adjustment, which is key to ensure high accuracy of the deformation field, had not been discussed in detail in previous studies related to registration of CT scan images, to the extent of our knowledge. In this work, we provided experimental results about parameter selection for performance optimization. $\sigma_{LK}$ and $n_{layers}$ were the parameters having the most significant impact on the registration performance. Carefully selecting $\sigma_{LK}$ and $n_{layers}$ respectively led to a decrease in the minimum registration error of 31.3\% and 36.2\%. On our dataset, we found optimal results with $\sigma_{LK}=2.0$ and $n_{layers} = 3$ or $n_{layers} = 4$. It was confirmed that using only one layer was hampering the registration performance, which correlates with the observations in \cite{xu2008lung, zhang2008use}. This is due to the high amplitude of the lung motion in the CT scan images used, relative to the image resolution. 

This study is a step forward in lung radiotherapy because better compensation of the treatment system latency will entail more accurate tumor targeting. In addition, it will enable reducing the radiation margin around the tumor for compensation of unexpected motion, leading thus to a decrease in the irradiation of surrounding healthy tissue, and in turn to less undesirable side effects such as radiation pneumonitis. Further research about prediction of more irregular breathing patterns will bring more insights into the capabilities of online learning methods such as RTRL to adapt to unexpected temporal events. This study of the RTRL algorithm could be further enriched by investigating the variation of the prediction performance as a function of the prediction horizon. Finally, we could extend this work by tracking more accessible surrogate signals such as points on the diaphragm recorded using kV imaging, or external markers placed on the skin \cite{ehrhardt20134d}.

Some of the Matlab source code used in this research is available online under the 3-clause BSD license \cite{pohl_michel_2020_4011750, pohl_michel_2021_4548433, pohl_michel_2021_4452210}. The data of patients 2,3 and 4 has been retrieved from the 4D-Lung data collection \cite{Lung4DCTdatabase, hugo2017longitudinal, balik2013evaluation, roman2012interfractional} in the Cancer Imaging Archive open-access database\footnotemark \cite{clark2013cancer}.

\footnotetext{Patients 2, 3, and 4 correspond respectively to the patients' IDs 111\_HM10395, 117\_HM10395, and 118\_HM10395 in the 4D-Lung collection. The sequences used were acquired respectively on December 16th, 1999, December 4th, 2000, and December 7th, 2000.}


\bibliographystyle{model1-num-names}

\section*{Conflicts of interest statement}
The authors declare no conflict of interest.

\section*{Acknowledgments}
The authors thank Dr. Stephen Wells (Department of Nuclear Engineering and Management, The University of Tokyo) who proofread the article.

\bibliography{cas-refs}

\onecolumn
\appendix

\section{Appendix : Displacement vector fields obtained with the pyramidal and iterative Lucas-Kanade optical flow algorithm}%
\label{appendix:DVF_display}
\begin{figure}[h!]%
    \captionsetup[subfigure]{labelformat=empty}
    \centering
    \subfloat[Patient 1 \text{\normalsize $t=t_{2209}$}]{{\includegraphics[height=3cm]{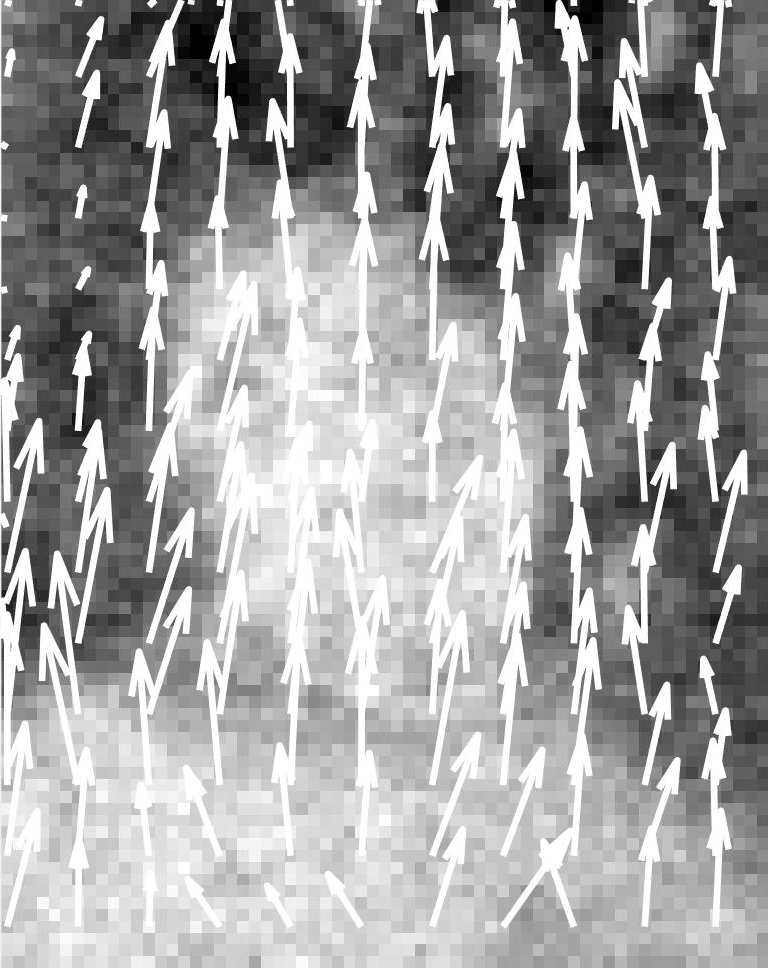} }}%
	\qquad  
    \subfloat[Patient 2 \newline \text{\normalsize $t=t_{2209}$}]{{\includegraphics[height=3cm]{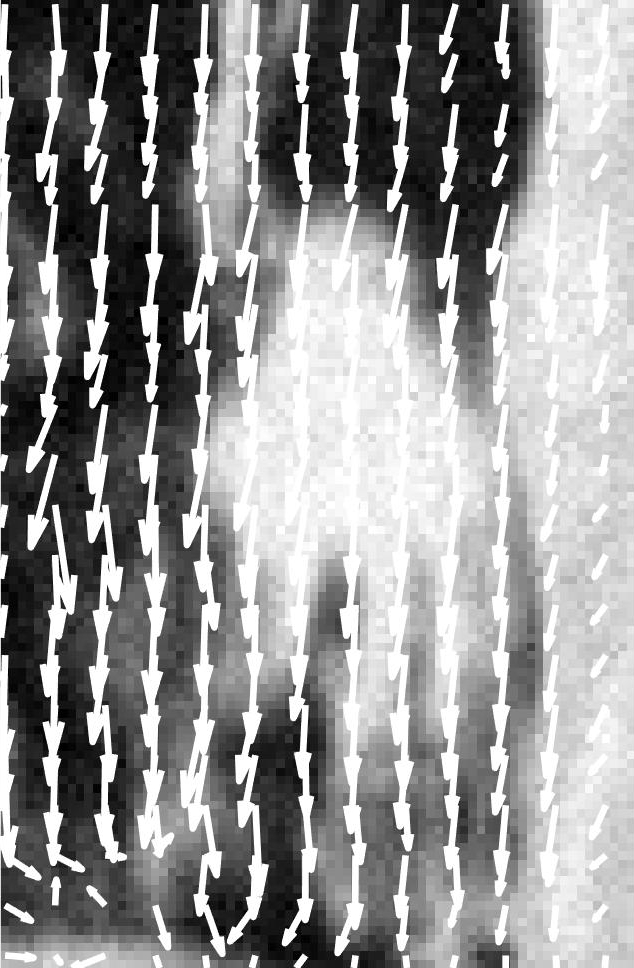} }}%
	\qquad     
    \subfloat[Patient 3 \newline \text{\normalsize $t=t_{2209}$}]{{\includegraphics[height=3cm]{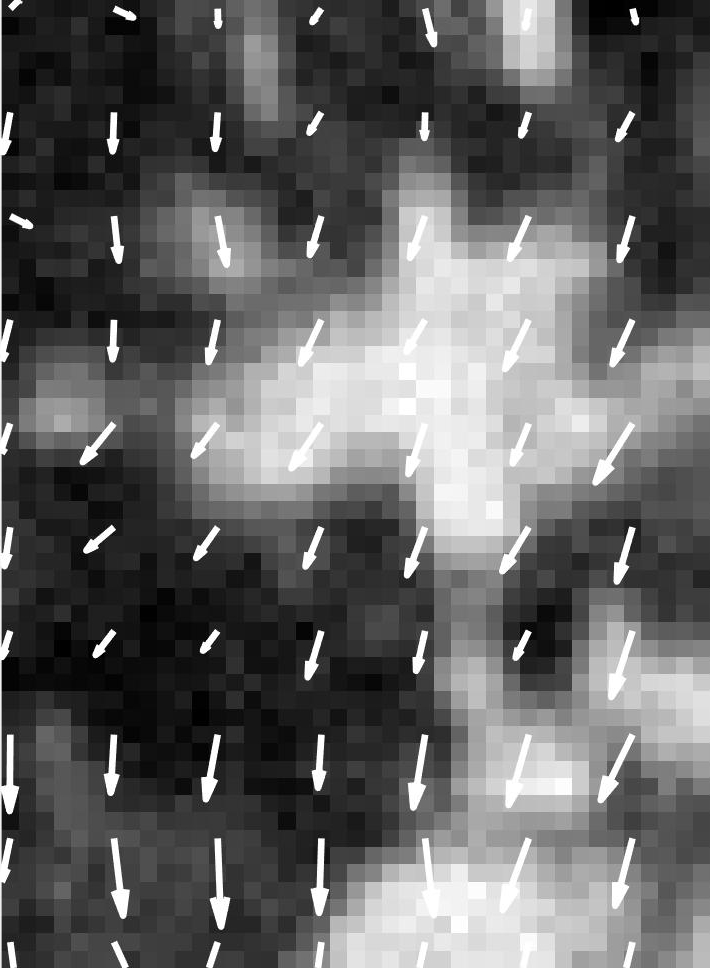} }}%
	\qquad      
    \subfloat[Patient 4 \text{\normalsize $t=t_{2209}$}]{{\includegraphics[height=3cm]{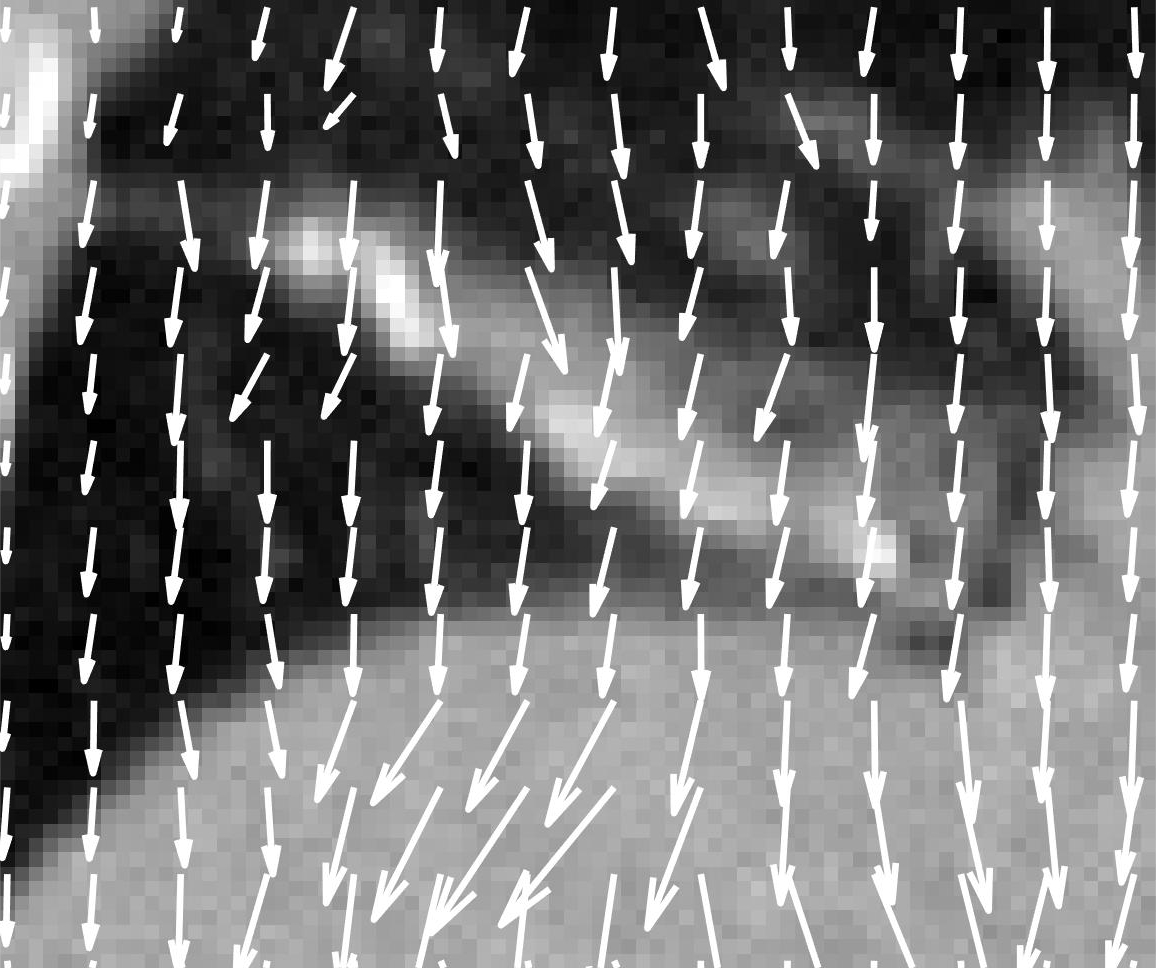} }}%
  
    \subfloat[Patient 1 \text{\normalsize $t=t_{2374}$}]{{\includegraphics[height=3cm]{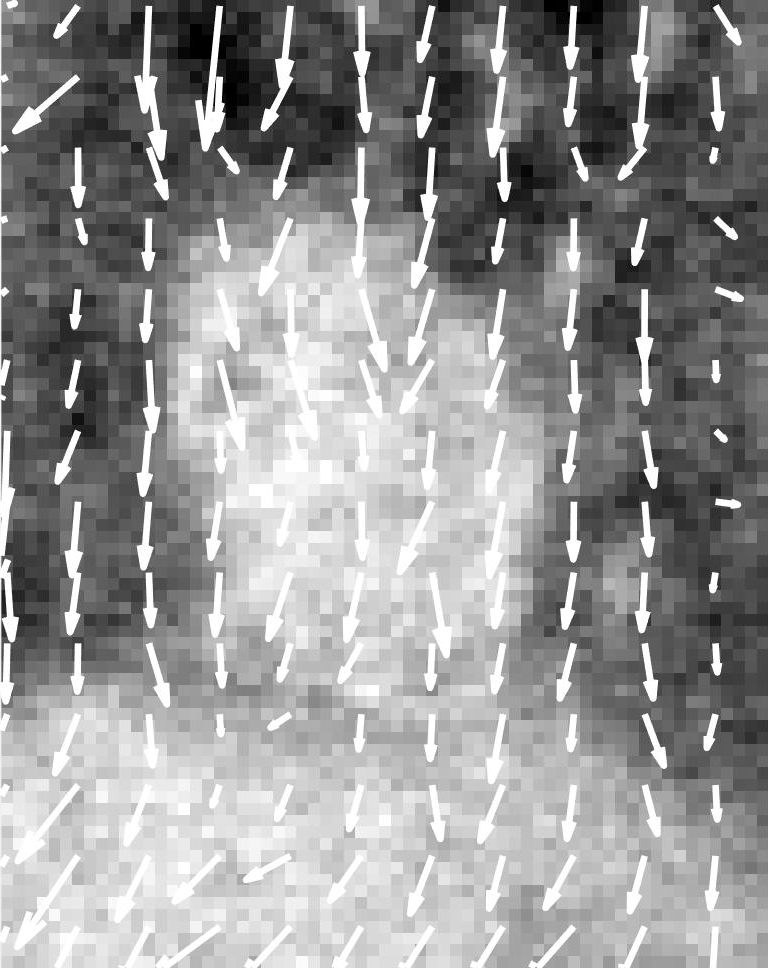} }}%
	\qquad      
    \subfloat[Patient 2 \newline \text{\normalsize $t=t_{2374}$}]{{\includegraphics[height=3cm]{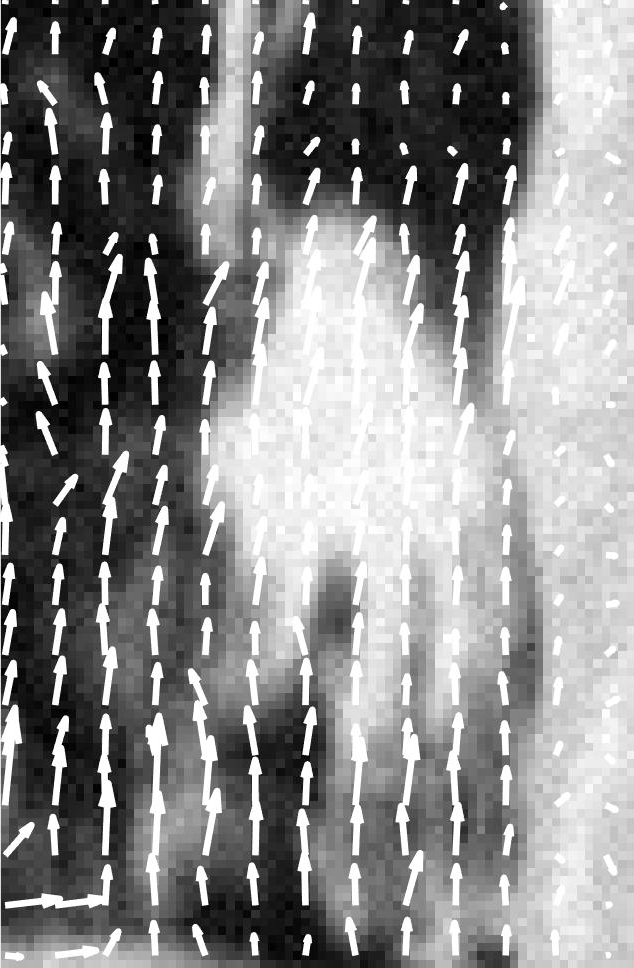} }}%
	\qquad     
    \subfloat[Patient 3 \newline \text{\normalsize $t=t_{2374}$}]{{\includegraphics[height=3cm]{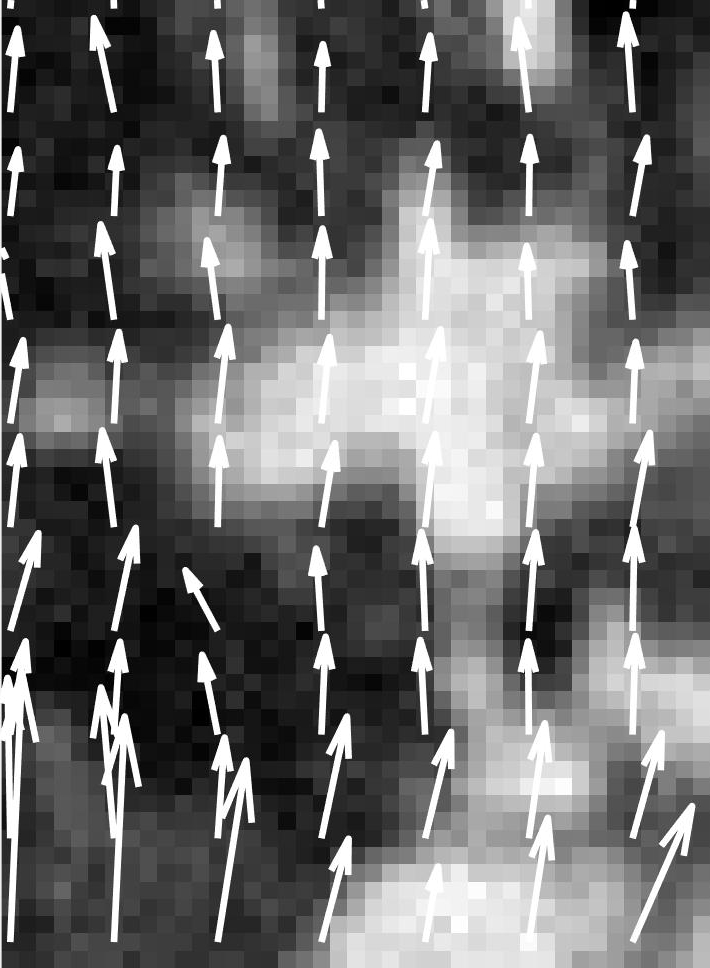} }}%
	\qquad      
    \subfloat[Patient 4 \text{\normalsize $t=t_{2374}$}]{{\includegraphics[height=3cm]{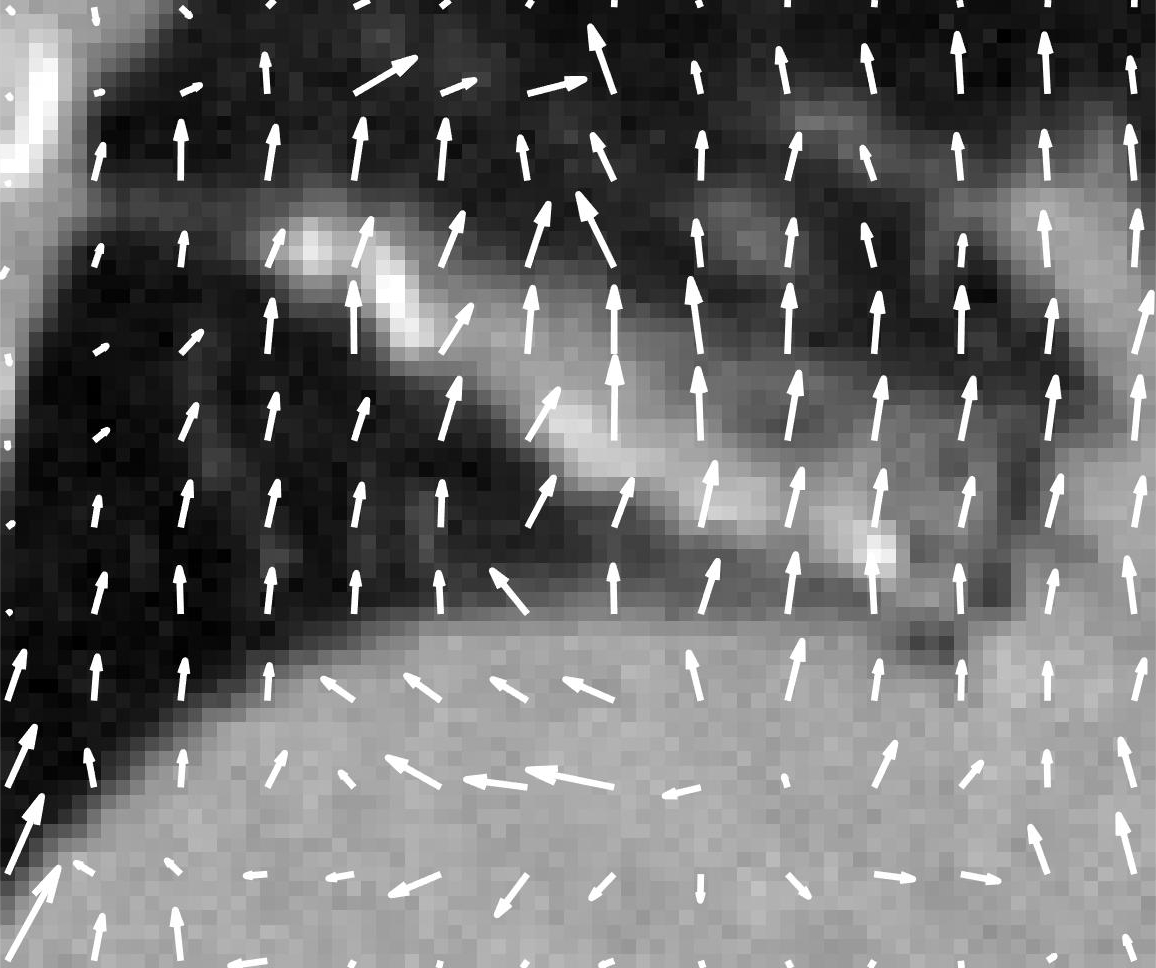} }}%
    \caption{Displacement vector field in the ROI for each patient at \text{\normalsize $t=t_{2209}$} (end of expiration for patient 1 and end of inspiration for the other patients) and \text{\normalsize $t=t_{2374}$} (opposite case) projected in a coronal plane (same coordinates as in Fig. \ref{fig:org_im}). The corresponding coronal cross-section at \text{\normalsize $t=t_1$} is displayed in the background. The origins of each of the displayed 2D displacement vectors are separated from each other by 6 voxels.}
\end{figure}%

\section{Appendix : Trajectories of the selected internal points}%
\label{appendix:traj_display}
\begin{figure}[h!]%
    \captionsetup[subfigure]{labelformat=empty}
    \centering
    \subfloat[Patient 1 sagittal]{{\includegraphics[height=3cm]{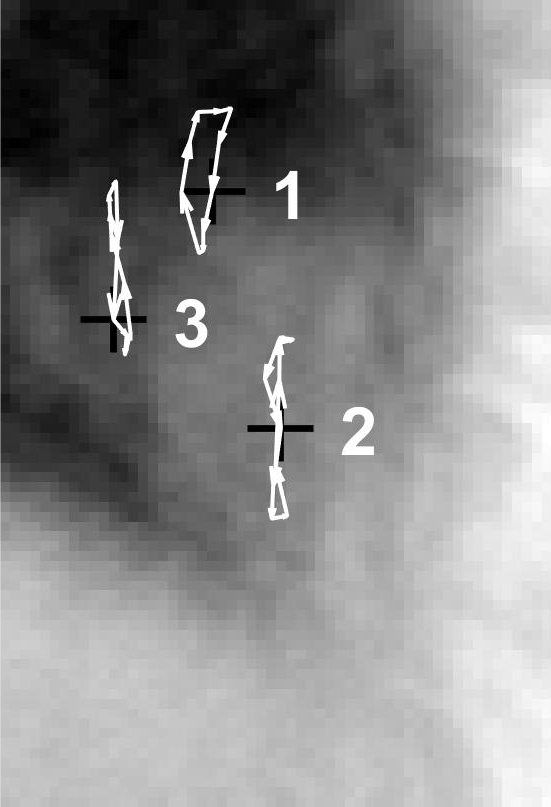} }}%
    \qquad
    \subfloat[Patient 2 sagittal]{{\includegraphics[height=3cm]{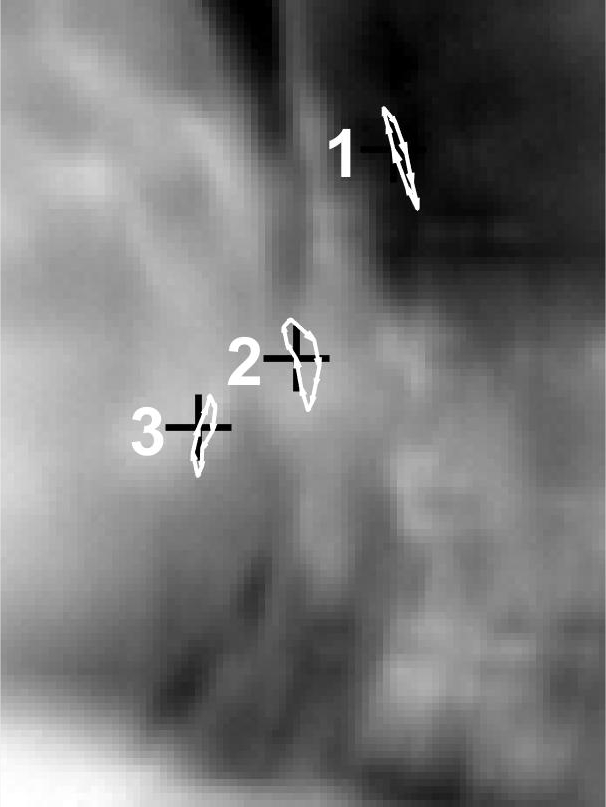} }}%
    \qquad  
    \subfloat[Patient 3 sagittal]{{\includegraphics[height=3cm]{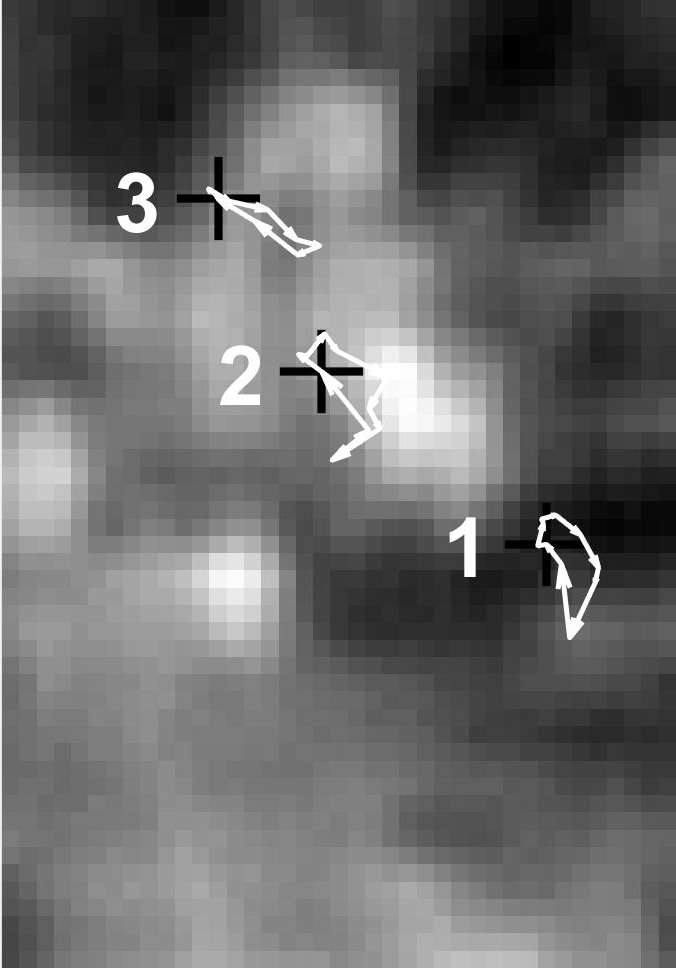} }}%
	\qquad    
    \subfloat[Patient 4 sagittal]{{\includegraphics[height=3cm]{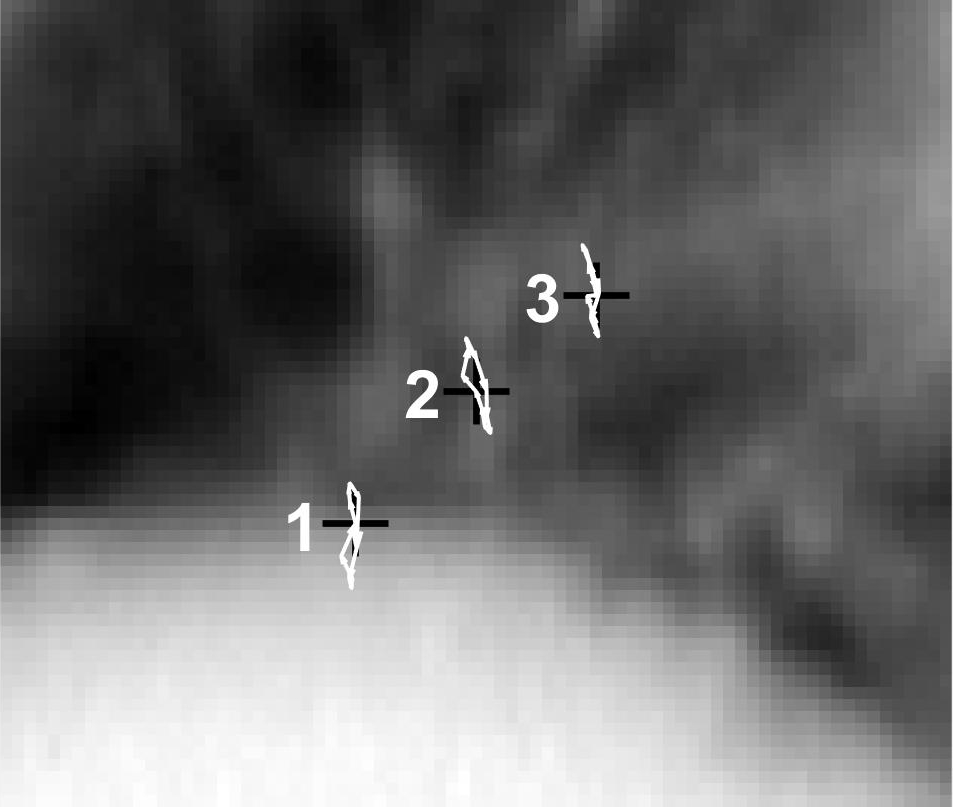} }}%
  
    \subfloat[Patient 1 coronal]{{\includegraphics[height=3cm]{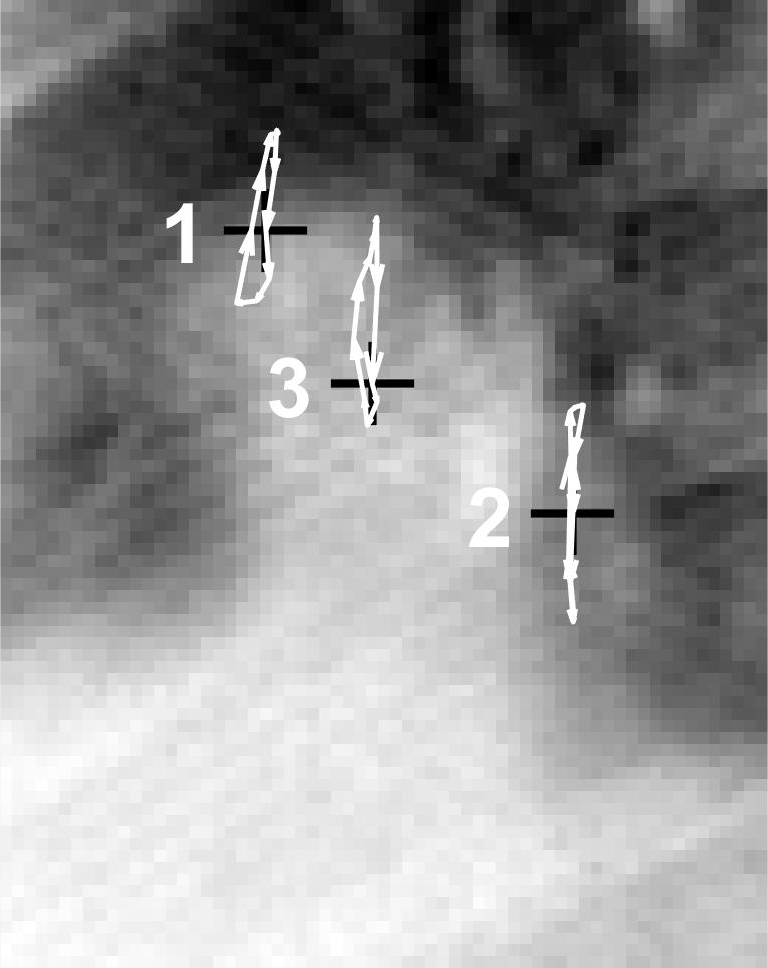} }}%
	\qquad    
    \subfloat[Patient 2 coronal]{{\includegraphics[height=3cm]{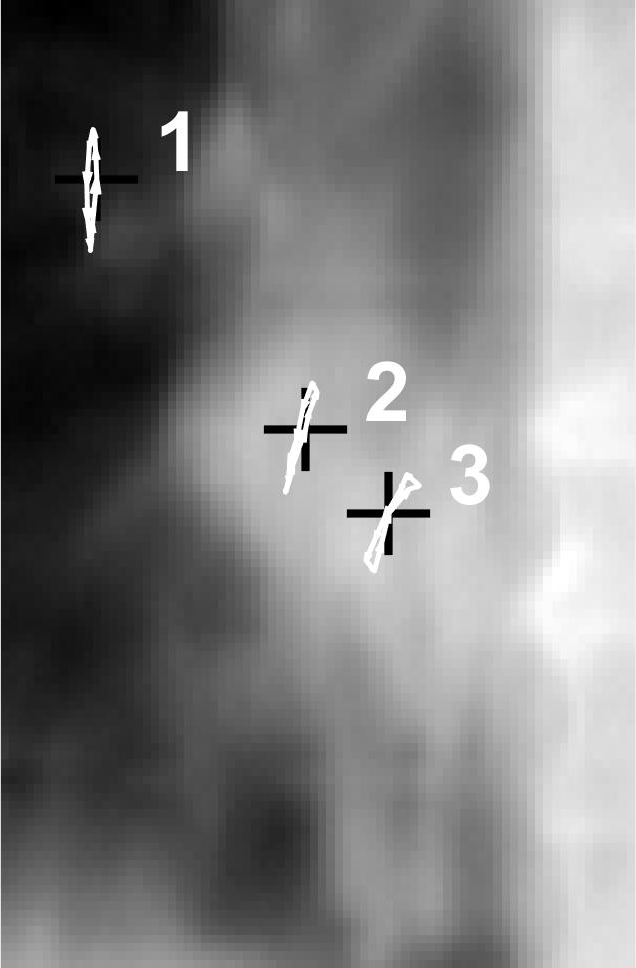} }}%
	\qquad     
    \subfloat[Patient 3 coronal]{{\includegraphics[height=3cm]{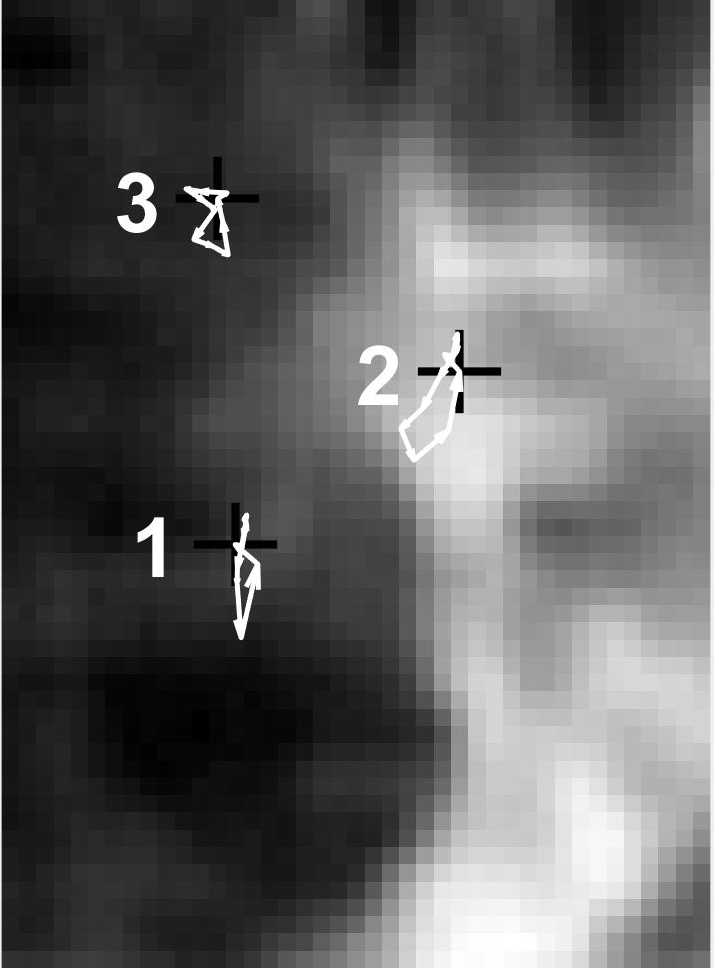} }}%
	\qquad      
    \subfloat[Patient 4 coronal]{{\includegraphics[height=3cm]{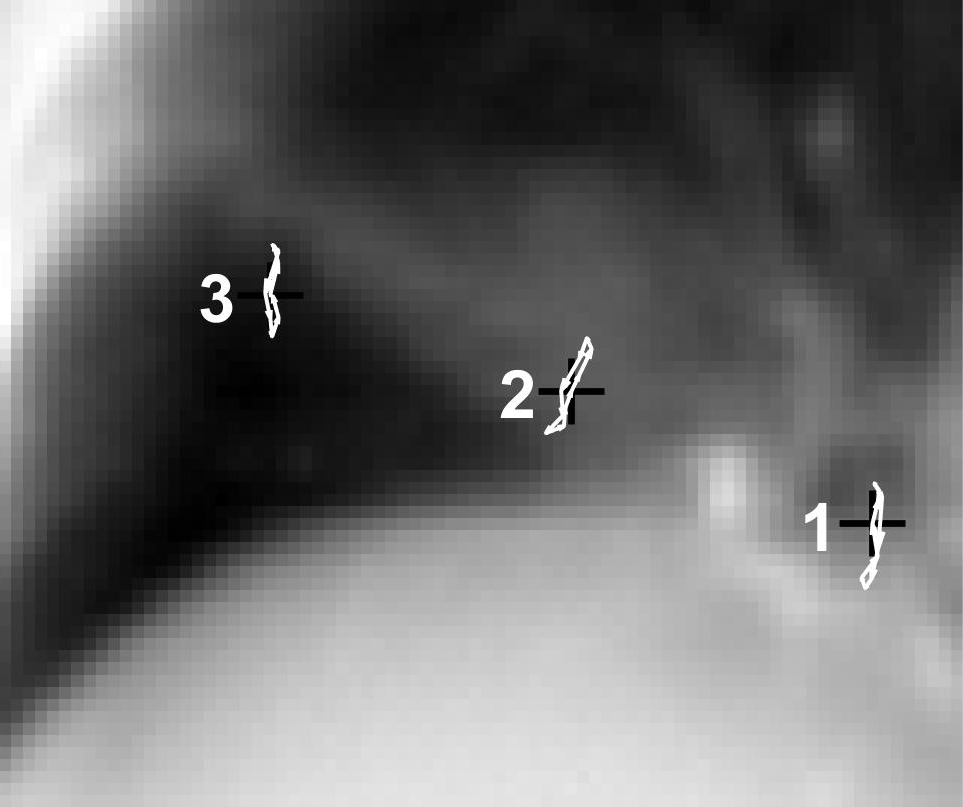} }}%
    \caption{Trajectories of the internal points between \text{\normalsize $t=t_1$} and \text{\normalsize $t=t_{10}$} for each patient, calculated using the pyramidal Lucas-Kanade optical flow algorithm and displayed on top of the average intensity projection (AIP) of the ROI at \text{\normalsize $t=t_1$}. The position of these internal points at \text{\normalsize $t=t_1$} is denoted by a black cross marker.}
\end{figure}%

\newpage 

\section{Appendix : Motion of the markers of patient 3}%
\label{appendix:entire_motion_dataset_patient3}
\begin{figure}[h!]%
    \centering
    \includegraphics[height=3.1cm]{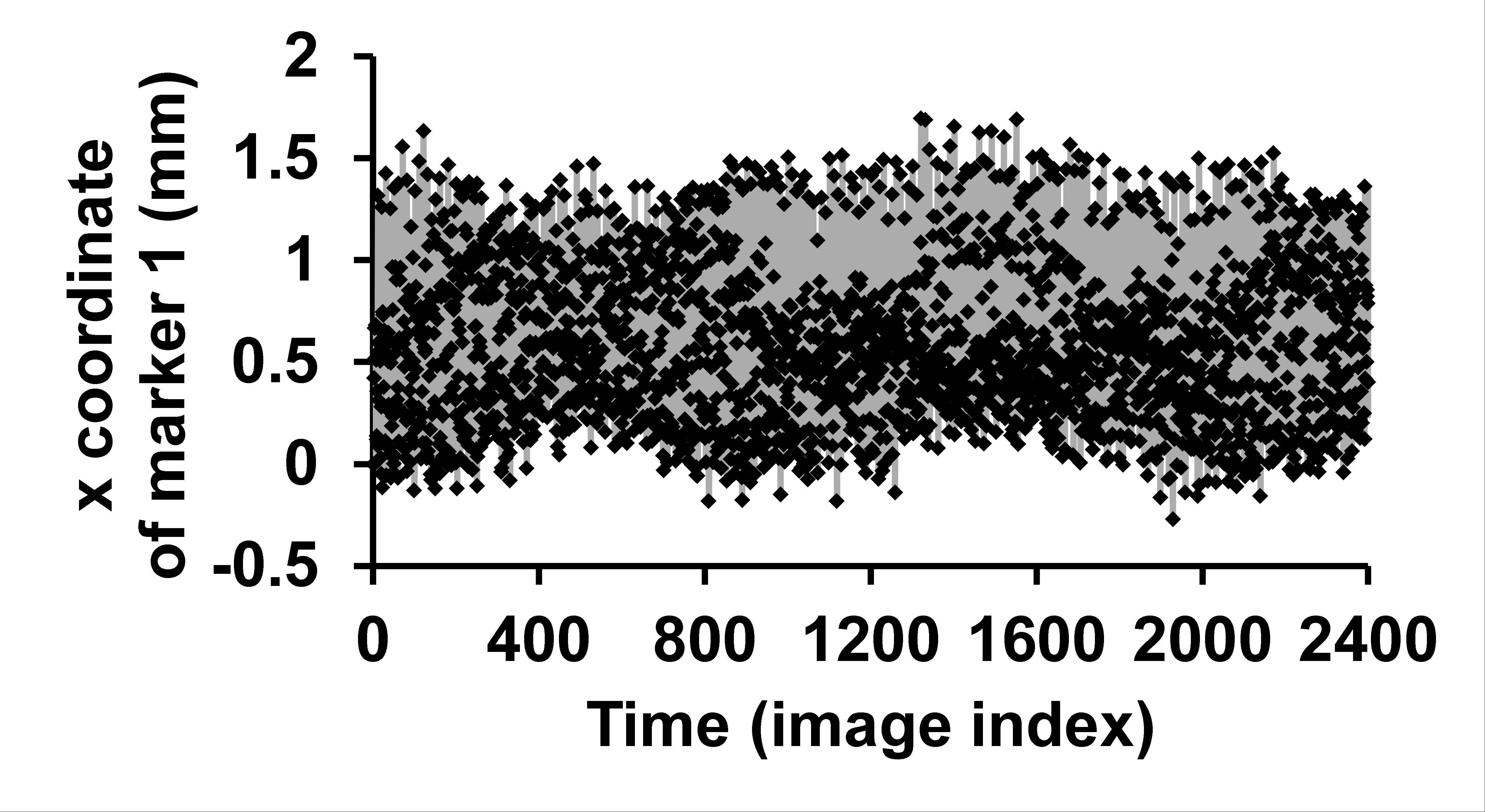}
    \includegraphics[height=3.1cm]{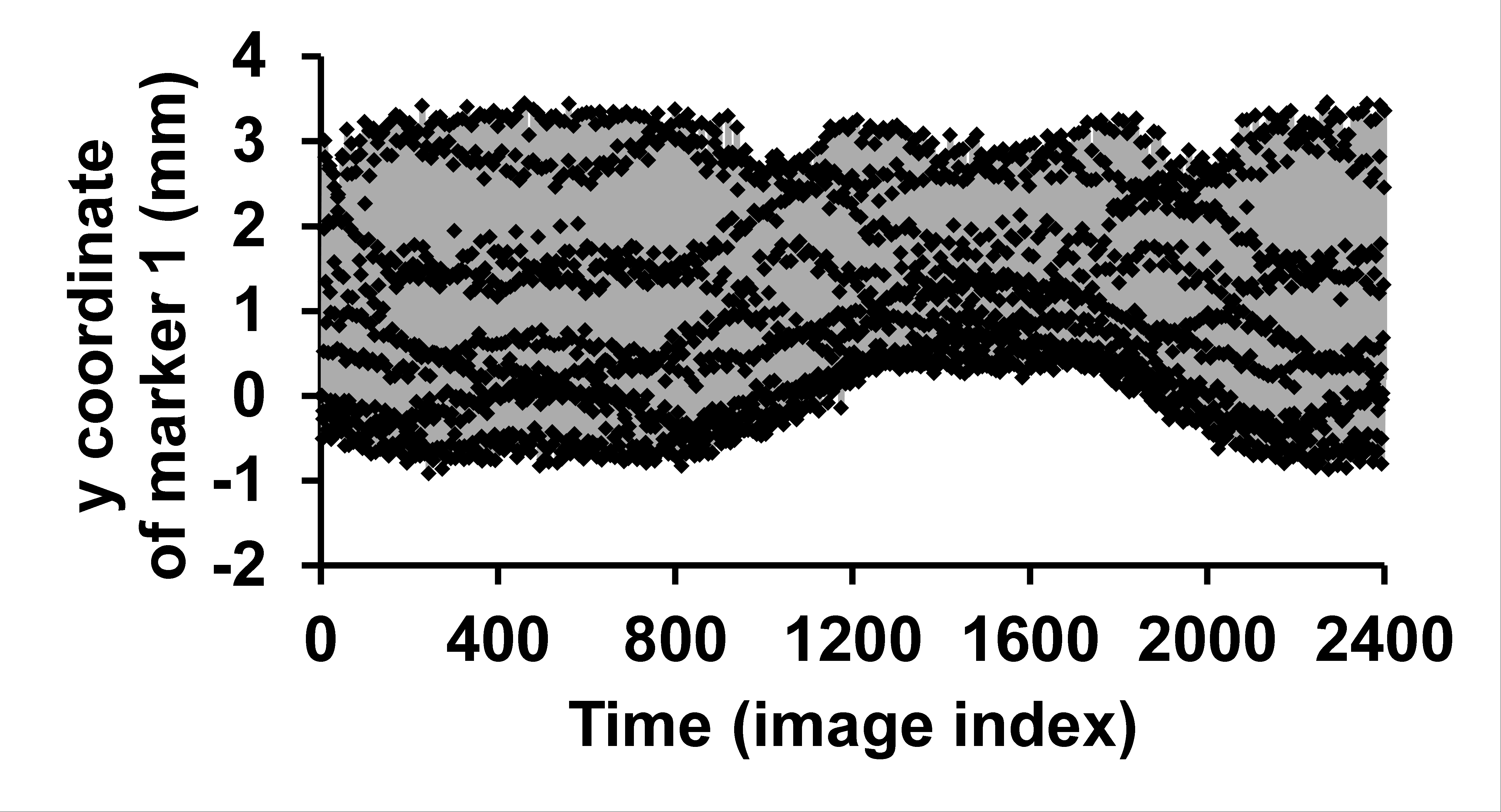}
    \includegraphics[height=3.1cm]{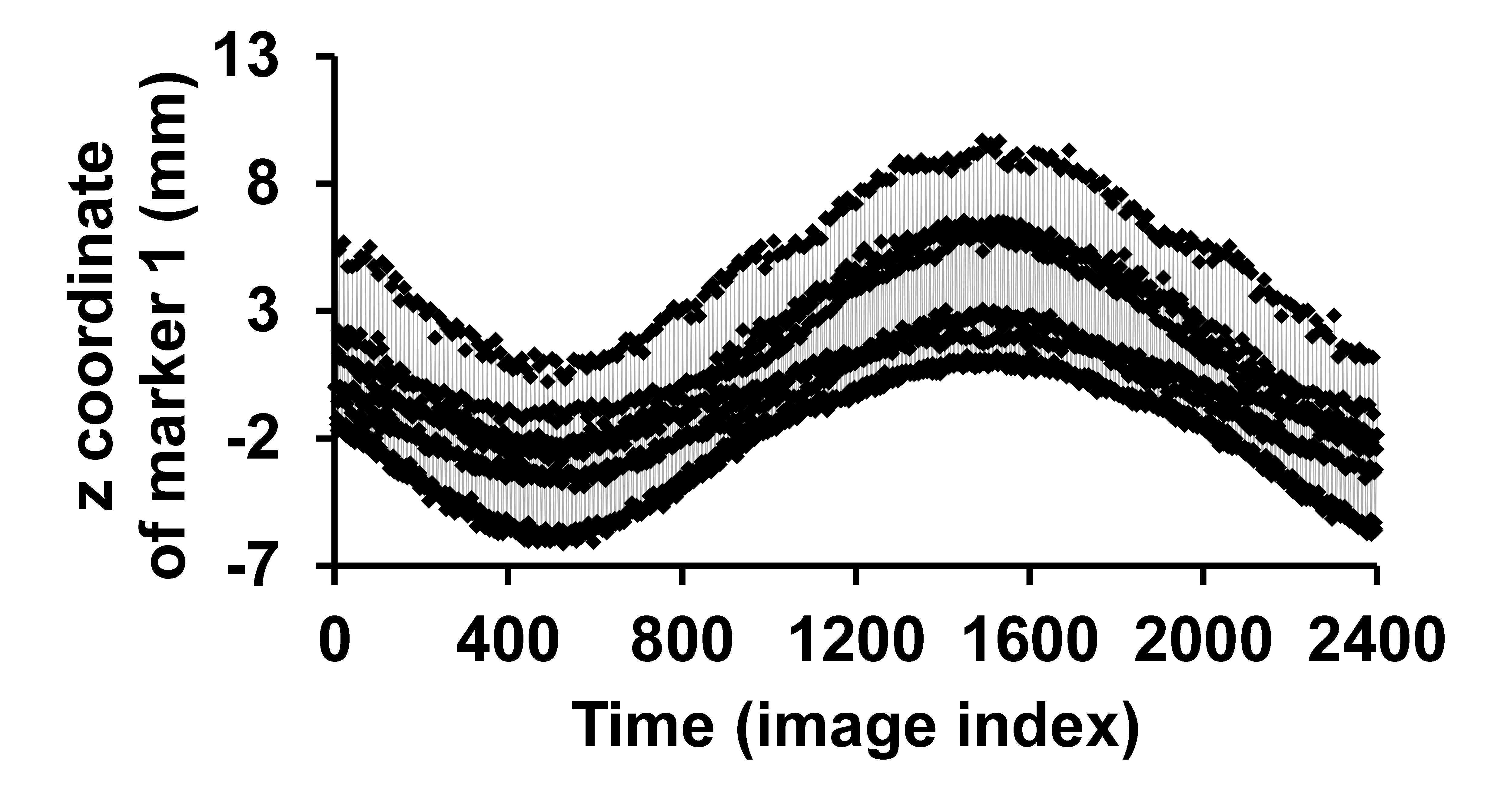}
    \quad
    \includegraphics[height=3.1cm]{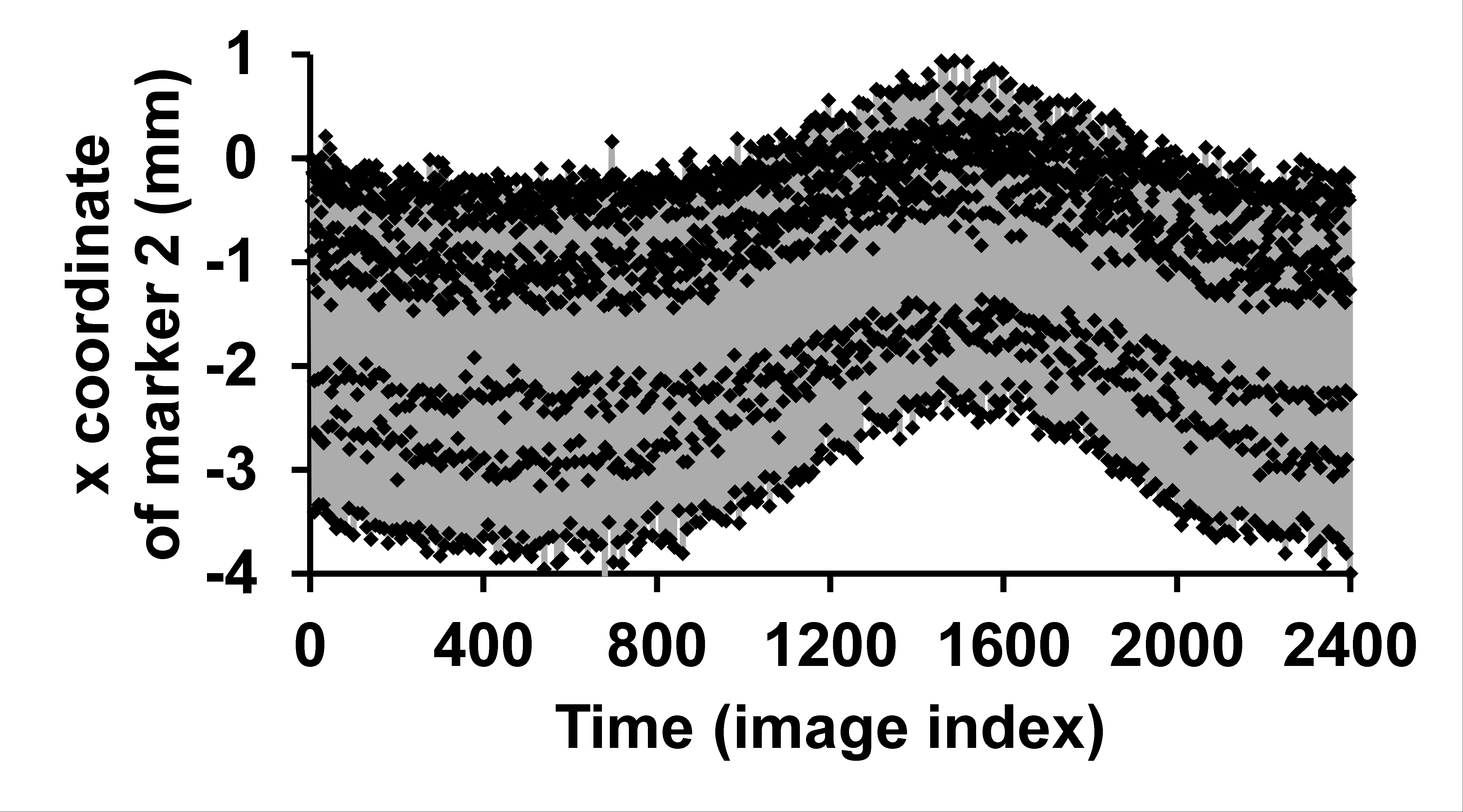}
    \includegraphics[height=3.1cm]{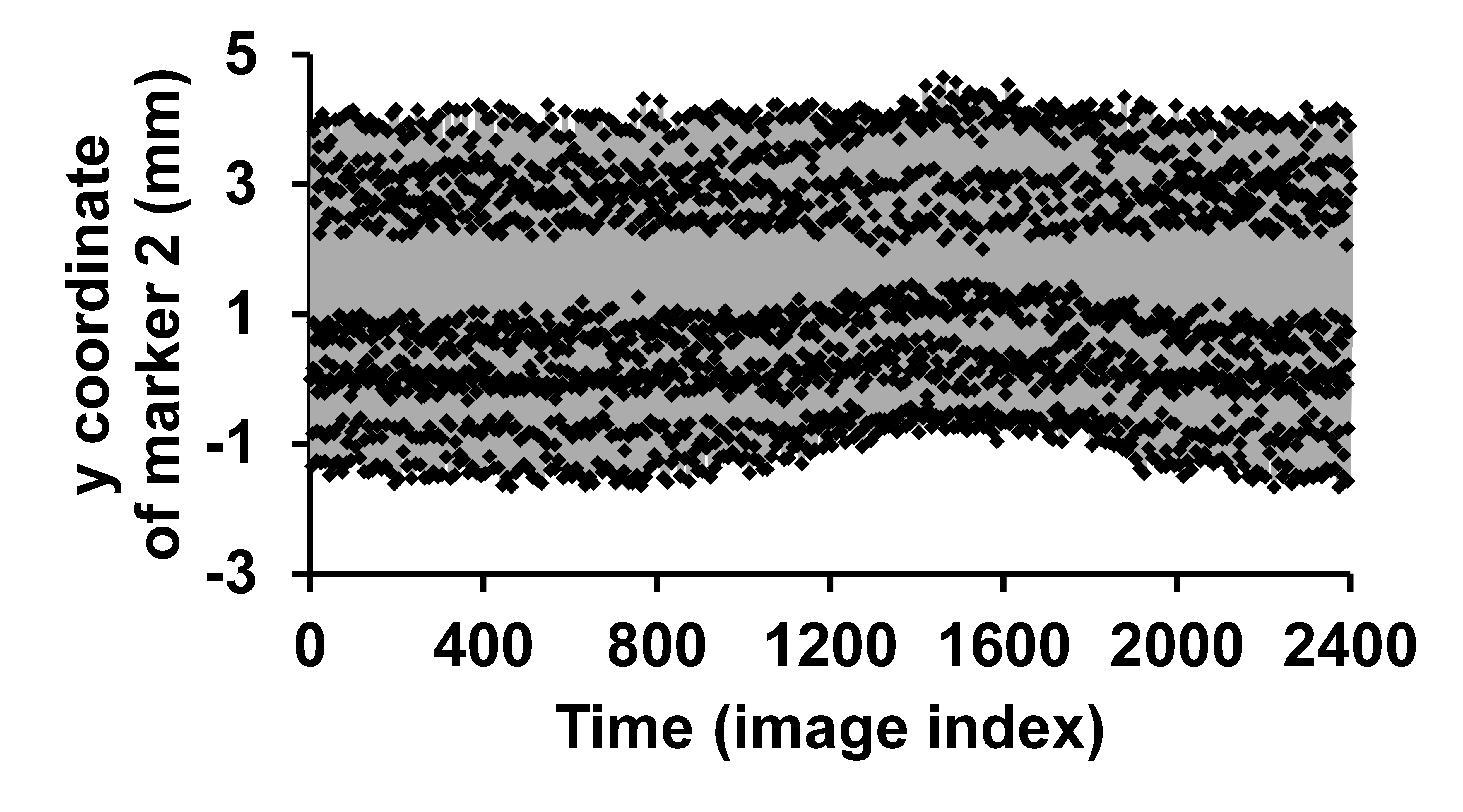}
    \includegraphics[height=3.1cm]{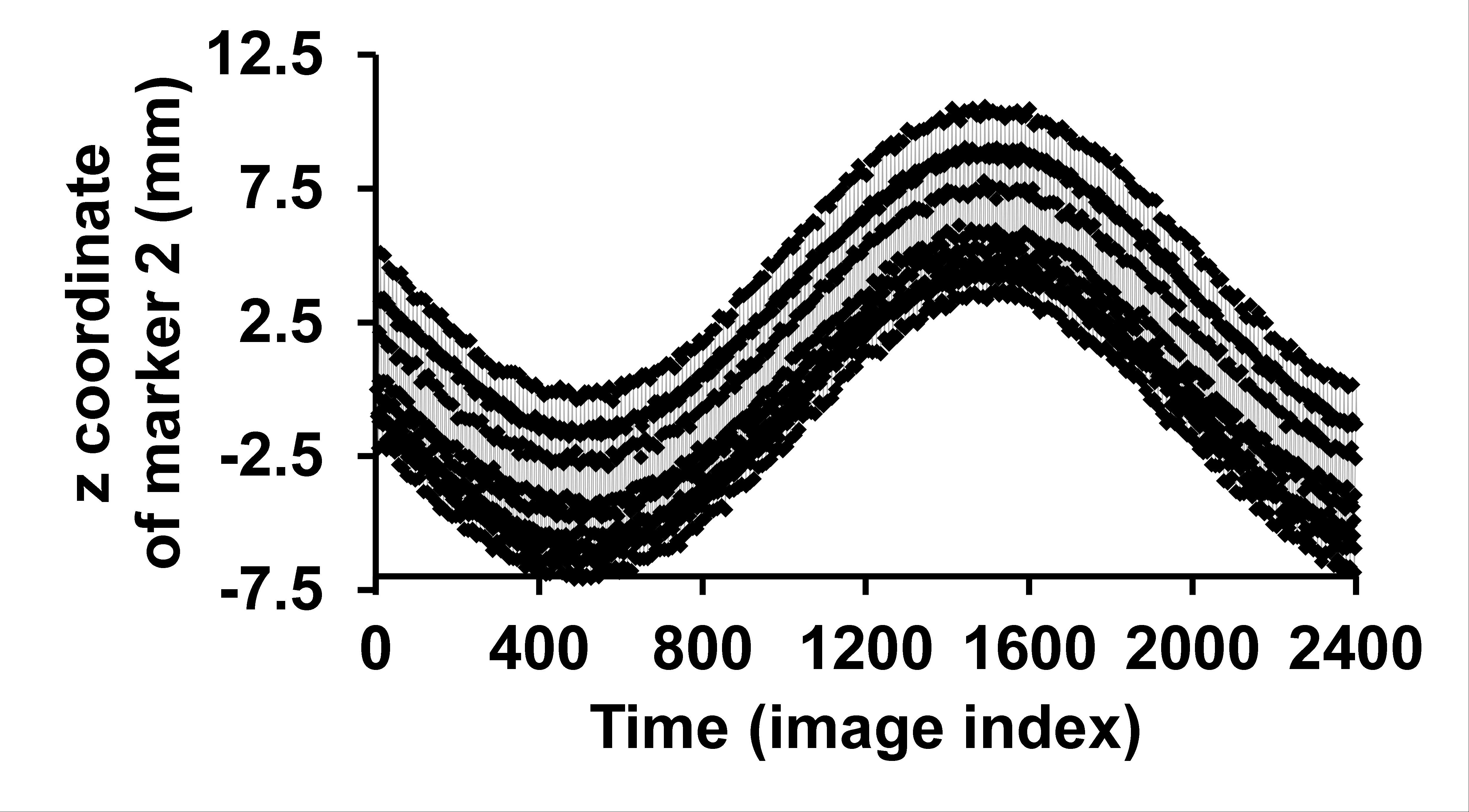}
    \quad
    \includegraphics[height=3.1cm]{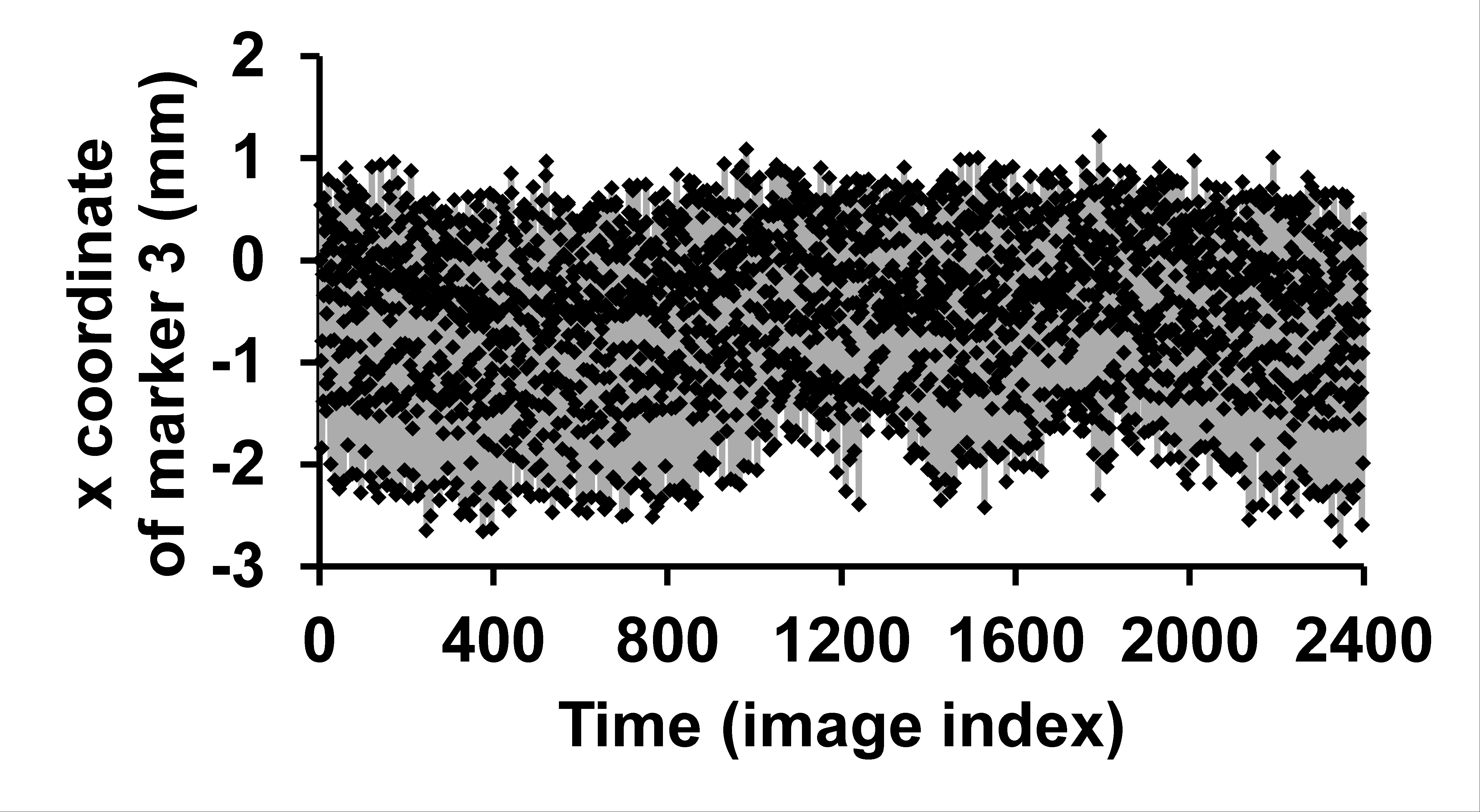}
    \includegraphics[height=3.1cm]{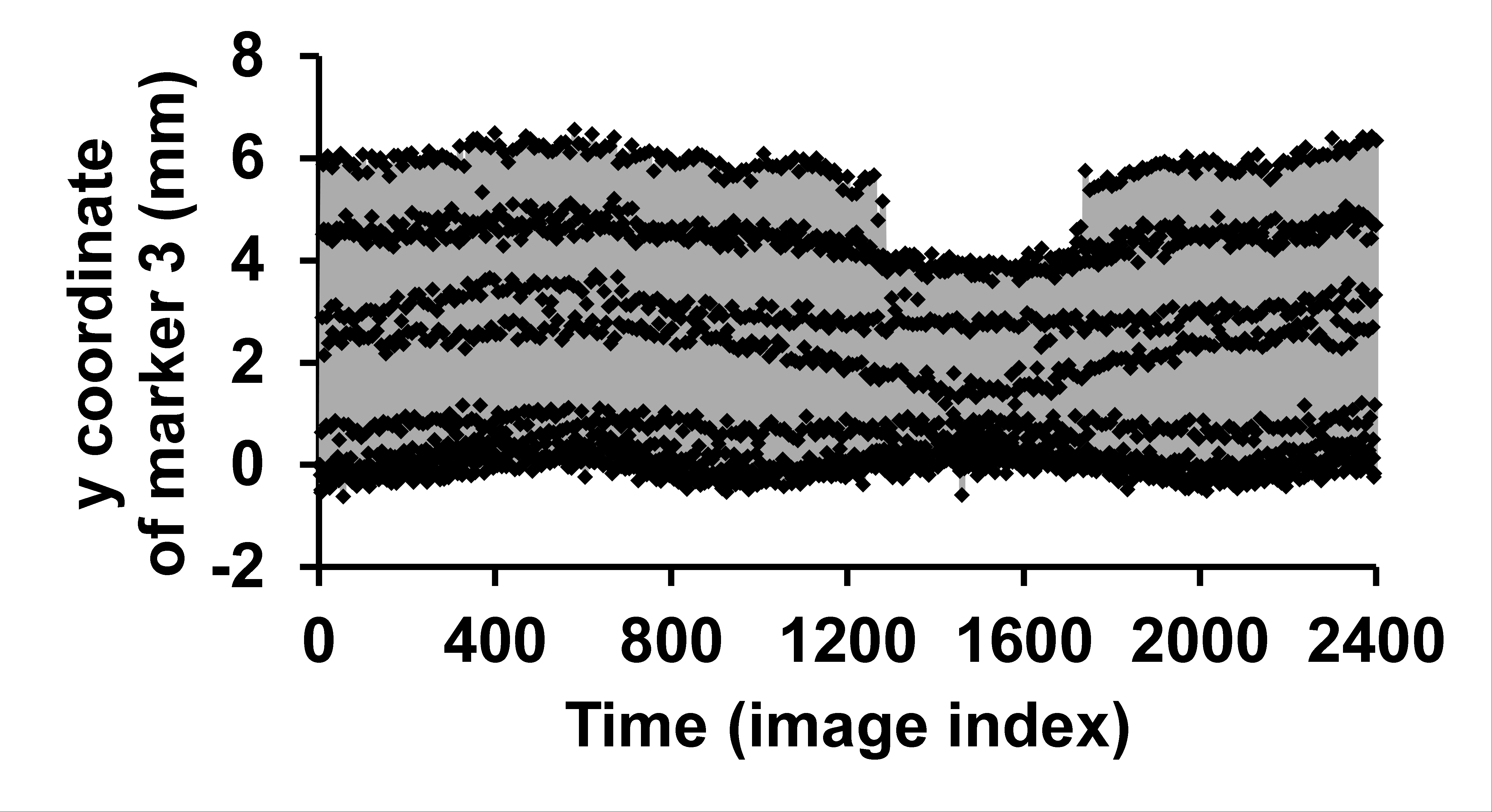}
    \includegraphics[height=3.1cm]{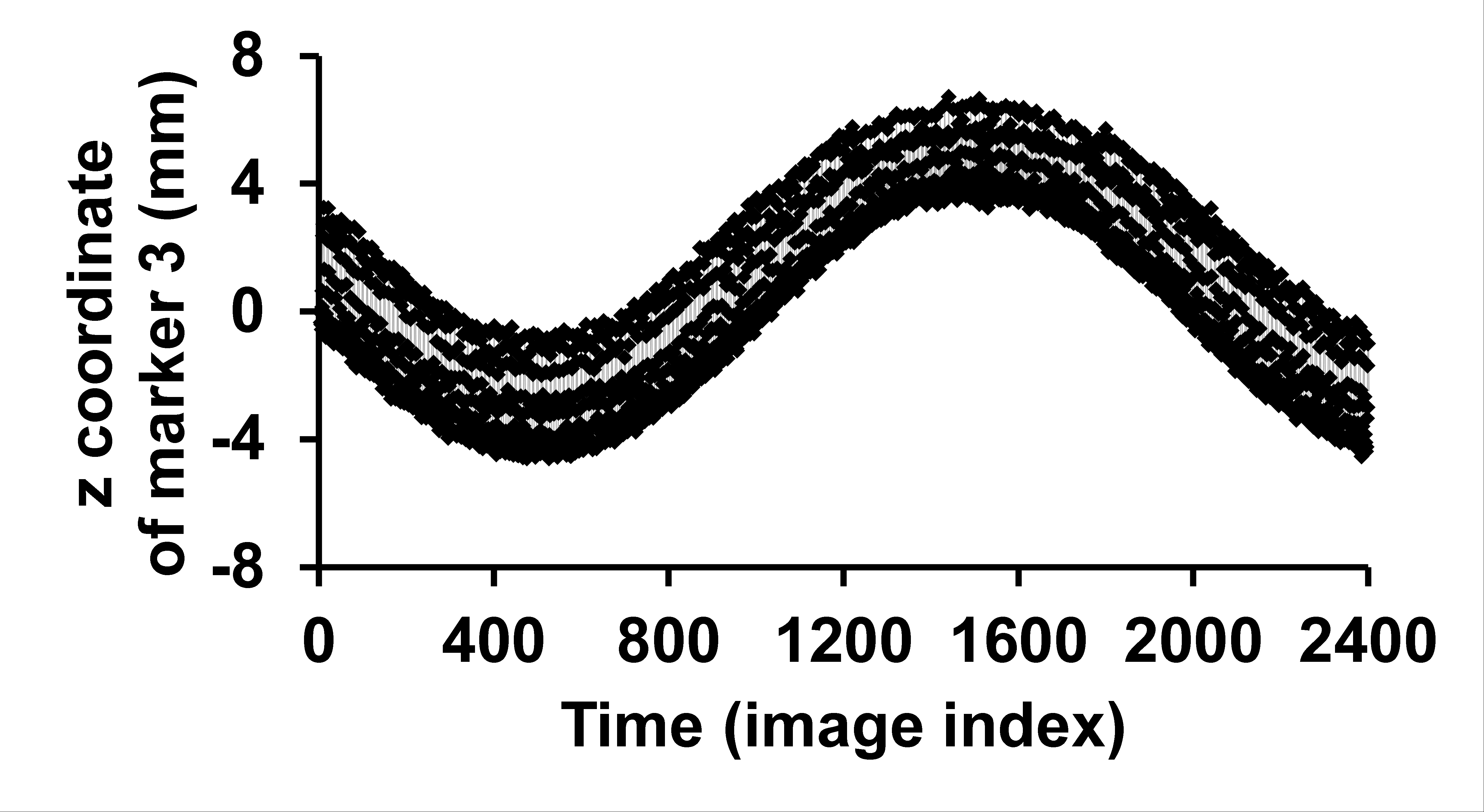}       
    \caption{Motion of the markers of patient 3. The dot at time t corresponds to the signal sampled at time t.  The axes are the same as in Fig. \ref{fig:traj}. The data is divided into 3 sets, namely the training set, between \text{\normalsize $t=t_{1}$} and \text{\normalsize $t=t_{2000}$}, the cross-validation set, between \text{\normalsize $t=t_{2001}$} and \text{\normalsize $t=t_{2200}$}, and the test set, between \text{\normalsize $t=t_{2201}$} and \text{\normalsize $t=t_{2400}$}.}%
\end{figure}%

\section{Appendix : Predicted images}%
\label{appendix:predicted_images}
\begin{figure}[h!]%
    \centering
    \captionsetup[subfigure]{justification=centering} 
    \captionsetup[subfigure]{labelformat=empty}
    \subfloat[Patient 1 \\* predicted \\* \text{\normalsize $t=t_{2209}$}]{{\includegraphics[height=2.8cm]{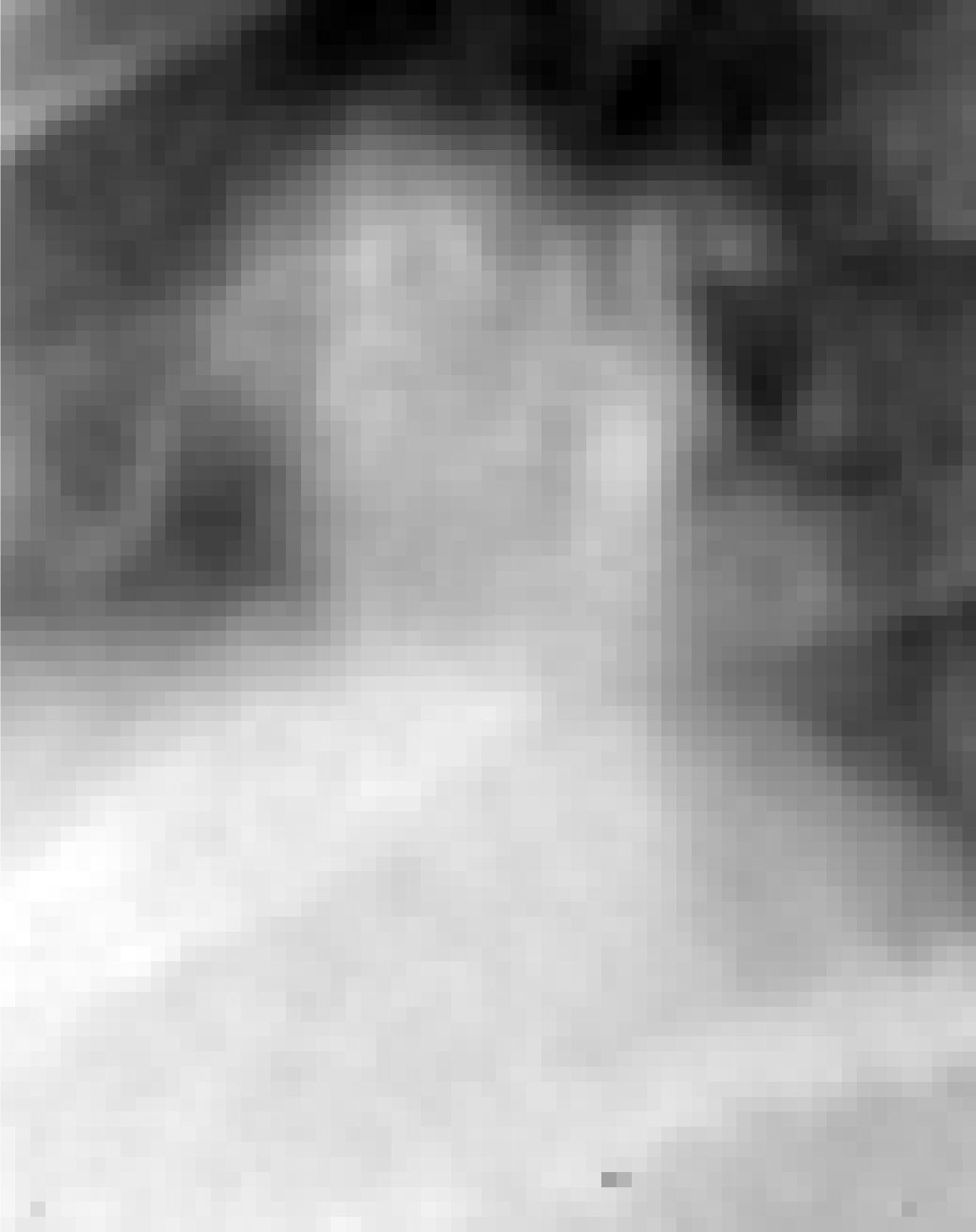} }}%
    \subfloat[Patient 1 \\* original \\* \text{\normalsize $t=t_{2209}$}]{{\includegraphics[height=2.8cm]{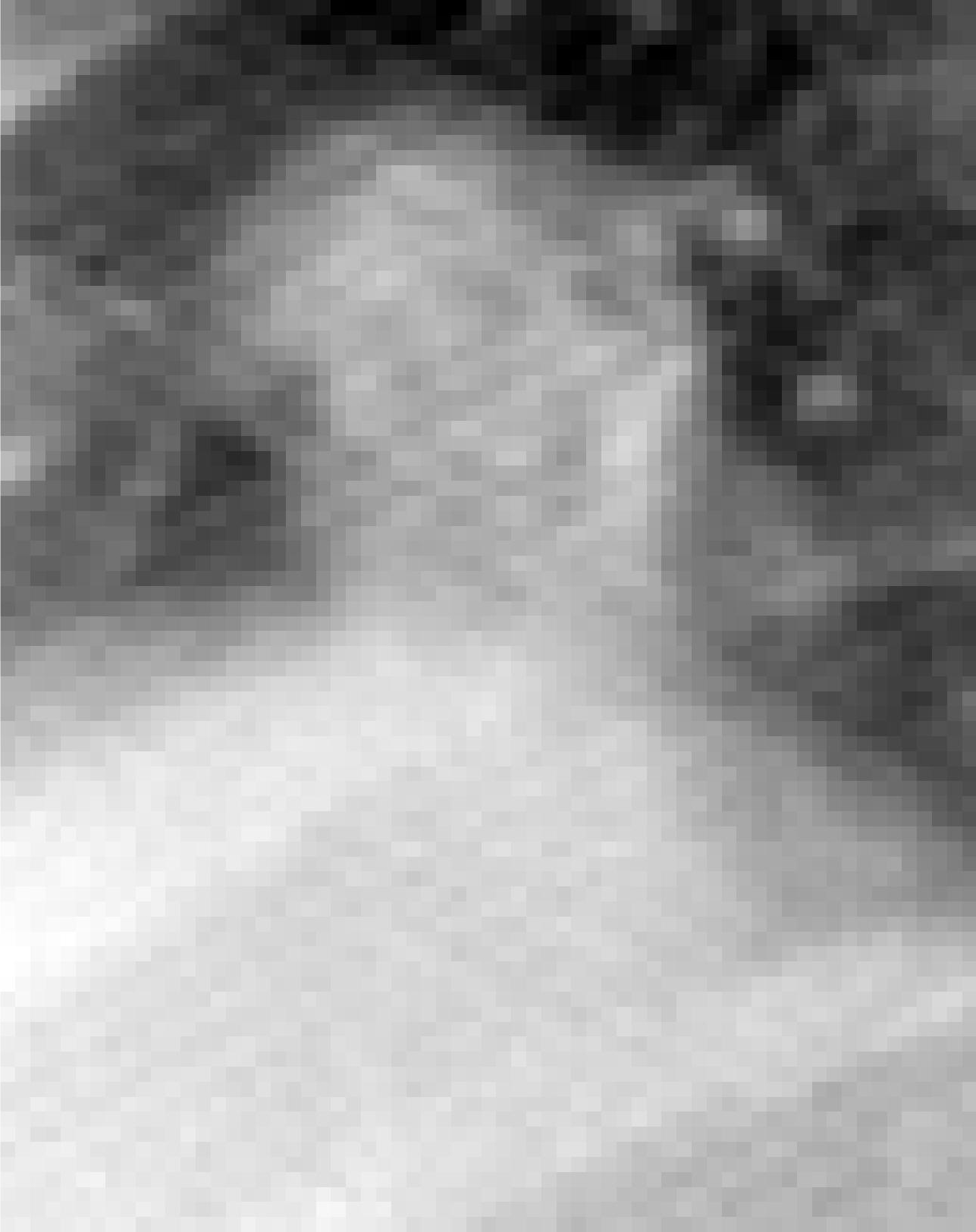} }}%
    \subfloat[Patient 1 \\* predicted \\* \text{\normalsize $t=t_{2374}$}]{{\includegraphics[height=2.8cm]{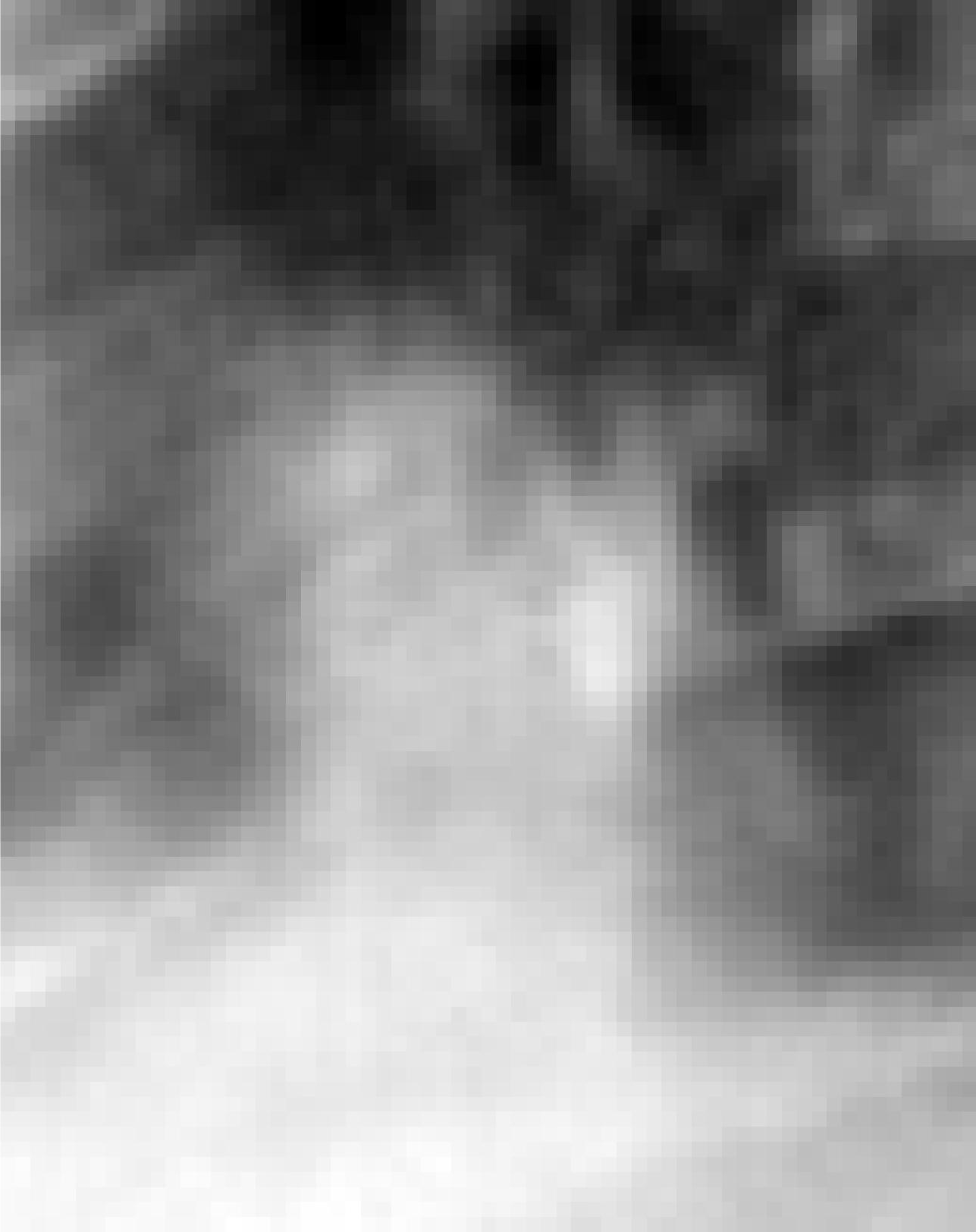} }}%
    \subfloat[Patient 1 \\* original \\* \text{\normalsize $t=t_{2374}$}]{{\includegraphics[height=2.8cm]{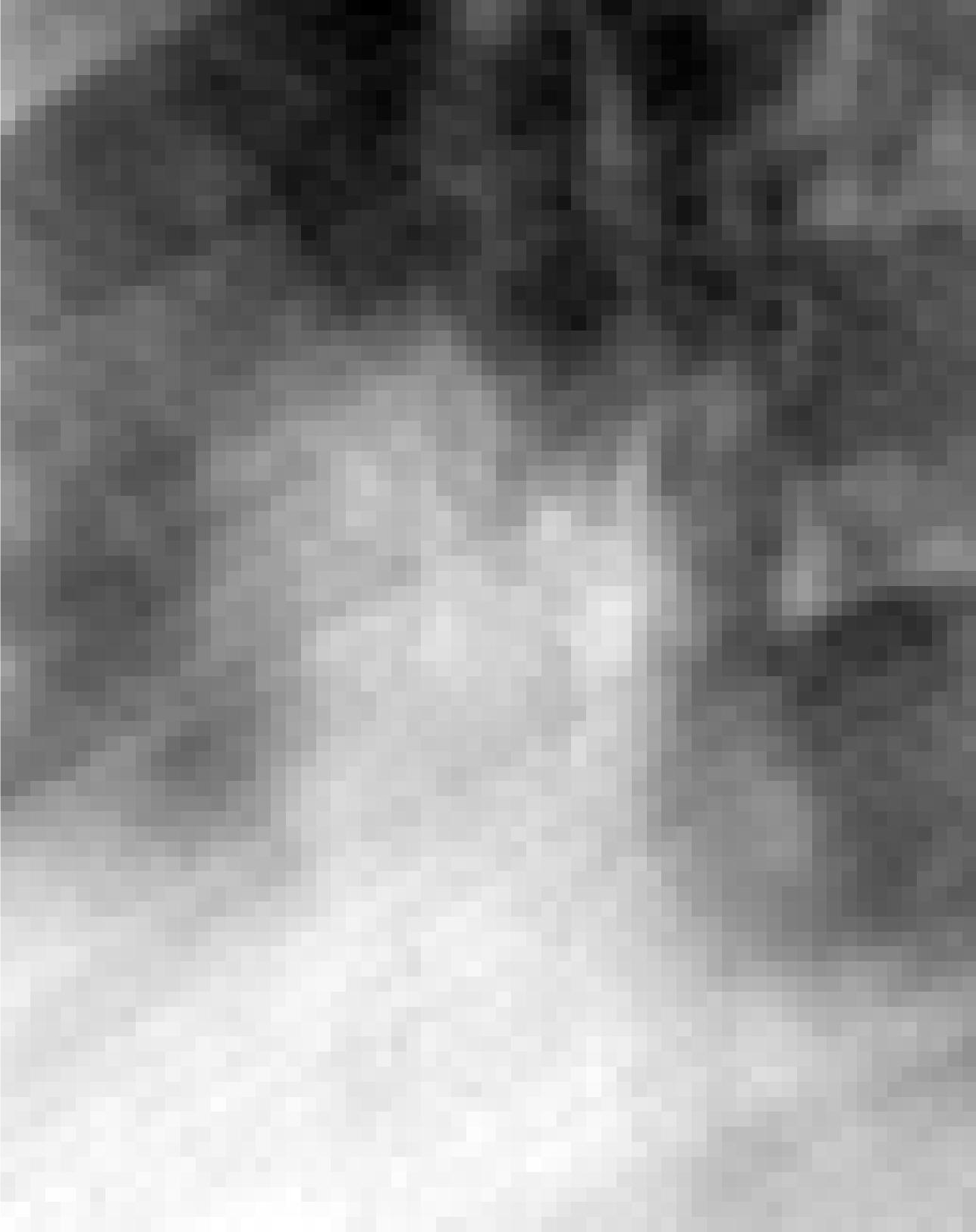} }}%
    \subfloat[Patient 2 \\* predicted \\* \text{\normalsize $t=t_{2209}$}]{{\includegraphics[height=2.8cm]{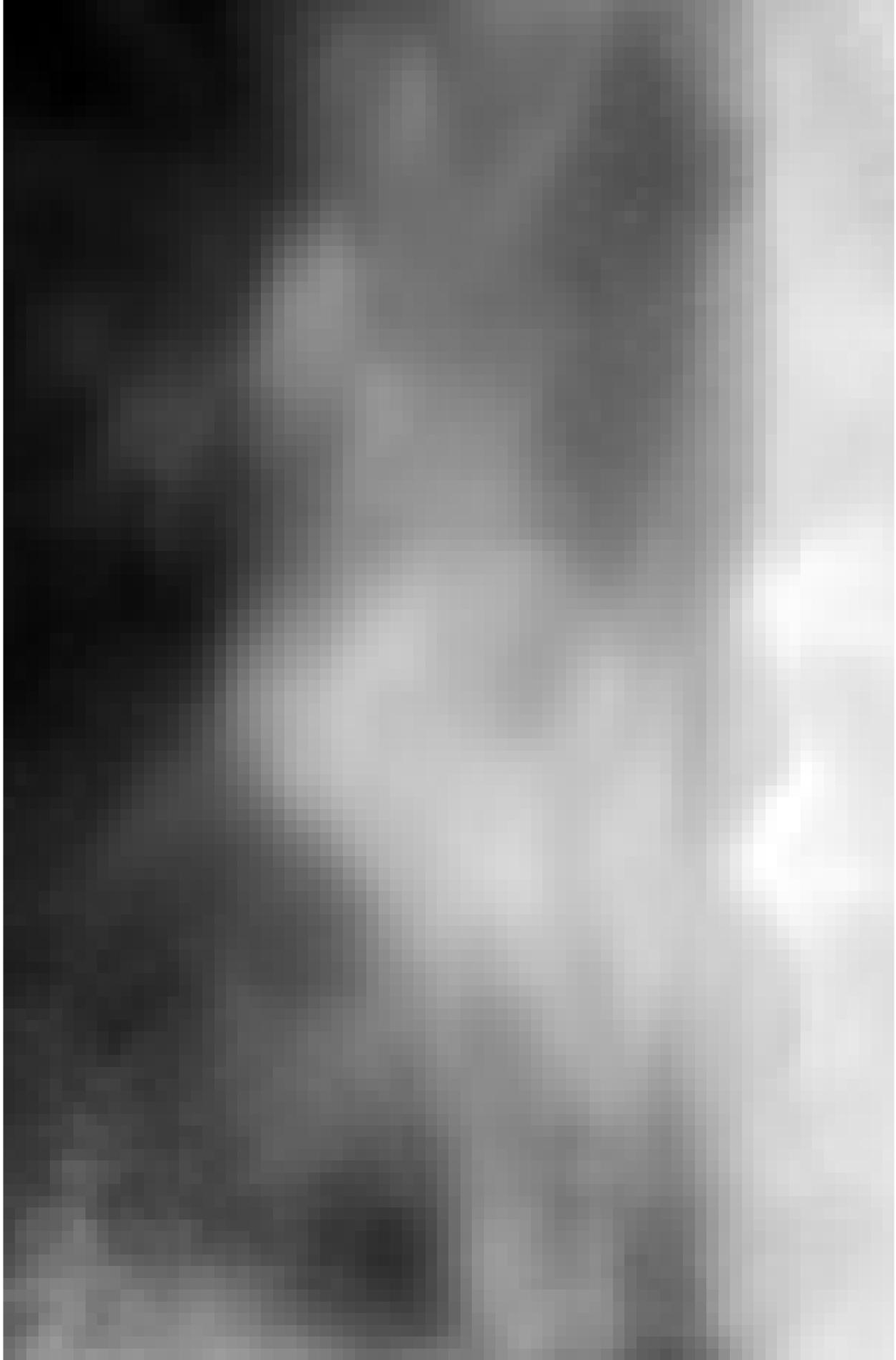} }}%
    \subfloat[Patient 2 \\* original \\* \text{\normalsize $t=t_{2209}$}]{{\includegraphics[height=2.8cm]{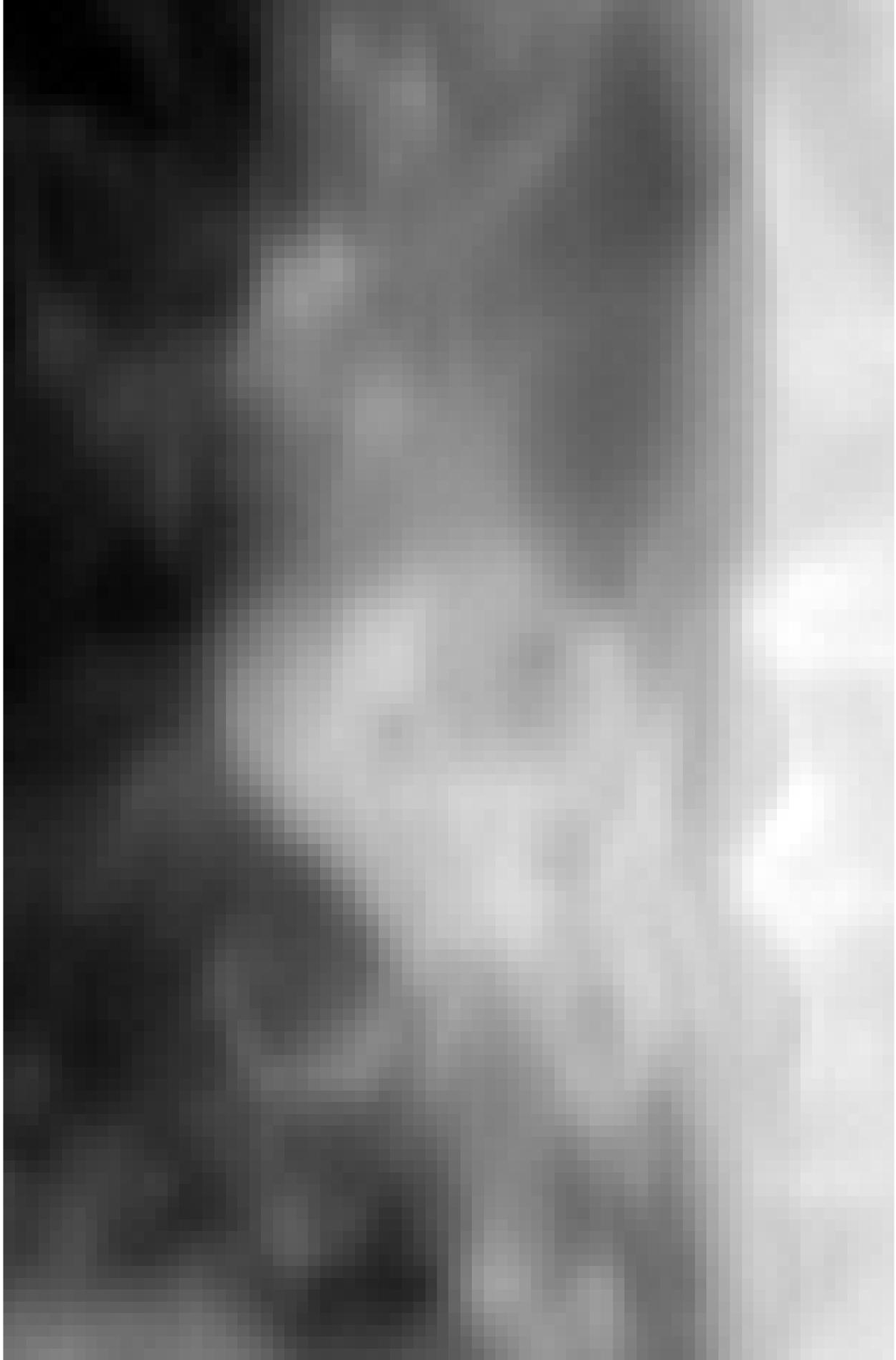} }}%
    \subfloat[Patient 2 \\* predicted \\* \text{\normalsize $t=t_{2374}$}]{{\includegraphics[height=2.8cm]{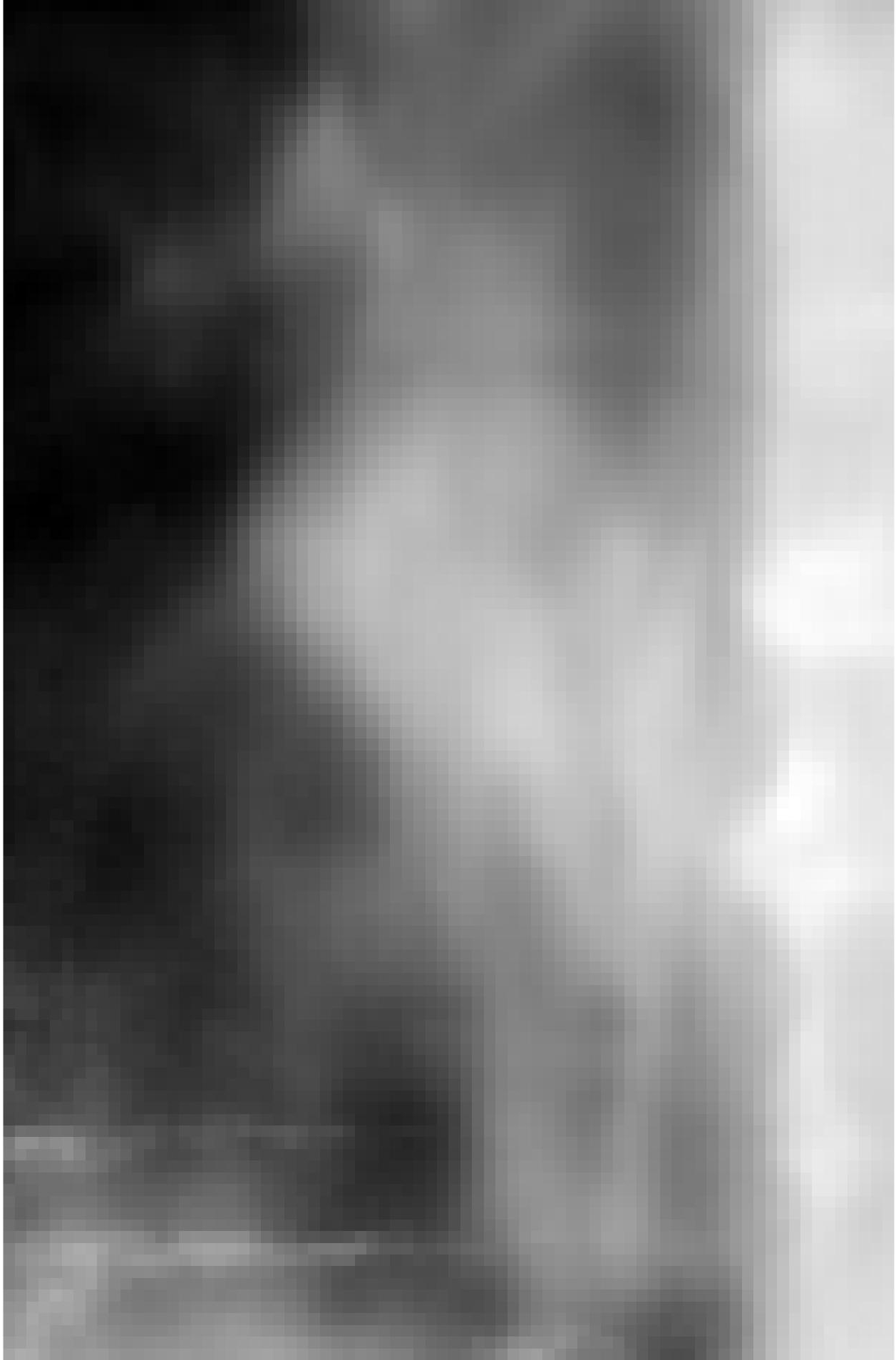} }}%
    \subfloat[Patient 2 \\* original \\* \text{\normalsize $t=t_{2374}$}]{{\includegraphics[height=2.8cm]{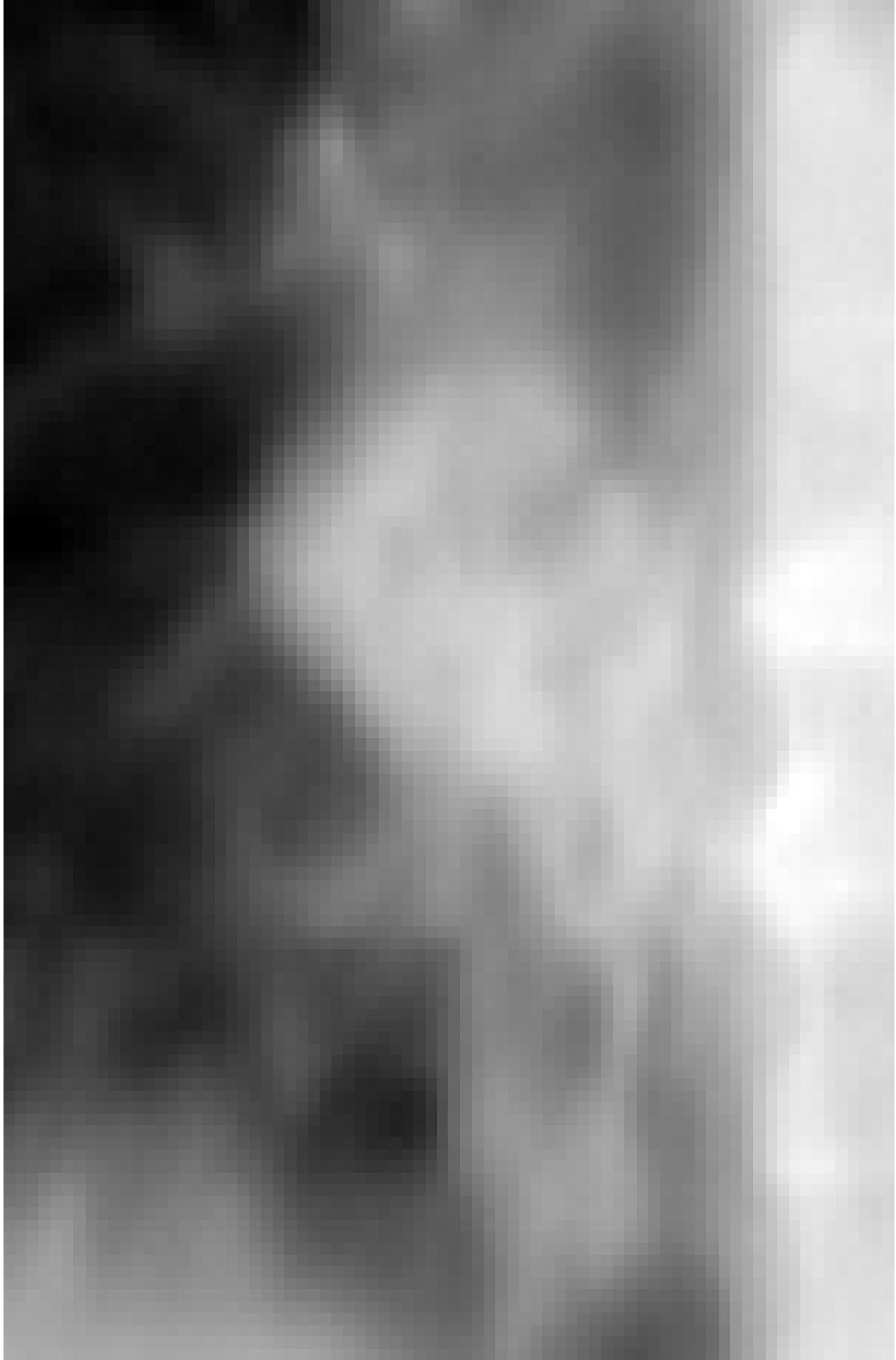} }}%
    \quad
    \subfloat[Patient 3 \\* predicted \\* \text{\normalsize $t=t_{2209}$}]{{\includegraphics[height=2.1cm]{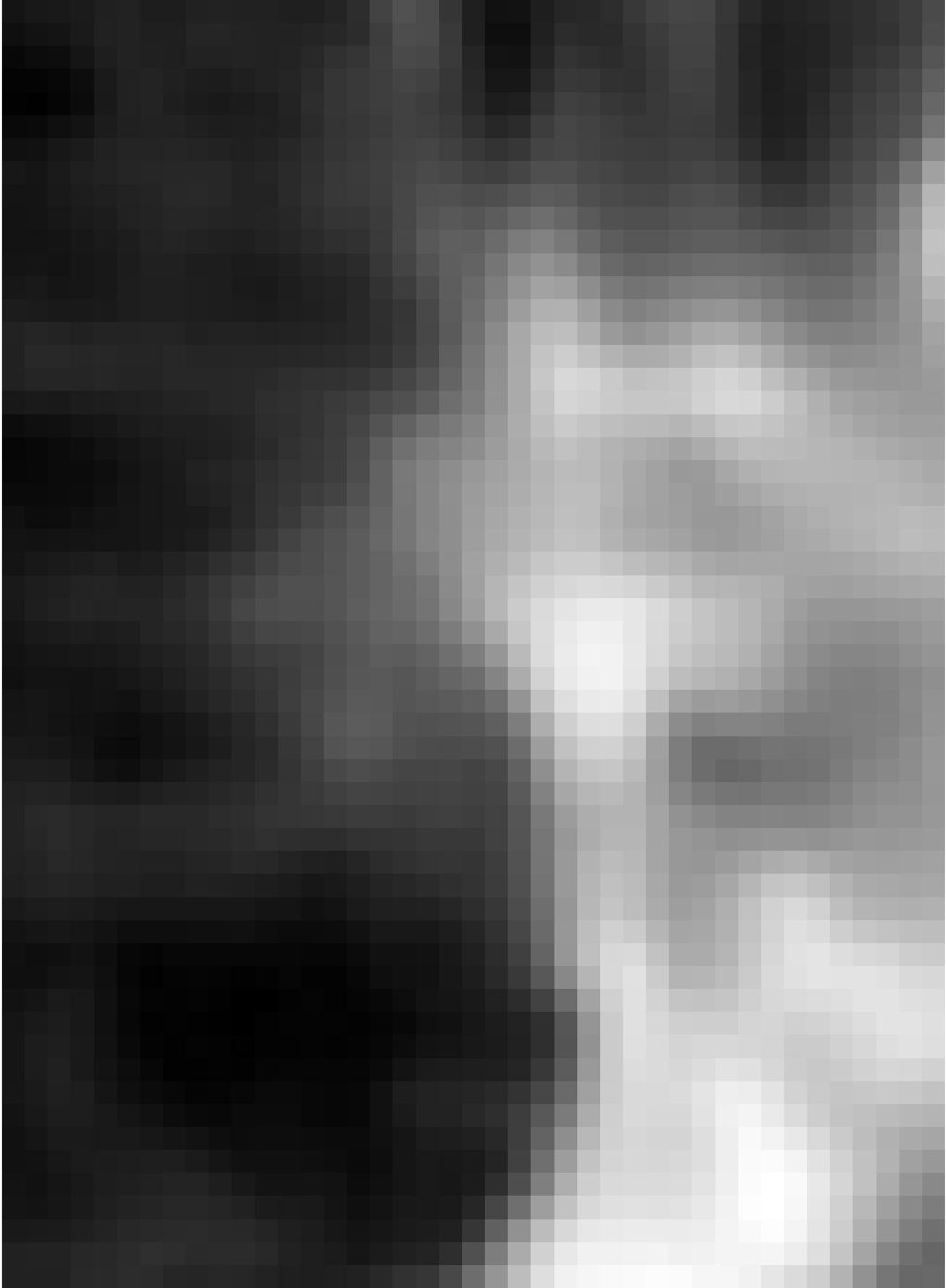} }}%
    \subfloat[Patient 3 \\* original \\* \text{\normalsize $t=t_{2209}$}]{{\includegraphics[height=2.1cm]{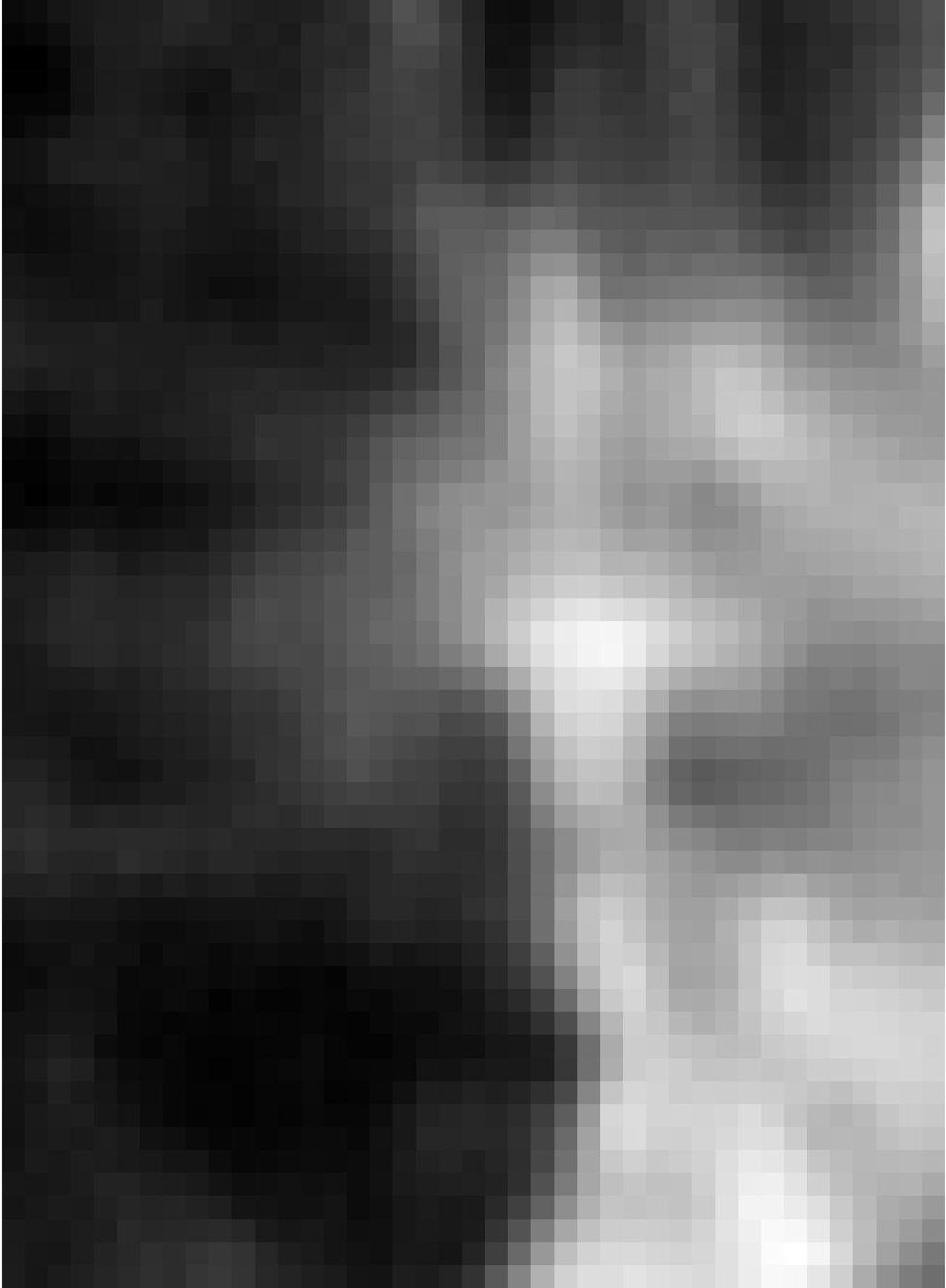} }}%
    \subfloat[Patient 3 \\* predicted \\* \text{\normalsize $t=t_{2374}$}]{{\includegraphics[height=2.1cm]{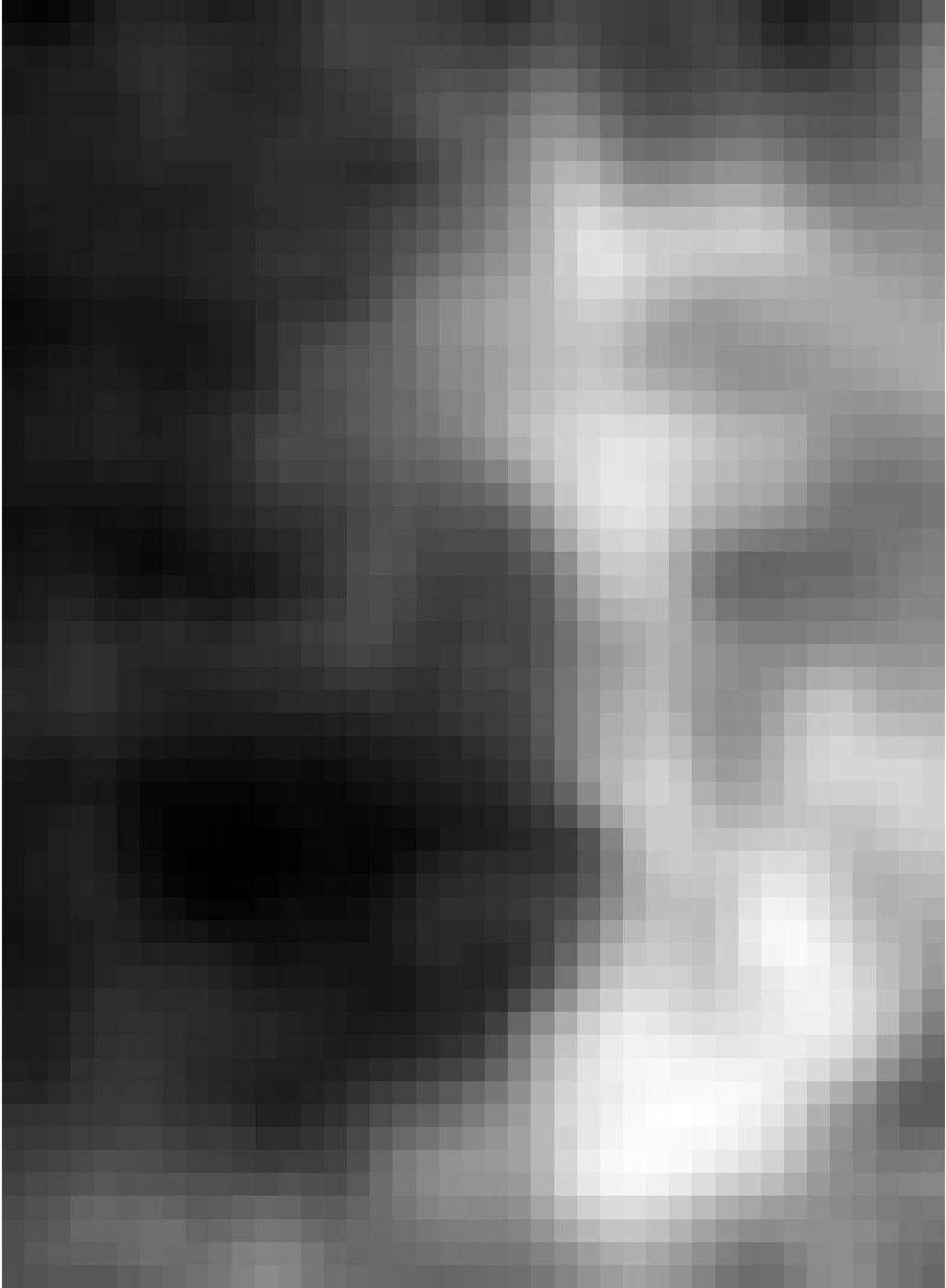} }}%
    \subfloat[Patient 3 \\* original \\* \text{\normalsize $t=t_{2374}$}]{{\includegraphics[height=2.1cm]{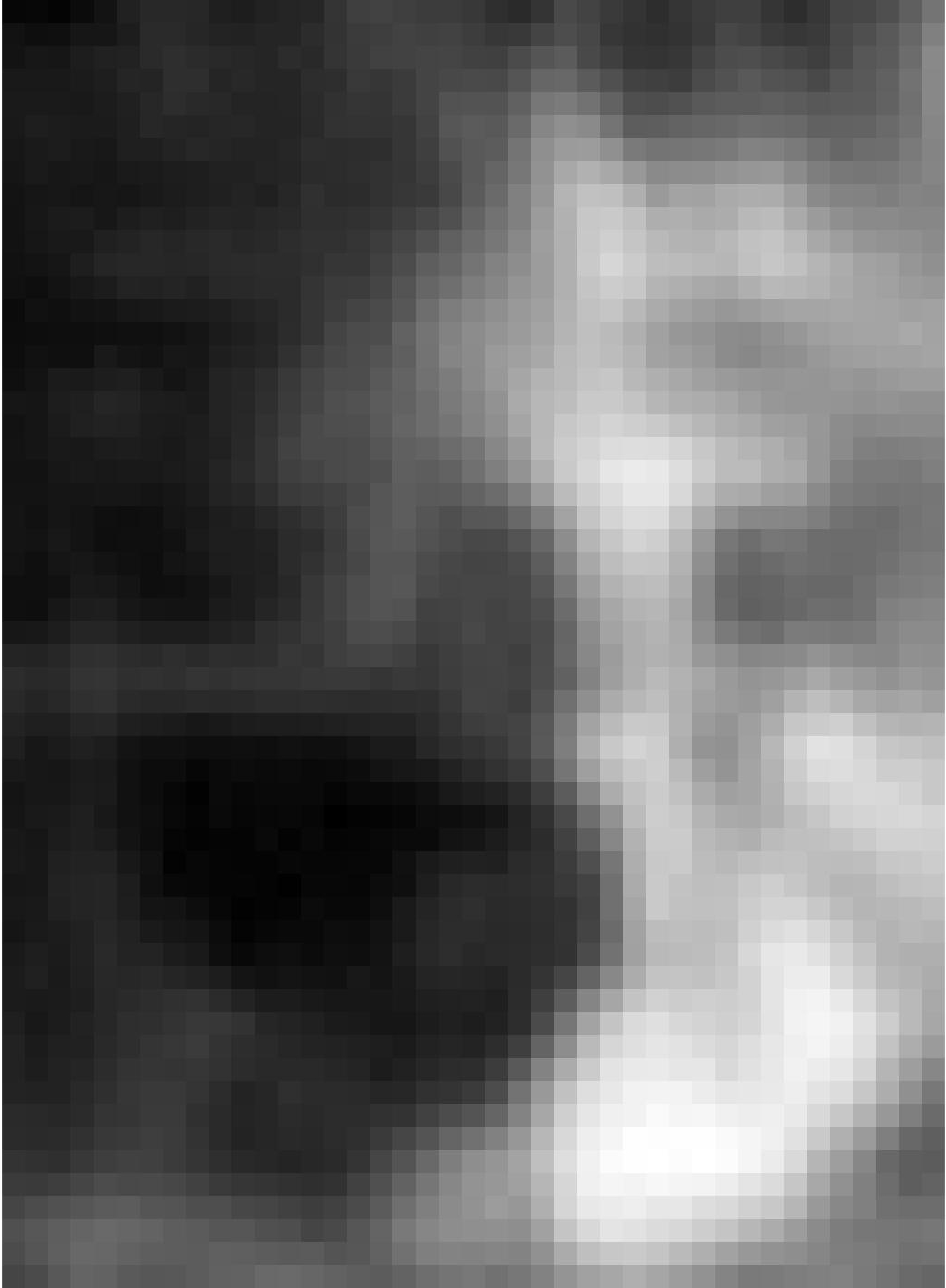} }}%
    \subfloat[Patient 4 \\* predicted \\* \text{\normalsize $t=t_{2209}$}]{{\includegraphics[height=2.1cm]{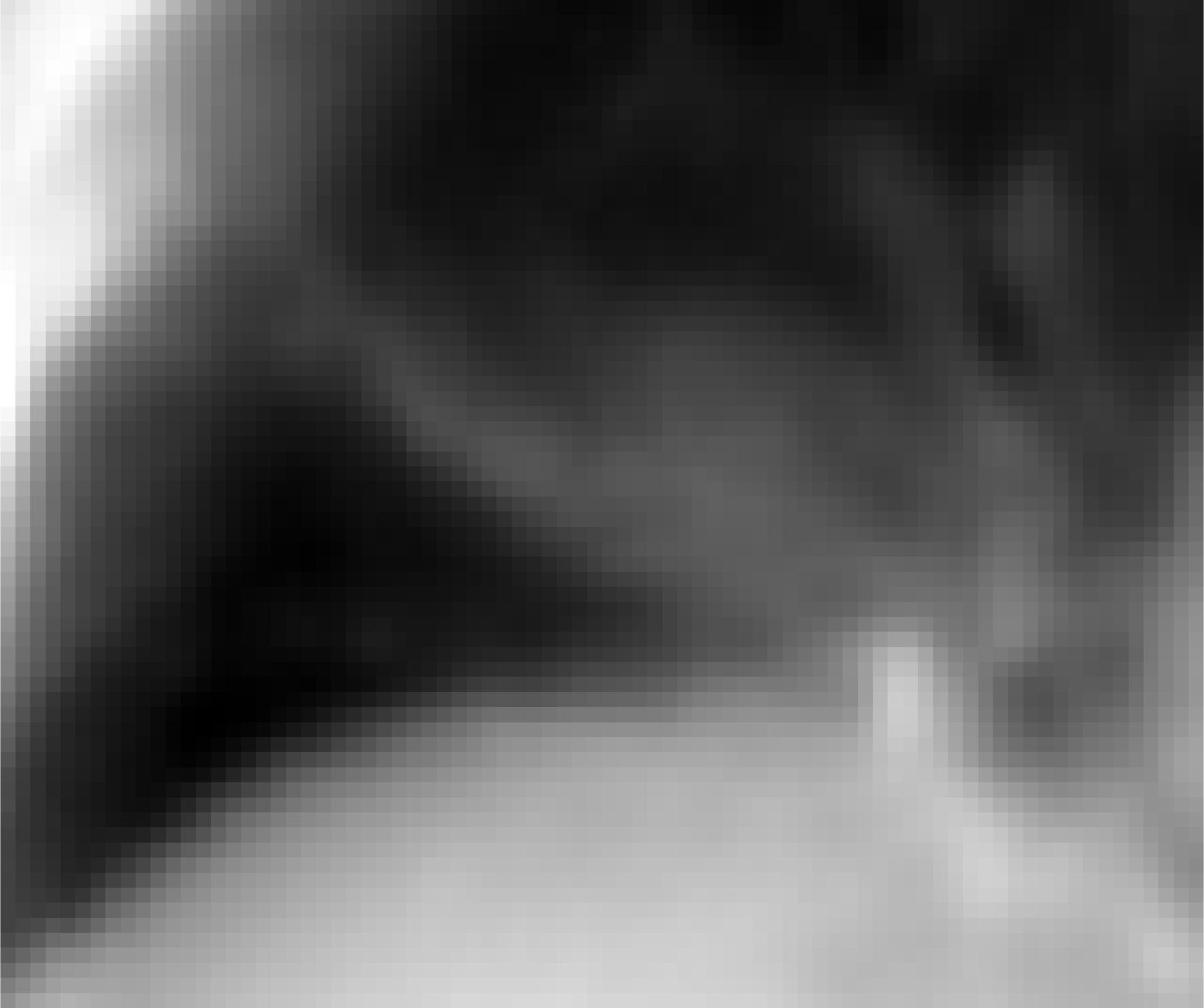} }}%
    \subfloat[Patient 4 \\* original \\* \text{\normalsize $t=t_{2209}$}]{{\includegraphics[height=2.1cm]{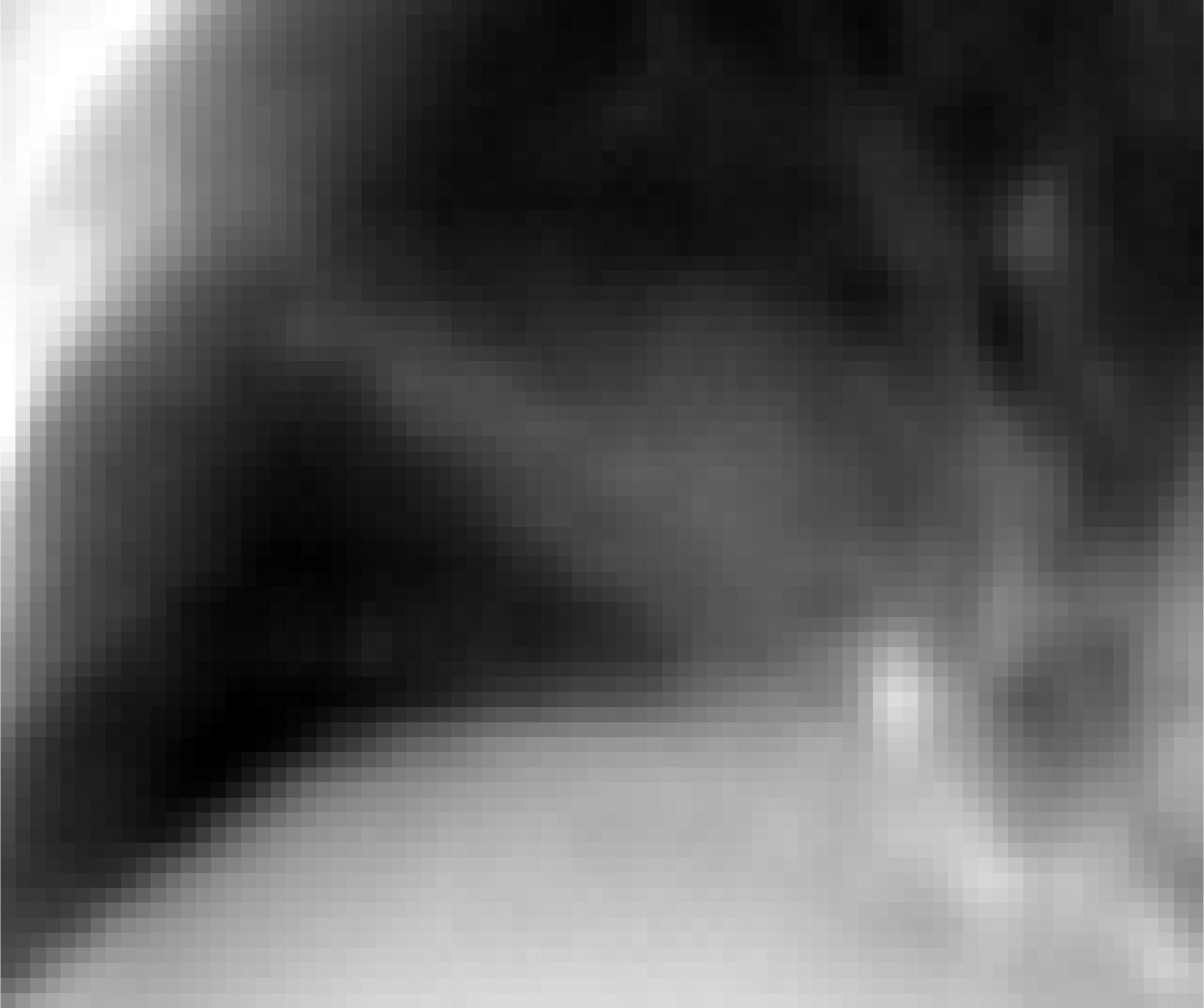} }}%
    \subfloat[Patient 4 \\* predicted \\* \text{\normalsize $t=t_{2374}$}]{{\includegraphics[height=2.1cm]{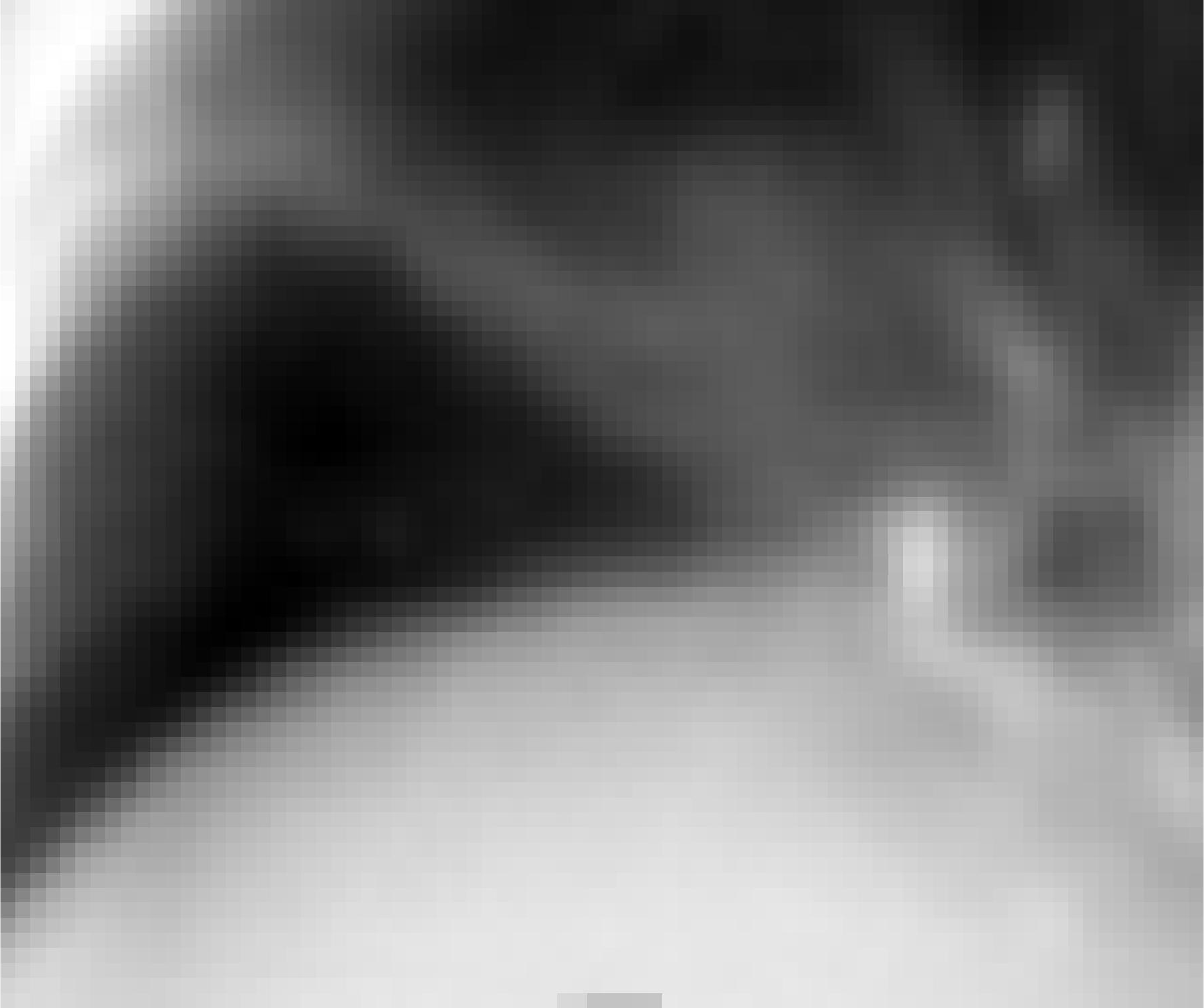} }}%
    \subfloat[Patient 4 \\* original \\* \text{\normalsize $t=t_{2374}$}]{{\includegraphics[height=2.1cm]{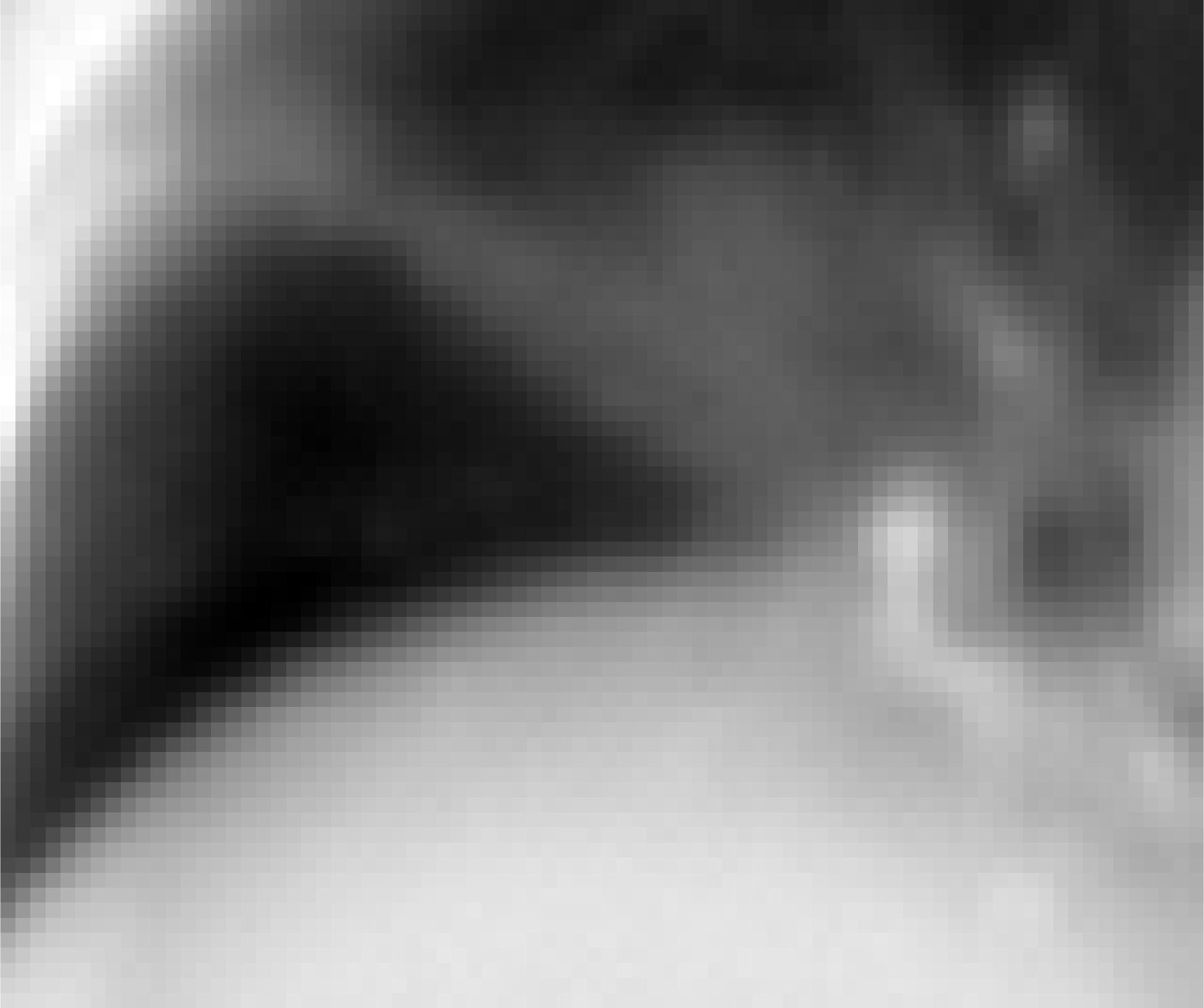} }}%
    \caption{Original and predicted ROI coronal AIP, at an end-of-exhale and an end-of-inhale positions.}%
    \label{fig:pred_cor_aip}%
\end{figure}%

\newpage 

\begin{figure}[h!]%
    \centering
    \captionsetup[subfigure]{justification=centering}   
    \captionsetup[subfigure]{labelformat=empty} 
    \subfloat[Patient 1 \\* predicted \\* \text{\normalsize $t=t_{2209}$}]{{\includegraphics[height=2.8cm]{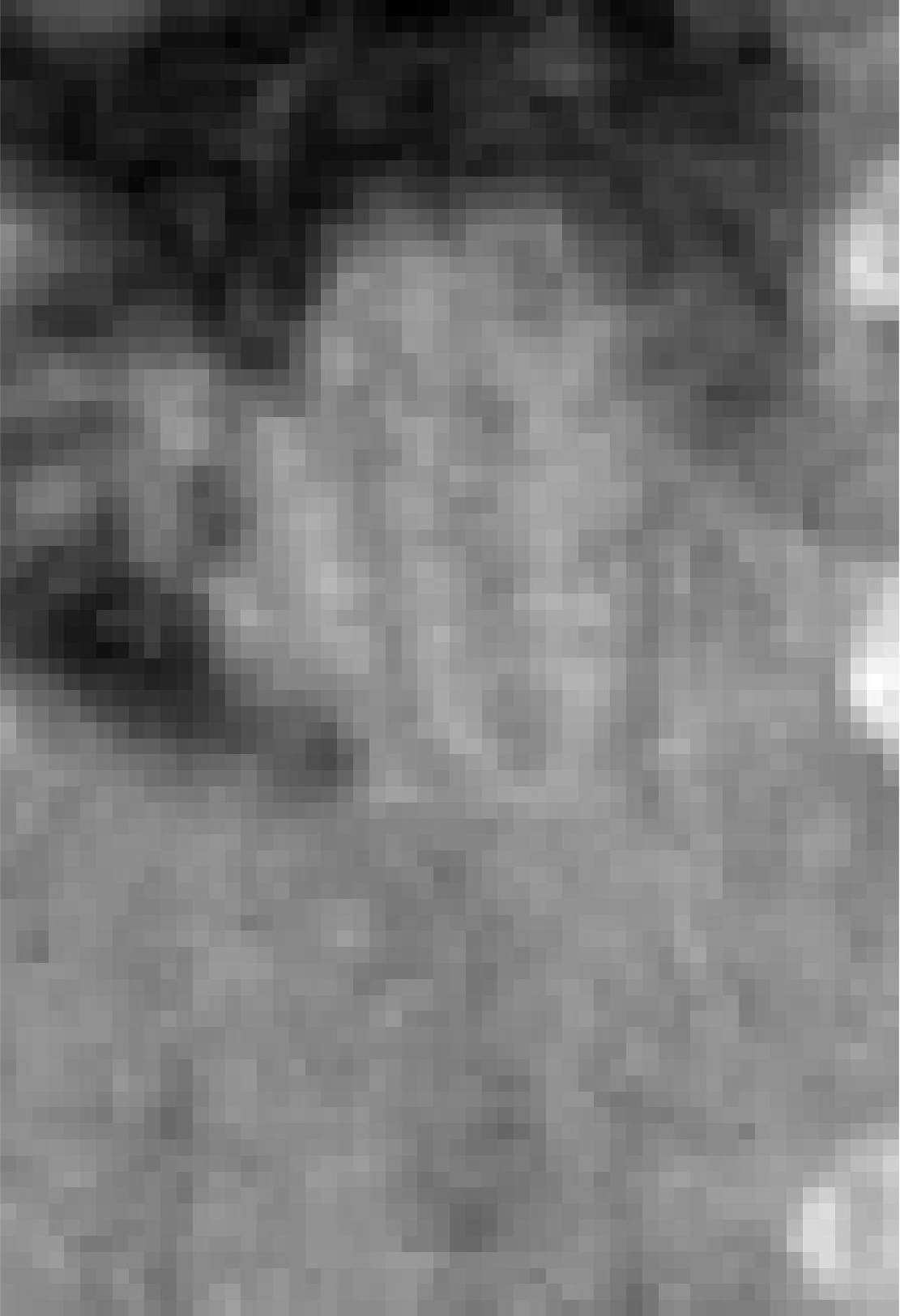}}}%
    \subfloat[Patient 1 \\* original \\* \text{\normalsize $t=t_{2209}$}]{{\includegraphics[height=2.8cm]{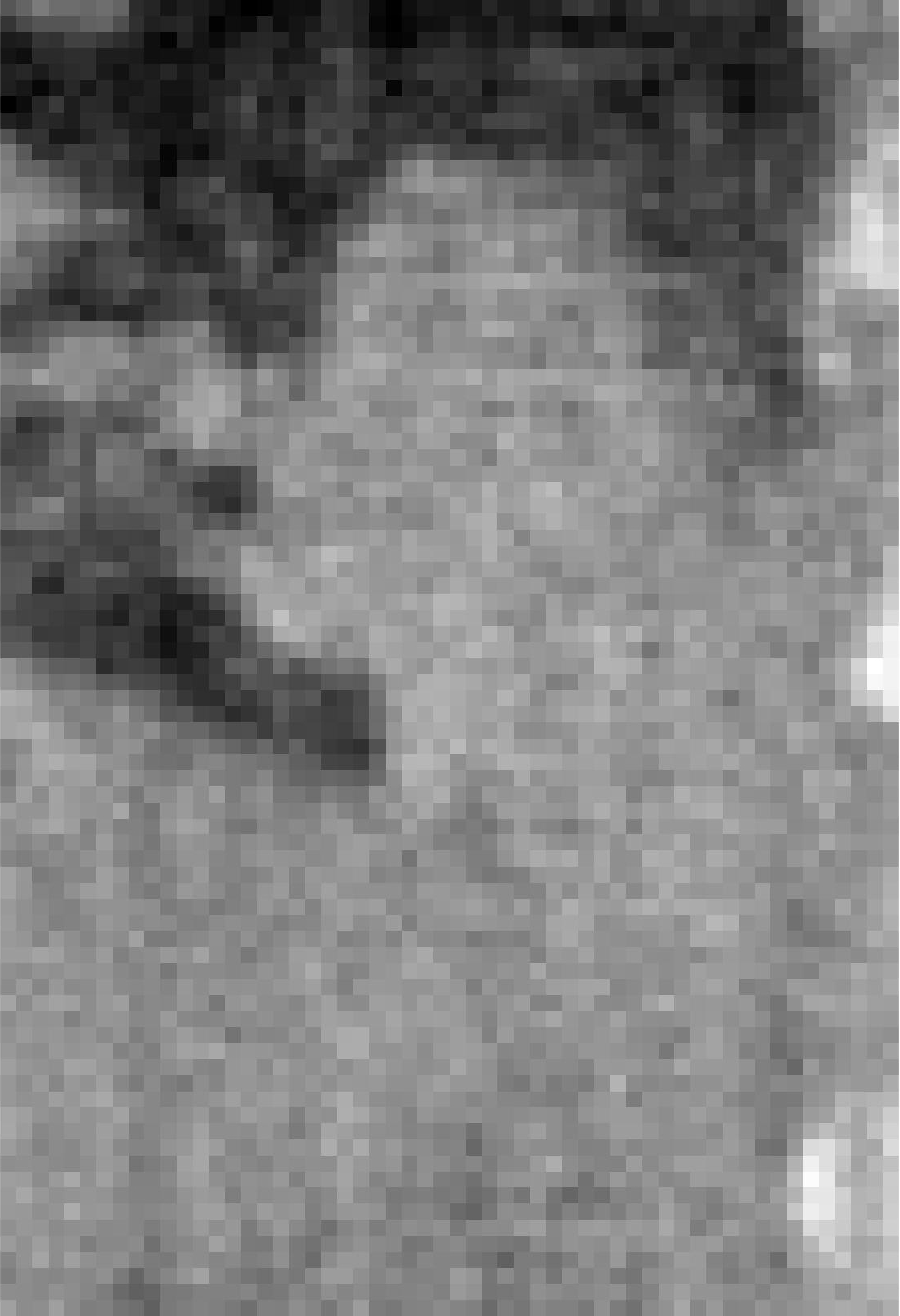} }}%
    \subfloat[Patient 1 \\* predicted \\* \text{\normalsize $t=t_{2374}$}]{{\includegraphics[height=2.8cm]{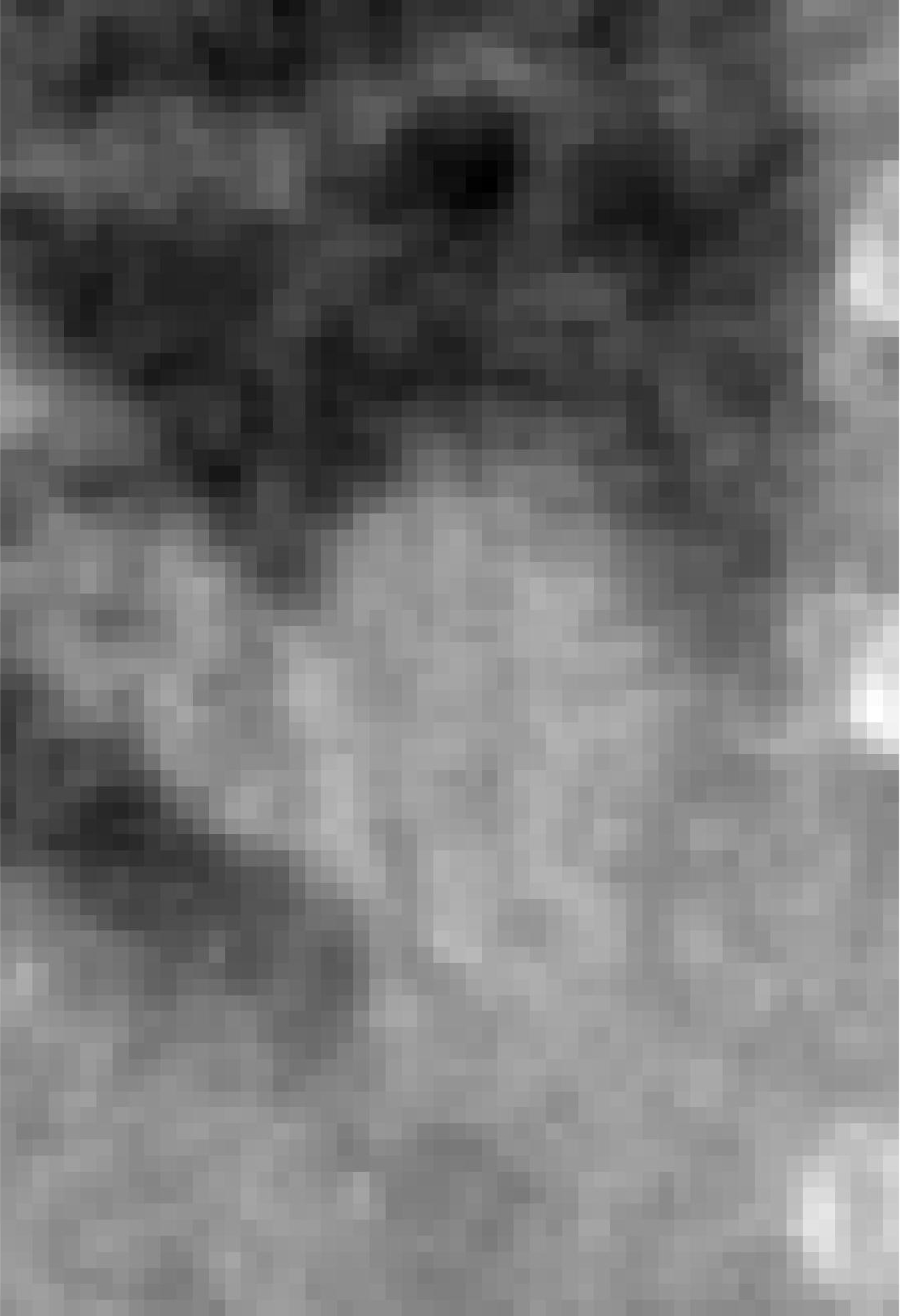} }}%
    \subfloat[Patient 1 \\* original \\* \text{\normalsize $t=t_{2374}$}]{{\includegraphics[height=2.8cm]{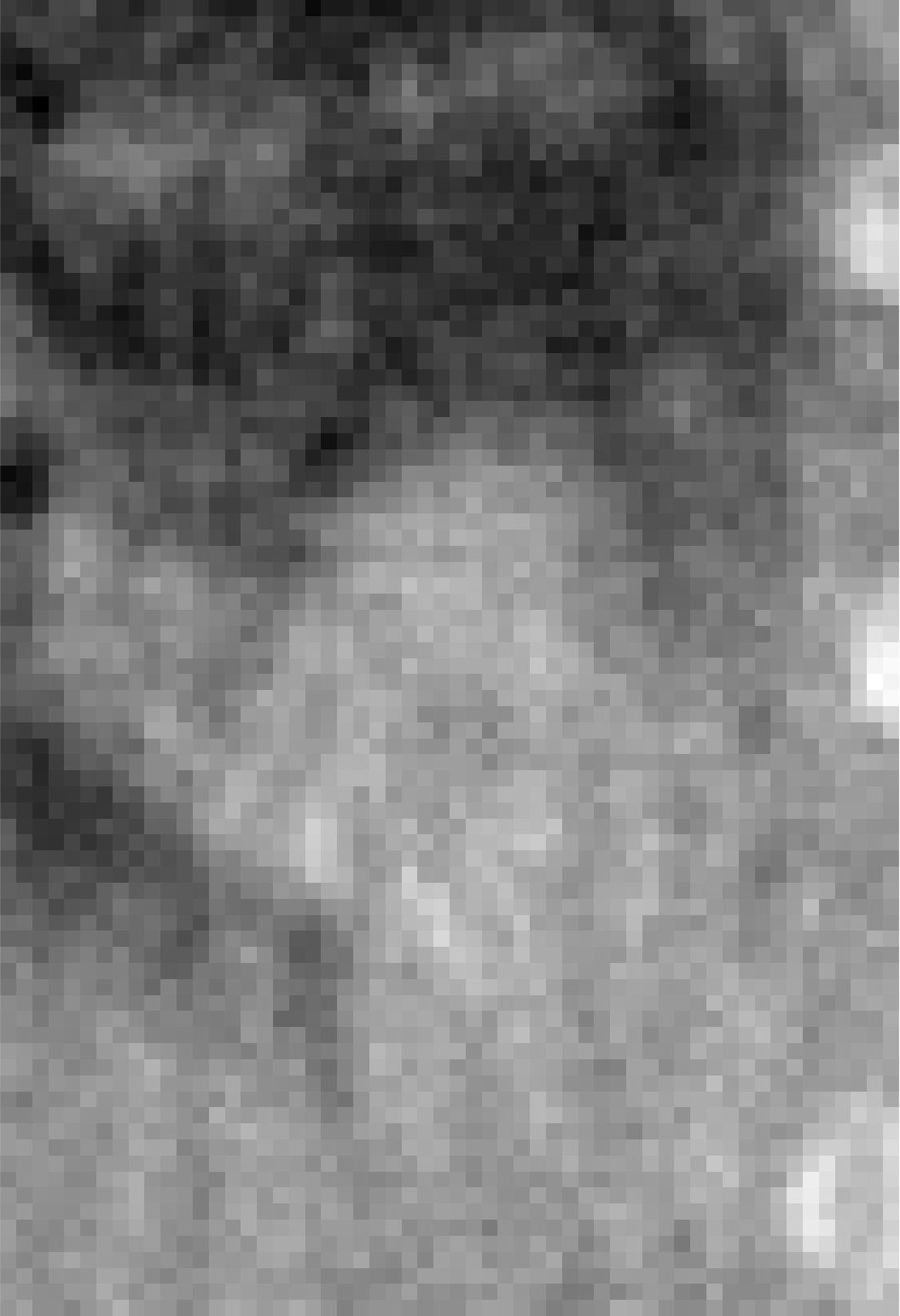} }}%
    \subfloat[Patient 2 \\* predicted \\* \text{\normalsize $t=t_{2209}$}]{{\includegraphics[height=2.8cm]{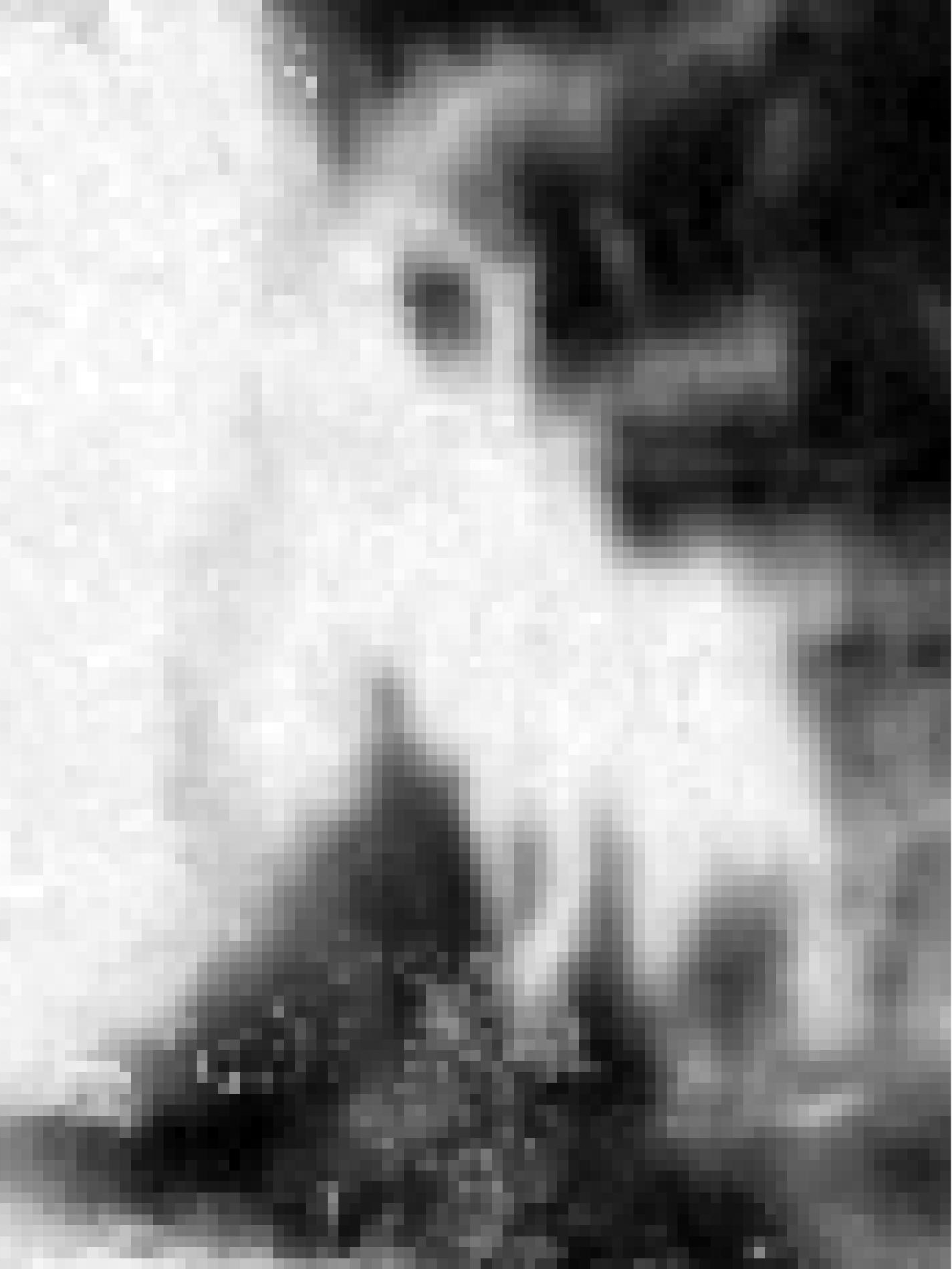} }}%
    \subfloat[Patient 2 \\* original \\* \text{\normalsize $t=t_{2209}$}]{{\includegraphics[height=2.8cm]{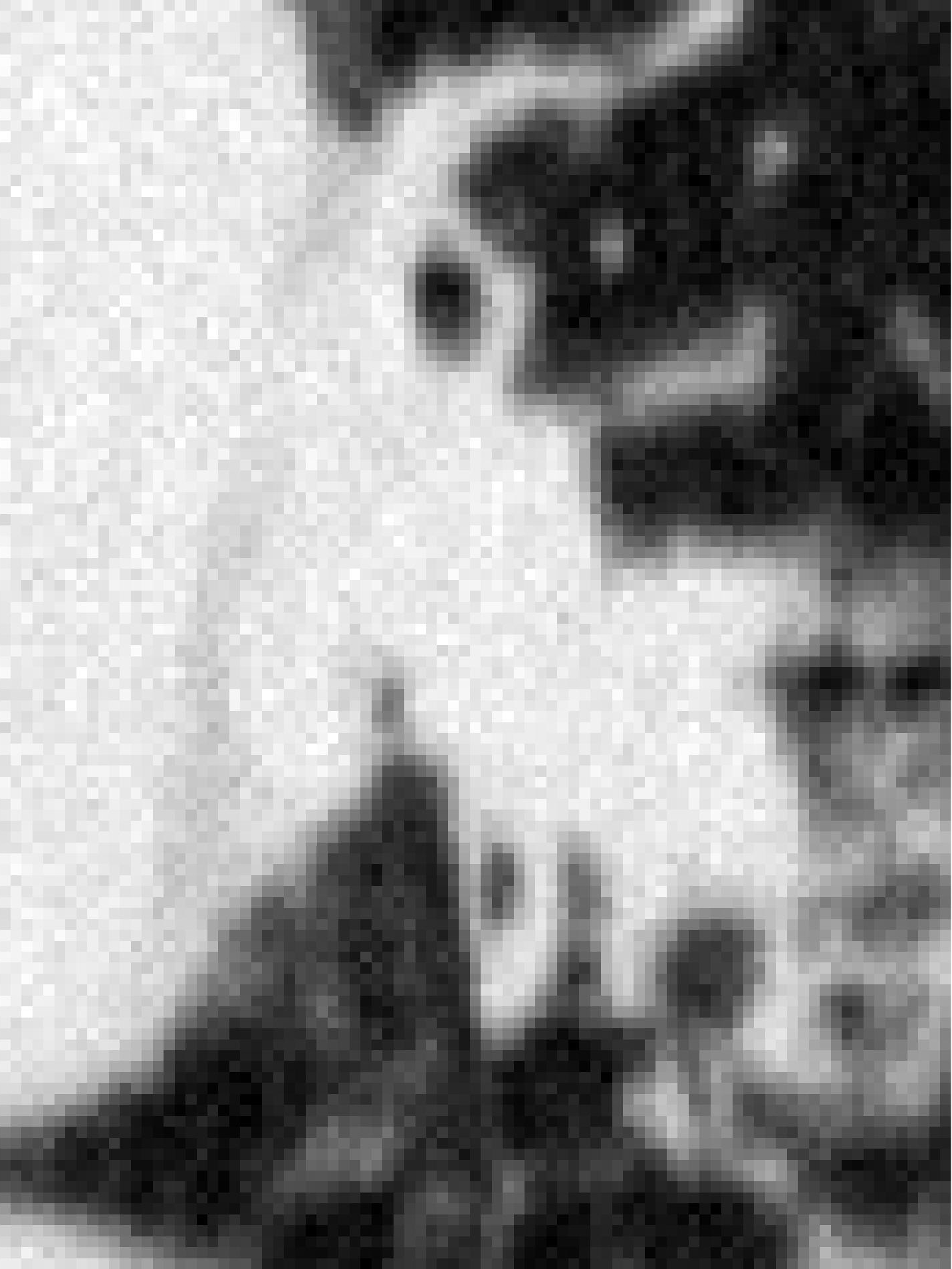} }}%
    \subfloat[Patient 2 \\* predicted \\* \text{\normalsize $t=t_{2374}$}]{{\includegraphics[height=2.8cm]{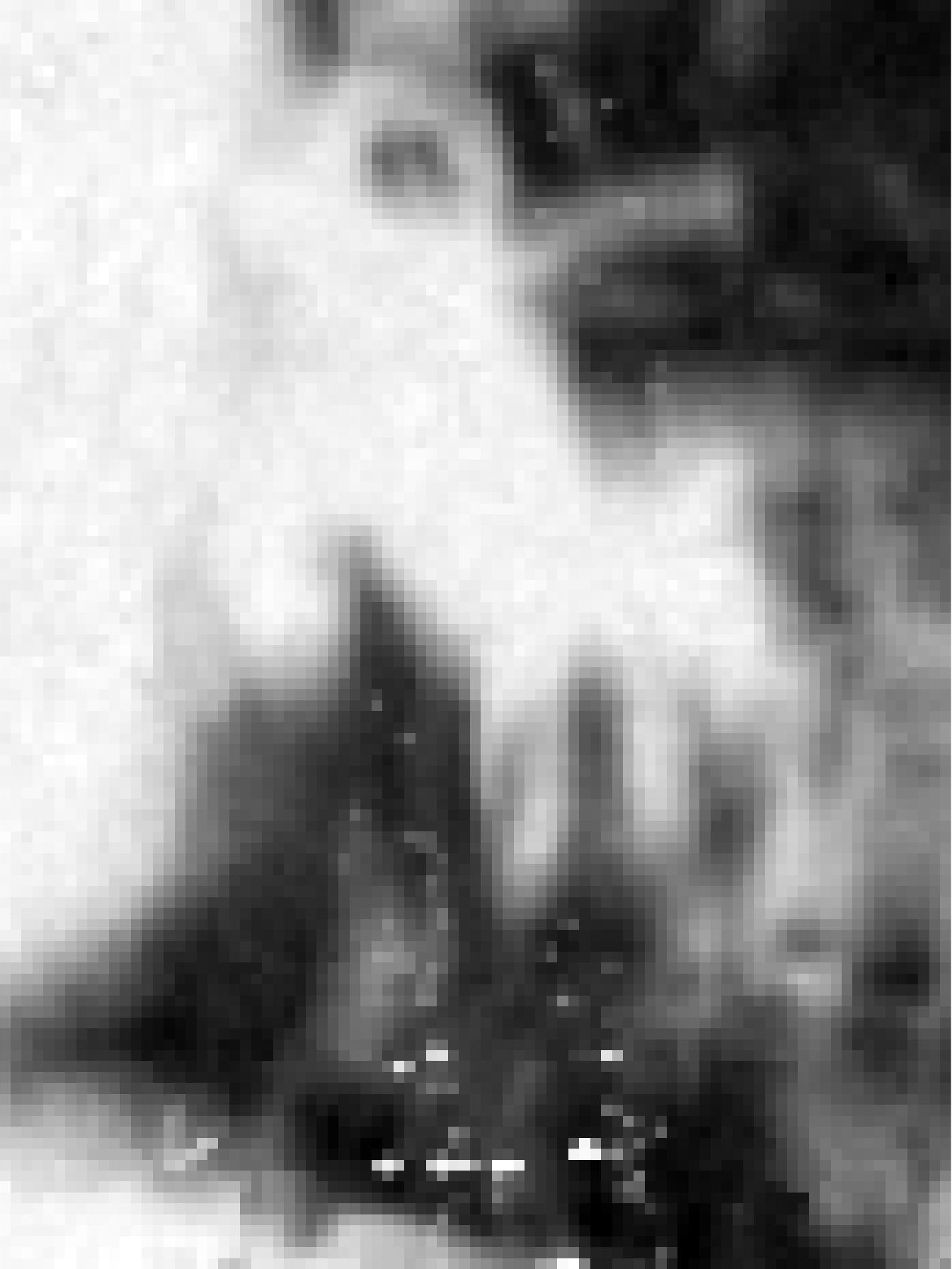} }}%
    \subfloat[Patient 2 \\* original \\* \text{\normalsize $t=t_{2374}$}]{{\includegraphics[height=2.8cm]{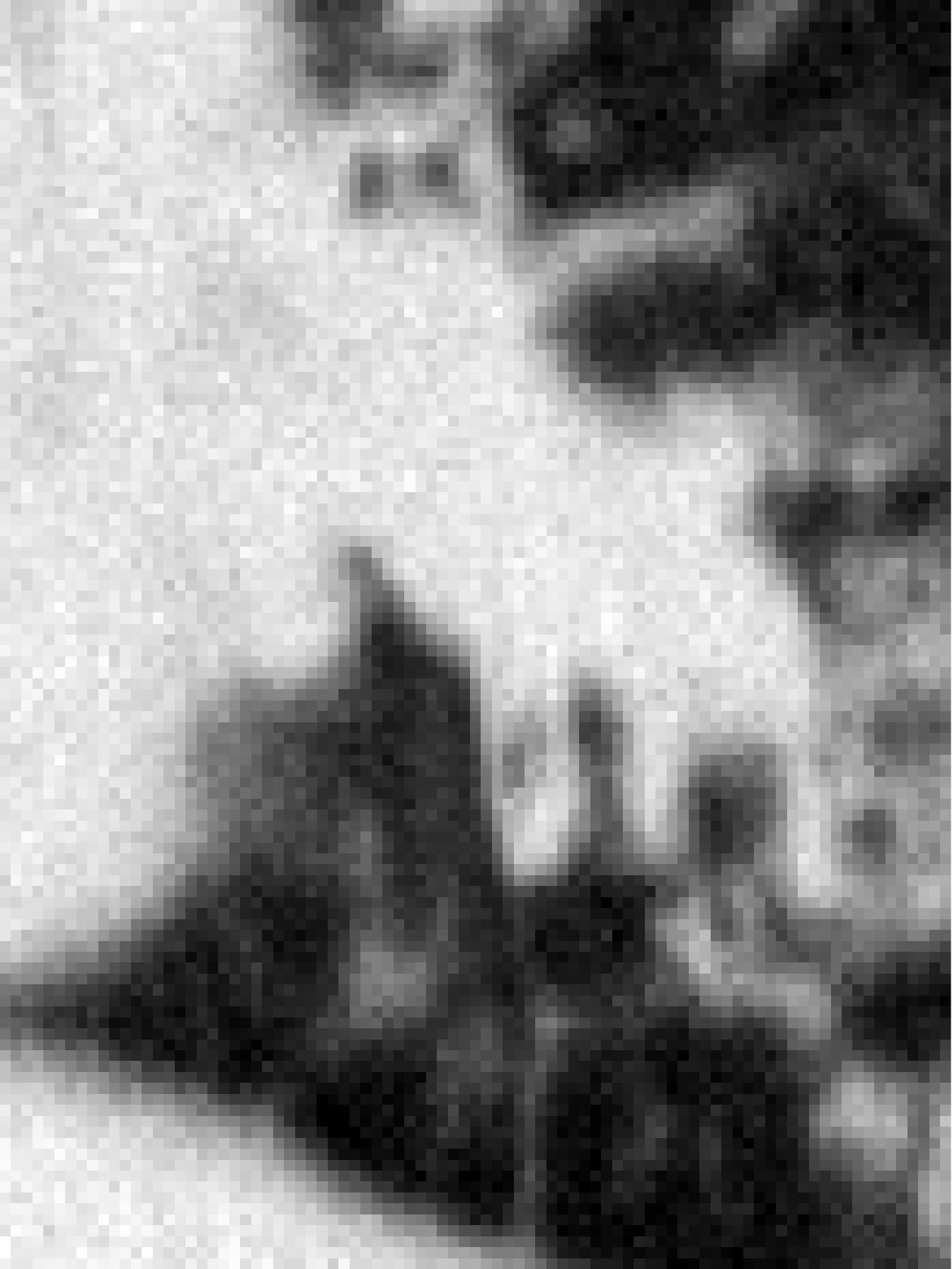} }}%
    \quad
    \subfloat[Patient 3 \\* predicted \\* \text{\normalsize $t=t_{2209}$}]{{\includegraphics[height=2.1cm]{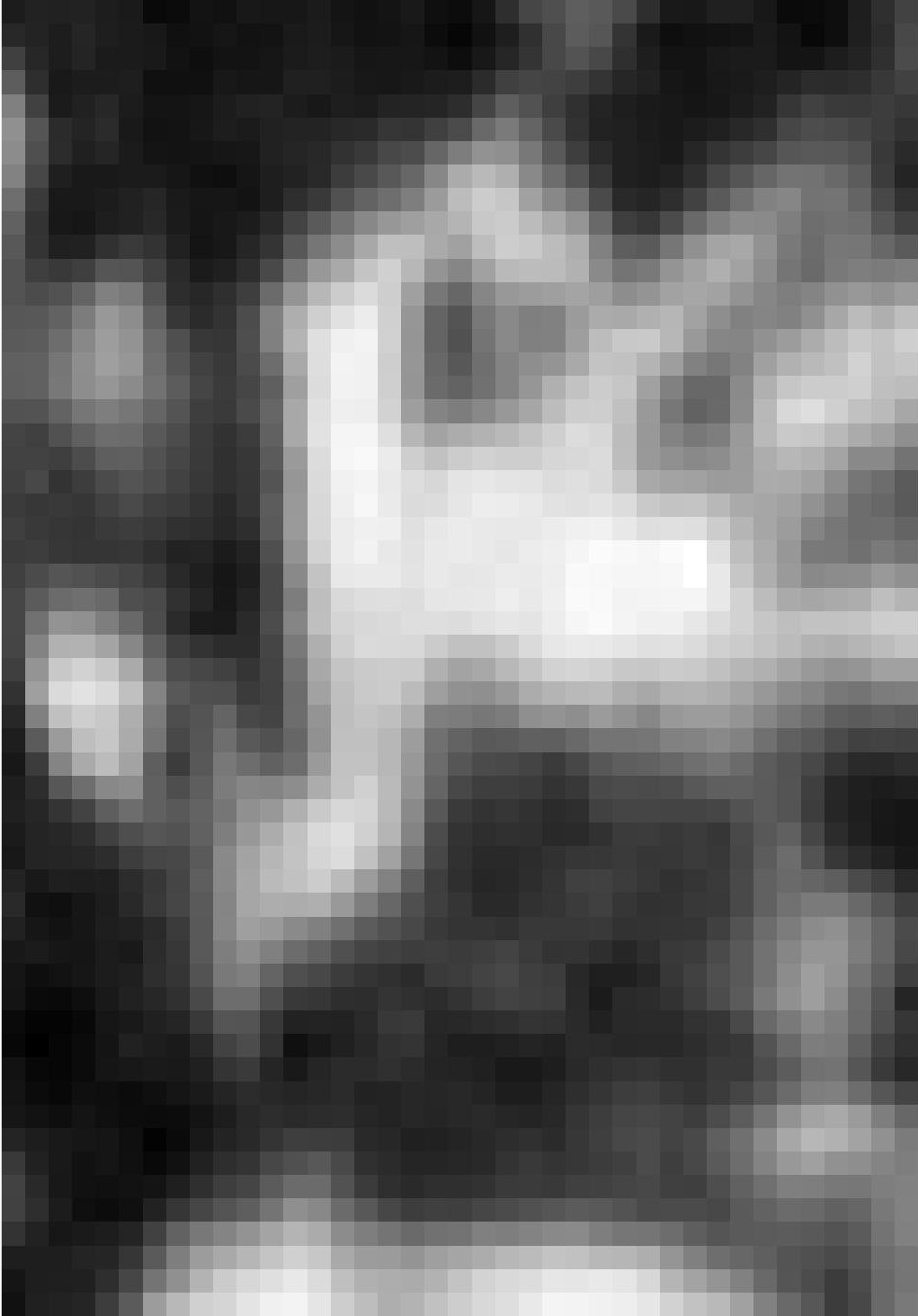} }}%
    \subfloat[Patient 3 \\* original \\* \text{\normalsize $t=t_{2209}$}]{{\includegraphics[height=2.1cm]{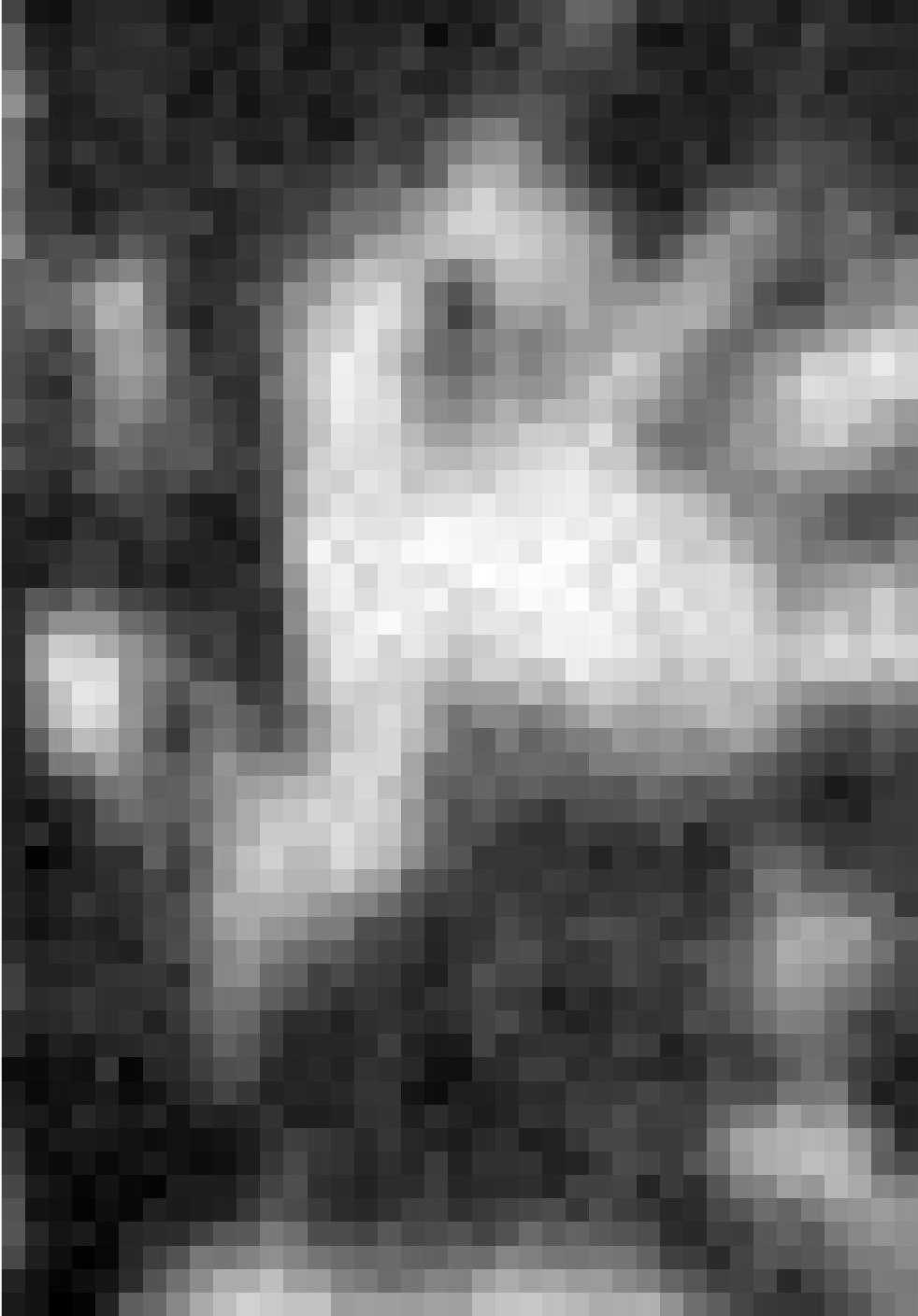} }}%
    \subfloat[Patient 3 \\* predicted \\* \text{\normalsize $t=t_{2374}$}]{{\includegraphics[height=2.1cm]{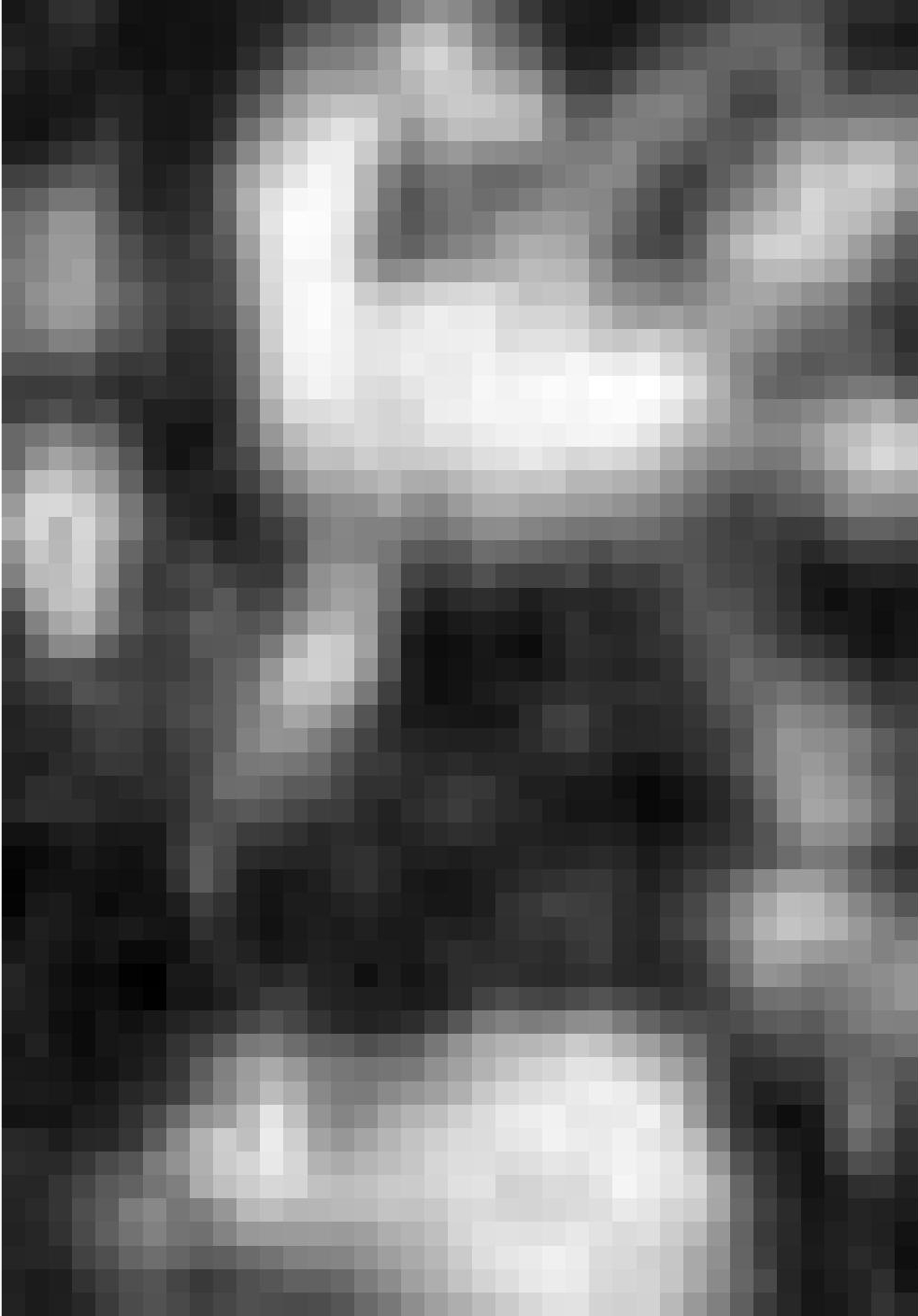} }}%
    \subfloat[Patient 3 \\* original \\* \text{\normalsize $t=t_{2374}$}]{{\includegraphics[height=2.1cm]{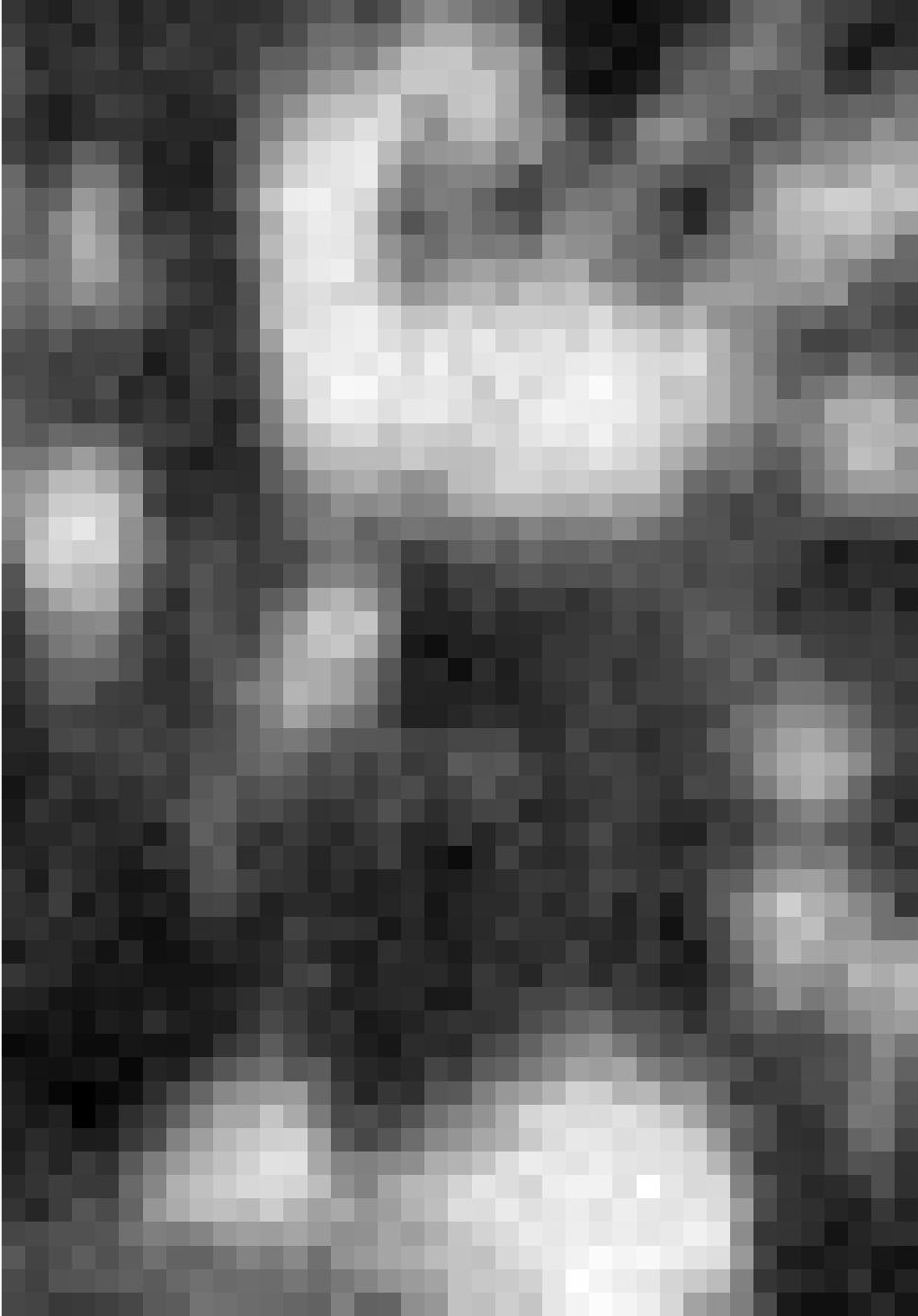} }}%
    \subfloat[Patient 4 \\* predicted \\* \text{\normalsize $t=t_{2209}$}]{{\includegraphics[height=2.1cm]{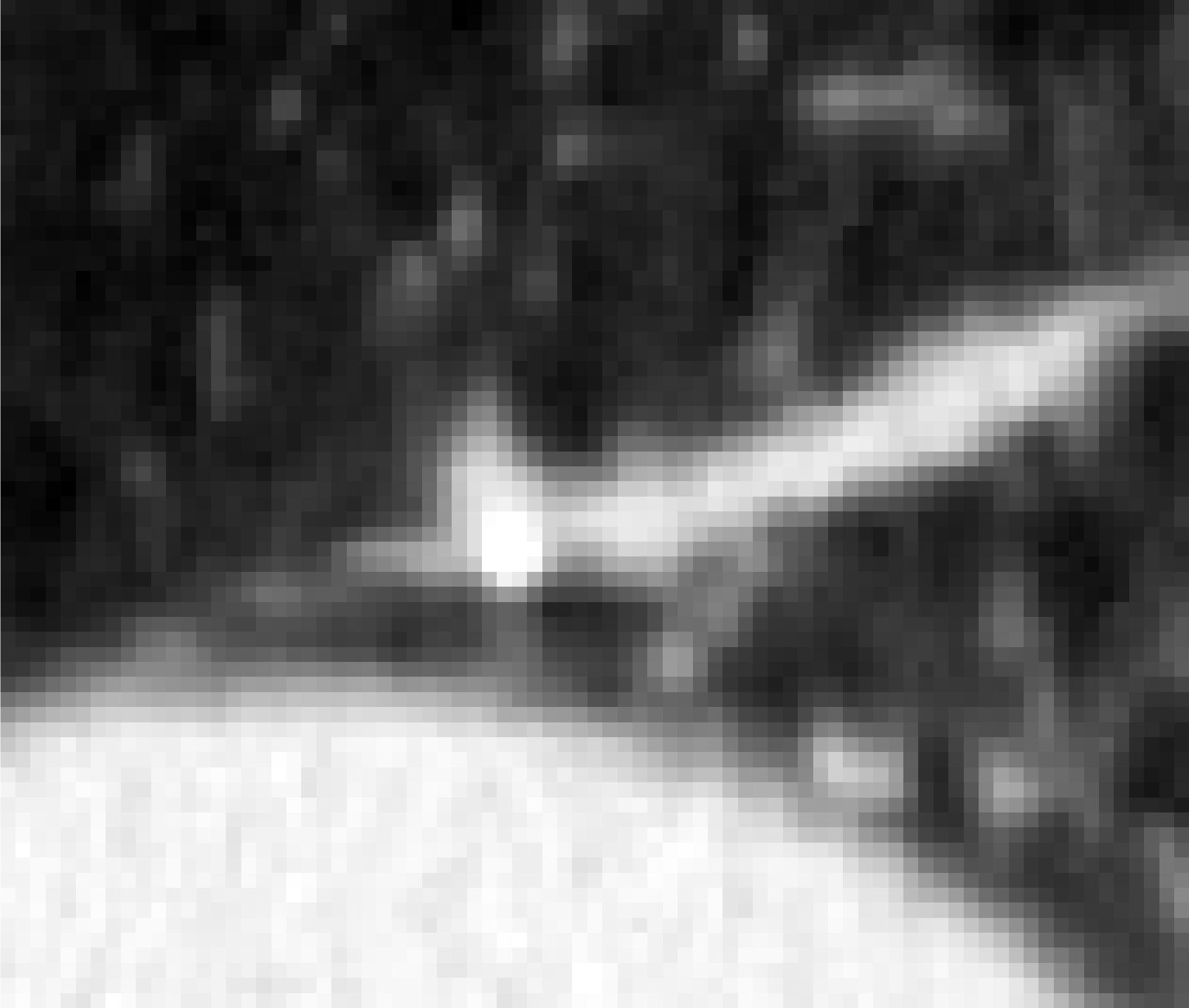} }}%
    \subfloat[Patient 4 \\* original \\* \text{\normalsize $t=t_{2209}$}]{{\includegraphics[height=2.1cm]{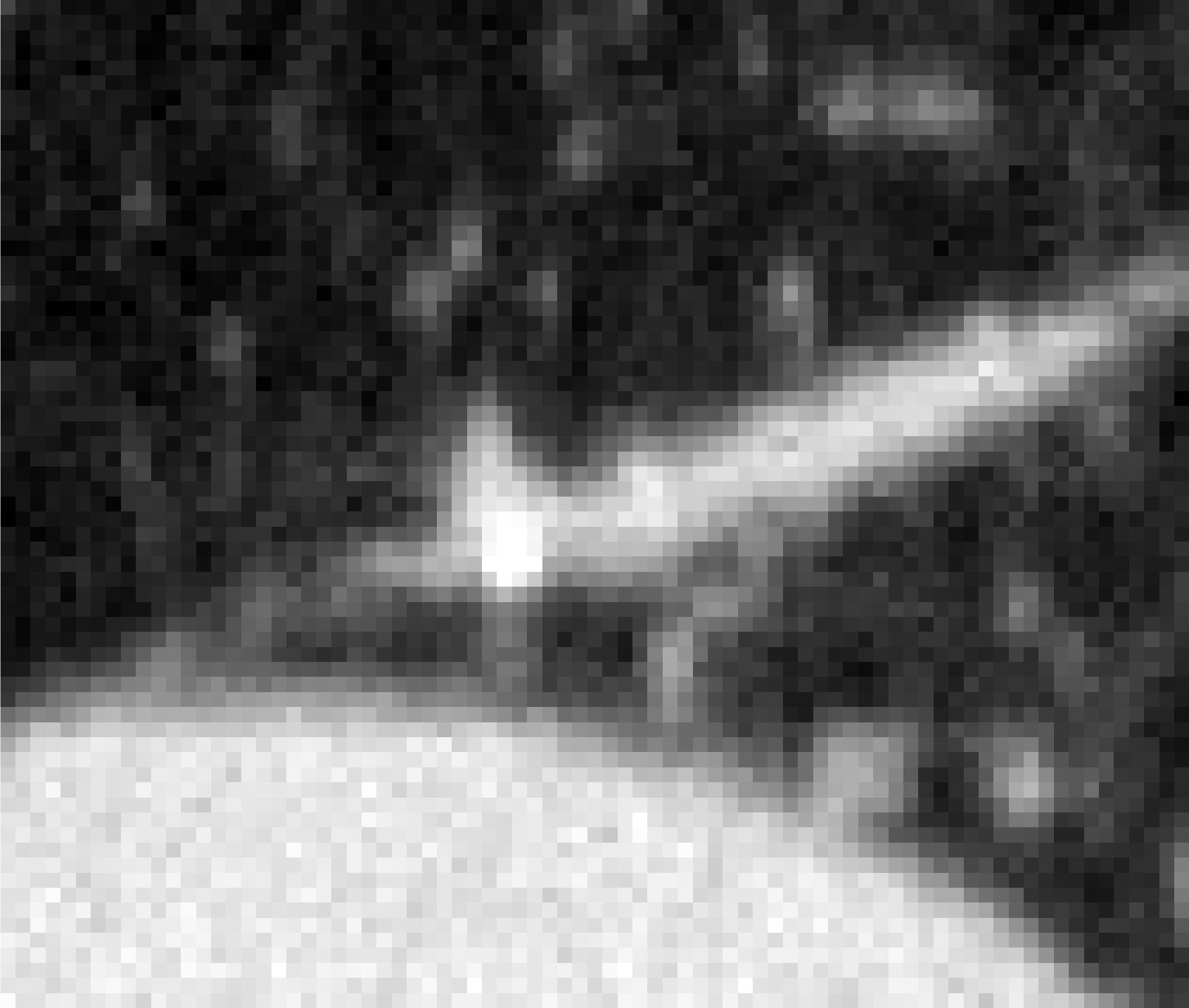} }}%
    \subfloat[Patient 4 \\* predicted \\* \text{\normalsize $t=t_{2374}$}]{{\includegraphics[height=2.1cm]{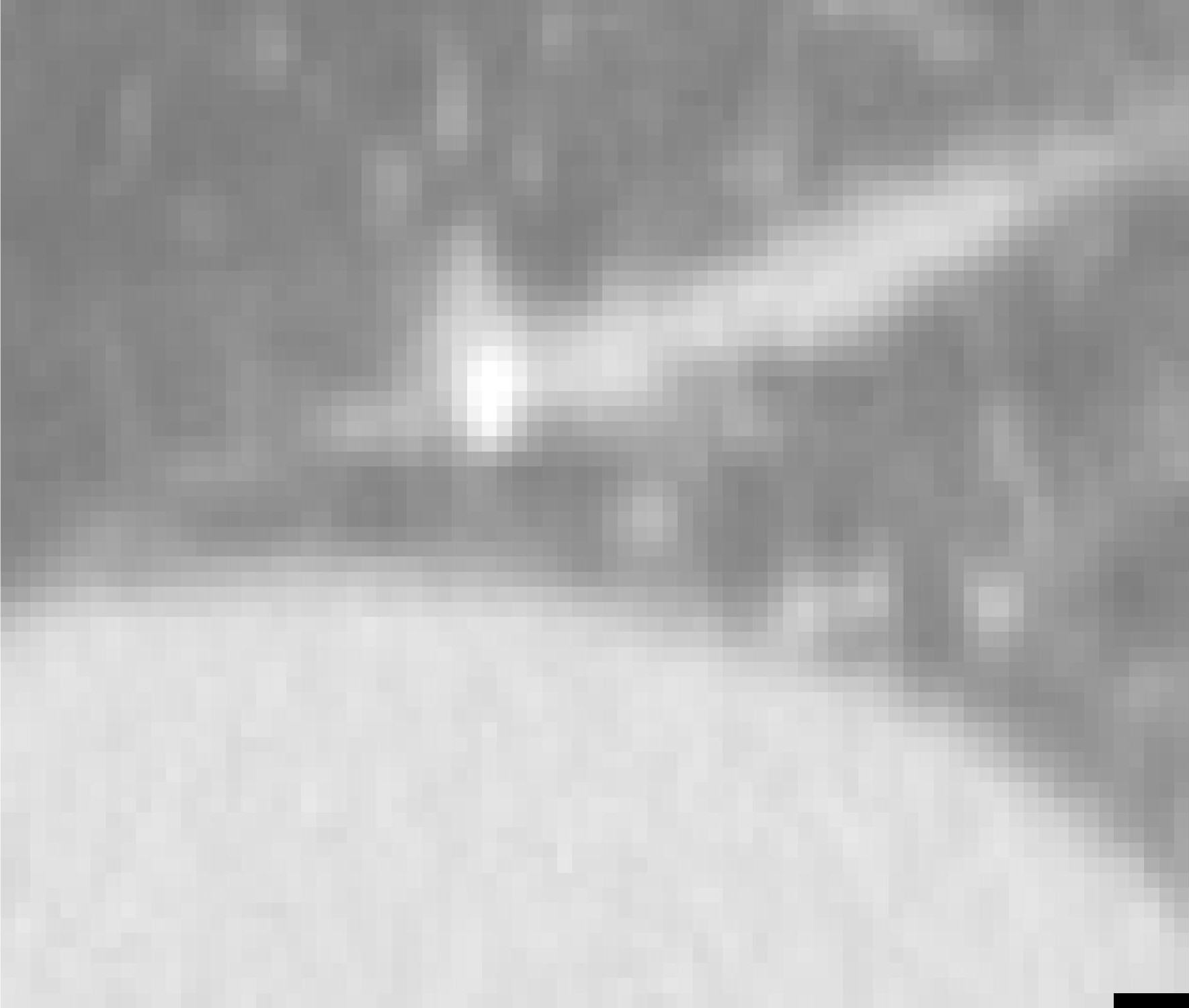} }}%
    \subfloat[Patient 4 \\* original \\* \text{\normalsize $t=t_{2374}$}]{{\includegraphics[height=2.1cm]{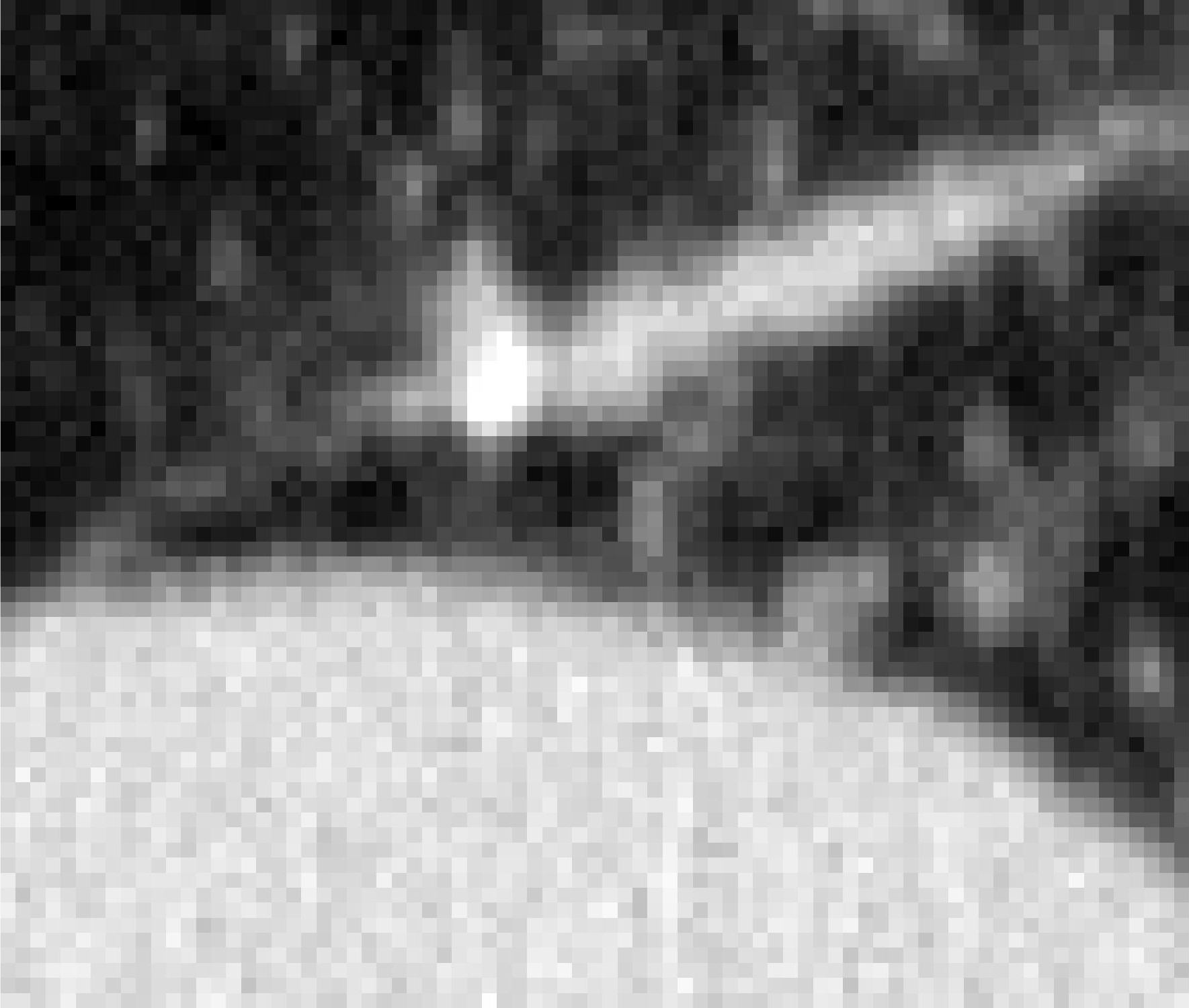} }}%
    \caption{Original and predicted ROI sagittal cross-sections (same coordinates as in Fig. \ref{fig:org_im}), at an end-of-exhale and an end-of-inhale positions. The predicted image at \text{\normalsize $t=t_{2374}$} for patient 4 seems to have high voxel intensity values but this is in fact due to post-processing with contrast enhancement, which takes into account the black voxels appearing on the lower right corner when displaying the image.}%
    \label{fig:pred_sag_cc}%
\end{figure}%

\begin{figure}[h!]%
    \centering
    \captionsetup[subfigure]{justification=centering} 
    \captionsetup[subfigure]{labelformat=empty}
    \subfloat[Patient 1 \\* predicted \\* \text{\normalsize $t=t_{2209}$}]{{\includegraphics[height=2.8cm]{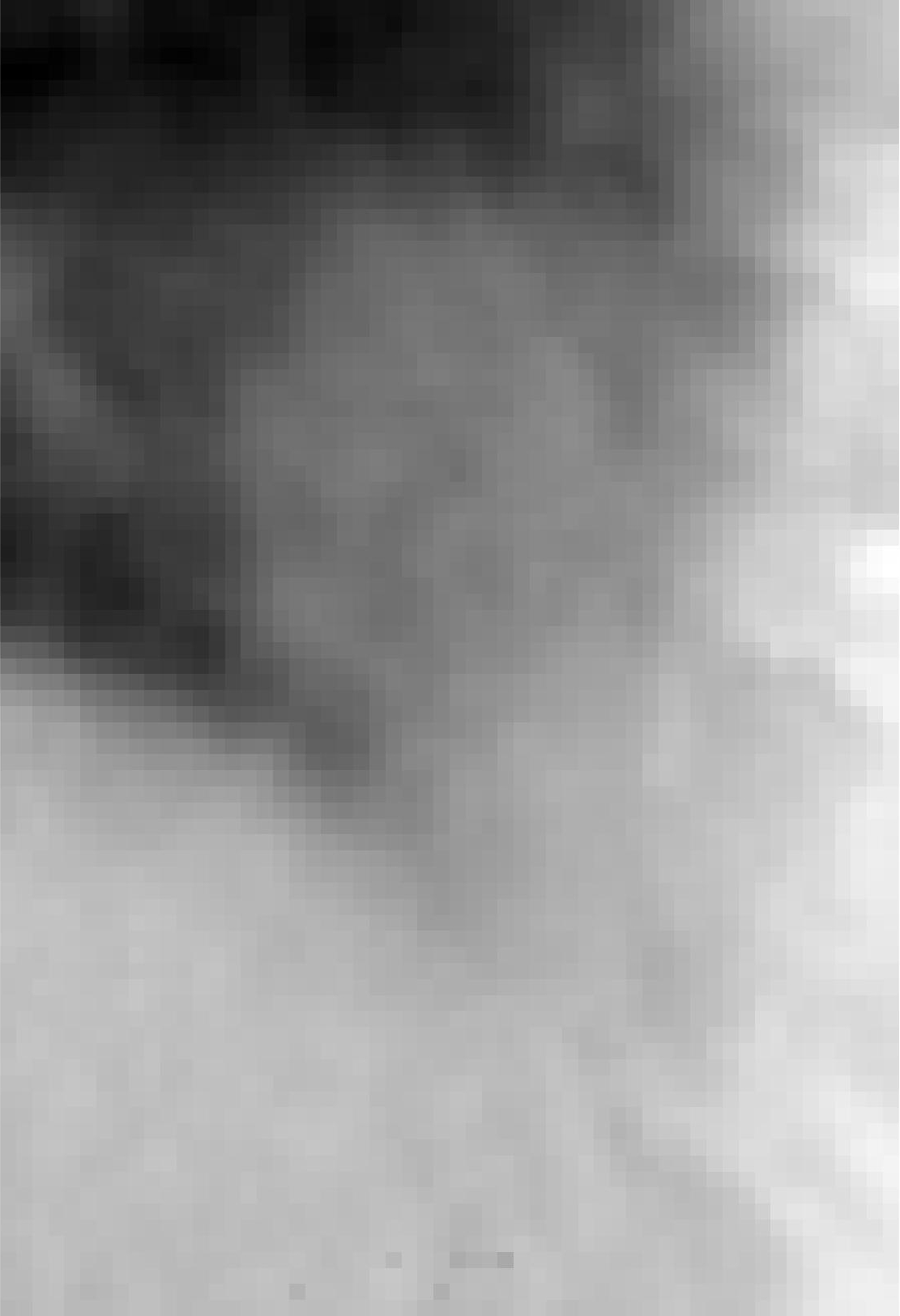} }}%
    \subfloat[Patient 1 \\* original \\* \text{\normalsize $t=t_{2209}$}]{{\includegraphics[height=2.8cm]{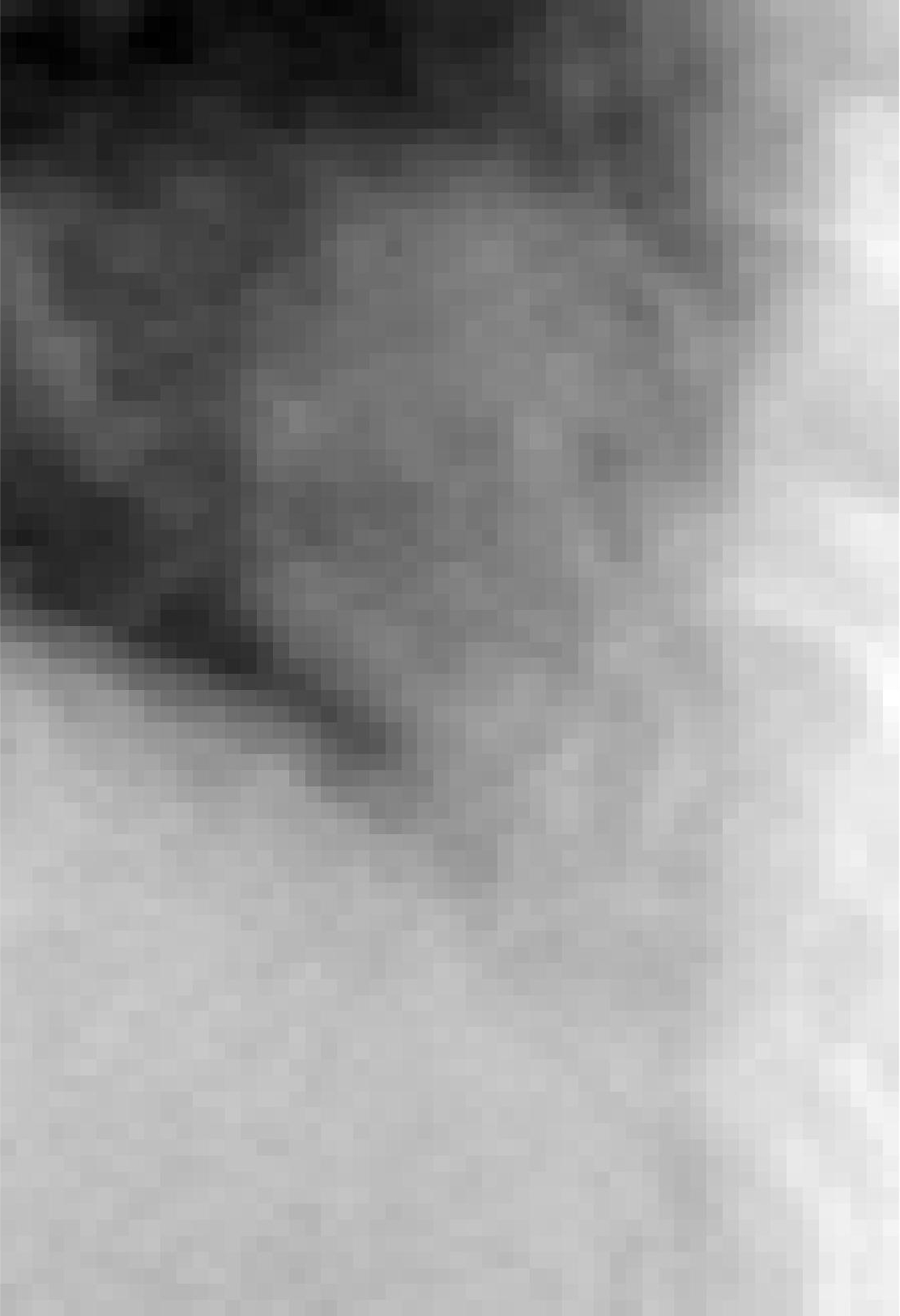} }}%
    \subfloat[Patient 1 \\* predicted \\* \text{\normalsize $t=t_{2374}$}]{{\includegraphics[height=2.8cm]{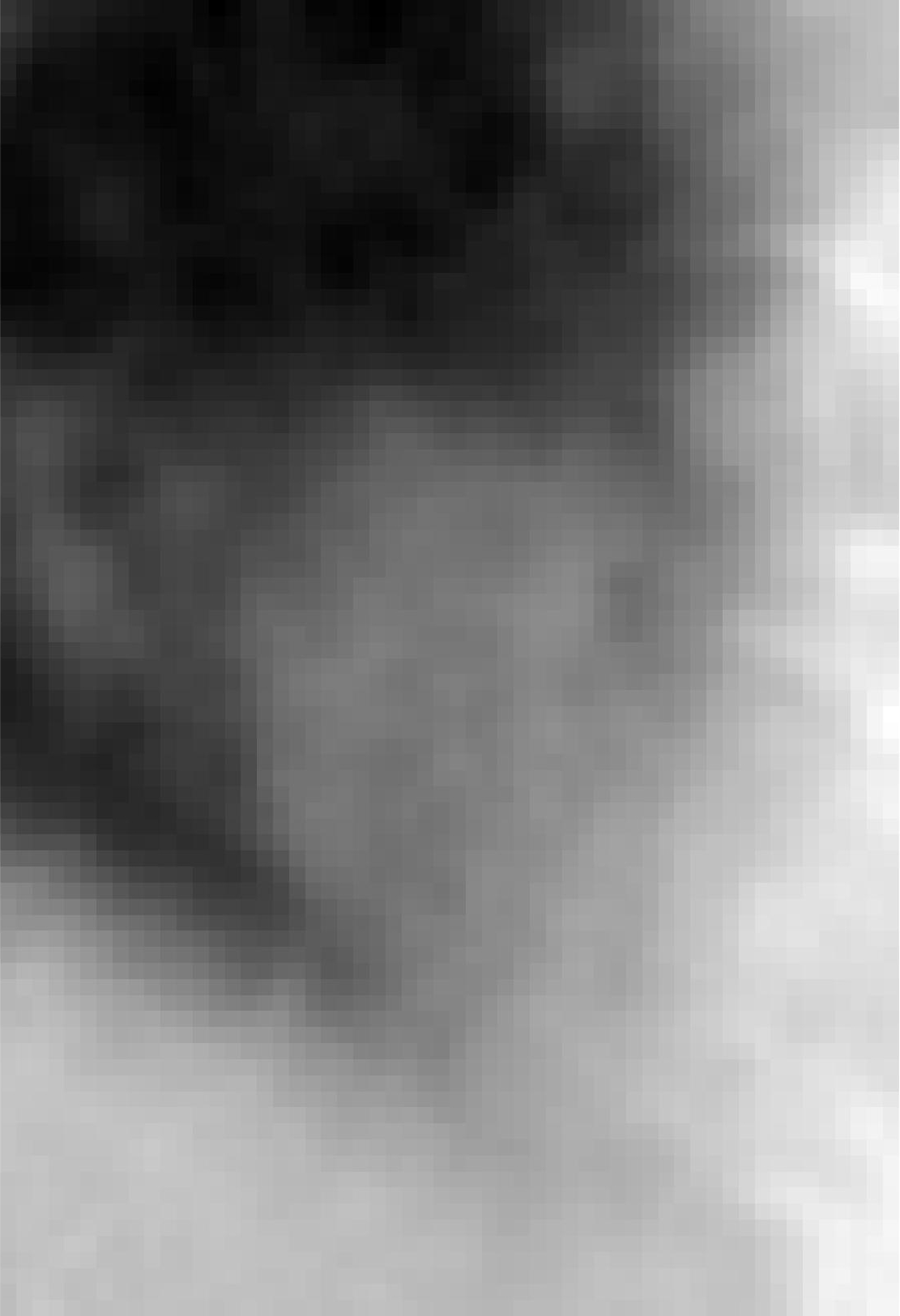} }}%
    \subfloat[Patient 1 \\* original \\* \text{\normalsize $t=t_{2374}$}]{{\includegraphics[height=2.8cm]{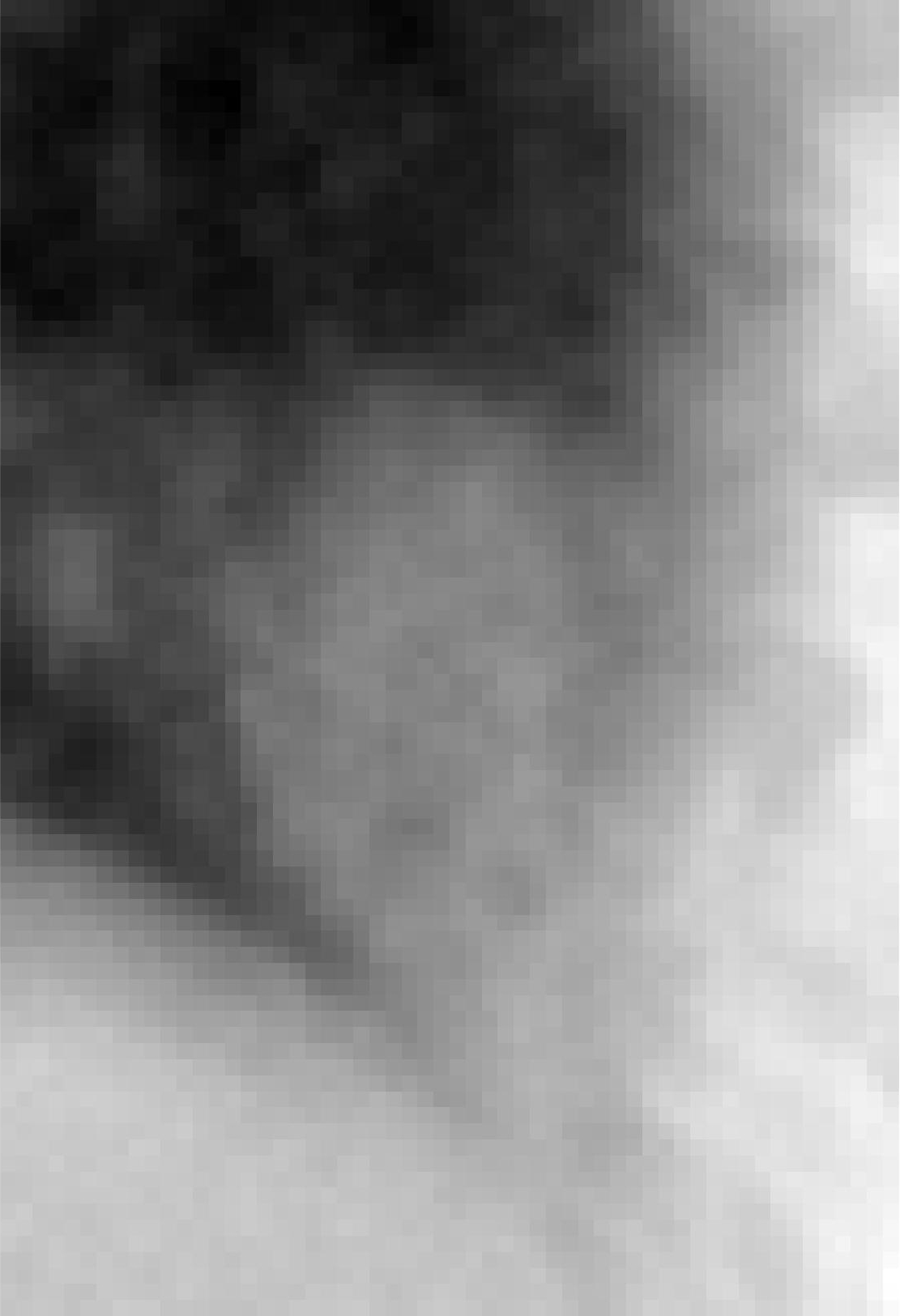} }}%
    \subfloat[Patient 2 \\* predicted \\* \text{\normalsize $t=t_{2209}$}]{{\includegraphics[height=2.8cm]{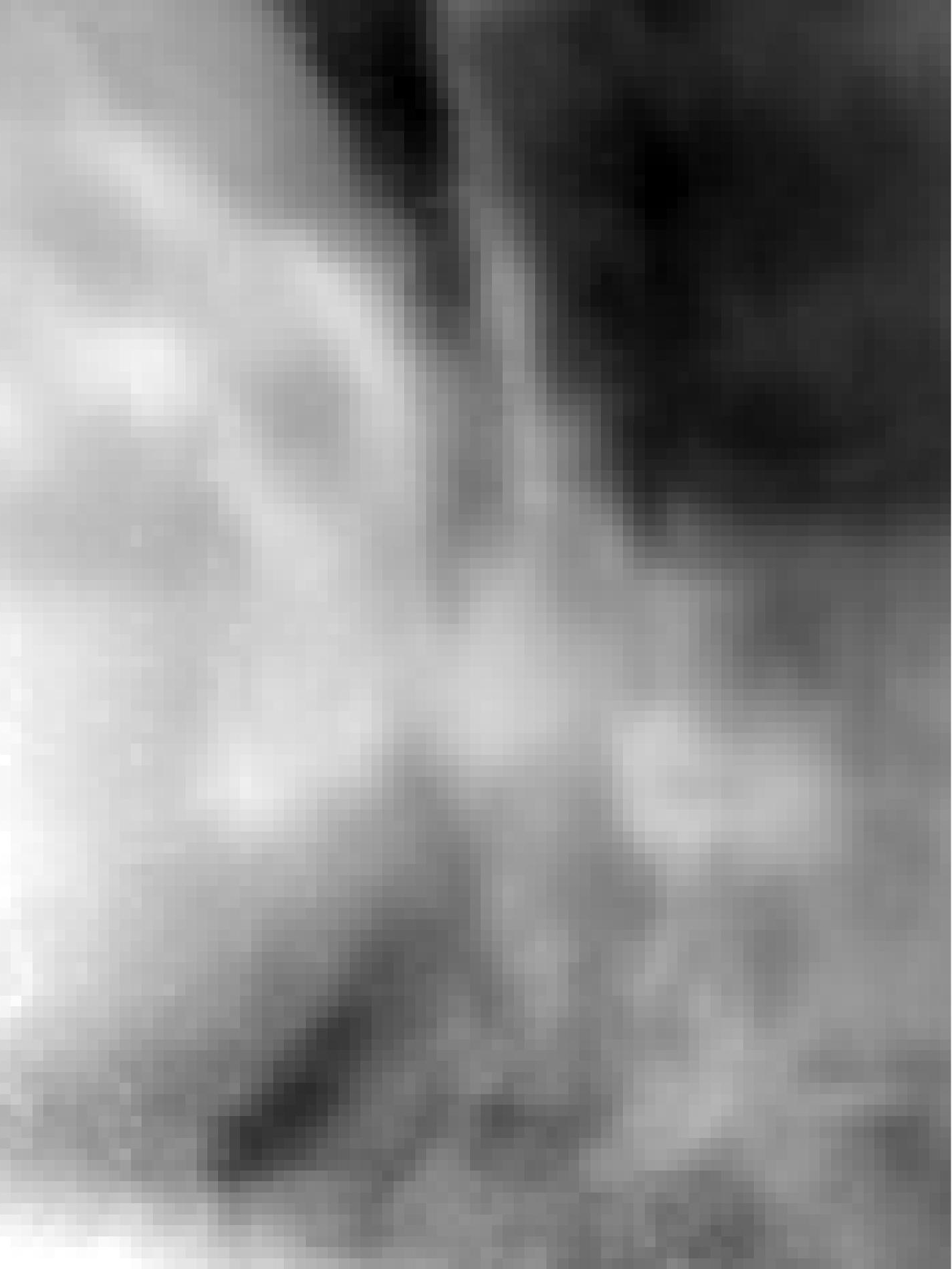} }}%
    \subfloat[Patient 2 \\* original \\* \text{\normalsize $t=t_{2209}$}]{{\includegraphics[height=2.8cm]{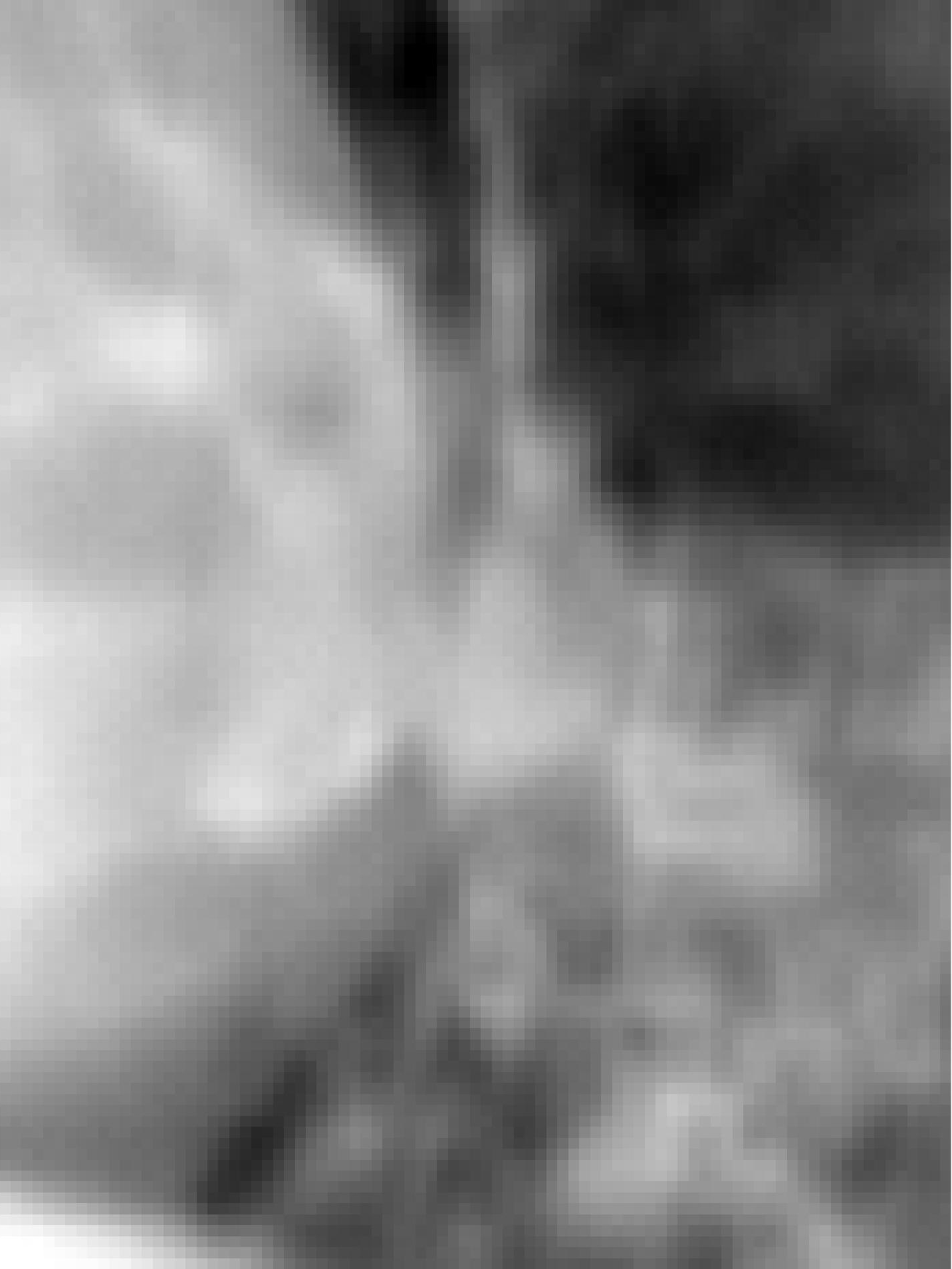} }}%
    \subfloat[Patient 2 \\* predicted \\* \text{\normalsize $t=t_{2374}$}]{{\includegraphics[height=2.8cm]{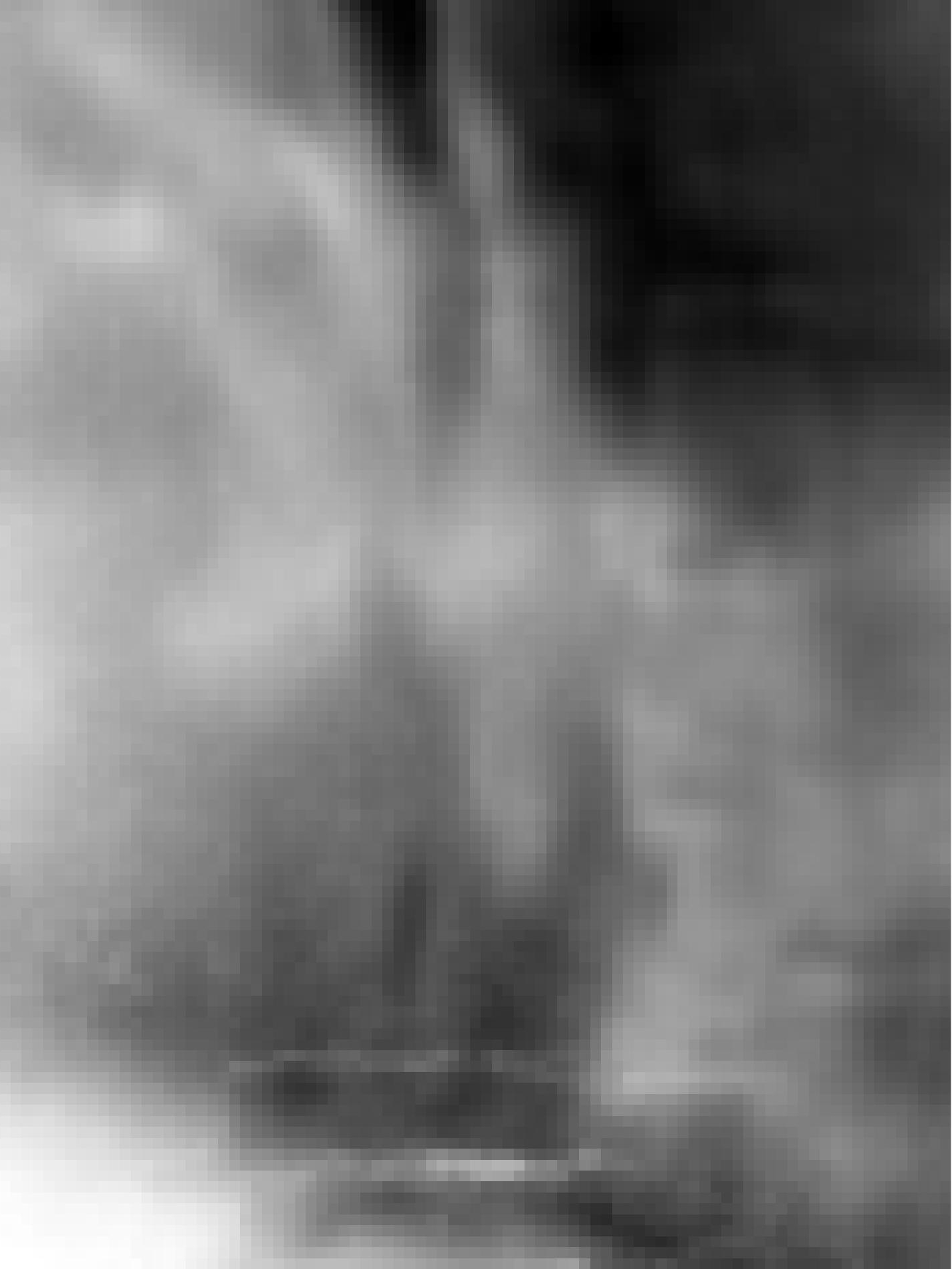} }}%
    \subfloat[Patient 2 \\* original \\* \text{\normalsize $t=t_{2374}$}]{{\includegraphics[height=2.8cm]{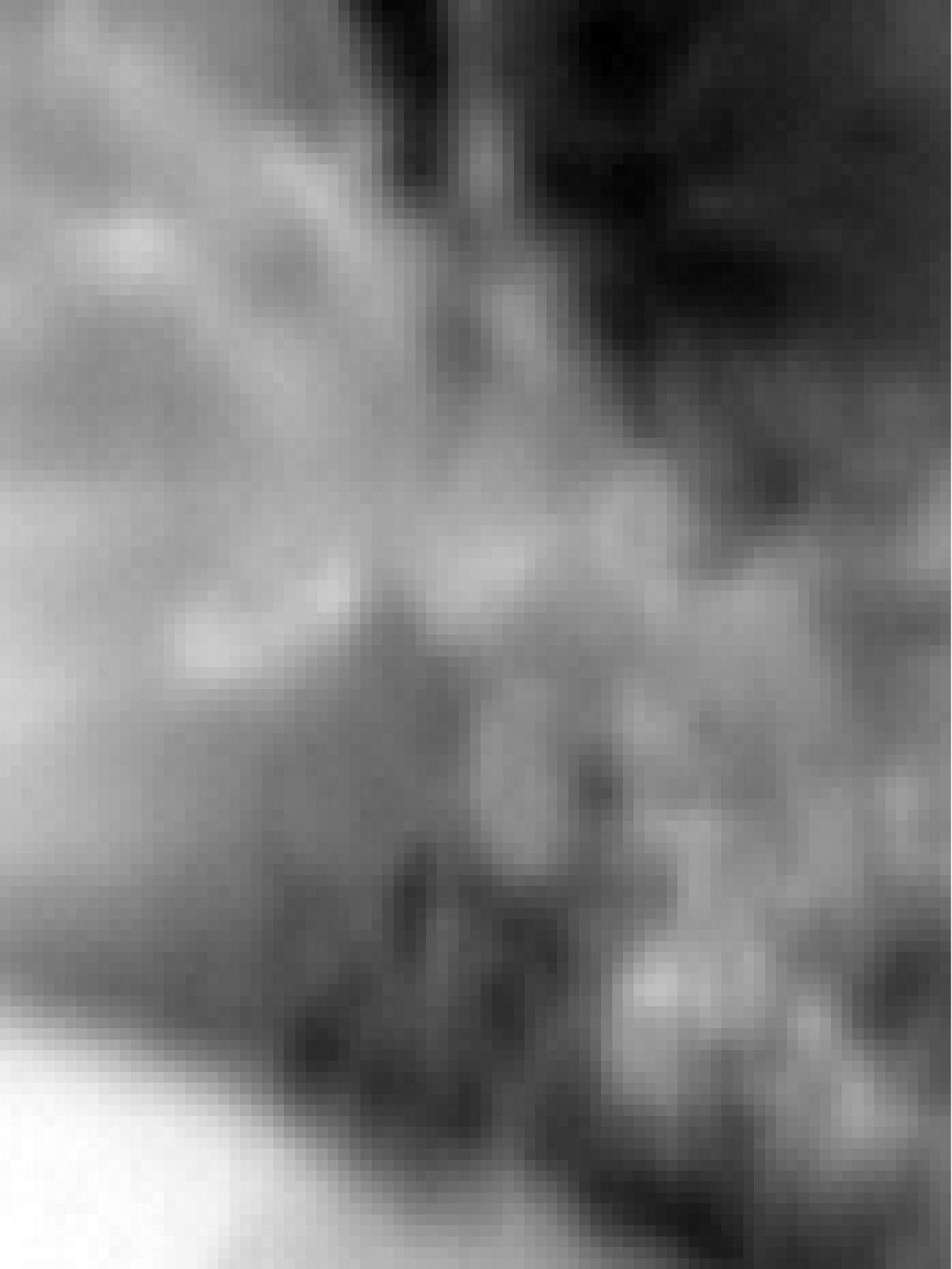} }}%
    \quad
    \subfloat[Patient 3 \\* predicted \\* \text{\normalsize $t=t_{2209}$}]{{\includegraphics[height=2.1cm]{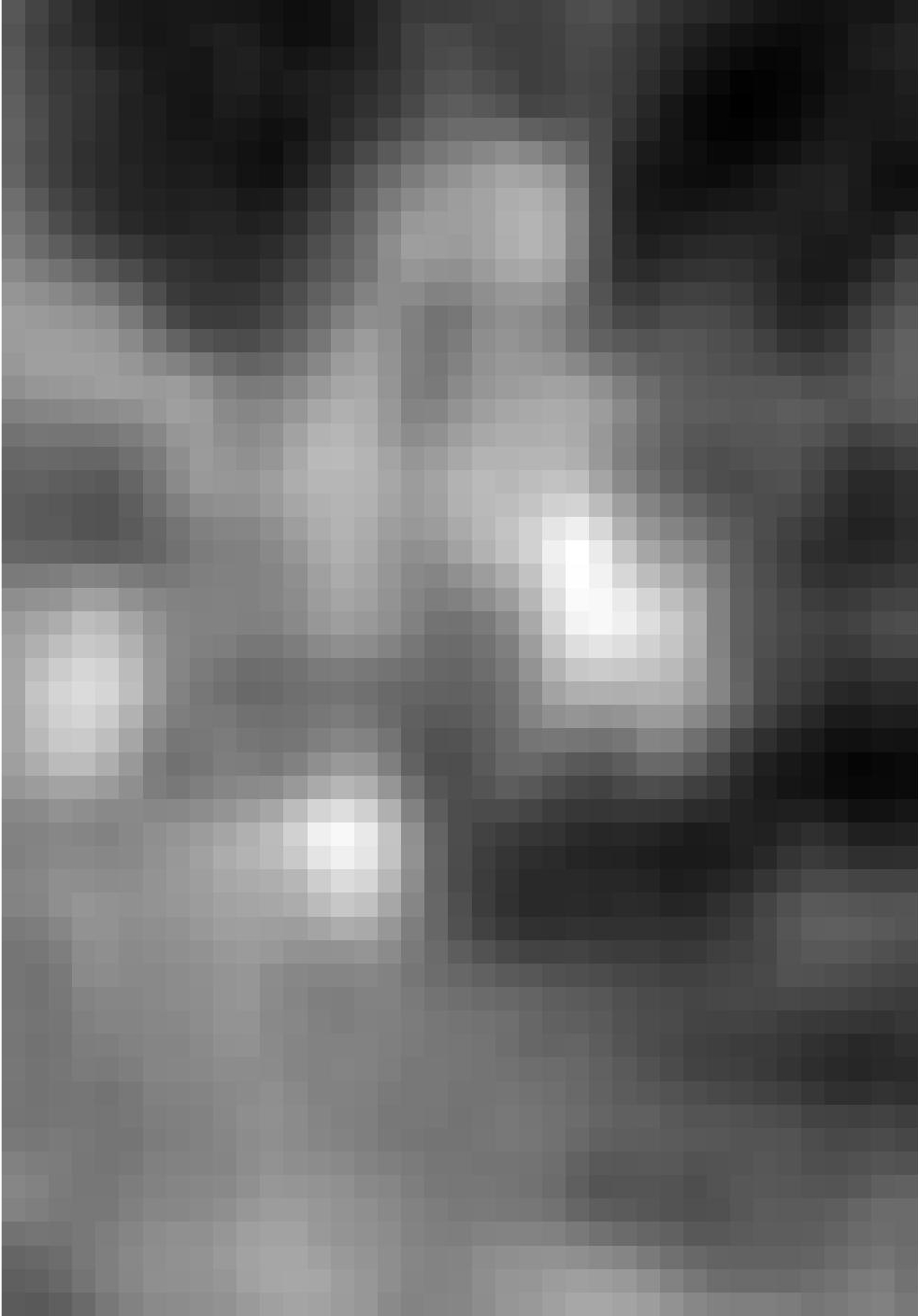} }}%
    \subfloat[Patient 3 \\* original \\* \text{\normalsize $t=t_{2209}$}]{{\includegraphics[height=2.1cm]{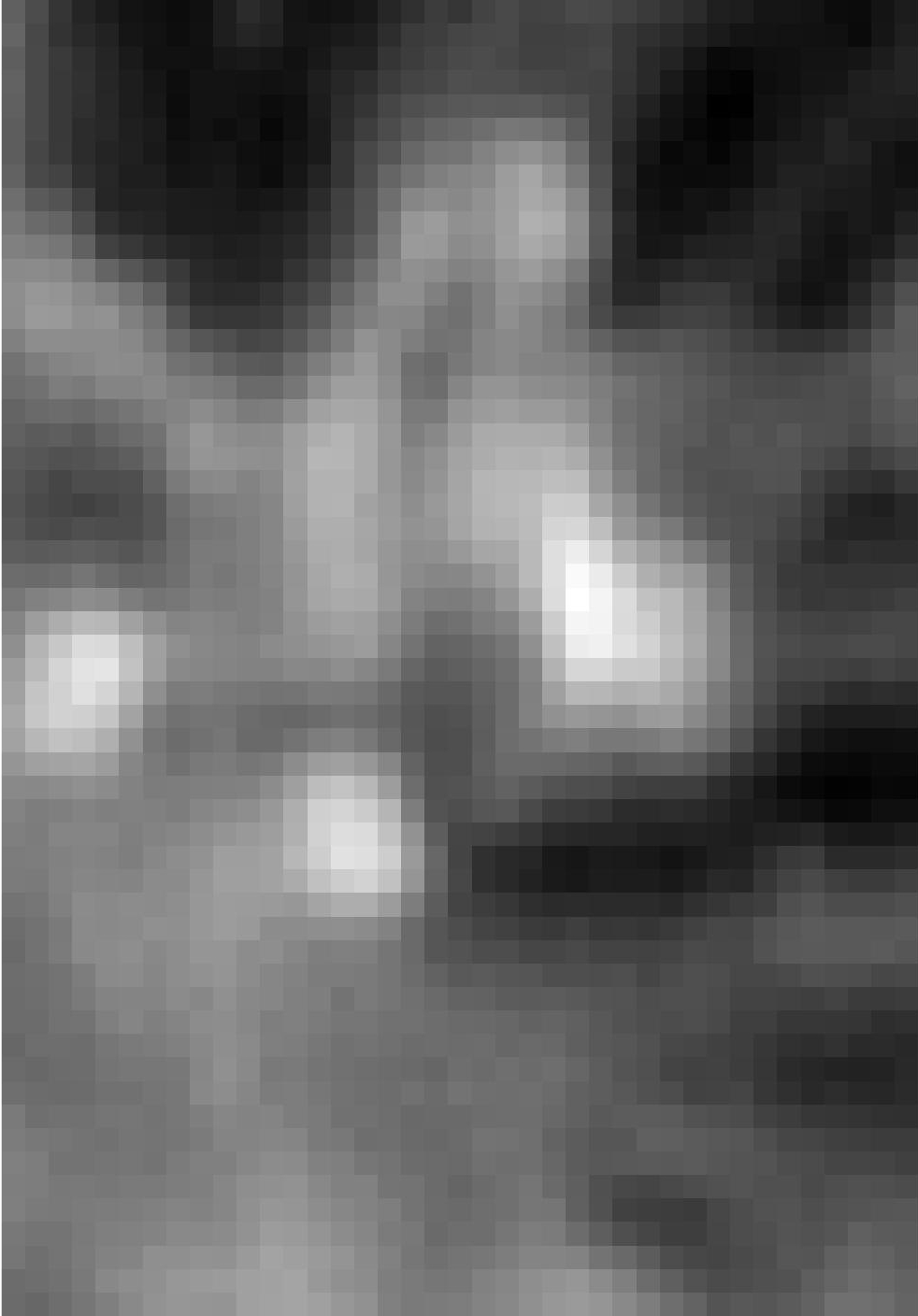} }}%
    \subfloat[Patient 3 \\* predicted \\* \text{\normalsize $t=t_{2374}$}]{{\includegraphics[height=2.1cm]{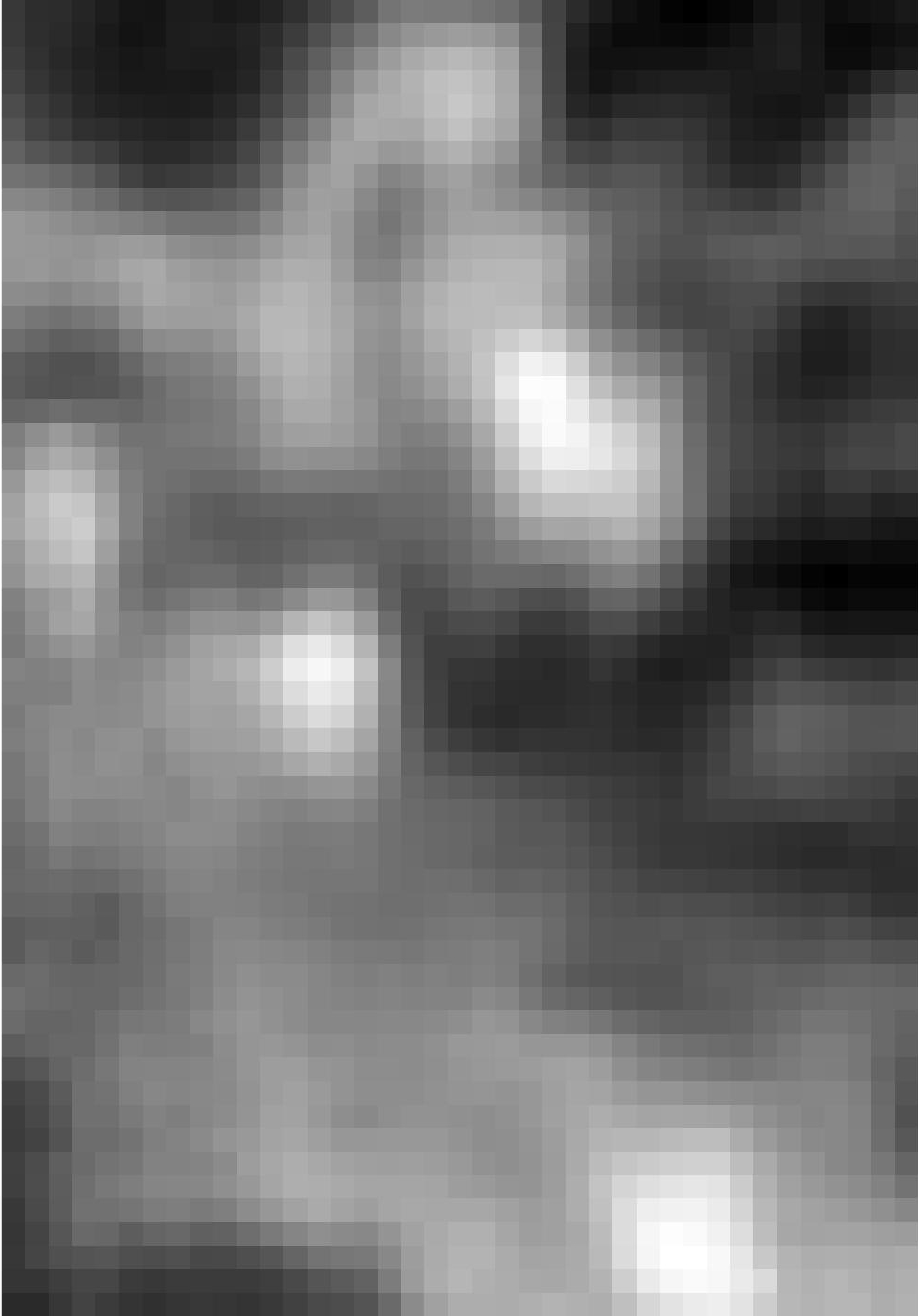} }}%
    \subfloat[Patient 3 \\* original \\* \text{\normalsize $t=t_{2374}$}]{{\includegraphics[height=2.1cm]{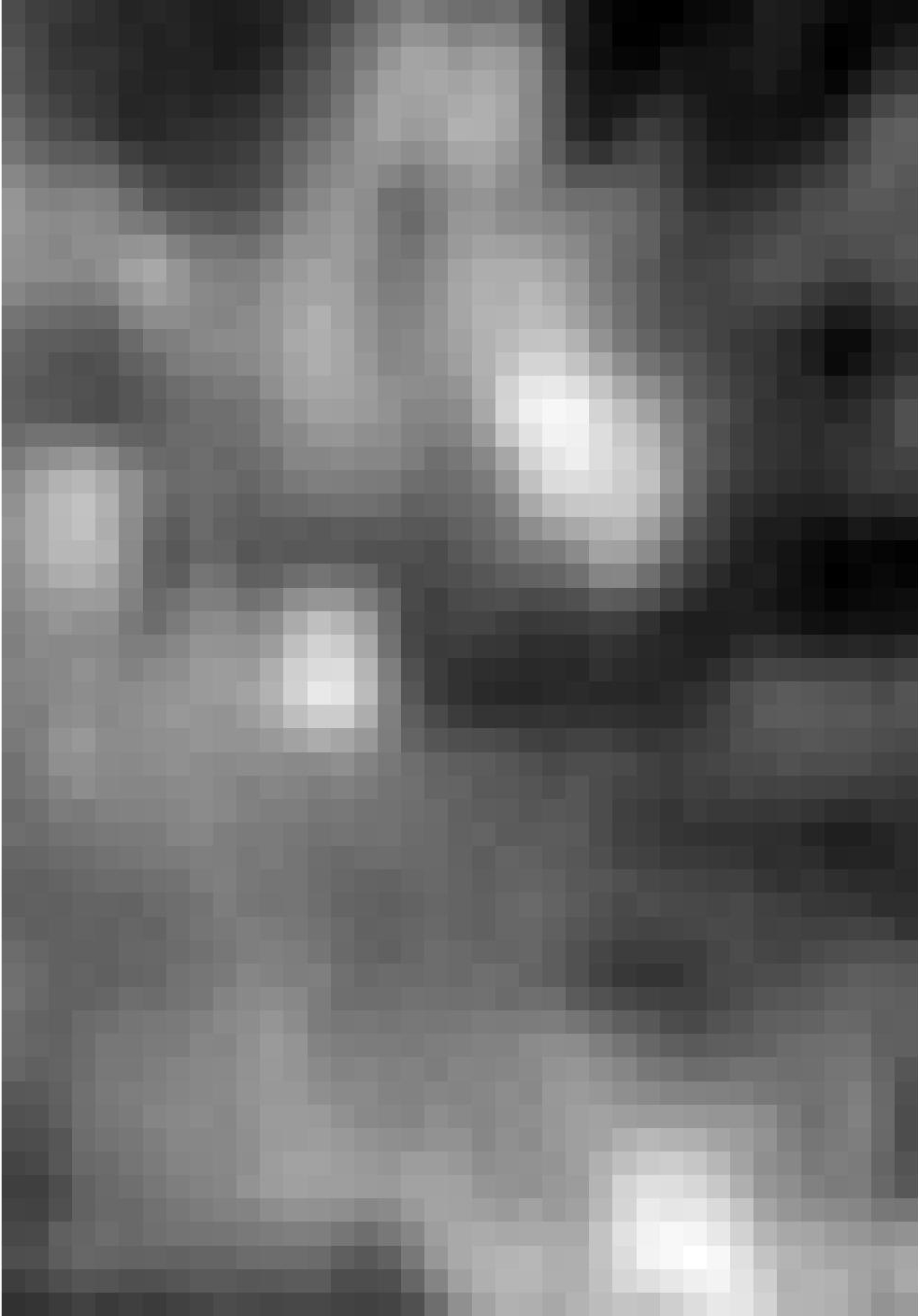} }}%
    \subfloat[Patient 4 \\* predicted \\* \text{\normalsize $t=t_{2209}$}]{{\includegraphics[height=2.1cm]{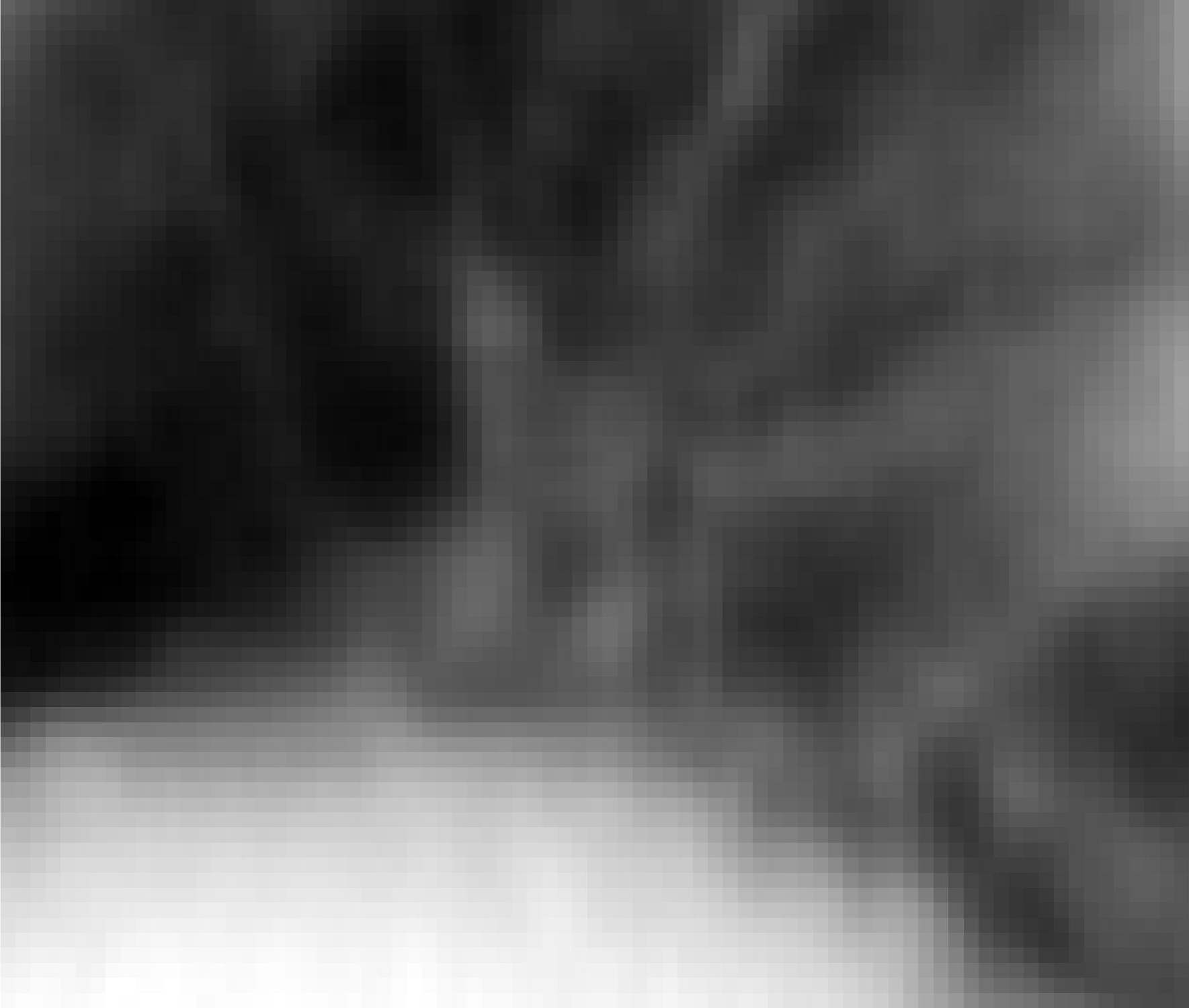} }}%
    \subfloat[Patient 4 \\* original \\* \text{\normalsize $t=t_{2209}$}]{{\includegraphics[height=2.1cm]{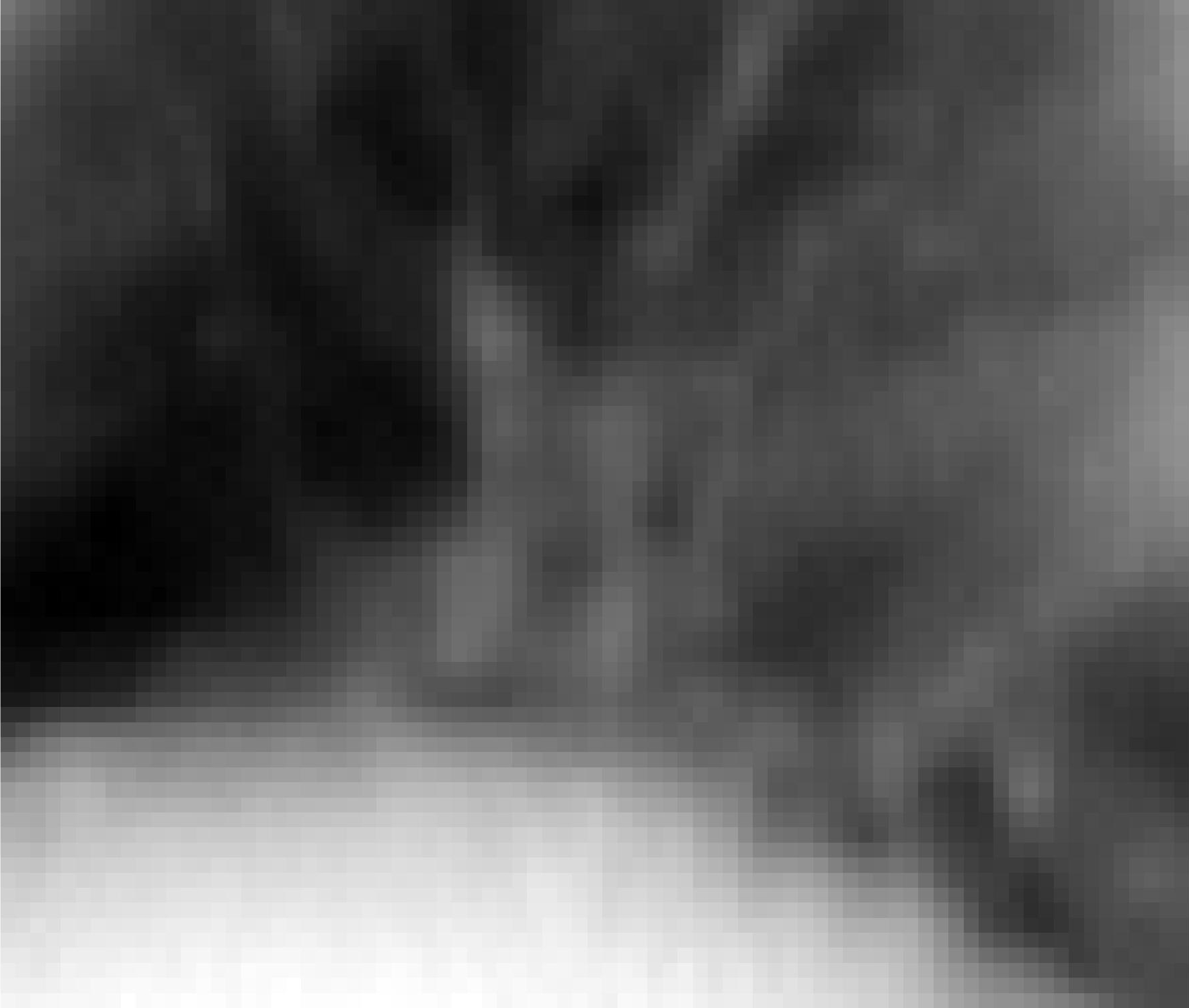} }}%
    \subfloat[Patient 4 \\* predicted \\* \text{\normalsize $t=t_{2374}$}]{{\includegraphics[height=2.1cm]{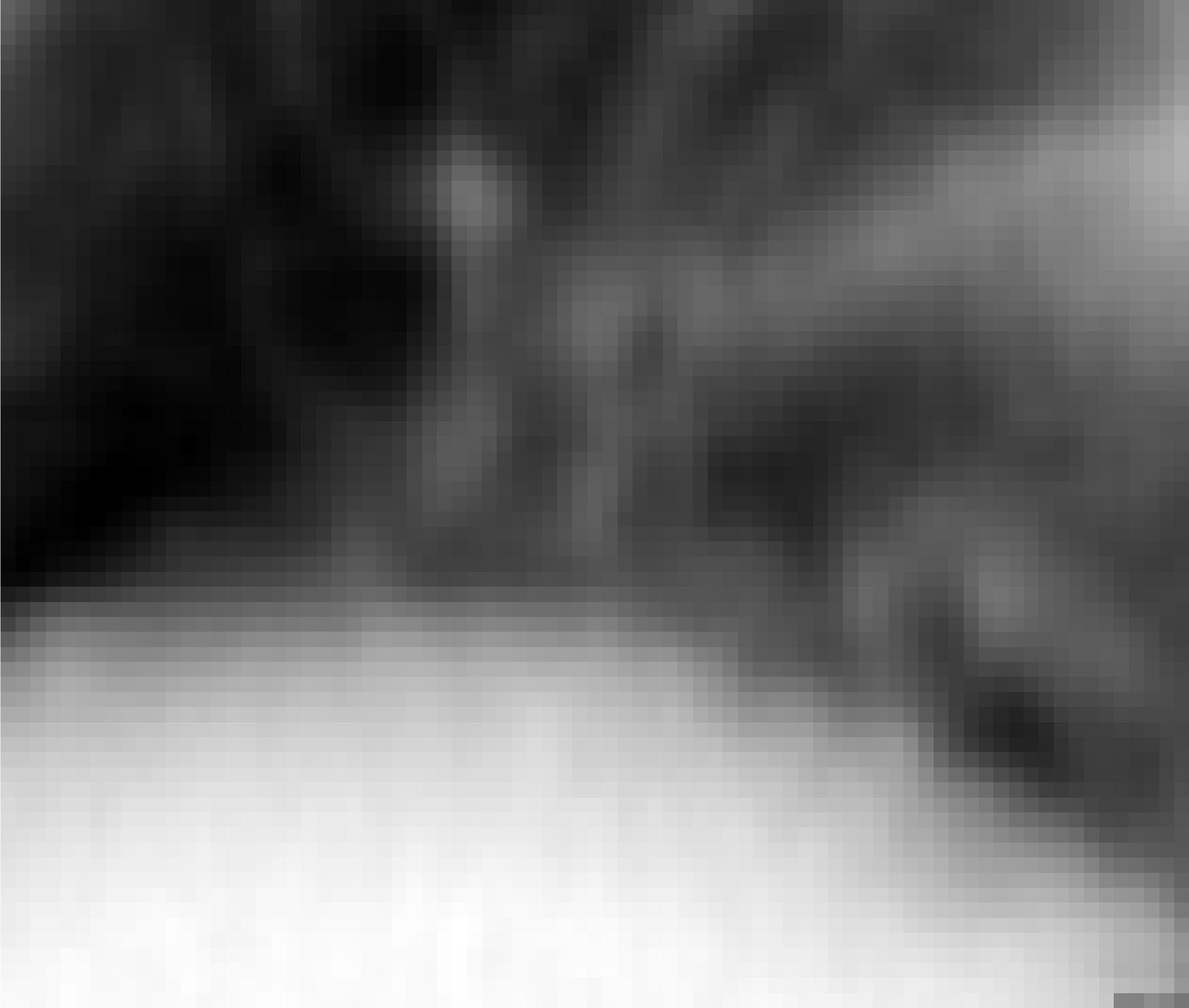} }}%
    \subfloat[Patient 4 \\* original \\* \text{\normalsize $t=t_{2374}$}]{{\includegraphics[height=2.1cm]{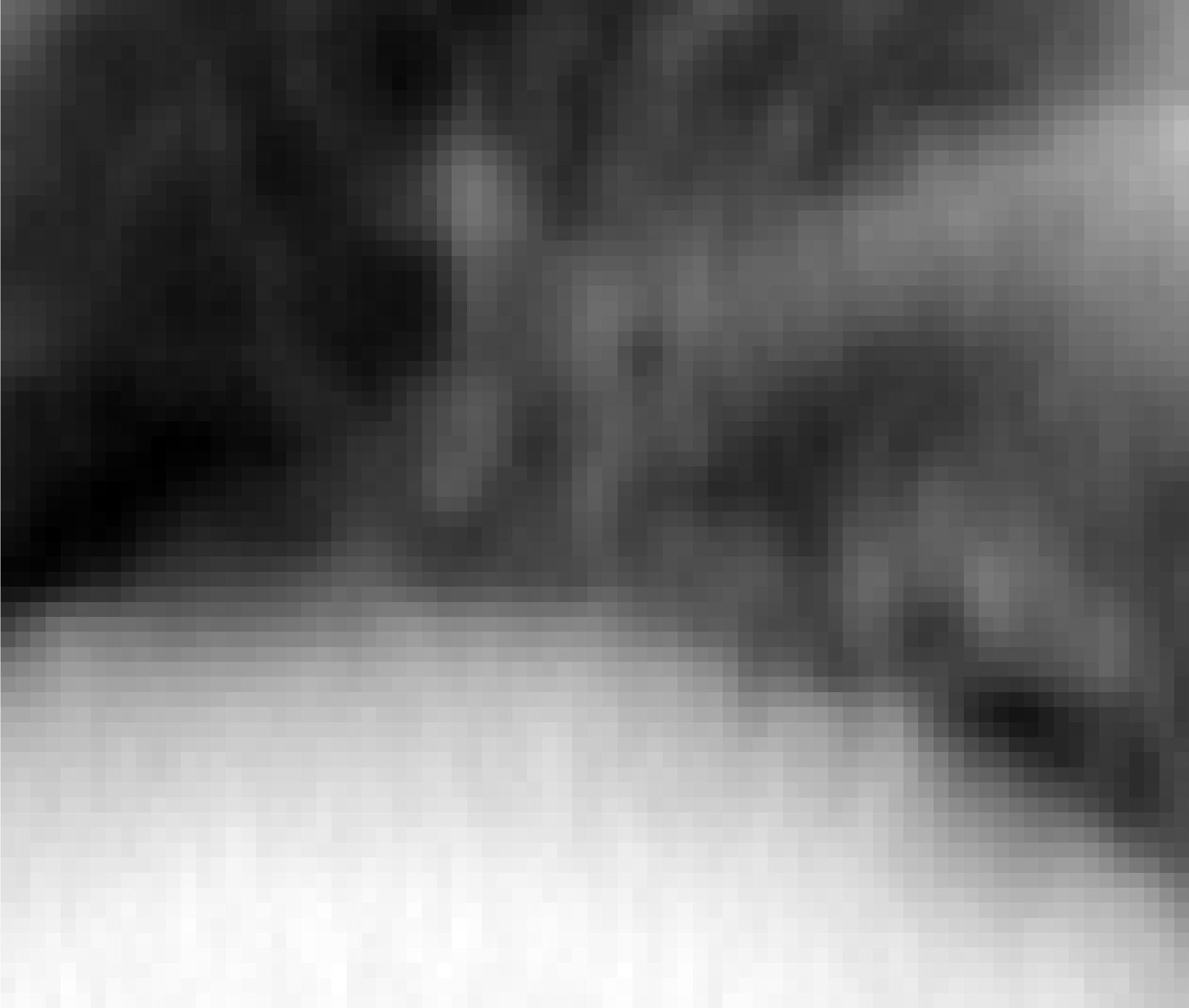} }}%
    \caption{Original and predicted ROI sagittal AIP, at an end-of-exhale and an end-of-inhale positions.}%
    \label{fig:pred_sag_AIP}%
\end{figure}%


\end{document}